\documentclass[a4paper,12pt]{article}
\usepackage{color}
\ifx\TeXtures\undefined
\usepackage{amsfonts}%
\usepackage{amsmath}%
\else
\usepackage{amsmath}%
\usepackage{amsfonts}%
\usepackage{times}
\fi
\usepackage{amsthm} 
\usepackage{epsfig}                
\newcommand{\DATUM}{}              
\newcommand{\change}
{{\marginpar{\#}}}        
\newcommand{\comma}{\: ,}              
\newcommand{\period}{\: .}             
\newcommand{\eps}{{\varepsilon}}        
\newcommand{\vphi}{{\varphi}}           
\newcommand{\Om}{\Omega}                
\newcommand{\om}{\omega}                
                
\newcommand{\la}{\langle}                
\newcommand{\ra}{\rangle}

\newcommand{\boQ}{{\bf Q}}
\newcommand{\boa}{{\bf a}}
\newcommand{\boC}{{\bf C}}
\newcommand{\boG}{{\bf G}}
\newcommand{\boP}{{\bf P}}
\newcommand{\bob}{{\bf b}}
\newcommand{\boe}{{\bf e}}
\newcommand{\boH}{{\bf H}}
\newcommand{\boV}{{\bf V}}
\newcommand{\boA}{{\bf A}}

\newcommand{\boE}{{\bf E}}         

\newcommand{\one}{{\bf 1}}
\newcommand{\cA}{{\mathcal{A}}}
\newcommand{\cB}{{\mathcal{B}}}

\newcommand{\cE}{{\mathcal{E}}}
\newcommand{\cF}{{\mathcal{F}}}
\newcommand{\cG}{{\mathcal{G}}}
\newcommand{\cH}{{\mathcal{H}}}

\newcommand{\cK}{{\mathcal{K}}}
\newcommand{\cO}{{\mathcal{O}}}         
\newcommand{\cP}{{\mathcal{P}}}         

\newcommand{\cS}{{\mathcal{S}}}

\newcommand{\cW}{{\mathcal{W}}}

\newcommand{\RR}{\mathbb{R}}            
\newcommand{\ZZ}{\mathbb{Z}}            
\newcommand{\NN}{\mathbb{N}}            
\newcommand{\CC}{\mathbb{C}}            
\newcommand{\fh}{\mathfrak{h}}         
\newcommand{\bP}{{\overline P}}        

\newcommand{\btheta}{\bar{\theta}}

\newcommand{\bcP}{{\overline \cP}}
\newcommand{\hC}{\widehat{C}}

\newcommand{\hkappa}{\hat{\kappa}}

\newcommand{\tH}{\widetilde{H}}        
\newcommand{\tP}{\widetilde{P}}   
\newcommand{\tR}{\widetilde{R}}         

\newcommand{\vnabla}{{\vec{\nabla}}}
\newcommand{\veps}{{\vec{\varepsilon}}}

\newcommand{\vk}{{\vec{k}}}

\newcommand{\vv}{{\vec{v}}}

\newcommand{\vx}{{\vec{x}}}

\newcommand{\od}{{\rm od}}

\newcommand{\dom}{{\rm dom}}
\newcommand{\rIm}{{\rm Im}}
\newcommand{\spec}{\sigma}                

\newcommand{\discspec}{\sigma_{\rm disc}}
\newcommand{\essspec}{\sigma_{\rm ess}}
\newcommand{\cirS}{\mathop{\bigcirc\kern -.73em {\scriptstyle{\rm S}}}}

\newcommand{\dist}{{\rm dist}}

\newcommand{\at}{{at}} 
 
\newcommand{\gs}{{\mathrm{gs}}}

\newcommand{\hf}{{\check{H}}}

\newcommand{\Rem}{\mathrm{Rem}}

\newcommand{\pf}{P-F}

\newcommand{\fF}{\mathfrak{F}}          
          
\newcommand{\fH}{\mathfrak{H}}          

\newcommand{\QED}{\phantom{blablabla}\hfill\qed\newline}  
\renewcommand{\thesection}
{\arabic{section}}                      
\renewcommand{\theequation}
{\thesection.\arabic{equation}}        
\newcommand{\secct}[1]{\section{#1}
\setcounter{equation}{0}}              

\newtheorem{assumptions}{Assumptions}[section]                                      
\newtheorem{hypothesis}{Hypothesis}[section]
\newtheorem{theorem}{Theorem}[section]         
\newtheorem{lemma}[theorem]{Lemma}             
\newtheorem{corollary}[theorem]{Corollary}     
\newtheorem{definition}[theorem]{Definition}   
\newtheorem{remark}[theorem]{Remark}           
\newtheorem{proposition}[theorem]{Proposition} 
\theoremstyle{plain}
\newcommand{\el}{{\mathrm{el}}}

\def\beq{\begin{equation}}
\def\ene{\end{equation}}
\newcommand{\p}{{\bf P}}
\newcommand{\bp}{\overline{{\bf P}}}

\begin{document}
\bibliographystyle{plain}
\setcounter{page}{0}
\thispagestyle{empty}

\title{Existence and Construction of Resonances for Atoms Coupled 
to the Quantized Radiation Field}

\author{
Volker Bach \\
\small{Institut f\"ur Analysis und Algebra}\\
\small{Technische Universit\"at Braunschweig } \\[-1ex]
\small{Germany (vbach@tu-bs.de)}
\and
Miguel Ballesteros \\
\small{Institut f\"ur Analysis und Algebra}\\
\small{Technische Universit\"at Braunschweig} \\[-1ex]
\small{Germany (m.ballesteros@tu-bs.de)}
\and
Alessandro Pizzo\\
\small{Department of Mathematics} \\
\small{University of California, Davis}\\[-1ex]
\small{ USA (pizzo@math.ucdavis.edu)} \\ 
}
\date{\DATUM}

\maketitle

\begin{abstract}
  For a nonrelativistic atom, which is minimally coupled to the
  quantized radiation field, resonances emerging from excited atomic
  eigenstates are constructed by an iteration scheme inspired by
  \cite{Pizzo2003} and \cite{BachFrohlichPizzo2006}. This scheme
  successively removes an infrared cut off in momentum space and yields
  a convergent algorithm enabling us to calculate the resonance
  eigenvalues and eigenstates, to arbitrary order in the feinstructure
  constant $\alpha \sim 1/137$, and is thus an alternative method of
  proof of a similar result obtained in \cite{Sigal2010}.
 \end{abstract}

\thispagestyle{empty}

\tableofcontents

\newpage
\setcounter{page}{1}

\secct{Introduction and Main Result} \label{sec-I}

\subsection{Resonances, Time-Decay, and Metastable States} 
\label{subsec-I.1}
%

The general framework of quantum mechanics is given by a Hilbert space
$\fH_S$, whose normalized elements $\psi_S(t) \in \fH_S$ - called wave
functions - represent the state of the system at any time $t \in \RR$,
and a self-adjoint Hamiltonian $H_S = H_S^*$ acting on (a dense domain
contained in) $\fH_S$. The subscript $S$ stands for system. The dynamics of the system evolving from a state
$\psi_S(0) \in \fH_S$ is determined by Schr\"odinger's equation
\begin{align} \label{eq-I.1}
\forall t > 0: \quad \dot\psi_S(t) \; = \; -i H_S \psi_S(t), 
\end{align}
which is solved by $\psi_S(t) = e^{-iH_St} \psi_S(0)$, showing that the
dynamical information of \eqref{eq-I.1} is fully contained in the
spectral decomposition of $H_S$.

In case that $\psi_S(0)$ is an eigenvector of $H_S$ corresponding to an
eigenvalue $E_S \in \RR$, the dynamics \eqref{eq-I.1} leaves $\psi_S(0)$
unchanged, up to multiplication by a phase factor, i.e., $\psi_S(t) =
e^{-iE_St} \psi_S(0)$. For many physical systems, most of the eigenvectors
of $H_S$ are unstable in the sense that a generic perturbation of $H_S$ -
no matter how small it may be - lets the eigenvalue disappear and
turns it into a resonance.

There are several definitions of a resonance (see, e.g.,
\cite{Simon1983}), but with varying degree of mathematical rigor and
no clear logical hierarchy of stronger assumptions implying weaker
ones. In this paper we define resonances as eigenvectors and
corresponding eigenvalues of a complex deformation $H_S(\theta)$ of the
Hamiltonian $H_S \equiv H_S(\theta =0)$, following
\cite{a-c,Hunziker1986}. Here, $\theta \in \CC$ is a
deformation parameter, and $\theta \mapsto H_S(\theta)$ defines a type-A
analytic family in a neighbourhood of $\theta =0$, i.e., $D_r \ni
\theta \mapsto H_S(\theta) [H_S+i]^{-1} \in \cB[\fH_S]$ is an analytic
$\cB[\fH_S]$-valued map, for some $r >0$. The usual choice of complex
deformation, which we also make here, is defined by $H_S(\theta) :=
U(\theta) \, H_S \, U^{-1}(\theta)$, where $U(\theta) := e^{i\theta M}$
and $M = M^*$ is a self-adjoint operator on $\fH_S$ such that $H$ is
sufficiently $M$-regular.

The relation of \eqref{eq-I.1} to $H_S(\theta)$ is given through an
application of the Laplace transform and analytic continuation in
$\theta$,

\begin{align} \label{eq-I.2}
\la \vphi |  e^{-iH_St} \psi \ra
\ & = \  
\frac{-1}{2\pi i} \int_{\RR+i\eps} e^{-izt} 
\Big\la \vphi \Big|  \; (H_S - z)^{-1} \, \psi \Big\ra 
\; dz
\\[1ex] \notag 
\ & = \  
\frac{-1}{2\pi i} \int_{\RR+i\eps} e^{-izt} 
\Big\la \vphi_{\btheta} 
\Big|  \; 
\big( H_S(\theta) - z \big)^{-1} \, \psi_\theta \Big\ra 
\; dz ,
\end{align}

where $\vphi$ and $\psi$ belong to a suitable dense domain (e.g.,
analytic vectors of $M$) in $\fH_S$ and $\theta$ varies in a range for
which the spectrum of $H_S(\theta)$ doesn't intersect $\CC^+$.

Now suppose that the Hamiltonian is of the perturbative form $H_{S,\alpha}
= H_{S,0} + W_{S, \alpha}$, where $W_{S,\alpha} \to 0$, as $\alpha \to 0$, and that
$E_{S, 0} \in \RR$ is an eigenvalue of $H_{S,0}$. Then, $E_{S, 0}$ is also an
eigenvalue of $H_{S, 0}(\theta)$, for all $\theta \in D_r \cap \RR$,
because $H_{S, 0}(\theta)$ is unitary equivalent to $H_{S, 0}$. The analyticity
of $\theta \mapsto H_{S, 0}(\theta)$ then implies that $E_{S, 0}$ is, in fact,
an eigenvalue of $H_{S, 0}(\theta)$, for all $\theta \in D_r$. This implies
that $z \mapsto \big(H_{S, 0}(\theta) - z \big)^{-1}$ is singular, as $z
\to E_{S, 0}$, and hence, close to $E_{S, 0}$, the integration contour in
\eqref{eq-I.2} cannot be deformed from the upper half-plane into the
lower half-plane. If $\alpha \neq 0$, however, then the eigenvalue
$E_{S, 0}$ often turns into an eigenvalue $E_{S, \alpha} \in \CC^-$ of
$H_{S, \alpha}(\theta)$ in the lower half-plane, and the real axis about
$E_{S, 0}$ is free of spectrum of $H_{S, \alpha}(\theta)$. This, in turn,
implies that, for $\alpha \neq 0$, the integration contour in
\eqref{eq-I.2} can, indeed, be deformed from the upper half-plane into
the lower half-plane, yielding a local exponential decay in time. The
eigenvalue $E_{S, \alpha} \in \CC^-$ of $H_{S, \alpha}(\theta)$ is then called a
\emph{resonance} and the corresponding eigenvector of
$H_{S, \alpha}(\theta)$ a \emph{metastable state}.

We remark that complex deformations leave isolated eigenvalues of
finite or infinite multiplicity unchanged. In particular, $\sigma_{\rm
  disc}[H_S(\theta)] \supset \sigma_{\rm disc}[H_S]$, for all $\theta \in D_r$.
Furthermore, the generator $M$ of the deformation group is necessarily
unbounded in order to have the deforming effect on the spectrum of $H_S$
described above.

The relation of the spectral information on $H_S(\theta)$ to the
time-decay properties of $e^{-itH_S}$ is rather indirect. Exponential
decay of the form $|\la \vphi | e^{-iH_St} \psi \ra| \leq e^{-\kappa
  |t|}$, for some $\kappa >0$, for instance, necessarily requires that
$\sigma(H_S) = \RR$, for otherwise a contradiction to the Paley-Wiener
theorem would emerge. Therefore, for a semibounded Hamiltonian $H_S$,
one cannot expect better decay estimates than
\begin{align} \label{eq-I.3}
|\la \vphi | e^{-iH_St} \psi \ra| 
\ \leq \ 
C_1 e^{-\kappa |t|} \: + \: C_2(t),
\end{align}
with $|C_1| \gg |C_2(0)|$, but $|C_1| e^{-\kappa |t|} \ll |C_2(t)|$,
for $t \gg 1$. Under mild further assumptions involving functions of
Gevrey type \cite{KleinRama2008}, rather than analytic functions, the
error term obeys the estimate $|C_2(t)| \leq \hC_\beta
\exp[-\hkappa_\beta |t|^{\beta}]$, where $\beta <1$ may be chosen
arbitrarily close to $1$. 
Accepting that an estimate of the form \eqref{eq-I.3} is optimal, the
question arises in which sense there is a natural choice for the decay
rate $\kappa$. For many models which allow for a perturbative
formulation as $H_{S, \alpha}$ above, this natural choice for $\kappa$ is
to leading order in $\alpha$ determined by second-order perturbation
theory - leading to the \emph{weak-coupling limit} of the dynamics.
Many physical processes show a varying decay rate on different time
scales, however, and the adequate mathematical description of these
phenomena remains a challenge.

\subsection{Excited Eigenvalues turning into Resonances}
\label{subsec-I.2}
%
In the present paper, we study a model for a one-electron atom in
interaction with the quantized radiation field. The Hilbert space of
this system is

\begin{equation} \label{eq-I.4}
\cH^0 \ = \ \cH_\at \otimes \cF^0,
\end{equation}
where $\cH_\at = L^2(\RR^3)$ is the Hilbert space of one spinless
particle, and $\cF^0 \equiv \fF_b[\mathfrak{h}^0]$ is the boson Fock space of
photons, with $\mathfrak{h}^0:= L^2(\RR^3 \times \ZZ_2)$ being the corresponding
one-photon Hilbert space. The dynamics is generated by the Hamiltonian
\beq \label{eq-I.5} 
\boH \ = \ 
\big( \frac{1}{i}\nabla +  \boA^0 \big)^2 \; - \; \boV 
\; + \; \hf^0,
\ene
where the potential $\boV \geq 0$ is a multiplication operator in the electro position $x \in \mathbb{R}^3$ which is assumed to be an infinitesimal 
perturbation of $-\Delta$ and to decay to zero at infinity.
The field energy is represented by
\beq \label{eq-I.6}
\hf^0 \ = \ \int \om(k) \: a^*(k) a(k) \, dk,
\ene
with $k = (\vk, \mu) \in \RR^3 \times \ZZ_2$, $\int f(k) \, dk :=
\sum_{\mu \in \mathbb{Z}_2} \int f(\vk,\mu) \, d^3k$, and $\om(k) :=
|\vk|$. The fine structure constant $\alpha >0$ is assumed to be a
sufficiently small coupling constant, ignoring that its physical value
is about $1/137$. The vector potential of the quantized magnetic field
in Coulomb gauge is
\beq \label{eq-I.7}
\boA^0(x)  := \ 
\frac{\alpha^{3/2}}{(2 \pi)^{3/2}}\int_{\RR^3 \times \ZZ_2} 
\frac{\kappa(k)}{(2\om(k))^{1/2}} 
\big\{ \veps(k) \, e^{-i \alpha \vk \cdot x} \, a^*(k) \: + \:
\veps(k)^* \, e^{i \alpha \vk \cdot x} \, a(k) \big\} \, dk,
\ene
where the $a^*(k)$ and $a(k)$ are the creation and annihilation
operators representing the canonical commutation relations on $\cF^0$ as
\beq \label{eq-I.8}
[a(k), a^*(k')] \: = \: \delta(k-k') \cdot \one,
\quad
[a(k), a(k')] \: = \: [a^*(k), a^*(k')] \: = \: 0,
\quad a(k) \Om \: = \: 0,
\ene
for all $k, k' \in \RR^3 \times \ZZ_2$, in the sense of
operator-valued distributions, i.e., it is understood that
\eqref{eq-I.8} is integrated against sufficiently regular test
functions. The ultraviolet cut-off $\kappa$ is (the restriction
to the real line of) an entire function and of sufficiently rapid
decay. Analyticity of $\kappa$ is important because we use the
method of complex deformations, and the characteristic function
of a set would not be an
admissible choice. A natural choice would be $\kappa(k) :=
\exp[- \vk^2]$. Furthermore, the transversal polarization
vectors $\veps(\vk,\mu) \equiv \veps(\vk/|\vk|,\mu)$ are measurable
maps on  $\mathbb{S}^2$ which, together with $\vk/|\vk|$, constitute an
orthonormal basis (\emph{Dreibein}) in $\RR^3$, for (almost) all $\vk
\neq 0$. Hence, $\veps(r \vk,\mu) = \veps(\vk,\mu)$, 
for all $r > 0$.

The atomic potential $\boV$ is assumed to be dilation analytic and compact relatively to the kinetic energy $-\Delta$ of the electron.
Consequently, the essential spectrum $\essspec(\boH_\at) = \RR_0^+$ of
$\boH_\at := -\Delta + \boV$ is the positive half-axis, and its discrete
spectrum $\discspec(\boH_\at) = \{\boe_0, \boe_1, \boe_2, \ldots, \boe_M\}
\subseteq \RR^-$ lies on the negative axis, where $M \in \NN$ or $M =
\infty$ and \ $\boe_0 < \boe_1 < \cdots < \boe_M$. We assume in
this paper that $M \geq 2$, i.e., that $\boH_\at$ has at least two 
isolated eigenvalues of finite multiplicity. 

Under these assumptions it is easy to see that the spectrum
$\sigma(\boH_0) = [\boe_0, \infty)$ of $\boH_0 = \boH_\el \otimes \one + \one
\otimes \hf^0$ covers the half-axis above $\boe_0$ and is entirely
essential, $\sigma(\boH_0) = \essspec(\boH_0) = [\boe_0, \infty)$. The atomic
eigenvalues $\{\boe_0, \boe_1, \boe_2, \ldots, \boe_M\}$ are eigenvalues of $\boH_0$,
as well, but now embedded in a continuum. Fixing $j \in
\{1,2,\ldots,M\}$, it is expected that $\boe_j$ is unstable under
perturbations and that $\boH$ has no (real) eigenvalue in the
vicinity of $\boe_j$ - no matter how small $\alpha >0$ may be.

In the past two decades, a lot of research has been carried out on the
model defined by \eqref{eq-I.4}-\eqref{eq-I.8} to establish this
picture. The first basic step is to establish the existence of the
model, as it is defined by an unbounded operator. After several works
of increasing generality, it was established in
\cite{Hiroshima1998,HaslerHerbst2008} that $\boH$ is a
semibounded, self-adjoint operator with domain $\dom(\boH_0)$, for all
$\alpha >0$. Its ground state energy

\beq \label{eq-I.9}
\boE_\gs(\alpha) \ = \ \inf\big\{ \sigma(H_\alpha) \big\}
\ene
is a simple eigenvalue, which was first shown for small $\alpha >0$ in
\cite{BachFroehlichSigal1999} and for all $\alpha >0$ in
\cite{GriesemerLiebLoss2000}. Its simplicity is a consequence of the
assumption that the electron is spinless. In reality, electrons are
spin-$\tfrac{1}{2}$ fermions, and the ground state is two-fold
degenerate \cite{MyaoSpohn2007}. 

The half-line above $\boE_\gs(\alpha)$
is filled with essential spectrum of $\boH$. In particular,
$\boH$ has no isolated eigenvalues.  Absence of eigenvalues,
positive commutator estimates, and limiting absorption principles in the interval $\big( \boE_\gs(\alpha), \Sigma(\alpha) \big)$ have
been established, where
\beq \label{eq-I.10}
\Sigma(\alpha) \ := \ 
\inf\Big\{ \sigma\big( 
( -i\nabla -  \boA^0 )^2 + \hf^0 \big) \Big\},
\ene
which, in particular, proves the instability of $\boe_j$ for $j \geq 1$
\cite{BachFroehlichSigalSoffer}.

Our main result is to further establish resonances for $j = 1$ (or
any other $j \geq 2$) and show that the eigenvalue $\boe_j$ is not only
absent from the real axis but pushed into the lower half-plane. 

The coupling function (often called ``form factor'' which, however, is a misnomer)
$\frac{1}{(2\om(k))^{1/2}}$ (see (\ref{eq-I.7})) yields major difficulties for the study of resonances,
indeed, in this case the perturbation is (superficially) \emph{marginal} in the renormalization group sense.
On
the contrary,  we can control the renormalization scheme in presence of
any
regularization of the form factor, i.e., $1/\sqrt{|k|^{1/2-\mu}}$, as $k\to
0$,
 with $\mu>0$.  We cannot handle $\boH$ directly but instead we can handle another Hamiltonian which is unitarily equivalent (and therefore physically equivalent). This new Hamiltonian is obtained by a change of gauge which is called the Pauli-Fierz transformation. In the new Hamiltonian the interaction with the photons is
modeled by a less singular form factor, but the long-distance behavior in the electron variable is more difficult to handle.
The relation between small momenta and long distances is manifest in the oscillating phase factors ($e^{-i \alpha \vk \cdot x}$) that appear in Eq. (\ref{eq-I.7}). Roughly speaking, we trade small momenta for long distances by applying the Pauli-Fierz transformation. The long-distance behavior in the electron variable can be controlled by exploiting the exponential localization of the eigenfunctions. The use of a Pauli-Fierz transformation similar to the one we use was first done by Sigal in \cite{Sigal2010}. 

Now we describe the Pauli-Fierz transformation and the Hamiltonian that we analyze. We select a function $\eta \in C_0^\infty(\mathbb{R}) $ that is identically $1$ in a neighbourhood of zero. We define the self-adjoint operator

\begin{align} \label{eq-I.11} 
\lambda^0_{PF} \ : & = \ 
\frac{\alpha^{3/2}}{(2 \pi)^{3/2}}\int_{\RR^3 \times \ZZ_2} 
\frac{\kappa(k)}{(2\om(k))^{1/2}} 
\Big \{ \veps(k) \, \cdot \eta\big(|x||k|\big) x \, a^*(k) \: \\\\ & + \:
\veps(k)^* \, \cdot \eta(|x||k|) x \, a(k) \Big \} \, dk.
\end{align}

The Pauli-Fierz transformation is the unitary operator $e^{- i \lambda^0_{PF}}$ and the Hamiltonian that we study is the conjugation of 
$\boH$ with respect to this unitary operator: 
\beq \label{eq-I.12}
H^0 : = e^{- i \lambda^0_{PF}} \boH e^{ i \lambda^0_{PF}}.
\ene
The operator $H^0$ is actually the energy operator in a different representation, which is described by the Pauli-Fierz transformation. The Pauli-Fierz transformation can be regarded as a change of gauge, as formula (\ref{eq-I.12}) suggests.    

Thanks to our analyticity
assumptions on $\boV$ and $\kappa$, we can trace the eigenvalue and
associate to the Hamiltonian a natural analytic family
$\{H^0(\theta) \} $ of operators which are complex deformations of
$H^0$. If $\rIm\, \theta$ is strictly positive (and small enough), and $\alpha $ is small, we prove that
 $H^0(\theta)$ possess a simple eigenvalue $E_\infty$ in a neighbourhood of $\boe_1$ (the same proof holds for $\boe_j$ with $j > 1$). 
We prove furthermore that the imaginary part of $E_\infty$ is strictly negative and that it is, therefore, a resonance. We prove additionally that  locally,
in a neighbourhood of $E_\infty$ and $\boe_1$,  there is no point of the spectrum of $H^0(\theta)$ with imaginary part larger than $\rIm\, E_\infty$ and we estimate the norm of the  resolvent $\| \frac{1}{H^0(\theta) - z} \|  $ for such points $z$ in the resolvent set. We prove also that the eigenvector associated to $E_\infty$ is exponentially localized in the electron variable by a rather simple method.    
  
The precise formulation of our main result is given in
Section~\ref{Resonances}.
 This result has already been established in
\cite{Sigal2010} by methods different from those used in the present
paper. In \cite{Sigal2010} the
renormalization group based on the Feshbach-Schur map is used to
prove Section~\ref{Resonances},
 while our approach borrows ideas from
\cite{BachFrohlichPizzo2006} and is based on a successive reduction
of momentum slices known as \emph{Pizzo's method}. We include new ingredients in the method taking advantage of the maximum modulus principle 
for analytic functions and using the Feshbach-Schur map as a tool to restrict the domain of our Hamiltonian to functions that are exponentially localized in the electron variable. As we do not use renormalization based on the Feshbach-Schur map, we do not deal with the usual combinatoric problems of the theory of renormalization.

\subsection{Strategy of the Proof and Description of the Content of the Paper}

We describe the strategy of our proof. We restrict the momentum of the 
photon $|k|$ to be larger than a positive number $s$ and we implement this restriction in all operators and Hilbert spaces. 
The parameter $s$ is actually an infrared cut-off (when strictly positive). We take a decreasing sequence of positive numbers $\{ \sigma_n \}_{n \in \mathbb{N} \cup\{ 0\}}$ in which each $\sigma_n$ represents an infrared cut-off and $\sigma_n \to 0$, for $n \to \infty$. We denote by $\overset{n}{H}$ the Hamiltonian with the restriction $|k| \geq \sigma_n $. We can construct (see Theorem~\ref{princ-secc}) an eigenvalue $E_0$ of $\overset{0}{H}$  in a neighbourhood of $ \boe_1 $. This eigenvalue is actually isolated (by the infrared cut-off), which makes it simple to analyze. We can prove the existence of $E_0$, for sufficiently small $\alpha$. We can not proceed in this way to construct eigenvalues of $ \overset{n}{H} $ because the maximal magnitude of $\alpha$, that we could still admit, would tend to zero, as $n \to \infty$. To construct $E_n$, we rather proceed inductively. Assuming to have already constructed the eigenvalues 
$ E_m$, for $m\leq n$, we construct the eigenvalue $E_{n + 1}$ of $ \overset{n + 1}{H}$. After completing the induction step (see Section~\ref{induction-step}) we conclude that, for any $n \in \mathbb{N}_0 : = \mathbb{N} \cup\{ 0\}$, there is an eigenvalue $E_n$ of $ \overset{n}{H} $. We conclude by proving that the limit 
$$E_\infty : = \lim_{n \to \infty} E_n  $$         
exists and is an eigenvalue of $ \overset{\infty}{H} := H^0 $ (see  Section~\ref{Resonances}).

We now briefly describe the structure of our paper. In Section~\ref{definition-model} we describe the mathematical objects that we use and the physical model. In Section~\ref{estimates-electron-hamiltonian} we compile the results that we need concerning the atom Hamiltonian. The proofs of the statements in this section are given in Section~\ref{proofs-estimates-electron-hamiltonian}. In Section~\ref{s.rbapha} we state basic bounds on the interaction with respect to the non-interacting Hamiltonian and use this results to prove some basic estimates. The proofs of the statements in this section are given in Section 
 \ref{rbapha-proofs}. In Section~\ref{s.rirh} we prove the induction basis of our induction scheme. The main results of this section are summarized in Theorem~\ref{princ-secc}. We prove that $E_0$ is a simple eigenvalue of $\overset{0}{H} $, we estimate its imaginary part, we construct the projection onto the corresponding eigenspace, and we derive some estimates for the resolvent operator and the resolvent set. The proofs of these assertions are carried out in Section~\ref{s.rirh.p}, there we use similar estimates as in \cite{BachFroehlichSigal1999}. In our case the analysis is much simpler because our eigenvalues are isolated. Section~\ref{infrared-limit} is the heart of the paper.  In that Section we state and prove all our results. In Section~\ref{feshbach-infrared} we introduce some mathematical tools that we need and prove exponential decay of the eigenfunctions (see Theorem~\ref{T.il.fm.1}). In Section~\ref{il.notation} we define the sequence of Hamiltonians and fix some notations and assumptions. Section~\ref{induction-hypothesis} is the core of our proof. There we establish the inductive scheme and prove the induction step. We inductively construct  $E_n$ and prove that it is a simple eigenvalue of $\overset{n}{H} $, we construct the projection onto the corresponding eigen-space and we derive some estimates for the resolvent operator and the resolvent set. In the proof of the induction step we make use of the exponential decay of the eigenfunctions by restricting the domain of our operators to a space of functions which decay exponentially in the electron variable. The mathematical object that permits us to justify this restriction is the Feshbach-Schur map. Here we use it just as a mathematical tool that allows us to use the exponential decay of the eigenfunctions, and \emph{not} as a fundamental object on which a renormalization scheme is based. In this paper we do not use renormalization group techniques.  In Section~\ref{Resonances} we state and prove the main results using Section~\ref{induction-hypothesis}. We take the limit $n \to \infty$ and prove that $\overset{\infty}{H}$ has a simple eigenvalue ($E_\infty$) in a neighbourhood of $\boe_1$. 
We furthermore prove  that the imaginary part of $E_\infty$ is strictly negative and it is, therefore, a resonance. We  additionally prove that  locally,
in a neighbourhood of $E_\infty$ and $\boe_1$,  there is no point of the spectrum of $  \overset{\infty}{H} $ with imaginary part larger than $\rIm\, E_\infty$ and we estimate the norm of the  resolvent $\| \frac{1}{H^0(\theta) - z} \|  $ for such points $z$ in the resolvent set.

\newpage

\section{Definition of the Model}\label{definition-model}

\subsection{The Atom Hamiltonian}\label{atom-hamiltonian}
\subsubsection{The Group of Dilation Operators}

For every $\theta \in \RR$,
we denote by $u(\theta): L^2(\mathbb{R}^3) \to L^2(\mathbb{R}^3)$ 
 the group of 
 dilation operators.
 $u(\theta)$ is defined by the 
formula, 
\beq \label{d1}
(u(\theta)\phi)(\vx) := e^{3 \theta /2}\phi(e^{\theta} \vx), \, 
 \,\forall \phi  \in L^{2}(\RR^3) \period 
\ene

\subsubsection{The Electric Potential}\label{electric-potential}
The electric potential is a complex valued function defined in $\mathbb{R}^3$. It depends on the 
parameters $\alpha,  \beta $ and  $\theta$. $\alpha $ is a positive real number less than one and $\theta$ and $\beta$ 
are a complex numbers with $|\theta |, |\beta| < \frac{1}{2} $. We denote the electric potential by 
$$
V(\theta) : \mathbb{R}^3 \to \mathbb{C}
$$ 
and, in general, in the notation we omit the dependence on $ \alpha$ and $\beta$ 
\beq \label{ep-1}
  V(\theta, \alpha, \beta) : =  V(\theta).
\ene 

The electric potential is the sum of 3 operators,  
\beq \label{ep-2}
V(\theta) : = \boV(\theta) + \alpha^3 V_{PF}(\theta) +  \tilde V(\theta, \beta).   
\ene

The function $\boV(0)$ is the {\it physical} electric potential and $\boV(\theta)$ is an analytic continuation of $\boV(0)$ (see Section~\ref{pep}). The term $V_{PF}(\theta)$ is an effective potential that comes from the 
Pauli-Fierz transformation (see Sections~\ref{effective-potential} and	 \ref{pauli-fierz-transformation}) and the term $ \tilde  V(\theta,\beta)  $ is a potential yielded by the conjugation with the exponential $e^{\beta \la x \ra}$, see (\ref{potential.1}) and (\ref{geh.7.m1}). We use this conjugation 
to prove exponential decay for the eigenfunctions of the Hamiltonians (see Sections~\ref{potential}, \ref{S.ebe.atom} and Theorem  \ref{T.il.fm.1})  and to control the long distance behavior for the electron (see (\ref{ei.49}) for example). Adding the potential $\tilde V(\theta,\beta)$ allows us to collocate an exponential decreasing factor $ e^{ -\beta \la x \ra} $ in a convenient place using the exponential decay of terms that we can control to compensate the term 
$e^{\beta \la x \ra}$ (for real and positive $\beta$, see (\ref{ei.49})).      

\paragraph{The Physical Electric Potential} \label{pep}   $\;$\\
\\ 
We assume that the physical electric potential satisfies the following properties:
\begin{itemize} 
\item
For any $\theta  = \mu + i \nu $ 
$$
\lim_{|x| \to \infty} \sup_{ |\theta | < 1/2} |  \boV(\theta)(x)   |= 0, \:   \boV(\theta)^* = \boV(\overline{\theta}),\:
\boV(\theta) = u(\mu) \boV(i\nu) u(\mu)^{-1}.
$$
\item The function 
$$
\theta \to \boV(\theta) \frac{1}{- \Delta + 1},    \: |\theta | < \frac{1}{2},
$$
is an operator valued analytic function, where $\Delta $ is the Laplace operator. We suppose furthermore that 
\beq\label{l.10}
\lim_{ r  \to \infty} \sup_{ |\theta | < 1/2} \|  \boV(\theta) \frac{1}{ - \Delta + r}  \|= 0,
\ene
where the norm $ \| \cdot  \| $ is the operator norm. 

\end{itemize}

\paragraph{The Effective Potential $V_{PF}(\theta)$} \label{effective-potential} $\;$\\
\\ 
This potential comes from the Pauli-Fierz transformation (see Section~\ref{pauli-fierz-transformation}). In particular, it appears in Eq. (\ref{eph.1}).

\noindent Let $ \eta \in C^{\infty}_0 ([0,\infty))$ be a decreasing function such that
\beq \label{pft.1}
\eta(r) = \begin{cases} 1, \, \, \, \text{if}\, r \leq 1, \\ 0, \, \, \, \text{if}\,  r \geq 2  \period  \end{cases}
\ene 
\noindent We denote by $\veps$, a fixed function 
$\veps: = (\varepsilon_1, \varepsilon_2,\varepsilon_3) :
  \RR^3 \times \mathbb{Z}_{2} \to \CC^3 $  
that satisfies 
\begin{equation} \begin{array}{l} \label{i.1} 
\veps(\vk, \lambda)^* \cdot \veps(\vk, \mu) \; = \; \delta_{\lambda,  \mu} 
\comma \hspace{5mm} 
\vk \cdot \veps(\vk, \lambda) \; = \; 0 \\\ 
\hspace{5mm} \overline{\veps( -\vk, \lambda)} = \veps( \vk, \lambda), \hspace{5mm} \veps( r\vk, \lambda) =  \veps(\vk, \lambda), \, r > 0,   
\end{array}
\end{equation}
where $  \overline{\veps( -\vk, \lambda)} $ is the vector whose entries are the complex-conjugate of the entries of $ \veps( -\vk, \lambda) $ and 
$\delta_{\lambda, \mu}$ is the Kronecker symbol.

The effective potential is defined by the following:
\begin{align} \label{pft.15} 
 V_{PF}(\theta)(x) : =  &  e^{- \theta} \sum_{\lambda = 1}^2 
\frac{1}{ 2(2\pi) ^3} \int_{\mathbb{R}^3} \big[ \exp(- 2e^{-2 \theta} |\vk|^2) \\ \noindent &  \cdot
|\eta(|x||\vk|) \veps(\vk, \lambda) \cdot x|^2 d \big]
\vk_1 d\vk_2 d\vk_3,   
\end{align}
for any $\theta \in \CC$ with $|\theta| < 1/2$.

\paragraph{The Potential $\tilde V(\theta, \beta)$} \label{potential} $\;$\\
\\ 
We denote by 
\beq \label{geh.7.m1}
\la x \ra : = ( 1 +  |x|^2 )^{1/2}.  
\ene
We define
\beq \label{potential.1}
\tilde V(\theta, \beta)(x) : = -  e^{- 2  \theta}e^{ -\beta \la x \ra }\Delta  e^{ \beta \la x \ra} + e^{- 2\theta}\Delta,\:
\theta, \beta \in \CC, \: |\theta|,  |\beta| < \frac{1}{2},
\ene
where $\Delta$ is the Laplace operator.

\subsubsection{The Atom Hamiltonian}

The atom Hamiltonian is a closed operator with domain contained in the atom Hilbert space
$$
\cH_{\at} : = L^2(\mathbb{R}^3).
$$
It is defined by the formula
$$
H_{\at}(\theta) : = -e^{- 2 \theta} \Delta + V(\theta).
$$

We do not write explicitly the dependence of $ H_{\at}(\theta) $ on $\alpha$, $\beta $ (see \ref{ep-1}). If it is required we also write
$$
H_{\at}(\theta, \alpha, \beta) := H_{\at}(\theta), \:  \theta, \beta \in \CC, |\theta|, |\beta|< \frac{1}{2}, \: \:   \alpha \in [0, 1].  
$$  

\begin{hypothesis}[The Energy of the First Excited State of $\boH_\at$]
\label{hypothesis}
We denote by 
$$\boe_0 : = \inf(\sigma (H_{\at}(0,0,0))  ), \: \boe_1 : = \inf(\sigma (H_{\at}(0,0,0))  \setminus \{ \boe_0 \}), 
$$ 
where $\sigma(O)$ is the spectrum of the operator $O$.  

We assume that 
$$
\boe_0 < \boe_1 < 0,
$$ 
and that $\boe_1 $ is a non-degenerate eigenvalue. 
\end{hypothesis}
\begin{definition}\label{delta-el}
We denote by 
\beq  \label{e.delta-el}
\delta_\at : = \dist\big(\boe_1, \sigma(H_\at(0, 0, 0)) \setminus \{ \boe_1\}  \big).
\ene
\end{definition}

\subsection{The Photon Hamiltonian}\label{photon-hamiltonian}
\subsubsection{The Photon Hilbert Space}
The Hilbert space of one photon restricted to energies between $s $ and $t$ (with $0 \leq  s < t \leq \infty$) is denoted by 
\beq \label{extra.1}
\mathfrak{h}^{s, t} : = L^2(\cK^{s, t}),
\ene
where 
\beq \label{extra.2}
\cK^{s, t} : = \{ (\vk, \lambda) \in \mathbb{R}^3 \times \mathbb{Z}_2 \:  | \:  s \leq|\vk| < t \}.  
\ene
A pair $(\vk, \lambda ) \in  \mathbb{R}^3 \times \mathbb{Z}_2  $ is denoted by 
\beq\label{extra.3}
k := (\vk, \lambda). 
\ene
The modulus of an element $ k = (\vk, \lambda ) $ is
\beq\label{extra.4}
|k| : = |\vk|.
\ene
The integral of a function $  f :  \mathbb{R}^3 \times \mathbb{Z}_2  \to \CC $ is defined by 
\beq \label{extra.5}
\int f(k)dk  = \int f(\vk, 0)d\vk_1 d\vk_2 d\vk_3 + \int f(\vk, 1)d\vk_1 d\vk_2 d\vk_3,     
\ene
where $\mathbb{Z}_2 = \{ 0, 1 \}$.

The Hilbert space of $N $ photons with energies between $s$ and $t$ is 
\beq \label{eq-pg4}
\cF_N^{s, t} : = \cS_N \bigotimes_{l= 1}^N \mathfrak{h}^{s,t}, \:  \cF^{s,t}_0 : = \CC,
\ene
where $\cS_N$ is the projection onto the space of totally symmetric functions.
The photon Fock space is the direct sum
$$
\cF^{s,t} : = \bigotimes_{N= 0}^\infty \cF_N^{s,t}. 
$$   
If $t = \infty$ we omit the variable $t$ and write
$$
\cF^s : = \cF^{s, \infty}. 
$$
The vacuum vector is the element 
$$
\Omega^{s,t} : = (1, 0, 0, \cdots) \in \cF^{s,t}. 
$$

\begin{remark}
We use frequently the superscript $s,t$ in our formulas throughout the paper. We will use  the convention that if the variable $t$ is omitted then it is $\infty$. 
\end{remark}

\subsubsection{The Photon Hamiltonian}
The photon Hamiltonian is the operator that takes an element $ \phi = (\phi_j)_{j \in \mathbb{N}\cup \{ 0 \}} \in \cF^{s, t} $ to the element 
$ \hf^{s,t}(\theta)\phi  $ with
$$
  (\hf^{s,t}(\theta)\phi)_j(k_1, \cdots, k_j  ): = e^{- \theta} (|k_1| + \cdots + |k_j|) \phi_j(k_1, \cdots, k_j),\: j \geq 1,
$$
$$
(\hf^{s,t}(\theta)\phi)_0 = 0,
$$

for any  $  \theta \in \CC$ , with $ |\theta| <  \frac{1}{2} $. If $t = \infty$ we omit the variable $t$:
$$
\hf^{s,t}(\theta): = \hf^s(\theta).
$$

\subsection{The Atom-Photon Hamiltonian}\label{S-atom-photon-hamiltonian}
\subsubsection{The Atom-Photon Hilbert Space}
The atom-photon Hilbert space is the tensor product
$$
\cH^{s,t} : = \cH_{\at} \otimes \cF^{s,t}
$$
(If $t = \infty$ we omit it in our notations). 
\subsubsection{The Free Atom-Photon Hamiltonian}
The free atom-photon Hamiltonian is the operator
\beq \label{fatph.1}
H^{s, t}_0(\theta) : = H_{\at}(\theta)\otimes \one_{\cF^{s,t}}  + \one_{\cH_{\at}}\otimes \hf^{s,t}(\theta), 
\ene
where $\one$ denotes the identity operator. $ H^{s, t}_0(\theta) $ depends also on the variables $\alpha$ and $\beta$. If it is necessary we
write this dependence explicitly 
\beq \label{fatph.2}
H^{s, t}_0(\theta, \alpha, \beta) : = H^{s,t}_0(\theta), \:  \theta, \beta \in \CC, |\theta|, |\beta| < \frac{1}{2}, \: \: \alpha \in [0, 1].
\ene
(If $t = \infty$ we omit it in our notations).

\subsubsection{The Atom-Photon Hamiltonian}\label{atom-photon-hamiltonian}
The atom-photon Hamiltonian is a Schr\"odinger operator with a second quantized magnetic potential (see Section~\ref{electron-photon-physical-hamiltonian}). The singularity at k=0 of the form factor that models the
interaction between the electron and the photons does not allow to study these
resonances  directly. To overcome this difficulties we transform it to another one that is unitarily equivalent via the Pauli-Fierz transformation (see Section~\ref{pauli-fierz-transformation}), which is a gauge transformation. 
The new Hamiltonian has to be analytically continued in order to study the resonances (see Section~\ref{analytic-continuation-resonances}). 
The long distance behavior in the electron variable becomes worse after the Pauli-Fierz transformation, this is the price to pay for a less singular form factor, that in fact will turn out to be regular at k=0. To overcome this problem we include a new parameter ($\beta$) that is technically convenient to analyze the long distance regime for the electron (see Section~\ref{parameter-beta}). The parameter $\beta$ is also useful to prove exponential decay of the eigenfunctions (see Section~\ref{S.ebe.atom} and Theorem~\ref{T.il.fm.1}). Finally we introduce an infrared cut off into the Hamiltonian that will be eventually removed (see Section~\ref{infrared-cut off}). 
In Section~\ref{infrared-cut off} we present the final Hamiltonian that we study in the rest of this article, and we give a precise description of the steps mentioned above.

\paragraph{Creation and Annihilation Operators}$\\$
\\
For any function $f \in \fh^{s, t}$, the creation operator $ a^*(f) $ is the operator that takes an element 
$\phi \in \cF_{N }^{s,t}$ to the vector
$$
a^*(f)(\phi) : = \sqrt{N + 1} \cS_{N + 1} f \otimes \phi \in \cF_{N + 1}^{s,t},
$$ 
see (\ref{eq-pg4}).
We extend $a^*(f)$ below by linearity and take the closure to define the creation operator $ a^*(f) $ in $\cF^{s,t}$. The annihilation
operator $a(f)$ is the adjoint of $a^*(f)$. \\
The creation and annihilation operators satisfy the canonical commutation
relations: 
 \beq \label{ih.3}\begin{array}{l}
[a^*(f) \, , \, a^*(g)] = [a(f) \, , \, a(g)]  = 0,  \; \;
[a(f) \, , \, a^*(g)]  = \la f \: | \; g \ra \comma \; \; \\\\  a(f)\Om^{s,t} = 0  \comma 
\end{array}
\ene 
where $[\cdot, \cdot]$ and $\la \cdot | \; \cdot \ra$ denote the commutator and the scalar product, respectively.  

For any vector valued function $\vec{f} \in  (\fh^{s, t})^3 $  we define 
$ a^*(\vec{f}) $ and $a(\vec{f})$ component-wise.  
We use the isomorphism 
$$
\cH^{s,t} = \cH_{\at} \otimes \cF^{s,t} \cong L^2(\mathbb{R}^3; \cF^{s,t}) 
$$
to extend the definition of creation and annihilation operators to $\cH^{s,t}$. Let $f : \mathbb{R}^3 \times \cK^{s,t} \to \CC $
be such that for all $x \in \mathbb{R}^3$, the function $f_x(\cdot) : = f(x, \cdot) $ belongs to $\fh^{s,t}$. We define
$$
a(f)^*\phi (x) : = a^*(f_x) \phi(x), \:  \phi \in  L^2(\mathbb{R}^3; \cF^{s,t}).  
$$
The definition of $a(f)$ is similar. Both definitions can be extended to vector valued functions component-wise. For any operator 
$O$ defined in $\cH_\at$ and any function  $f : \mathbb{R}^3 \times \cK^{s,t} \to \CC $  as before we identify
\beq \label{op-creation}
a^*( O f): =  O\otimes \one_{\cF^{s,t}}a^*(f), \:  a( Of  ): =   a(f)O^*\otimes \one_{\cF^{s,t}}.
\ene
and accordingly we define $a^*(  fO)$, $a( fO)$.

It is convenient to define the creation and annihilation operators 
point-wise ((X.74) and (X.75) \cite{ReedSimonII1980}). They are 
operator-valued tempered distributions on the Fock space $\cF^{s,t}$ 
obeying the canonical commutation relations,
\beq \label{ih.8} \begin{array}{l}
[a^\#(\vk_1, \lambda ), a^\#(\vk_2, \mu ) ] = 0, \; [a(\vk_1, \lambda ), a^*(\vk_2, \mu ) ]= 
\delta_{\lambda, \mu}  \delta (\vk_1 - \vk_2) \comma \\\\
a(\vk_1, \lambda)\Om = 0 \comma
\end{array}
\ene
where $a^\# = a$ or $  a^* $. For any  $O$ and $f$ are as in (\ref{op-creation}) we write
\beq \label{ih.10}
a^*(O f) : = \int_{\cK^{s,t}} O f(x,k) a^*(k) dk, \; a(O f) : = \int_{\cK^{s,t}}  \overline{f(x,k)}  O^*  a(k) dk \comma
\ene
and accordingly we write $a^*(fO)$ and $ a(fO) $.

\paragraph{The Atom-Photon Hamiltonian: The Second-Quantized Magnetic Schr\"odinger Equation}\label{electron-photon-physical-hamiltonian} $\\$
\\
For any $\theta $ in $\CC$ with $|\theta | < \frac{1}{2}$ we define the function $G^{s,t}(\theta) : \mathbb{R}^3 \times \cK^{s,t} \to \CC^3  $ by 
\beq \label{i.2}
G^{s,t}(\theta)(x, k) : =  \frac{\alpha^{3/2}}{(2 \pi)^{3/2}} \frac{e^{- \theta}\exp(e^{- 2 \theta} |k|^2)}{\sqrt{2 |k|}} 
e^{-i \alpha  \vk \cdot x} \veps(k), 
\ene
where the polarization vector $ \veps(k) $ is defined in Section~\ref{effective-potential}. The constant $\alpha$ is the fine structure constant but it will be assumed to be sufficiently small. The exponential function $ \exp(e^{- 2 \theta} |k|^2)  $ plays the role of an ultraviolet cut off. The parameter $s$ is an infrared cut off if it is strictly positive.  

The second quantized magnetic potential is the operator 

\beq \label{i.6}
\boA^{s,t}(\theta) : = a^*(G^{s,t}(\theta)) + a(G^{s,t}(\overline{\theta})).
\ene
In the formula above we omit the parameter $t$ when it is infinity. 

$  \boA^0(\theta) $ is an analytic continuation of the physical second quantized magnetic potential
$\boA^0(0)$.  
The electron-photon Hamiltonian is the operator
\beq \label{number}
\boH : = ( -i \nabla \otimes \one_{\cF^{0}}  -  \boA^0(0) )^2 + V(0, 0, 0) \otimes \one_{\cF^{0}} + \one_{\cH_{el}}\otimes \hf^0 
\ene
(see (\ref{ep-1})), which is a second quantized version of the magnetic Schr\"odinger operator.

\paragraph{The Pauli-Fierz Transformation}\label{pauli-fierz-transformation} $\\$
\\
The form factor
$\frac{1}{\sqrt{|k|}}$ (see (\ref{i.2})) yields major difficulties in the study of resonances,
indeed, in this case the perturbation is (superficially) \emph{marginal}.
On
the contrary,  we can control the renormalization scheme in presence of
any
regularization of the form factor, i.e., $1/\sqrt{|k|^{1/2-\mu}}$  as $k\to
0$,
 with $\mu>0$.  We cannot handle directly $\boH$ but instead we can handle another Hamiltonian wich is unitarily equivalent (and therefore physically equivalent). This new Hamiltonian is obtained by a change of gauge which is called the Pauli-Fierz transformation. In the new Hamiltonian the interaction of the electron with the photons is
modeled by a less singular form factor, but the long distance behavior in the electron variable is more difficult to handle.
The relation between small momenta and long distances can be read from the oscillatory functions ($e^{-i \alpha \vk \cdot x}$). In some sense we can say that we trade small momenta for long distances via the Pauli-Fierz transformation. The long distance behavior in the electron variable can be controlled by exploiting the exponential localization of the eigenfunctions. 

We define the operator
\beq \label{lambda-pf}
\lambda^{s, t}_{PF}(\theta) : = a^*( G^{s,t}(\theta)(0, k) \cdot \eta (|x| |k|) e^{\theta} x) +  a( G^{s,t}(\overline{\theta})(0, k) \cdot \eta (|x| |k|)e^{\overline{\theta}} x)
\ene
(see (\ref{pft.1})). We omit the variable $t$ if it is infinity and we do not write the dependence on $\alpha$. 

The Pauli-Fierz transformed Hamiltonian is the operator 
\beq \label{PF}
H^0(0, \alpha, 0) : =   e^{- i \lambda^0_{PF}(0)}\boH e^{i \lambda^0_{PF}(0) }. 
\ene
Explicitly we have (see (\ref{ep-1}, \ref{ep-2}) and appendix \ref{ap-b}) 
\begin{align} \label{eph.1} 
H^0(0, \alpha, 0)  = & 
( -i \nabla \otimes \one_{\cF^{0}}  -  A^0(0) )^2 + b^0(0)  \\ \notag &
+ V(0, \alpha, 0) \otimes \one_{\cF^{0}} + \one_{\cH_{\at}}\otimes \hf^0,
\end{align}
where
\begin{align} \label{eph.2} 
A^{s, t}(\theta) := \ & \boA^{s, t}(\theta) - (e^{- \theta}\nabla \otimes \one_{\cF^{s,t}})  \lambda^{s, t}_{PF}(\theta), \\ \notag
b^{s, t}(\theta) := \ & a^*( ie^{-  \theta}|k|G^{s,t}(\theta)(0, k) \cdot \eta (|x| |k|) x) \\ \notag 
 & + a(ie^{- \overline{\theta}}|k|G^{s,t}(\overline{\theta})(0, k) \cdot \eta (|x| |k|) x)
\end{align}
and we do not write $t$ if it is infinity.

\paragraph{Analytic Continuation} \label{analytic-continuation-resonances} $\\$
\\
The Hamiltonian $H^0(0)$ is the object of our study. It represents the energy of the electron-photon system in the representation that follows from the change of gauge in formula (\ref{PF}). We are interested on the resonances of the Hamiltonian $H^0(0)$. To study the resonances we need an analytic continuation of $H^0(0)$. It is given by (\ref{ep-2}) and (\ref{eph.2}) as follows
\beq \label{eph.3} \begin{array}{l}
H^0(\theta,\alpha, 0)  = 
( -i e^{- \theta}\nabla \otimes \one_{\cF^{0}}  -  A^0(\theta) )^2 + b^0(\theta)  \\\\
+ V(\theta, \alpha, 0) \otimes \one_{\cF^{0}} + e^{- \theta}\one_{\cH_{\at}}\otimes \hf^0.
\end{array}
\ene

For real $\theta$ we define the unitary operator 
$$
U(\theta) : = u(\theta) \otimes \big( \one_{\CC} \oplus \bigoplus_{j = 1}^{\infty} u(- \theta)^{\otimes^j }\big)
$$
(see (\ref{d1})) acting on the Hilbert space $\cH^0$. It is clear from the definitions that
\beq \label{eph.1pirata}
H^0(\theta, \alpha, 0) : = U(\theta) H^0(0, \alpha, 0) U(\theta)^{-1}.
\ene
In Section~\ref{s.rbapha} we prove that $H^0(\theta, \alpha, 0)$ is closed and self-adjoint for real $\theta$. We prove furthermore that it is an analytic family of operators (see Theorem~\ref{L.inh.1}).   

\paragraph{Controlling the Long-Distance Behavior for the Electron Variable: The parameter $\beta$} \label{parameter-beta}$\\$
\\
The Pauli-Fierz transformation yields a form factor that is regular at $k=0$, but as a consequence the long distance behavior of the electronic part is worse. To control the electron at log distances we make use of the exponential decay of the eigenfunctions. This is done by introducing a new parameter $\beta \in \CC,\, |\beta| < \frac{1}{2}$, in the Hamiltonian. We define
\beq \begin{array}{l}
H^0(\theta, \alpha, \beta) : = e^{ - \beta \la x\ra   } H^0(\theta, \alpha, 0) e^{ \beta \la x\ra   }\\\\
= ( -i e^{- \theta} \nabla \otimes \one_{\cF^{0}}  -i e^{- \theta} \beta (\nabla   \la x\ra)\otimes \one_{\cF^0}   -  A^0(\theta) )^2 + b^0(\theta)  \\\\
+ V(\theta, \alpha, 0) \otimes \one_{\cF^{0}} + e^{- \theta}\one_{\cH_{\at}}\otimes \hf^0,
\end{array}
\ene
(see (\ref{geh.7.m1})). The parameter $\beta$ is also useful to prove exponential decay of the eigenfunctions (see Section~\ref{S.ebe.atom} and Theorem~\ref{T.il.fm.1}).

\paragraph{The Infrared cut off: The Final Hamiltonian}\label{infrared-cut off} $\\$
\\   
We conclude our construction of the Hamiltonian by including an infrared cut off (that will be eventually removed). We restrict the momentum of the 
photons to be larger or equal to $s$ and smaller than  $t$. We point out that if $s = 0$ there is no infrared cut off. Finally, we define the
Hamiltonian
\beq \begin{array}{l}\label{eph}
H^{s, t}(\theta, \alpha, \beta) : = 
( -i e^{- \theta} \nabla \otimes \one_{\cF^{s,t}} -  A^{s,t}(\theta) )^2  \\\\ + 
i e^{- \theta} \beta(\nabla   \la x\ra) \cdot  A^{s, t}(\theta) + 
   A^{s, t}(\theta) \cdot i e^{- \theta} \beta (\nabla \la x\ra) \\\\
+ b^{s, t}(\theta)  
+ V(\theta, \alpha, \beta) \otimes \one_{\cF^{s, t}} + e^{- \theta}\one_{\cH_{\at}}\otimes \hf^{s, t} \\\\ =
H_0^{s,t}(\theta, \alpha, \beta) + W^{s,t}(\theta, \alpha, \beta),
\end{array}
\ene
where
\beq \label{icfh.2}
W^{s,t}(\theta, \alpha, \beta)  : =  H^{s,t}(\theta, \alpha, \beta) - H_0^{s,t}(\theta, \alpha, \beta)
\ene 
is the interaction (see the definition of $H_0^{s,t}(\theta, \alpha, \beta)$ in (\ref{fatph.1})).

Whenever we do not use explicitly the parameters $\alpha$ and $\beta$ we simply write 
\beq \label{icfh.1}
H^{s,t}(\theta) : = H^{s,t}(\theta, \alpha, \beta), \: \:  W^{s,t}(\theta)  : =  W^{s,t}(\theta, \alpha, \beta).  
\ene
(If $t = \infty$ we omit it in our notation).

Using (\ref{ih.8}) and (\ref{ih.10}) we can write the interaction in the form 
\begin{align} \label{i.10.4} 
W^{s,t}(\theta, \alpha, \beta) := & W^{s,t}_{1, 0}(\theta, \alpha, \beta) + 
  W^{s,t}_{0, 1}(\theta, \alpha, \beta)  + W^{s,t}_{2, 0}(\theta, \alpha, \beta) \\ \notag & +  
W^{s,t}_{0, 2}(\theta, \alpha, \beta) + W^{s,t}_{1, 1}(\theta, \alpha, \beta) + W^{s,t}_{0, 0}(\theta, \alpha, \beta)
\period 
\end{align}
where
\begin{align} \label{i.10.3}
W^{s,t}_{m, n}(\theta, \alpha, \beta) := & \int_{ (\cK^{s,t})^{(m + n)}} dk_1\cdots dk_{m} d \tilde{k}_1
\cdots d\tilde{k}_n \\ & \notag \big[ w^{s,t}_{m,n}(\theta, \alpha, \beta)(k_1, \cdots, k_{m}, \tilde{k}_1 \cdots \tilde{k}_n) \\ & \notag
 a^*(k_1)\cdots a^*(k_m)a(\tilde{k}_{1}) \cdots a(\tilde{k}_{n})\big] \period
\end{align}
The functions $w^{s,t}_{m, n}(\theta, \alpha, \beta)$ can be derived from (\ref{icfh.2}).  We report them in 
(\ref{i.10.3.0})-(\ref{i.10.3.6}). The
operator on the r-h-s of Eq. (\ref{i.10.3}) is understood in the sense of quadratic forms (see Theorem X.44 and (X.74, X.75) of \cite{ReedSimonII1980}).

\section{Estimates for the Atom Hamiltonian} \label{estimates-electron-hamiltonian}
In this section we state some results for the electron Hamiltonian that are used later on. The proofs are deferred to Section~\ref{proofs-estimates-electron-hamiltonian}. 

\begin{lemma} \label{L.eeh.m1}
For every $\epsilon > 0$  
 there exists a constant $b_\epsilon$ such that,  $ \forall  |\theta| < \frac{1}{2} $ : 
\beq \label{geh.5}
\| V(\theta, \alpha, \beta)  \phi \|_{L^2(\mathbb{R}^3)} \leq \epsilon \| - \Delta \phi  \|_{L^2(\mathbb{R}^3)} + b_{\epsilon} \|   \phi \|_{L^2(\mathbb{R}^3)}
\ene 
for any $\phi \in H^2(\mathbb{R}^3) $, where 
$ H^2(\mathbb{R}^3) $ is the Sobolev space of functions in $L^2(\mathbb{R}^3)$ with distributional derivatives up to order $2$ in $ L^2(\mathbb{R}^3)  $.
\end{lemma}
\emph{Proof:}
Eq. (\ref{geh.5}) follows from  (\ref{l.10}), (\ref{pft.15}) and (\ref{potential.1}). 
\QED

\begin{definition} \label{notation-electron}
The following constants are repeatedly used through the paper: 
\begin{align} \label{geh.6.1}
\boa : = & \ a(\frac{1}{19}( \frac{32}{\delta_\at})^{-1})^{1/3} \comma \\ \notag
\bob : = & \ \frac{1}{19} ( 50 \cdot ( 4  + (2|\boe_0| + 2 b_{1/2} + 1) 
 \frac{32}{\delta_\at} ) )^{-1} \period
\\ 
\label{an.0.7}
C_{\ref{an.0.7}} : =  & \ \sup_{\alpha \leq 1, |\beta| \leq 1  }  \max_{|\theta| = \frac{1}{60}} \| V(\theta, \alpha, \beta) \frac{1}{1 - \Delta} \| \comma
\\  \label{an.0.9}
C_{\ref{an.0.9}} : =  & \ (\frac{1}{2 (1/60 - 1/120)^2}C_{\ref{an.0.7}} +
 2e^{1/15}) \cdot \frac{4}{3}\big(  1  + 2 \frac{|\boe_0| + \delta_\at/16   + b_{1/4} + 1}{ \delta_\at/16  }  \big)\period
\end{align}

\end{definition}

\begin{theorem} \label{p.eeh.1}

We suppose that $\alpha \leq \boa$, $|\beta| \leq \bob$ and that
\beq \label{an.0.14}
| \theta   | \leq \min\Big((32 C_{\ref{an.0.9}})^{-1}, \frac{1}{120}\Big)\period
\ene
The following holds true:
\begin{itemize}
\item 
There are only two points ($\{ e_0(\theta, \alpha, \beta),  e_1(\theta, \alpha, \beta) \}$) in the spectrum of $H_\at(\theta, \alpha, \beta)$ with real part less than $\boe_1 + \frac{15}{16} \delta_\at$ (see Definition \ref{delta-el}). 
They are simple eigenvalues and they do not depend on $\beta$ and $\theta$ (they are therefore real). They satisfy 
\beq \label{e.p.eeh.1}
| e_j(\theta,\alpha, \beta) - \boe_j | \leq \frac{\delta_\at}{16}, \: j \in \{0, 1\}.  
\ene
It follows that 
\beq \label{e.p.eeh.1.1}
e_j(\theta,\alpha, \beta)  = e_{j}(0, \alpha, 0): = e_{j}(\alpha), \: j \in \{0, 1\}.  
\ene
We omit writing the dependence on $\alpha$ when it is not strictly necessary. 
\item Let 
\beq \label{e.p.eeh.2}
P_{\at, j}(\theta, \alpha, \beta),\:   j \in \{ 0, 1 \} 
\ene
be the projection onto the eigen-space corresponding to $ e_j  $, $j  \in \{ 0, 1  \}$ respectively. 
It follows that 
\beq \label{e.p.eeh.3} \begin{array}{l}
\| P_{\at, j}(0, \alpha, \beta)  - P_{\at, j}(0, 0, 0) \| \leq \frac{1}{8}\comma \\\\
 \| P_{\at, j}(\theta, \alpha, \beta)  - P_{\at, j}(0, \alpha, \beta) \| \leq \frac{1}{8}.
\end{array}
\ene
\end{itemize}

\end{theorem}

\begin{definition} \label{d.eeh.1}
We denote by 
\beq\label{eprima}
e_i' := e_i - \frac{\delta_\at}{16},\: i \in \{1, 2\}. 
\ene
and 
\beq\label{delta}
\delta =\frac{ 7 \delta_\at}{8}.
\ene
We notice that
\beq
\delta \leq e_1 - e_0, \: \:   e_i' \leq e_i, \: \: i \in \{0, 1\}\comma
\ene 
for any $\alpha \leq \boa$.

\end{definition}

\noindent We define the projection operator $P_{disc}(\theta)$ by

\beq \label{eeh.3.0.0}
P_{disc}(\theta, \alpha, \beta) : = P_{\at, 0}(\theta, \alpha, \beta) + P_{\at, 1}(\theta, \alpha, \beta) \period
\ene
For any projection $P$ we define 
\beq \label{eeh.3}
\overline{P}: = 1- P \period
\ene

\begin{remark}
From the proof of Theorem~\ref{p.eeh.1} (more precisely from Eq. (\ref{an.0.6})) it follows that  
\beq \label{an.0.10}
 \|(1 - \Delta) \frac{1}{|H_{\at}(0, \alpha, \beta) - z| }  \| \leq C^2_{\ref{an.0.11}}  \comma
\ene
where 
\beq \label{an.0.11}
C^2_{\ref{an.0.11}} : =  \frac{4}{3}\big(  1  + 2 \frac{|\boe_0| + \delta_\at/16   + b_{1/4} + 1}{ \delta_\at/16  }  \big)   \comma
\ene
for any $z \in \CC$ with ${\rm Re } (\, z) \leq 0$ and  $\dist(z, \sigma(H_\at(0, 0,0))) \geq \frac{\delta_\at}{16}$. 
\end{remark}

\begin{theorem}\label{L.eeh.1} Suppose that $\theta$
 satisfies (\ref{an.0.14}). Suppose furthermore that $\alpha \leq \boa$ and $|\beta| \leq \bob$. 
Let $z \in \CC   $  with $\mathrm{Re (z)} < \boe_1 + \frac{7}{8}\delta_\at $. Then,  
\beq \label{L.eeh.1.e1}\begin{array}{l}
\|  (H_\at(\theta, \alpha, \beta) - z )^{-1} \overline{P}_{disc}(\theta, \alpha, \beta)  \|
\leq 54   \frac{1}{ |z - \boe_{\ref{an.0.16.1}}| } \\\\ 
\leq C_{\ref{an.0.16.2}} \frac{1}{| z - e_{0}| }     \comma
\end{array}
\ene
where
\beq \label{an.0.16.1}
\boe_{\ref{an.0.16.1}}  : =  \boe_1 + \frac{15}{16}\delta_\at      \comma
\ene
and 
\beq \label{an.0.16.2}
 C_{\ref{an.0.16.2}} := 972 \frac{|\boe_0|}{\delta_\at}     \period
\ene

\end{theorem}

\begin{corollary} \label{C.eeh.1} Suppose that $\theta$
 satisfies (\ref{an.0.14}). Suppose furthermore that $\alpha \leq \boa$ and $|\beta| \leq \bob$. 
Let $z \in \CC \setminus \{ e_{0}, e_{1}  \}   $  with $\mathrm{Re (z)} < \boe_1 + \frac{7}{8}\delta_\at $. The following estimates 
hold true,
\beq \label{C.eeh.1.e1}
\| (H_\at(\theta, \alpha, \beta) - z)^{-1}  \| \leq 4 C_{\ref{an.0.16.2}} ( \frac{1}{|z-e_{0}|} +   
\frac{1}{|z - e_{1}|} ) \period
\ene
\beq \label{C.eeh.1.e2}
\| (H_\at(\theta, \alpha, \beta) - z)^{-1} \overline{P}_{\at, 1}(\theta,  \alpha, \beta) 
 \| \leq  4 C_{\ref{an.0.16.2}}   \frac{1}{|z-e_{0}|}    \period
\ene

\end{corollary}

\begin{corollary}\label{L.eeh.2}
Suppose that $\theta$
 satisfies (\ref{an.0.14}). Suppose furthermore that $\alpha \leq \boa$ and $|\beta| \leq \bob$.  
Let $z, \; \mu \in \overline{B_{\delta_\at/6}(e_{1})}  $, it follows that
\beq \label{L.eeh.2.e1}
\|  (H_{\at}(\theta, \alpha, \beta)- z) \frac{\overline{P}_{\at, 1}(\theta, \alpha, \beta)}{ (H_{\at}(\theta, \alpha,\beta) - \mu)   } 
  \| \leq 4 + 4 C_{\ref{an.0.16.2}} \period 
\ene
\end{corollary}

\subsection{Exponentially Boundedness of the Eigenfunctions} \label{S.ebe.atom} 

It is well known that the eigenfunctions of the electron Hamiltonian are exponentially bounded (see \cite{CombesThomas1973}). 
But for our proposes we need uniform bounds in $\theta $ and $\alpha$. We provide below some explicit bounds for the exponentially
boundedness of the eigenvalues as well as explicit uniform estimates for the exponential decay rate.   
   
\begin{theorem}\label{T.ebe.1}
Suppose that $\theta$
 satisfies (\ref{an.0.14}). Suppose furthermore that $\alpha \leq \boa$ and $|\beta| \leq \bob$.  Then 
the range of $ P_{\at, 1}(\theta, \alpha, 0)  $ is contained in the domain of $e^{\beta \la x \ra}$ (see (\ref{geh.7.m1})) 
and
\beq \label{T.eb.1.e1} \begin{array}{l}
 \|  e^{\beta \la x \ra}  P_{\at, 1}(\theta, \alpha, 0) \|  \leq
 8 \| e^{\beta \la x\ra}  \phi   \|_{\cH_\at}  \period
\end{array}
\ene
In particular the following uniform bounds (in $\theta$ and $\alpha$) hold true
\beq \label{T.eb.1.e3} \begin{array}{l}
 \|  e^{\beta \la x \ra}  P_{\at, 1}(\theta, \alpha, 0) \| \leq  C_{\ref{T.eb.1.e2}} \comma
\end{array}
\ene
and 
\beq \label{T.eb.1.e4} \begin{array}{l}
 \| (1 + |x|^2) P_{\at, 1 }(\theta, \alpha, 0) \| \leq  C_{\ref{T.eb.1.e2.1}}  \comma
\end{array}
\ene
where
\beq \label{T.eb.1.e2}
C_{\ref{T.eb.1.e2}}  = 8 \| e^{\bob \la x \ra}  \phi   \|_{\cH_\at}   \comma
\ene
\beq \label{T.eb.1.e2.1}
C_{\ref{T.eb.1.e2.1}}  =   8 \|  e^{\bob \la x \ra}  \phi   \|_{\cH_\at} (1 + e^{-2}(1 + 4/\bob^2))  \comma
\ene
and $  \phi $ is any unit eigenvector of $  P_{\at, 1, 0, 0}(0)  $.  

\end{theorem}

\section{Relative Boundedness of the Atom-Photon Hamiltonian and Analyticity.} \label{s.rbapha}
In this section we establish some basic properties of the atom-photon Hamiltonian as well as some useful estimates. The proofs of this sections 
are deferred to Section~\ref{rbapha-proofs}.

First we recall a basic result the proof of which can be found for example in Lemma 1 
of \cite{HaslerHerbst2008}.

\begin{lemma}\label{L.ih.1}

Let  $g_i : \mathbb{R}^3 \to  \mathfrak{h}^{s,t} $, $i \in \{1, 2 \}$, be uniformly bounded functions. We suppose furthermore 
that the functions $ x \to |k|^{-1/2} g_i(x)  $ are uniformly bounded with values in $ \mathfrak{h}^{s,t} $ (see (\ref{extra.1})-(\ref{extra.5})).  

For any $   a^\#, \tilde a^\# \in \{ a, a^* \} $, it follows that 
$$ \dom (\one_{\cH_\at} \otimes (\hf^{s, t} )^{1/2} ) \subset \dom(a^\#(g_i)) $$
and $$  a^\#(g_i)  [ \dom(  \one_{\cH_\at} \otimes \hf^{s,t}) ] 
\subset  \dom(  a^\#(g_j) ).  $$
It follows furthermore that for every $\rho > 0$ 	and every $\phi \in \dom(\one_{\cH_\at}  \otimes (\hf^{s,t} )^{1/2}  )$, 
$ \psi \in \dom  (  \one_{\cH_\at}  \otimes \hf^{s,t}    ) $,
\beq
\begin{array}{l}
 \big\| a^\#(g_i) \, \phi \big\|_{\cH^{s,t}} 
 \leq 
\|  g_i \|_{\rho} \|  \one_{\cH_\at}  \otimes(\hf^{s,t} + \rho)^{1/2}  \phi  \|_{\cH^{s,t}} \comma
\\\\ 
\big\| \tilde a^\#(g_2) \, a^\#(g_1) \psi \,  \big\|_{\cH^{s,t}}   
 \leq 
  \| g_2 \|_{\rho}   \|  g_1 \|_{ \rho} 
      \|  \one_{\cH_\at}  \otimes(\hf^{s,t} + \rho)  \psi  \|_{\cH^{s,t}}   \comma
\end{array}
\ene
where
\beq \label{norm-rho}
 \| g_i \|_{\rho} := \sup_{x \in \mathbb{R}^3} \frac{1}{\rho^{1/2}} \|  g_i(x)  \|_{\mathfrak{h}^{s,t}} + \sup_{x \in \mathbb{R}^3} \| |k|^{-1/2} g_i(x)  \|_{\mathfrak{h}^{s,t}}.
\ene

\end{lemma}

In the following remark we fix some notations that will be used later. 
\begin{remark}\label{remark-chafa}
Let $s' \leq s < t  $ be positive real or infinite numbers. \\
The union
\beq \label{ilprima.6}
\cK^{s',t} = \cK^{s', s}\cup \cK^{s, t} 
\ene
gives rise to the isomorphisms
\beq \label{ilprima.7}
\cF^{ s',t  } \cong    \cF^{s,t} \otimes \cF^{ s', s } \comma
\ene 
\beq \label{ilprima.8}
\cH^{ s',t} \cong    \cH^{ s,t } \otimes  \cF^{ s', s }\comma
\ene 
\beq \label{ilprima.8.1}
\hf^{s',t} =   \hf^{ s,t } \otimes  \one_{ \cF^{ s', s }   }  +    \one_{  \cF^{s,t}  }  \otimes \hf^{  s', s} \period 
\ene 

Given an operator $O$ on $   \cH^{s,t}  $, we use the same symbol to denote the 
operator 
\beq  \label{ilprima.9}
O : = O \otimes 1_{  \cF^{ (s', s) }  }\period
\ene 

\end{remark}

The next lemma establishes the relative boundedness of the interaction with respect to the free atom-photon Hamiltonian (see (\ref{eph})-(\ref{icfh.1})),
which implies that the atom-photon Hamiltonian is closed (for small $\alpha $ and $|\beta|$) and furthermore that it is self-adjoint for $ \theta \in \RR $ and ${\rm Re}\, \beta = 0$  ( see Theorem 1.1 page 190 of \cite{kato1976} and the Kato Rellich Theorem (Theorem X.12 of 
\cite{ReedSimonII1980}). For the proofs see Section~\ref{rbapha-proofs}.

\begin{lemma}\label{L.i.1}
 Let $ s' \leq s  < t$ be positive real numbers, possibly $t$ is infinity. Suppose that $\alpha \leq \boa $ and $|\beta| \leq \bob $ ( see \ref{geh.6.1}). 

For every $\rho > 0$ and every $\theta, \eta \in \CC$ with $|\theta| \leq \frac{1}{120}$ and $ |\theta + \eta| \leq \frac{1}{120} $ there exist constants $C_{\ref{i.13.4}}$ and 
$C_{\ref{i.17}}$ such that the following  
estimates hold true 
\begin{align}\label{L.i.1.e1}
\| W^{s,t}(\theta) \phi \|_{\cH^{s',t}} \leq &  C_{\ref{i.13.4}} ( (1 + \frac{1}{\rho^{1/2} })\alpha^{3/2} +  (1 + \frac{1}{\rho^{1/2} })^2\alpha^{3})\cdot \\ \notag & 
(\|(H_0^{s',t}(\theta) + \rho)\phi \|_{\cH^{s',t}} +  \| \phi \|_{\cH^{s',t}}) \comma
\end{align}

moreover, if $-\rho $ belongs to the resolvent set of $H_{0}^{s,t}(0, \alpha, \beta)$,
\begin{align}
\label{L.i.1.e2}
\big\| \big(W^{s,t}(\theta + \eta) - W^{s,t}(\theta)\big) \frac{1}{ H^{s',t}(0) + \rho}   \big\| 
\leq &  |\eta|  C_{\ref{i.17}} \big( (1 + \frac{1}{\rho^{1/2} })\alpha^{3/2} +  \\  \notag (1 + \frac{1}{\rho^{1/2} })^2 & \alpha^{3}\big) 
(1 + \|(H_0^{s',t}(0) + \rho)^{-1} \| ) \period
\end{align}
and the operator-valued function $\theta \to W^{s,t}(\theta)
\frac{1}{ H^{s',t}(0) + \rho} $ is analytic for $|\theta| < \frac{1}{120}$.\\
Explicit values for the constants are given in (\ref{i.13.4}) and (\ref{i.17}). 
\end{lemma}
\beq 
\ene
In the next theorem we establish the analyticity of the Hamiltonian. 

\begin{theorem} \label{L.inh.1}
Suppose that $\alpha \leq \boa $ and $|\beta| \leq \bob $ (see \ref{geh.6.1}).
For every $\rho > 0$ such that $- \rho$ belongs to the resolvent set 
of $H^{s,t}_0(0)$, the function $\theta  \to H^{s,t}( \theta) \frac{1}{H^{s,t}(0) + 
\rho}$ is an operator-valued analytic function for $|\theta| < \frac{1}{120}$.
 Moreover, for every 
$\theta, h \in \CC$ such that $  |\theta| < \frac{1}{120}, |\theta + h| < \frac{1}{120} $
 the following estimate holds

\begin{align} \label{L.inh.1.e1} 
\|(H^{s,t}(\theta + h) - H^{s,t}(\theta )) \frac{1}{H_0^{s,t}(0) + \rho}\| \leq 
C_{\ref{L.inh.1.e2}}  |h|(1 + \| \frac{1}{H_0^{s,t}(0) + \rho}  \|) \comma
\end{align}
where
\begin{align} \label{L.inh.1.e2}
C_{\ref{L.inh.1.e2}}(\alpha,\rho):= C_{\ref{L.inh.1.e2}} : = & \big( C_{\ref{i.17}} ( (1 + \frac{1}{\rho^{1/2} })\alpha^{3/2}  
+  (1 + \frac{1}{\rho^{1/2} })^2\alpha^{3}) 
 \\ & \notag + \frac{4}{3} (  1 + b_{1/4} )(\frac{C_{\ref{an.0.7}}}{2(1/60 - 1/120)^2} +  3 e^{1/15} )\big) 
\end{align}

\end{theorem}

Using Lemma~\ref{L.i.1} and Theorem~\ref{L.inh.1} we can prove the following basic estimates that are useful for the next chapters.

\begin{corollary}\label{L.an.1}
Suppose that $\alpha \leq \boa $ and $|\beta| \leq \bob $ (see \ref{geh.6.1}). Suppose furthermore that $\alpha^{3/2} \leq \frac{1}{128 C_{\ref{i.17}} }$ and that $z \in \CC$ is such that $ \rm{Re}(z) \leq \frac{3}{2} e_0 - \frac{3}{4}  $. Then 
$H^{s,t}(0, \alpha, \beta) - z$ is invertible and 
\beq \label{L.an.1.e1}
\|  \frac{1}{H^{s,t}(0, \alpha, \beta) - z}  \| \leq \Big(\frac{4}{3}\Big)^2 \frac{1}{| z - \boe_0  |} \comma
\ene
\beq \label{L.an.2.e1}
\| (H_{0}^{s,t}(0, \alpha, \beta) - z)  \frac{1}{H^{s,t}(0, \alpha, \beta) - z}  \| \leq  2 \period
\ene
Suppose furthermore that $\theta, h \in \CC$ are such that $|\theta| < \frac{1}{120}$ and $|\theta + h| < \frac{1}{120}$, then
\begin{align} \label{L.an.3.e1} 
\| (H^{s,t}(\theta + h, \alpha, \beta) - H^{s,t}(\theta, \alpha, \beta))  \frac{ 1  }{H^{s,t}(0, \alpha, \beta) - z}  \| &  \leq 
|h|\cdot 2 \Big( 1 +   \Big( \frac{4}{3}\Big)^2 \Big) \\ \notag  \cdot \big( \frac{1}{16} +   \frac{4}{3} (  1 + b_{1/4} )  (\frac{C_{\ref{an.0.7}}}{2(1/60 -  1/120)^2} &   +  3 e^{1/15} )\big)  \period
\end{align}
In particular if  
$$ |h|  \leq   \frac{1}{4}    \Big( 2 \Big( 1 +  \Big( \frac{4}{3}\Big)^2 \Big) \cdot \Big( 
  \frac{1}{16} +   \frac{4}{3} (  1 + b_{1/4} )(\frac{C_{\ref{an.0.7}}}{2(1/60 - 1/120)^2} +  3 e^{1/15} ) \Big) \Big)^{-1},  
$$ it follows that 
\beq \label{L.an.3.1.e1}
\|  \frac{ 1  }{H^{s,t}(\theta, \alpha, \beta) -  z}  \| \leq \Big(\frac{4}{3 }\Big)^3 \cdot \frac{1}{| z - \boe_0  |}   \period
\ene

\end{corollary}

\begin{corollary}\label{L.an.4}

Suppose that $\alpha \leq \boa $ and $|\beta| \leq \bob $ ( see (\ref{geh.6.1})). Suppose furthermore that  $\alpha^{3/2} \leq \frac{1}{128 C_{\ref{i.17}} }$. Then
for every $\theta \in B^\CC_{1/120}(0)$, $h \in B^\CC_{1/120}(0) \cap(i \RR)$   with 

\beq
\begin{array}{l}\label{theta} 
 |\theta| <  \frac{1}{16}    \Big( 2 \Big( 1 +  \Big( \frac{4}{3}\Big)^2 \Big) \cdot
 \Big( \frac{1}{16} +   \frac{4}{3} (  1 + b_{1/4} )(\frac{C_{\ref{an.0.7}}}{2(1/60 - 1/120)^2} +  3 e^{1/15} )\Big)  \Big)^{-1}  $$
\end{array}
\ene
 the following estimates hold

\beq \label{L.an.4.em1}
\|  (H^{s,t}_0(0, \alpha, \beta) -    \rho )\frac{1}{  H^{s,t}(\theta, \alpha, \beta) - \rho   }    \| \leq 2 \comma
\ene

\beq \label{L.an.4.e1}
\|  (H^{s,t}_0(0, \alpha, \beta) -    \rho )\frac{1}{  H^{s,t}(\theta, \alpha, \beta) - e^{h} (\rho + r)  }    \| \leq 8 \comma
\ene

where $r \in \RR $ is any negative number and $\rho \leq 2 e_0 - 1 $.



\end{corollary}

\section{Resonances for the Infrared-Regularized Hamiltonian} \label{s.rirh}
In this section we construct an eigenvalue of the infrared regularized Hamiltonian in a neighbourhood of $e_1$. We can estimate the 
imaginary part of it and  give conditions for it to be strictly negative (which implies that the eigenvalue is actually a resonance). The main results are collected in Theorem~\ref{princ-secc}. Theorem~\ref{princ-secc} is the starting point of an inductive argument that takes the infrared cut off to zero and proves the existence of resonances for the non-regularized Hamiltonian. The latter is the main result of this paper and it is proved in the next section (Section~\ref{infrared-limit}). 

\noindent In this section we fix the parameter $\theta $, 

\beq \label{rirh.1}
\theta =  i\nu    
\ene
with
\beq \label{rirh.2}
\nu \in \RR \setminus \{0 \}.
\ene
We suppose that it satisfies (\ref{an.0.14}) and (\ref{theta}).

\noindent We recall that the parameter $s$ in the Hamiltonian denotes an infrared cut off. In this section 
we suppose that 
\beq \label{rirh.3}
s \leq \frac{1}{2} \delta_\at
\ene
(see (\ref{e.delta-el})).
  
\noindent We start by setting some notations.

\noindent 
We define the following subsets of the complex 
plane
\begin{align} \label{neri.0.1}
\cE^{s,t}(\theta) =  \mathcal{E}^{s}(\theta): = &  [  e_{1} - s/4 , e_{1} + s/4 ] \\ \notag & +
i [- (|\sin(\nu)|/2) s, (|\sin(\nu)|/2) s]    
- e^{- i \nu}[0, \infty) \comma
\end{align}

\beq \label{neri.1}
\cA^{s, t}(\theta): = \cA^{s}(\theta): =  \mathcal{E}^{s}(\theta) \setminus B^{\CC}_{(|\sin(\nu)|/2) s/4}( e_{1})\comma
\ene
where $  B^{\CC}_{\rho}(z) $ is the ball of radius $ \rho$ and center $z$ in the complex plane. 
The factor $|\sin(\nu)/2|$ ensures that
$e_{0}$ is not contained in  $ \mathcal{A}^{s}(\theta)  $  and that     
\beq \label{neri.1.1}
\mathcal{A}^{s}(\theta) - re^{-i \nu} \subset \mathcal{A}^{s}(\theta) \comma 
\ene
for every $r  \in \{ 0 \} \cup [s, \infty) $.

\noindent For every $z$ in the resolvent set of $ H^{s,t}_{0}(\theta, \alpha, \beta)  $ we
 denote by $ R^{s,t}_{0}( \theta, \alpha, \beta)(z) $ the resolvent of $H^{s,t}_{0}( \theta, \alpha, \beta)$ at $z$: 
\beq \label{neri.2}
R^{s,t}_{0}(\theta, \alpha, \beta)(z) := \frac{1}{  H^{s,t}_{0}(\theta, \alpha, \beta) - z  } \period
\ene
We identify (see \ref{e.p.eeh.2})
\beq \label{neri.2.1}
P_{\at}(\theta, \alpha, \beta)  \equiv P_{\at}(\theta, \alpha, \beta)\otimes \one_{\cF^{s,t}} ,\: 
\overline{P}_{\at}(\theta, \alpha, \beta)  := \overline {P}_{\at}(\theta, \alpha, \beta)\otimes \one_{\cF^{s,t}}  \period
\ene 
We will prove that the set $\cA^s(\theta)$ is contained in the resolvent set of the Hamiltonian 
$H^{s,t}(\theta, \alpha, \beta)$.  We prove furthermore that $\cE^{s}(\theta)$ contains only one point of the spectrum of 
$H^{s,t}(\theta, \alpha, \beta)$. This point is a simple eigenvalue.

\begin{lemma}\label{L.neri.1}
Suppose that $\theta =\pm i  \nu $, where  $\nu \in \RR$ is different
from zero. Suppose furthermore that it satisfies (\ref{an.0.14}) and (\ref{theta}) and that $\alpha \leq \boa$ and $|\beta| \leq \bob$. Then,  
for every $\mu \geq 0$ and every $z \in \cE^{s}(\theta) \setminus \{e_{1}\}$, 
\beq \label{L.neri.1.e0}
\|   R^{s,t}_{0}(\theta, \alpha, \beta)(z)     \| 
\leq \frac{ 50 |e_0| C_{\ref{an.0.16.2}}  }{\delta |\sin(\nu)|} 
 \frac{1}{|z - e_{1}| }  \comma 
\ene
\beq \label{L.neri.1.e1}
\| |  R^{s,t}_{0}(\theta, \alpha, \beta)(z)|  (\hf^{s,t} + \mu)   \| 
\leq \frac{50 |e_0| C_{\ref{an.0.16.2}}}{\delta|\sin(\nu)|} (1  +
 \frac{\mu}{|z - e_{1}| }  ) \comma
\ene
\beq \label{L.neri.1.e2}
\|  | R^{s,t}_{0}(\theta, \alpha, \beta)(z)|  ( H_\at \pm i \delta)   \| \leq 
\frac{  C_{\ref{L.neri.1.e3} } }{|\sin(\nu)|} ( 1 +
  \frac{1}{|z- e_{1}|}  )) \period
\ene
For any $ z \in \cE^{(1/2 \delta_\at)} $,
\beq \label{L.neri.1.e0.1}
\| \overline{P}_{\at, 1}(\theta, \alpha, \beta)  R^{s,t}_{0}(\theta, \alpha, \beta)(z)     \| 
\leq \min \Big(  \frac{ 16 C_{\ref{an.0.16.2}}  }{|\sin(\nu)|\delta} ,
\frac{ 20 |e_{0}| C_{\ref{an.0.16.2}}  }{|\sin(\nu)|\delta} \frac{1}{|z - e_{0}|} \Big)  \comma 
\ene

where 
\beq \label{L.neri.1.e3}
C_{\ref{L.neri.1.e3} } :=    2\big(1+ 8 C_{\ref{an.0.16.2}}  + \frac{8 C_{\ref{an.0.16.2}}  }{|\sin(\nu)|}(1 + \frac{|e_0'|}{\delta}) + 
 \frac{8C_{\ref{an.0.16.2}} }{| \sin{\nu} |}(\delta + |e_1'|)\big)
 \period 
\ene

\end{lemma}

\noindent In the next Lemma we prove the basic estimates on the interaction that are necessary to study the spectral properties of our Hamiltonian.

For the next Lemma recall (\ref{i.10.4}), (\ref{i.10.3}), (\ref{ilprima.7})-(\ref{ilprima.9}), and (\ref{i.10.3.0})-(\ref{i.10.3.6}). 
\begin{lemma}\label{L.neri.2}

Suppose that $ s' \leq  s < t  $ and that $\theta = \pm i \nu $, with $\nu \in \RR \setminus \{0 \}$. 
Suppose furthermore that  $ \theta  $ satisfies (\ref{an.0.14}) and (\ref{theta}) and that $\alpha \leq \boa$ and $|\beta| \leq \bob$. For every 
$z \in \mathcal{E}^{s}(\theta) \setminus \{ e_{1} \}$ there is a constant $  C_{\ref{neri.15.2}} $ such that  
\beq \label{L.neri.2.e0}\begin{array}{l}
\|  |R^{s,t}_{0}(\theta, \alpha, \beta)(z)|^{1/2} W^{s,t}_{m, n}(\theta, \alpha, \beta)   |R^{s,t}_{0}(\theta, \alpha, \beta)(z)|^{1/2}  \| \\\\
\leq   \frac{C_{\ref{neri.15.2}}}{|\sin(\nu)|^{2}}\alpha^{3(m+n)/2} (1 + \frac{1}{|z- e_{1}|})^{1/2}  \comma m + n \geq 1\comma
\end{array}
\ene
\beq \label{L.neri.2.e0.1}\begin{array}{l}
\|  |R^{s,t}_{0}(\theta, \alpha, \beta)(z)|^{1/2} W^{s,t}_{0, 0}(\theta, \alpha, \beta)   |R^{s,t}_{0}(\theta, \alpha, \beta)(z)|^{1/2}  \| \\\\
\leq   \frac{C_{\ref{neri.15.2}}}{|\sin(\nu)|}\alpha^{3} (1 + \frac{1}{|z- e_{1}|})  \comma
\end{array}
\ene
\beq \label{L.neri.2.e0.2}\begin{array}{l}
\|  |R^{s,t}_{0}(\theta, \alpha, \beta)(z)|^{1/2} W^{s,t}(\theta, \alpha, \beta)   |R^{s,t}_{0}(\theta, \alpha, \beta)(z)|^{1/2}  \| \\\\
\leq    \frac{C_{\ref{neri.15.2}}}{|\sin(\nu)|^{2}}\alpha^{3/2} (1 + \frac{1}{|z- e_{1}|})^{1/2}   +    
 \frac{C_{\ref{neri.15.2}}}{|\sin(\nu)|}\alpha^{3} (1 + \frac{1}{|z- e_{1}|})  \comma
\end{array}
\ene
For any $z \in \cE^{1/2 \delta_\at}$,
\beq \label{L.neri.2.e0.2.1}\begin{array}{l}
\|  W^{s,t}(\theta, \alpha, \beta)  
 \overline{P}_{\at}(\theta, \alpha, \beta)R^{s',t}_{0}(\theta, \alpha, \beta)(z)  \| \\\\
\leq    C_{\ref{neri.15.2.1}}\alpha^{3/2}     \period
\end{array}
\ene
The explicit values of the constants  $  C_{\ref{neri.15.2}} $ and $ C_{\ref{neri.15.2.1}} $  are given in (\ref{neri.15.2}) and ({\ref{neri.15.2.1}}). 
\end{lemma}

For any $z $ in the resolvent set of $H^{s,t}(\theta, \alpha, \beta)$, we define by
\beq \label{rrp.1}
R^{s,t}(\theta, \alpha, \beta)(z) : = \frac{1}{H^{s,t}(\theta, \alpha, \beta) - z} \period
\ene

Lemma~\ref{L.neri.2} provides the key ingredient to construct 
the resolvent of $ H^{s,t}(\theta, \alpha, \beta) $ for any $z \in \mathcal{A}^{s} (\theta)$
(see (\ref{neri.1})) in terms of a Neumann series.

\begin{theorem}\label{L.rrp.1}
Suppose that  $\theta = \pm i \nu $, with $\nu \in \RR \setminus \{0 \}$, and that  $ \theta  $ satisfies (\ref{an.0.14}) and (\ref{theta}).
Suppose furthermore that $\alpha $ is such that 
\beq \label{L.rrp.1.m10}\begin{array}{l}
 \frac{C_{\ref{neri.15.2}}}{|\sin(\nu)|^{2}}\alpha^{3/2} (1 + \frac{1}{(|\sin(\nu)|/2) s /4  })^{1/2}   +    
 \frac{C_{\ref{neri.15.2}}}{|\sin(\nu)|}\alpha^{3} (1 + \frac{1}{ (|\sin(\nu)|/2) s /4 }) \\\\
  \leq \frac{\delta|\sin(\nu )|^2}{250 |e_0| C_{\ref{an.0.16.2}}}.
 \end{array}
\ene 
and  that $\alpha \leq \boa$ and $|\beta| \leq \bob$. \\\\
Then $ \mathcal{A}^{s}(\theta)$ is contained in the resolvent set of $H^{s,t}(\theta, \alpha, \beta)$ and for every $z \in \mathcal{A}^{s}(\theta) $
\beq \label{L.rrp.1.e1}
\| R^{s,t}(\theta, \alpha, \beta)(z)  \| \leq \frac{60 |e_0| C_{\ref{an.0.16.2}}}{\delta|\sin(\nu)|}\frac{1}{|z - e_{1}|}\period
\ene
\end{theorem}

\noindent

\emph{Proof.}\\
We construct $R^{s,t}(\theta, \alpha, \beta)(z)$ by a norm-convergent  Neumann series 
\beq \label{rrp.2}
R^{s,t}(\theta, \alpha, \beta)(z) = \sum_{n= 0}^{\infty} R_{0}^{s,t}(\theta,\alpha, \beta)(z) \big[- W^{s,t}(\theta, \alpha, \beta) R_{0}^{s,t}(\theta,\alpha, \beta)(z) \big]^n
\ene
We estimate the $n^{th}$ order term using Lemma~\ref{L.neri.2} 
\beq \label{rrp.3} \begin{array}{l}
 \big\|R_{0}^{s,t}(\theta,\alpha, \beta)(z) \big[- W^{s,t}(\theta, \alpha, \beta) R_{0}^{s,t}(\theta,\alpha, \beta)(z) \big]^n \big\| \\\\ \leq 
\|   |R_{0}^{s,t}(\theta,\alpha, \beta)(z)|^{-1}  R_{0}^{s,t}(\theta,\alpha, \beta)(z)  \|^{n+1}  \cdot \|   |R_{0}^{s,t}(\theta,\alpha, \beta)(z)|   \| \\\\ \cdot  
\| |R_{0}^{s,t}(\theta,\alpha, \beta)(z)|^{1/2}  W^{s,t}_{\theta, \varsigma, \beta}  |R_{0}^{s,t}(\theta,\alpha, \beta)(z)|^{1/2}    \|^n
 \\\\
 \leq  \|   R_{0}^{s,t}(\theta,\alpha, \beta)(z)   \| \frac{(\delta|\sin(\nu)|^2)^n}{(250 |e_0| C_{\ref{an.0.16.2}})^n} \period 
\end{array}
\ene
This proves the convergence of the Neumann series and, by Lemma~\ref{L.neri.1}, the bound
in (\ref{L.rrp.1.e1}).

\QED

\noindent Once we know that  $R^{s,t}(\theta, \alpha, \beta)(z)$ exists for $ z \in \mathcal{A}^{s}(\theta) $, 
we can define the projections

\beq\label{rrp.4} \begin{array}{l}
P^{s,t}(\theta, \alpha, \beta): = \frac{i}{2 \pi} \int_{\gamma_\theta^{s,t}} R^{s,t}(\theta, \alpha, \beta)(z) dz \comma \\\\
P_{0}^{s,t}(\theta, \alpha, \beta): = \frac{i}{2 \pi} \int_{\gamma_\theta^{s,t}} R_{0}^{s,t}(\theta, \alpha, \beta)(z) dz \comma 
\end{array}
\ene 
where $ \gamma_\theta^{s,t}: [0, 2\pi) \to \CC $ is the curve given by
\beq\label{rrp.5}
\gamma_\theta^{s,t} (t): = e_{1} + (|\sin(\nu)|/2) s e^{ -i t} \period   
\ene 
Lemma~\ref{L.neri.1} and Eqs. (\ref{rrp.2}) and  (\ref{rrp.3})imply that 
\beq \label{rrp.6.0}
\|  R^{s,t}(\theta, \alpha, \beta)(z) -  R_{0}^{s,t}(\theta,\alpha, \beta)(z)  \| \leq \frac{|\sin(\nu)|}{4}   \frac{1}{|z- e_{1}|}\period
\ene
Integrating over $\gamma_\theta^{s,t}$ we obtain 
\beq \label{rrp.6}
\|  P^{s,t}(\theta, \alpha, \beta)  -  P_0^{s,t}(\theta, \alpha, \beta)  \| \leq \frac{|\sin(\nu)|}{4} \comma
\ene 
As  
\beq \label{rrp.6.2.1}
P_0^{s,t}(\theta, \alpha, \beta) = P_{\at, 1}(\theta, \alpha, \beta) \otimes \Omega^{s,t}, 
\ene
 (\ref{e.p.eeh.3}) implies 
\beq \label{rrp.6.2.2}
1 - \frac{2}{8} \leq \| P_0^{s,t}(\theta, \alpha, \beta) \| \leq 1 + \frac{2}{8}\period
\ene
Hence, using (\ref{rrp.6}) and that $|\nu| < \frac{1}{120}$, we have that
\beq \label{rrp.6.2.3}
1 - \frac{2}{8} - \frac{1}{120}   \leq  \|P^{s,t}(\theta, \alpha, \beta) \| \leq1 + \frac{2}{8} + \frac{1}{120}\period
\ene
For $\beta = 0$,  $P_{\at}(0, \alpha, 0)$ is an orthogonal projection and its norm is, therefore, $1$. Using 
(\ref{e.p.eeh.3}) we can conclude that 
\beq \label{in.10.1.gtilde}
1 - \frac{1}{8} - \frac{1}{120}   \leq  \|P^{s,t}(\theta, \alpha, 0) \| \leq1 + \frac{1}{8} + \frac{1}{120}\period
\ene

\begin{remark}\label{R.neri.1}
By Theorem~\ref{p.eeh.1} and (\ref{rrp.6.2.1}),  $e_{1}$ is the only 
point in the spectrum of $  H_0^{s,t}(\theta, \alpha, \beta)  $ that belongs to  $ \mathcal{E}^{s}(\theta) $ 
(see (\ref{neri.0.1})) and it is a simple
eigenvalue. \\  
It follows from (\ref{rrp.6}) that the range of the
projection $ P^{s,t}(\theta, \alpha, \beta)  $ is one-dimensional. Thus, by  (\ref{L.rrp.1.e1}), there exists 
a unique point in the spectrum of  $  H^{s,t}(\theta, \alpha, \beta)  $ that belongs to  $ \mathcal{E}^{s}(\theta) $ 
and it is a simple eigenvalue, we denote it by $E^{s,t}(\theta)$. Clearly, by Theorem~\ref{L.rrp.1}, $E^{s,t}(\theta)$ is 
contained in the disk $B^\CC_{(|\sin(\nu)|/2) s /4}(e_{1})$.
\end{remark}

\begin{corollary}\label{C.rrp.1}
Suppose that  $\theta = \pm i \nu $, with $\nu \in \RR \setminus \{0 \}$, and that  $ \theta  $ satisfies (\ref{an.0.14}) and (\ref{theta}).
Suppose furthermore that $\alpha $ satisfies (\ref{L.rrp.1.m10}) and that $\alpha \leq \boa$ and $|\beta| \leq \bob$.
Then for every $ z \in  \mathcal{E}^{s}(\theta) $, 
\beq \label{c.rrp.1.e1}
\| R^{s,t}(\theta, \alpha, \beta)(z) \overline{P}^{s,t}(\theta, \alpha, \beta)  \| \leq  \frac{100 |e_0| C_{\ref{an.0.16.2}}}{\delta |\sin(\nu)|} 
\frac{1}{(|\sin(\nu)|/2) s}\comma
\ene
where $  \overline{P}^{s,t}(\theta, \alpha, \beta):= 1 - P^{s,t}(\theta, \alpha, \beta)  $. 

\end{corollary}

\noindent

\emph{Proof.}\\
Let $\psi, \tilde{\psi}$ belong to $\cH^{s,t}$. The function 
$$
f(z) : = \la \psi | \;  R^{s,t}(\theta, \alpha, \beta)(z) \overline{P}^{s,t}(\theta, \alpha, \beta)   \tilde{\psi}  \ra
$$
is analytic in $\cE^{s}(\theta)$.
Hence, by the maximum modulus principle and Theorem~\ref{L.rrp.1}, we have that 
$$
\max_{z \in B_{s|\sin(\nu)|/2}(0)} |f(z)| \leq \frac{60 |e_0| C_{\ref{an.0.16.2}}}{\delta |\sin(\nu)|}\frac{1}{(|\sin(\nu)|/2) s} \|\psi \|_{\cH^{s,t}} 
\| \overline{P}^{s,t}(\theta, \alpha, \beta)   \tilde{\psi}\|_{\cH^{s,t}}, 
$$
which proves (\ref{c.rrp.1.e1})  for  $  z \in B^\CC_{s|\sin(\nu)|/2}(0)  $ (we use (\ref{rrp.6.2.3})). The rest follows from Theorem~\ref{L.rrp.1}.
\QED

In Remark \ref{R.neri.1} we establish the existence of an eigenvalue $E^{s,t}(\theta)$ of the infrared-regularized Hamiltonian 
$H^{s,t}(\theta, \alpha, 0)$. We claim that this eigenvalue is a resonance, i.e., that its imaginary part is strictly negative. This is the content of the next Theorem which is proven in Section~\ref{S.rrp.p}.       

\begin{theorem} \label{T.rrp}
Suppose that  $\theta = \pm i \nu $, with $\nu \in \RR \setminus \{0 \}$, and that  $ \theta  $ satisfies (\ref{an.0.14}) and (\ref{theta}) and that 
$\beta =0$.
Suppose furthermore that $\alpha \leq \boa $ satisfies (\ref{L.rrp.1.m10}) and that  $ (|\sin(\nu)|/2) s \leq 1 $ and $ s = \alpha^{\upsilon}$ for some 
$\upsilon \in (0, 2)$. 

Let (see (\ref{i.10.3.0}))
\begin{align} \label{imaginary-tilde}
\tilde E_I : = & - \pi \int_{\mathbb{S}^2}  dS \| P_{\at, 0}(0, \alpha, 0)|e_1 - e_0| \\ \notag & \cdot  w_{1,0}(0, \alpha, 0)(x, \frac{k}{|k|}|e_1 - e_0|)
 \psi_0  \|^2, 
\end{align} 
where $\mathbb{S}^2$ is the sphere and $\psi_0$  is a unit eigenvector of $ P_{\at, 1}(0, \alpha, 0)  $. 

There is a constant $C_{\ref{corr.16}}$ such that 
\begin{align} \label{imaginary-basis}
|\tilde E_I - { \, Im} (E^{s,t}(\theta)) | \leq & 4 C_{\ref{corr.16}} \alpha^3 
 (\alpha^\upsilon |\log(\alpha^\upsilon)|  \\ \notag & + \alpha^\upsilon + \alpha^{2\upsilon} +  \alpha^{(3-\upsilon)/2} (1 + \alpha^{(3 - \upsilon)/2})^3)
\end{align}

The explicit value of the constant $C_{\ref{corr.16}}$ is written in (\ref{corr.16}).  

\end{theorem}

We conclude this section by collecting the main results. They establish the existence of resonances for the infrared-regularized Hamiltonian as well as some estimates on the resolvent. In the next section we remove the infrared cut off. We do it step by step using an inductive argument. The induction basis is the content of the next theorem, where we make use of the previous results.       
  
\begin{theorem} \label{princ-secc}
Suppose that  $\theta = \pm i \nu $, with $\nu \in \RR \setminus \{0 \}$, and that  $ \theta  $ satisfies (\ref{an.0.14}) and (\ref{theta}).
Suppose furthermore that $\alpha $ satisfies (\ref{L.rrp.1.m10}) that s satisfies (\ref{rirh.3}) and  that $\alpha \leq \boa$ and $|\beta| \leq \bob$.\\ We assume also that 
 $ (|\sin(\nu)|/2) s \leq 1 $ and $ s = \alpha^{\upsilon}$ for some 
$\upsilon \in (0, 2)$. Then there is only one point ($E^{s,t}(\theta)$) in the spectrum of $H^{s,t}(\theta)(\theta, \alpha, \beta)$ included in the set $\cE^{s}(\theta)$ (see (\ref{neri.0.1})). It is a simple eigenvalue and for $\beta = 0$ its imaginary part is estimated by (\ref{imaginary-basis}). The distance of $ E^{s,t}(\theta) $ to $e_1$ is estimated by 
\beq \label{E-e1}
|E^{s,t}(\theta) - e_1| \leq \frac{ (|\sin(\nu)|/2) s}{4},
\ene
see (\ref{neri.1}).\\
The Riesz projection (see (\ref{rrp.4})) of the Hamiltonian corresponding to the eigenvalue $E^{s,t}(\theta)$ satisfies the Bounds (\ref{rrp.6}) 
(\ref{rrp.6.2.3}) and (\ref{in.10.1.gtilde}).
 
For any $z \in \cE^s(\theta)$ the resolvent operator (\ref{rrp.1}) satisfies the following bound:

\beq \label{bound-basis-resolvent}
\| R^{s,t}(\theta, \alpha, 0)(z) \overline{P}^{s,t}(\theta, \alpha, 0)  \| \leq  \frac{1000 |e_0| C_{\ref{an.0.16.2}}}{\delta |\sin(\nu)|} 
\frac{1}{(|\sin(\nu)|/2) s + |z - E^{s,t}(\theta)|}\comma
\ene
where $  \overline{P}^{s,t}(\theta, \alpha, 0):= 1 - P^{s,t}(\theta, \alpha, 0)  $.

\end{theorem}
\noindent {\it Proof:}
The only results not proven yet are  Eqs. (\ref{E-e1}) and (\ref{bound-basis-resolvent}). Eq. (\ref{E-e1}) follows from (\ref{neri.1}) and (\ref{L.rrp.1.e1}). 
Eq. (\ref{bound-basis-resolvent}) follows from Theorem~\ref{L.rrp.1}, (\ref{in.10.1.gtilde}) and (\ref{E-e1})  for $|z - e_1| \geq (|\sin(\nu)|/2) s$, and from Corollary~\ref{C.rrp.1} and (\ref{E-e1})  for $|z - e_1| \leq (|\sin(\nu)|/2) s$. 
\QED

\secct{Resonances for the Non-Regularized Hamiltonian: The Infrared Limit} \label{infrared-limit}

This is the main section of the paper. Here we prove the existence of resonances (see Section~\ref{existence-resonant-eigenvalue}), we provide an explicit formula for the imaginary part of the resonant eigenvalue 
up to order $\alpha^3$ (see Section~\ref{approximations-imaginary}). We prove also that the resonant eigenvalue is non-degenerate (see Section~\ref{non-degeneracy}), we give estimates for the resolvent operator and the resolvent set (see Section~\ref{estimates-resolvent}) and prove exponential decay of eigenfunctions (See Theorem~\ref{T.il.fm.1}). 
  
Most of the conclusive results of this section are stated and proved in Sections~\ref{existence-resonant-eigenvalue}-\ref{estimates-resolvent}, but the core of our proofs relies on the results of Section~\ref{induction-step}, which is the main technical ingredient of our paper.     
  
To accomplish our proofs we use an inductive argument. The induction basis is Theorem~\ref{princ-secc}. We take the infrared cut off to zero step by step inductively (Section~\ref{induction-step}). 

\subsection{The Feshbach Map}\label{feshbach-infrared}

In all statements of this section we suppose that $\alpha \leq \boa$ and $|\beta| \leq \bob$.

We present here an appropriate Feshbach map that is a useful tool for our proofs. The proof that it is well defined as well as some basic estimates are done in a similar way as in Section~\ref{S.rrp.p}, where complete detailed proofs are included. Some of the proofs of the basic results are based on the proofs of Section~\ref{S.rrp.p}. In these cases we refer to Section~\ref{S.rrp.p} to follow the argument.            
We fix some parameters $s',\, s$ and $t$, with $s' \leq s < t \leq \infty$. 
We define the following projection on $ \mathcal{H}^{s,t}$ (see (\ref{e.p.eeh.2})) 
\beq \label{il.fm.1} \begin{array}{l}
\cP  : = \cP_\beta : = \cP(\theta) : = P_{\at, 1}(\theta, \alpha, \beta) \otimes 1_{\cF^{s,t}} \comma \\\\
 \bcP: = \bcP_\beta 
: = \bcP(\theta) : = 1 -  \cP (\theta)  \period
\end{array}
\ene
We denote by 
\beq \label{il.fm.1.1}
\dot{H} := \dot{H}_\beta  : =  H^{s,t}(\theta)\otimes 1_{\cF^{s', s}}  +   1_{ \cF^{s,t} }  \otimes \hf^{s', s}
\ene
and by 
\beq \label{il.fm.2.0.0} 
\dot{H}_{\bcP}: =  \dot{H}_{\bcP, \beta}   : = \bcP \dot{H} \bcP \period
\ene
We define (formally) the Feshbach map corresponding to the projection (\ref{il.fm.1})
\beq \label{il.fm.2} \begin{array}{l}
\cF_\cP: = \cF_{\cP, \beta} := \cF_{\cP}(\dot{H} - z)  : = \cP (\dot{H} - z) \cP \\\\ -  
\cP \dot{H}
\bcP  (\dot{H}_{\bcP} - z)^{- 1} \bcP \dot{H} \cP \period
\end{array}
\ene
If $\alpha$ is sufficiently small, $ \cF_\cP  $ defines a closed operator. Its importance lies 
in the facts that  $ z \in \sigma(H^{s,t}(\theta) ) $ if and only if 
$   0 \in \sigma (\cF_\cP)  $ and that there are explicit formulae for the resolvent of each one of those operators 
in terms of the other as well as for the eigenvectors. We will state precisely and  prove this facts in the remaining of this sub-section.

\begin{lemma}\label{L.il.fm.1}
Suppose that  $\theta = \pm i \nu $, with $\nu \in \RR \setminus \{0 \}$, and that  $ \theta  $ satisfies (\ref{an.0.14}) and (\ref{theta}).
Suppose furthermore that $   \alpha^{3/2}   \leq \frac{1}{2 C_{\ref{neri.15.2.1}}}  $. 
Then $ \mathcal{E}^{\delta_\at /2}(\theta)$ is contained in the resolvent set of $\dot{H}_\bcP$ and  for every 
$\mu \in   \mathcal{E}^{\delta_\at /2}(\theta)$,
\beq \label{L.il.fm.1.e1}
  \|  (\dot{H}_{\bcP} - \mu)^{-1} \bcP \|  \leq \min(  \frac{32 C_{\ref{an.0.16.2}}}{|\sin(\nu)| \delta}, 
\frac{40 |e_{0}| C_{\ref{an.0.16.2}}}{|\sin(\nu)| \delta} \frac{1}{| z - e_{0} |})
 \period
\ene

\end{lemma}

\emph{Proof:}\\
The proof is similar to the one of Lemma~\ref{L.fm.1}, here we use (\ref{L.neri.1.e0.1}) and (\ref{L.neri.2.e0.2.1}).  

\QED

\begin{lemma}\label{L.il.fm.2}
Suppose that  $\theta = \pm i \nu $, with $\nu \in \RR \setminus \{0 \}$, and that  $ \theta  $ satisfies (\ref{an.0.14}) and (\ref{theta}). Suppose furthermore that $   \alpha^{3/2}   \leq \frac{1}{2 C_{\ref{neri.15.2.1}}}  $. 
 Then for every $z \in \cE^{\delta_\at /2} (\theta)$, 
\beq \label{L.il.fm.1.e2} \begin{array}{l}
\|    \bcP (\dot{H}_{\bcP} - z)^{- 1} \bcP \dot{H} \cP \| <  
 4 \alpha^{3/2}  C_{\ref{neri.15.2.1}}     \comma \\\\ 
\| \cP  \dot{H} \bcP (\dot{H}_{\bcP} - z)^{- 1} \bcP    \| < 4 \alpha^{3/2}  C_{\ref{neri.15.2.1}} \period
 \comma
\end{array} \ene
\end{lemma}

\emph{Proof:}\\ The proof of (\ref{L.il.fm.1.e2}) is similar to the one of (\ref{L.il.fm.1.e1}), here we use
Lemma     \ref{L.neri.2}   instead of Lemma~\ref{L.neri.1}.

\QED

\noindent Using  (\ref{e.p.eeh.3}), (\ref{L.i.1.e1}) and (\ref{L.il.fm.1.e2}) we obtain for $\rho =   1$ 
\beq \label{il.fm.3} \begin{array}{l}
\| \big(  \cP W^{s,t}(\theta, \alpha, \beta) \cP -  
\cP  W^{s,t}(\theta, \alpha, \beta) 
\bcP  (\dot{H}_{\bcP} - z)^{- 1}   W^{s,t}(\theta, \alpha, \beta)  \cP\big) \phi \|_{\cH^{s',t}} \\\\
\leq ( 2 + 4 \alpha^{3/2} C_{\ref{neri.15.2.1}})6 C_{\ref{i.13.4}} \alpha^{3/2} \\\\ \cdot  \big( \| \cP   (H^{s',t}_{0}(\theta, \alpha, \beta)  + 1)
 \cP \phi \|_{\cH^{s',t}} + \| \cP  \phi  \|_{\cH^{s',t}} \big) \period
\end{array}
\ene
As $ \cP   (H^{s',t}_{0}(\theta, \alpha, \beta)  + 1)  \cP $ is closed,
 we conclude (see Theorem 1.1 page 190 of \cite{kato1976}) that  $ \cF_\cP  $ is closed for any $\alpha $ such that 
$  ( 2 + 4 \alpha^{3/2} C_{\ref{neri.15.2.1}})6 C_{\ref{i.13.4}} \alpha^{3/2} < 1  $. Actually the domain of $ \cF_\cP $ is 
the same as the domain of $  \cP   H^{s',t}_{0}(\theta, \alpha, \beta)  \cP   $.   

\begin{lemma}\label{L.il.fm.2.1}
Suppose that  $\theta = \pm i \nu $, with $\nu \in \RR \setminus \{0 \}$, and that  $ \theta  $ satisfies (\ref{an.0.14}) and (\ref{theta}). 
Suppose furthermore that $   \alpha^{3/2}   \leq \frac{1}{2 C_{\ref{neri.15.2.1}}}  $. 
 Then for every $z \in \cE^{\delta_\at /2} (\theta)$ and every $\rho \in \CC$, 
\beq \label{L.il.fm.2.1.e1} \begin{array}{l}
\|  ( H^{s',t}_{0}(\theta, \alpha, \beta) + \rho )  \bcP (\dot{H}_{\bcP} - z)^{- 1} \bcP    \| \\\\
 < 6 \min \Big( ( 1  + |\rho + z| \frac{32 C_{\ref{an.0.16.2}}}{|\sin(\nu)| \delta} )\\\\ , 
 1 +  ( |\rho |  + | e_0|)  \frac{32 C_{\ref{an.0.16.2}}}{|\sin(\nu)| \delta}  +
   \frac{40  |e_0|C_{\ref{an.0.16.2}}}{|\sin(\nu)| \delta}  \Big)   \period 
\end{array} \ene

\end{lemma}
\noindent \emph{Proof:} 
The result follows from Eq. (\ref{L.neri.2.e0.2.1}), the Neumann expansion (\ref{re.5}) with $\cP$ instead of $\bp$ and (\ref{L.il.fm.1.e1}).

\QED

\begin{lemma}\label{L.il.fm.3}
Suppose that  $\theta = \pm i \nu $, with $\nu \in \RR \setminus \{0 \}$, and that  $ \theta  $ satisfies (\ref{an.0.14}) and (\ref{theta}). 
Suppose furthermore that $   \alpha^{3/2}   \leq \frac{1}{2 C_{\ref{neri.15.2.1}}}  $ and  that
$  ( 2 + 4 \alpha^{3/2} C_{\ref{neri.15.2.1}})6 C_{\ref{i.13.4}} \alpha^{3/2} < 1  $. Then  for any 
$z \in  \cE^{\delta_\at /2} (\theta)$, the operator $\cF_\cP (\dot{H}  - z )$ (see \ref{il.fm.2}) is closed and the 
following holds true:   
\begin{itemize}
\item[(i)]
 $\cF_\cP ( \dot{H}  - z )$ is invertible on $ \cP\cH^{s',t} $ if and only if 
$ H^{s,t}(\theta)  - z   $ is invertible on $\cH^{s',t}$ and the following formulas hold 
\beq \label{L.il.fm.3.e1}
(\cF_\cP)^{-1} : = \cP (   \dot{H} - z   )^{-1}\cP
\ene

\beq \label{L.il.fm.3.e2}\begin{array}{l}
(   \dot{H} - z)^{-1} =  \big[  \cP - \bcP ( \dot{H}_{\bcP} - z)^{-1} W^{s,t}(\theta, \alpha, \beta) \cP  \big] 
 \\\\ \cdot \cF_\cP^{-1} \big[ \cP - \cP W^{s,t}(\theta, \alpha, \beta) \bcP   (  \dot{H}_{\bcP} - z)^{-1}   \big] 
+ \bcP   ( \dot{H}_{\bcP} - z)^{-1}  \bcP \comma
\end{array}
\ene

\item[(ii)] If $ \dot{H} \psi = z \psi  $ for some eigenvector $ 0 \neq \psi \in\cH^{s',t} $ and z $\in \cE^{1/2\delta_\at} (\theta)$,
 then $ 0 \neq  \cP\psi   \in \cP  \cH^{s',t}  $ solves $\cF_\cP  \cP \psi = 0$ and 
\beq \label{L.il.fm.3.e2.1}
\|  \cP\psi  \| \geq \frac{\| \psi \|}{1 +  \|    \bcP (\dot{H}_{\bcP} - z)^{- 1} \bcP \dot{H} \cP \| }\period
\ene 
\item[(iii)]
If $  \cF_\cP \phi = 0 $ for some eigenvector $ 0 \neq \phi  = \cP \phi \in \cP \cH^{s',t} $, then the vector $ 0 \neq  \psi \in \cH^{s',t} $, 
defined by $\psi : =   [\cP -  (\dot{H}_\bcP - z )^{-1}  \bcP W^{s,t}(\theta, \alpha, \beta) \cP ]\phi $, 
solves $   \dot{H}\psi = z \psi   $. 

\item[(iv)]
\beq \label{ker}
\dim \ker  (  \dot{H} - z  ) = \dim \ker \cF_\cP \period
\ene

\end{itemize}

\end{lemma}

\noindent{\it Proof:} 
See the proof of Theorem II.1 \cite{BachFroehlichSigal1998b}. 

\QED

\begin{theorem}[Exponential Boundedness of the Eigenvalues] \label{T.il.fm.1}
Suppose that  $\theta = \pm i \nu $, with $\nu \in \RR \setminus \{0 \}$, and that  $ \theta  $ satisfies (\ref{an.0.14}) and (\ref{theta}).
Suppose furthermore that $   \alpha^{3/2}   \leq \frac{1}{16 C_{\ref{neri.15.2.1}}}  $,
 $  ( 2 + 4 \alpha^{3/2} C_{\ref{neri.15.2.1}})6 C_{\ref{i.13.4}} \alpha^{3/2} < 1  $, that $\alpha \leq \boa$ and $ \beta = 0$ (see \ref{geh.6.1}). If  $z \in \cE^{\delta_\at /2} (\theta)$
is an eigenvalue of  $\dot{H}_0$ and $\psi \ne 0 $ a corresponding eigenfunction, then 

\beq \label{T.il.fm.1.e1}
\frac{\|   e^{\beta \la   x  \ra } \psi  \|}{  \| \psi  \| }  \leq 12    C_{\ref{T.eb.1.e2}} \period
\ene
The constant $  C_{\ref{T.eb.1.e2}} $ defined in (\ref{T.eb.1.e2}) does not depend on $\alpha$, 
$\beta$  and $\theta$, nor on $s'$ and $s$ and $t$ (see the beginning of this section). 

\end{theorem}

\noindent{\it Proof:} 

By Lemma~\ref{L.il.fm.3}, the eigenvectors on the kernel of  $ \dot{H}_0  - z $  are of the form 
 $\psi  =   [\cP_0 -  (\dot{H}_{\bcP_0, 0} - z )^{-1}  \bcP_0 W^{s,t}(\theta, \alpha, 0) \cP_0 ]\phi $, where 
$\cF_{\cP_0} \phi = 0$ (see (\ref{il.fm.1})-(\ref{il.fm.2})). We have that 
\beq \label{eb-qed.1} \begin{array}{l}
\| e^{\beta \la   x  \ra } \psi  \|  = \|  e^{\beta \la   x  \ra }   [\cP_0 -  (\dot{H}_{\bcP_0, 0} - z )^{-1}  \bcP_0 W^{s,t}(\theta, \alpha, 0) \cP_0 ]
  e^{ - \beta \la   x  \ra } e^{ \beta  \la   x  \ra }  \phi \| \\\\
 =  \| [\cP_{-\beta} -  (\dot{H}_{\bcP_{-\beta}, -\beta} - z )^{-1}  \bcP_{-\beta} W^{s,t}(\theta, \alpha, -\beta) \cP_{-\beta} ]  e^{ \beta \la   x  \ra } \cP_0 \phi \| \\\\
\leq \|   [\cP_\beta -  (\dot{H}_{\bcP_{-\beta}, -\beta} - z )^{-1}  \bcP_{-\beta} W^{s,t}(\theta, \alpha, -\beta) \cP_{-\beta} ]   \| \cdot \| e^{\beta \la   x  \ra } \cP_0\|\cdot 
\|  \phi  \| \\\\  \leq 3  C_{\ref{T.eb.1.e2}}\|  \phi  \| 
\end{array}
\ene     
where we used (\ref{e.p.eeh.3}), (\ref{T.eb.1.e3}) and (\ref{L.il.fm.1.e2}). Notice that $\phi \in \cP_0 \cH^{s',t}$ implies that 
$\phi = \cP_0  \phi$. By 
(\ref{L.il.fm.1.e2}) we have that 
\beq \label{eb-qed.2}
\|  \psi \| \geq ( 1 - \frac{1}{4})\| \phi \| \geq \frac{1}{4}\| \phi \|\period
\ene
(\ref{T.il.fm.1.e1}) is a direct consequence of (\ref{eb-qed.1}) and (\ref{eb-qed.2}).

\QED

\subsection{The Sequence of Infrared-Regularized Hamiltonians (Notation, Definitions and Assumptions)} \label{il.notation}

\begin{assumptions}\label{parameters-basis}
In this section we suppose that the parameters $\theta$, $\alpha $ and $\beta$ satisfy the following: 
\begin{itemize}
\item   $\theta = \pm i \nu $, with $\nu \in \RR \setminus \{0 \}$ and   $ \theta  $ satisfies (\ref{an.0.14}) and (\ref{theta}).
\item  $\alpha \leq \boa$ (see \ref{geh.6.1}) and it satisfies (\ref{L.rrp.1.m10}) with $\sigma_0$ instead of s (see $\sigma_0$ below), \\
\begin{equation*} \begin{array}{l}
\alpha^{3/2}   \leq  \frac{1}{16 C_{\ref{neri.15.2.1}}} + \frac{1}{128 C_{\ref{i.17}}} +  \frac{1}{64(1 + 8 C_{\ref{an.0.16.2}} b_{1/4} )C_{\ref{R.a.2.1}}}
\\\\ +  \frac{1}{24 C_{\ref{i.13.4}} ( 1 +   8 C_{\ref{an.0.16.2}}  )}  +  \frac{1}{( 2 + 4 C_{\ref{neri.15.2.1}})6 C_{\ref{i.13.4}}  }. 
\end{array}
\end{equation*}
\item  $|\beta| \leq \bob$ (see \ref{geh.6.1}).
\end{itemize}

\end{assumptions}

Next we define an important constant that in some sense includes all the constants appearing in the error bounds in our estimations. This constant plays a fundamental role in our inductive procedure.  It depends only on the parameters of the physical atom Hamiltonian (without field) and $\theta$. It does not depend on $\alpha$ and $\beta$ satisfying Assumptions~\ref{parameters-basis}.      

\begin{definition}
We denote by $ \boC_{\ref{il.2.1}} $ the following constant 
\beq \label{il.2.1} \begin{array}{l}
\boC_{\ref{il.2.1}}  : =  \frac{1000 |e_0'| C_{\ref{an.0.16.2}}}{\delta |\sin(\nu)|} + C_{\ref{P.ei.1.e1}} +
C_{\ref{T.ei.1.e2}}  + C_{\ref{T.ee.1.e2}} \\\\ + C_{\ref{T.I.S.2}}  + C_{\ref{acs.1}} + C_{\ref{L.ei.10.e2}} + 
C_{\ref{P.ei.1.e2}}+ 1 \period 
\end{array}
\ene
Explicit values of the constants in the right hand side of Eq. (\ref{il.2.1}) are written in  (\ref{an.0.16.2}) and (\ref{P.ei.1.e1}), (\ref{T.ei.1.e2}),  (\ref{T.ee.1.e2}), 
(\ref{T.I.S.2}) and (\ref{acs.1}) below. They depend only on the physical atom Hamiltonian (with $\alpha = \beta = \theta = 0  $) and $\theta$, for 
$\alpha$ and $\beta$ satisfying Assumptions~\ref{parameters-basis}.  
\end{definition}

In the following definition we introduce a sequence of numbers that represent the infrared infrared-cut off. Taking the sequence index to infinity 
corresponds to removing the cut off in our induction scheme
\begin{definition} \label{sigmas-sequence}
We fix two constants ($\sigma_0$, $\mathcal{B}$) satisfying the following properties: 
\beq \label{in.10.1.3}\begin{array}{l}
\boC_{\ref{il.2.1}}^4 \sigma_0^{1/2} \leq \frac{1}{100},\: \: \boC_{\ref{il.2.1}}^2 \mathcal{B}^{1/2} \leq \frac{1}{100}, \: 
  \frac{\alpha^{3/2}}{\mathcal{B}} \leq 1 \comma \\\\
  \boC_{\ref{il.2.1}}^2 \sigma_0 \leq \frac{1}{10} (|\sin(\nu)|/2),  \: \: \sigma_0 = \alpha^{\upsilon}, \: \sigma_0 \leq \frac{1}{2}\delta_\at \comma
\end{array}
\ene
for some $\upsilon \in (0, 2) $.  
Notice that $\sigma_0 < 1$, $\mathcal{B} <1$ are consequences of (\ref{in.10.1.3}).  \\\\
We define a decreasing sequence $(\sigma_n)_{n = 0}^{\infty}$ by setting 
\beq \label{il.3}
\sigma_n : = \cB^n \sigma_0,  \; \; \; \; \; \sigma_\infty = 0.
\ene 
\end{definition}

\subsubsection{Scale of Hamiltonians}\label{scale-hamiltonians}

In this section we take $\beta = 0$ whenever we do not write $\beta$ explicitly.

For any $ \sigma_m < \sigma_n $, the disjoint union 
\beq \label{il.6}
\cK^{\sigma_m} = \cK^{\sigma_m, \sigma_n}\cup \cK^{\sigma_n} \period
\ene
Gives rise to the isomorphisms
\beq \label{il.7}
\cF^{\sigma_m} \cong   \cF^{\sigma_n}\otimes  \cF^{\sigma_m, \sigma_n} \comma
\ene 
\beq \label{il.8}
\cH^{\sigma_m} \cong   \cH^{\sigma_n}\otimes  \cF^{\sigma_m, \sigma_n} \period 
\ene 
We simplify our notation using 
\beq \label{ei.1} \begin{array}{l}
\overset{n}{H}: = \overset{n}{H}(\theta) := H^{\sigma_n}(\theta, \alpha,0) , \; \;  \overset{n}{R}(z) = \frac{1}{ \overset{n}{H} - z } \comma \\\\
\overset{n}{H}_0 := \overset{n}{H}_0(\theta) := H_0^{\sigma_n}(\theta,\alpha, 0) , \; \; \;  \overset{n}{R}_0(z): = \frac{1}{ \overset{n}{H}_0 - z }\comma\\\\
\end{array}
\ene
To compare the Hamiltonians $\overset{n}{H}$ and $\overset{n + 1}{H}$ at successive energy scales 
we introduce the Hamiltonians, 
\beq \label{il.8.1} \begin{array}{l}
\tH^{n} : = \overset{n}{H} \otimes 1_{\cF^{\sigma_{n + 1}, \sigma_n}} + e^{- \theta} 1_{\cH^{\sigma_n}} \otimes \hf^{\sigma_{n + 1}, \sigma_n}, \; \;  \tR^n(z) = \frac{1}{ \tH^n - z }  \comma \\\\
\tH_\infty^{n} : = \overset{n}{H} \otimes 1_{\cF^{\sigma_{\infty}, \sigma_n}} + e^{- \theta} 1_{\cH^{\sigma_n}} \otimes \hf^{\sigma_\infty, \sigma_n}, \; \;  \tR_\infty^n(z) = \frac{1}{ \tH_\infty^n - z } 
\period 
\end{array}
\ene
We introduce the velocity operator acting on  $\cH^{\sigma_n}$ by the formula 
\beq \label{il.1}
\vv^{\sigma_n}(\theta) : = \vv^{\sigma_n} : =  - i e^{- \theta} \vnabla  - A^{\sigma_n}(\theta),
\ene 
see (\ref{eph.2}), and the operator(we omit the variable $t$ if it is infinity) 
Then we have (see (\ref{eph})), 
\beq \label{il.2}
\overset{n}{H} = (\vv^{\sigma_n})^2 + b^{\sigma_n}   + e^{- \theta} \hf^{\sigma_n}  + V(\theta,\alpha, 0).  
\ene
We identify $\vv(\theta)^{\sigma_n}$ with $ \vv(\theta)^{\sigma_n} \otimes 1_{\cF^{\sigma_m, \sigma_n}}  $, which is 
acting on $\cH^{\sigma_m}$ for any $ m \geq n + 1  $. Then we have, 
\beq \label{il.9}
\vv(\theta)^{\sigma_m} = \vv(\theta)^{\sigma_n} - A^{\sigma_m, \sigma_n}(\theta) \period
\ene
We define the operator (for $m > n$)
\beq \label{il.11} \begin{array}{l}
W_m^n(\theta) : = W_m^n: =  (\vv(\theta)^{\sigma_m})^2 -  (\vv(\theta)^{\sigma_n})^2 +  b^{\sigma_m}(\theta) -   b^{\sigma_n}(\theta) \\\\
= - A^{\sigma_m, \sigma_n}(\theta) \cdot \vv(\theta)^{\sigma_n}  -   \vv(\theta)^{\sigma_n} \cdot  A(\theta)^{\sigma_m, \sigma_n}     +  (A(\theta)^{\sigma_m, \sigma_n})^2  
\\\\ +  b^{\sigma_m,\sigma_n}(\theta)  \period
\end{array}
\ene
With this notation we have that, 
\beq \label{il.13}
\overset{n + 1}{H} = \tH^{n} + W_{n+1}^n \period 
\ene
Now we define 
\beq \label{il.14} \begin{array}{l}
 \boG^{s,t}_\theta  :=  \boG^{s,t}_\theta (x) : =  (\boG^{s,t}_{\theta, 1},  \boG^{s,t}_{\theta, 2},   \boG^{s,t}_{\theta, 3} ) : =  
  (\boG^{s,t}_{\theta, 1}(x),  \boG^{s,t}_{\theta, 2}(x),   \boG^{s,t}_{\theta, 3}(x) )\\\\ : = G^{s,t}(\theta) (x, \cdot)  - e^{- \theta }\vnabla  \boQ^{s,t}(x, \cdot) \comma
\end{array}
\ene
where the central dot ($\cdot$) represents the variable $k$ and (see also \ref{pft.1.1})
\beq \label{pft.1.1prima} \begin{array}{l}
 \boQ^{s,t}(\theta)(x, k)  :=  G^{s,t}(\theta)(0, k) \cdot \Big( \eta\Big( |x||k|  \Big)  e^{\theta} x \Big)  \period
\end{array}
\ene
Then we have that 
\beq \label{il.15}
A^{s,t}(\theta) = a^*( \boG^{s,t}_\theta  ) + a( \boG^{s,t}_{\overline{\theta}}  )\period
\ene

As the variable $\theta$ is fixed, we will omit it in our notations unless we need to write it explicitly.  

\subsection{Inductive Scheme} \label{induction-hypothesis}

We remember that the energy $E^{s,t}(\theta)$ is defined in Remark \ref{R.neri.1} and the sets
$\cE^{s,t}(\theta)$ are defined in (\ref{neri.0.1}). For every $n \in \NN \cup \{ 0 \} \cup \{ \infty \}$,
  $\sigma_n$ is defined in (\ref{il.3}).

Our selection of  $\alpha$ and $\theta$ assure that there exist only one point on the spectrum of  $ \overset{0}{H}$ in the set 
$\cE^{\sigma_0}(\theta)$ and that it is a simple eigenvalue (see Remark \ref{R.neri.1}). This is  possible only for $\sigma_0$ because
the selection of the alpha that fulfills the hypothesis of Lemma~\ref{L.rrp.1} goes to zero as $s $ goes to 
infinity (see (\ref{L.rrp.1.m10})), thus only $\alpha = 0$ would assure the existence of a unique simple eigenvalue of  $ \overset{n}{H}$ 
in $\cE^{\sigma_n}(\theta)$. \\

 We wish to find simple eigenvalues of $\overset{n}{H}$ for every  $n \in \mathbb{N}$ and 
then take the limit when $n$ goes to infinity in order to find an eigenvalue for $ \overset{\infty}{H} $. \\

We find the eigenvalues recursively and prove by induction that they satisfy certain properties. \\

We construct inductively (and simultaneously) a sequence of numbers $ \{ E_j \}_{j \in \mathbb{N} \cup \{ 0 \}} $, a sequence of subsets of the
complex plane  $ \{ \cE_j \}_{j \in \mathbb{N} \cup \{ 0 \}} $ and a sequence of operators $ \{ P_j \}_{j \in \mathbb{N} \cup \{ 0 \}}   $
that satisfy certain properties that we specify below. 

\subsubsection{Induction Basis}\label{induction-basis}

Here we define the number $ E_0 $, the set $\cE_0 $ and the operator
$P_0$. They satisfy some properties that we list below (see Theorem~\ref{princ-secc}).  

\begin{itemize}
\item[1.]
\beq \label{in.1} 
E_0 = E^{\sigma_0}(\theta) \period
\ene
$E_0$ is a simple eigenvalue of $ \overset{0}{H}$.
Remember that the parameter $t$ does not appear if it is infinity. 

\item[2.]
\beq \label{in.2}
 \cE_0  := \begin{cases} \cE^{\sigma_0}(\theta)
 \setminus (E_0 + \RR - i ((|\sin(\nu)|/2) \sigma_0/2, \infty )), & {\rm if} \; {\rm Im}(\theta) > 0 \comma \\
\cE^{\sigma_0}(\theta) \setminus (E_0 + \RR + i ((|\sin(\nu)|/2) \sigma_0/2, \infty )), & {\rm if} \; {\rm Im}(\theta) < 0  \period  \end{cases}        
\ene 
By Theorem~\ref{princ-secc}, $E_0$ is the only point in the spectrum of $\overset{0}{H} $.

\item[3.]
\beq \label{in.2.1}
P_0 : = P^{\sigma_0}(\theta, \alpha, 0) \period
\ene

\item[4.]
Theorem~\ref{princ-secc} and (\ref{il.2.1}) imply that
\beq \label{in.3} \begin{array}{l}
  \| \overset{0}{R}(z)\overline{P}_0 \| \leq  \boC_{\ref{il.2.1}} \frac{1}{(|\sin(\nu)|/2) \sigma_0 + 
 |z - E_0 |}\comma 
\end{array}
\ene
where $ \overline{P}_0 = 1 - P_0 $. 

\end{itemize}

\begin{remark}
Theorem~\ref{princ-secc} implies that  $P_0$ satisfies 
\beq \label{in.10.1.6tilde} \begin{array}{l}
\|  P_{\at, 1}(\theta, \alpha, 0)  -  P_0  \| \leq \frac{|\sin(\nu)|}{4}, \\\\ 1 -   \frac{1}{8} - \frac{1}{120}   \leq \|P_{0}\| 
\leq  1 +  \frac{1}{8} +  \frac{1}{120}   \period 
\end{array}
\ene

\end{remark}

\subsubsection{Induction Hypothesis} \label{Induction-Hypothesis}

We suppose that we have already defined the numbers $ E_m$, the sets $\cE_m$ and the operators $P_m$ for every $m \leq n$. We suppose furthermore that they satisfy the properties listed below. 

\begin{itemize}
\item[1a.] 
 $E_m$ is a simple eigenvalue of $\overset{m}{H}$.
 
\item[1b.] 
 
\beq \label{in.5}
| E_m  -  E_{m - 1} | < \boC^{ m+ 1 }_{\ref{il.2.1}}  \alpha^{3/2} (\sigma_{m -1})^{2} \period
\ene

\item[2.] 
\beq \label{in.6}
\cE_m  := \begin{cases}  \cE^{\sigma_0}(\theta) \setminus (E_m + \RR  - i (    (|\sin(\nu)|/2) \sigma_m/2, \infty )), & {\rm if} \; {\rm Im}(\theta) > 0 \comma \\
 \cE^{\sigma_0}(\theta) \setminus (E_m + \RR  + i ( (|\sin(\nu)|/2) \sigma_m/2, \infty )), & {\rm if} \; {\rm Im}(\theta) < 0  \period  \end{cases}      
\ene 
$E_m$ is the only point in the spectrum of $\overset{m}{H} $ in $  \cE_m $.

\item[3a.]
\beq \label{in.7}
P_m : = \frac{i}{2 \pi } \int_{\gamma_m} \overset{m}{R}(z)   dz \comma 
\ene
where $ \gamma_m: [0, 2\pi] \to \CC $ is the curve given by
\beq\label{in.8}
\gamma_m (t): = E_m + (|\sin(\nu)|\sigma_m ) e^{ -i t} \period   
\ene

\item[3b.]
\beq \label{in.10.1.2}
\| P_{m}    -   P_{m - 1}\otimes P_{\Omega^{(\sigma_{m}, \sigma_{m - 1} )}} \| \leq  \boC_{\ref{il.2.1}}^{2m + 2} \sigma_{m -1}^{1/2} \comma
\ene
where  $ P_{\Omega^{\sigma_m, \sigma_{m - 1}}}   $ is  the projection on the vacuum state $ \Omega^{\sigma_m, \sigma_{m-1}} $.

\item[4.]
\beq \label{in.9}
\| \overset{m}{R}(z)\overline{P}_m \| \leq \boC_{\ref{il.2.1}}^{m + 1} \frac{1}{(|\sin(\nu)|/2) \sigma_m + 
 |z - E_m |}\comma 
\ene
where
 
\beq \label{in.10}
\overline{P}_m : = 1 - P_m \period
\ene

\end{itemize}

\begin{remark} From (\ref{in.10.1.3}),
(\ref{in.10.1.6tilde}), and (\ref{in.10.1.2}) it follows that
\beq 
\begin{array}{l} \label{in.8.1}
1 -   \frac{1}{8} - \frac{1}{120} - \frac{1}{100}\sum_{i=0}^{m - 1} (\frac{1}{100})^i  \leq  \|P_{m}\| 
\\\\  \| P_m \| \leq  1 +  \frac{1}{8} +  \frac{1}{120} + \frac{1}{100}\sum_{i=0}^{m - 1}(\frac{1}{100})^i   \period 
\end{array}
\ene
\end{remark}

{\bf Remarks.}\\

Items 1b. and 3b. do not appear in the induction basis, if we wish to have them then we could define $E_{-1} = E_0$, 
$ P_{-1} = P_0 $ and $\sigma_{-1} = \sigma_0$.\\   

Once we have defined the value $E_m$, we can construct the sets $ \cE_m $. If we know that the eigenvalue $E_m$ is the only 
spectral point of $\overset{m}{H}$ in $\cE_m$, then the projection $P_m$ is well defined. To accomplish the induction step, we have to 
construct $E_{n+1}$, to prove that it is simple and the only eigenvalue in   $ \cE_{n+1} $ and to verify that (\ref{in.5}), (\ref{in.10.1.2}) and 
(\ref{in.9}) are valid.

The following equation, that is a consequence of the induction basis and the induction hypothesis, is used below
\beq \label{in.10.1.5}
\|  P_{m}  - P_{\at, 1}(0, 0, 0) \otimes P_{\Omega^{(\sigma_m, \infty)}}  \| \leq \frac{1}{8} +  \frac{1}{120} + \frac{1}{100}\sum_{i=0}^{m - 1} 
(\frac{1}{100})^i \period
\ene

\subsubsection{Induction Step} \label{induction-step}
The main result in this section is the Theorem below. The proof of this Theorem is the content of this subsection. It is based in Theorems~\ref{T.ei.1}, \ref{T.ee.1} and \ref{T.Induction-Step}.  
\begin{theorem}\label{T-Princ-Step}
Suppose that for any $m \in \NN \cup \{ 0\}$, $m \leq n$, we have defined the number $ E_m$, the set $\cE_m$ and the operator $P_m$ that satisfy the properties stated in the induction hypothesis. Then there exists a number
$   E_{n + 1}$ a set $  \cE_{n + 1}$ and the operator $ P_{n + 1} $ satisfying the same properties. 
\end{theorem}

\noindent \emph{Proof:}
The proof is a consequence of Remarks \ref{r.ei.1}, \ref{r.ee.1}, and \ref{r.ee.2} (which describe Theorems~\ref{T.ei.1}, \ref{T.ee.1} and \ref{T.Induction-Step}).  
\QED

Once Theorem~\ref{T-Princ-Step} is proved, then we conclude that the numbers
$   E_{n}$ the sets $  \cE_{n}$ and an operators $ P_{n} $ are well defined for any $n\in \mathbb{N}$. Parallel to the sets $ \cE_{m}$ we define 
\beq \label{in.6prima}
 \cE_{(m, \infty)}  := \begin{cases}  \cE^{\sigma_0}(\theta) \setminus (E_m + \RR  + i (     \sigma_m, \infty )), & {\rm if} \; {\rm Im}(\theta) > 0 \comma \\
 \cE^{\sigma_0}(\theta) \setminus (E_m + \RR  - i (  \sigma_m, \infty )), & {\rm if} \; {\rm Im}(\theta) < 0  \comma  \end{cases}      
\ene 
\beq \label{ei.5}
\widetilde{\cE}_m  = \cE^{\sigma_0}(\theta) \setminus (E_m + \RR + i \frac{1}{2}\sin( - \nu) (\sigma_{m + 1} , \infty)) \period
\ene

\begin{corollary}[from the proof of Theorem~\ref{T-Princ-Step}] \label{C-Princ-Step}
Suppose that the induction hypothesis are valid. It follows that for every $z \in  \cE_{(n, \infty)}$
\beq \label{estimates-infinity-interaction}\begin{array}{l}
\| (1 + |x|^2)^{-1} W_\infty^n \tR_\infty^n(z)  \| \leq \boC_{\ref{il.2.1}}^{n +2} \sigma_n \comma  \\\\
\|  W_\infty^n \tR_\infty^n(z)  \| \leq \boC_{\ref{il.2.1}}^{n +2}  \period   
\end{array}
\ene
Furthermore, z belongs to the resolvent set of $\overset{\infty}{H}$ and
\beq \
\| (\overset{\infty}{H} - z)^{-1}  \| \leq \boC_{\ref{il.2.1}}^{n + 2} \frac{1}{ (|\sin(\nu)|/2) \sigma_{n + 1} + |z- E_n| } \period
\ene
\end{corollary}
\noindent \emph{Proof:}
It follows from Theorem~\ref{P.ei.1} and Theorem~\ref{C.ei.2}, see also (\ref{acs.1}). 
\QED

We remark that since Theorem~\ref{T-Princ-Step} closes the inductive scheme, then the conclusions of Corollary are valid for any $n \in \mathbb{N}$.

\paragraph{Proof of Theorem~\ref{T-Princ-Step}} $\\$

In this paragraph we prove Theorem~\ref{T-Princ-Step}. As in the case of the infrared-regularized Hamiltonian, the estimates for the interaction are very important. In the regularized case Theorem~\ref{L.neri.2} is the key ingredient for the proof of Theorem~\ref{princ-secc}, which establish the existence of resonances. In the first sub paragraph of this paragraph we do estimates for the interaction, which are stated in Theorem~\ref{P.ei.1}.\\

In Theorem~\ref{L.neri.2}, we estimate the interaction with respect to the free (non interacting) resolvent ($R_0^{s,t}(\theta, \alpha,\beta)$). This makes the analysis simple but it has the disadvantage that $\alpha$ depends on the infrared cut off parameter $s$ (see \ref{L.rrp.1.m10}) and that it goes to zero as $s$ goes to zero, which implies that only the non interacting case ($\alpha = 0$) is analyzable if we remove the infrared cut-off. \\     

Here we estimate the interaction (\ref{il.11}) with respect to the resolvent (\ref{il.8.1}). This analysis is more delicate than the one of Theorem~\ref{L.neri.2} because the resolvent  (\ref{il.8.1}) already has a piece of interaction. The resolvent (\ref{il.8.1}) is very singular and we cannot handle it as in the case of the free resolvent. We need extra (physical) information to control it. We have to make use of the exponential decay (in the atom variable) of the eigenfunctions. To do this, we restrict our operators to a space that is generated (in the atomic part) by an eigenvector corresponding to the first excited eigenvalue of the atom Hamiltonian (see (\ref{il.fm.1})). This is done using the Feshbach map (See (\ref{il.fm.2})). \\

In the second subparagraph we estimate the Feshbach map of our Hamiltonians. The key ingredient is Theorem~\ref{L.ei.10} which is the analogous result of Theorem~\ref{L.neri.2} and Theorem~\ref{P.ei.1}. Theorem~\ref{L.ei.10} (and Neumann expansions) imply the invertibility  of the Feshbach map applied to the Hamiltonian. In the third subparagraph we study the resolvent of the original Hamiltonian using (\ref{L.il.fm.3.e2}) and prove (\ref{in.10.1.2}) for the induction step as well as the existence of $E_{n +1}$ and the fact that it is non-degenerate (see Remark \ref{r.ei.1}). In the forth subparagraph we prove (\ref{in.5}) (see Remark \ref{r.ee.1}) and conclude the induction step by proving (\ref{in.9}) in Theorem~\ref{T.Induction-Step}, which finish the induction scheme (see Remark \ref{r.ee.2}).

\subparagraph{Estimates for the Interaction} $\\$

 As in the case of the infrared-regularized Hamiltonian, the estimates for the interaction are very important. In the regularized case Theorem~\ref{L.neri.2} is the key ingredient for the proof of Theorem~\ref{princ-secc}, which establish the existence of resonances. In this sub paragraph we do estimates for the piece of the interaction with photon energies between $\sigma_{n + 1} $ and $\sigma_n$ (see (\ref{il.11})). The main result is stated in Theorem~\ref{P.ei.1}. 

To study the operator $W_{n+ 1}^{n}$ we analyze many terms. To simplify our notation we organize some of these terms in different sets wich are introduced in the following definition together with some other notations.

\begin{definition}\label{sym} 
We define the following sets:
\beq 
\begin{array}{l} \label{sym-e1}
\mathcal W_{2}(\iota, m) = \Big \{ (1 + |x|^2)^{ -\iota} (A^{\sigma_m, \sigma_n})^2, (1 + |x|^2)^{-\iota}    (A_{ j}^{  \sigma_m, \sigma_n })^* A_{q}^{\sigma_m, \sigma_n},  \\\\ (1 + |x|^2)^{-\iota/2} ( A_{ i}^{\sigma_m, \sigma_n} )^* [(1 + |x|^2)^{-\iota/2} A_{j}^{\sigma_m, \sigma_n}, 
 \vv_{ q}^{\sigma_n} ],    \\\\  \, 
 [(1 + |x|^2)^{-\iota/2}(A_{j}^{\sigma_m, \sigma_n})^*,  \vv_{ q}^{\sigma_n} ]   (1 + |x|^2)^{-\iota/2}
A_{ i}^{\sigma_m, \sigma_n} \big | i, j, q \in \{1, 2, 3  \} \Big \}, 
\end{array}
\ene

\beq
\begin{array}{l}\label{sym-e2}
\mathcal W_{1}(\iota, m) = \Big \{ (1 + |x|^2)^{-\iota/2}  b^{\sigma_m, \sigma_n}, \ (1 + |x|^2)^{-\iota/2}  [A_{j}^{\sigma_m, \sigma_n},  \vv_{ q}^{\sigma_n} ] \\\\ \big  | j, q \in \{1, 2, 3  \}  \Big \}. 
\end{array}
\ene

The constant
\beq \label{sym-e3}
C_{\ref{sym-e3}} : = 40 C_{\ref{R.a.2.1}}^2 (1 + C_{\ref{T.eb.1.e2.1}}),
\ene
and the function 
\beq \label{sym-e4}
\cG_\alpha(\sigma,\rho) : =\alpha^{3/2}(\sigma^2 \rho^{-1/2} + \sigma^{3/2}) 
\ene
is repeatedly used in this section.

\end{definition}

\begin{remark}\label{us} {\rm
Eq. (\ref{il.8.1}) implies that $E_n$ is a simple eigenvalue of $\tH^n$. We denote by 
\beq \label{ei.9}
\tP_n = P_n \otimes  P_{\Om^{\sigma_{n +1},\sigma_n}} 
\ene  
the projection onto the corresponding eigenvector. It is given by the integral 
\beq 
\tP_n : = \frac{i}{2 \pi}\int_{\tilde{\gamma}_n} \tR^n(z)dz,  \comma
\ene 
where
$ \tilde{\gamma}_n: [0, 2\pi] \to \CC $ is the curve
\beq \label{ei.8}
\tilde{\gamma}_n (t): = E_n + (|\sin(\nu)|/2) \sigma_{n + 1}  e^{ -i t} \period   
\ene 
}
\end{remark}

\begin{lemma}\label{L.ei.1} For any $z \in \widetilde{\cE}_n  $ 
\beq \label{L.ei.1.e1}
\|  \tR^n(z)   \overline{\tP_n}  \| \leq  \frac{2}{(|\sin(\nu)|/2)}(\boC_{\ref{il.2.1}}^{n+ 1} + 2)  \frac{1}{ (|\sin(\nu)|/2) \sigma_{n + 1} + |z - E_n|   }\comma
\ene
for any $  z \in \cE_{(n, \infty)} $
\beq \label{L.ei.1.e1tilde}
\|  \tR_\infty^{n}(z)     \| \leq \frac{2}{(|\sin(\nu)|/2)} (\boC_{\ref{il.2.1}}^{n+1} + 2)  \frac{1}{ (|\sin(\nu)|/2) \sigma_{n + 1} + |z - E_n|   }\comma
\ene


\end{lemma}
\emph{Proof.}\\
We prove (\ref{L.ei.1.e1}) and suppose that $\rm{Im}(\theta) > 0$. The other cases are similar. \\ 
By functional calculus, 
\beq\label{ei.11} \begin{array}{l}
\|  \tR^n(z)   \overline{\tP_n}  \|  \leq \|  \tR^n(z)  \overline{P}_n \otimes 1_{\cH^{\sigma_{n + 1}, \sigma_n}}    \| +
 \|  \tR^n(z)  P_n \otimes  \overline{P}_{\Om^{\sigma_{n + 1}, \sigma_n}}   \| \\\\ \leq 
\sup_{r \in \spec(\hf^{\sigma_{n + 1}, \sigma_n})}    \| \overset{n}{R}(z - e^{-\theta} r )    \overline{P}_n   \|    + 
 \sup_{r \in [\sigma_{n + 1}, \infty)}   \| \frac{P_n}{ E_n - z  + r e^{- \theta} }   \| \\\\ \leq
\sup_{r \in \spec(\hf^{\sigma_{n + 1}, \sigma_n})} \boC_{\ref{il.2.1}}^{n+ 1} \frac{1}{(|\sin(\nu)|/2) \sigma_{n} + |z - E_n - r e^{- \theta} | }  \\\\ +
 \sup_{r \in [\sigma_{n + 1}, \infty)} 2  | \frac{1}{ z - E_n   - r e^{- \theta} }  |
\comma
\end{array}
\ene
where we used (\ref{in.9}) and (\ref{in.8.1}). \\
It is clear from the definition of $\tilde{\cE}_n$ that for $z$ in this set
\beq \label{ei.12}
(|\sin(\nu)|/2) \sigma_{n + 1 } \leq |z - E_n - re^{- \theta}|, \; \forall r \in [\sigma_{n + 1}, \infty) \period 
\ene
We denote by $\mathcal{C}_n$ the set
$$\mathcal{C}_n  : = \{ z = (z_1, z_2) \in \widetilde{\cE}_n : z_1 > \rm{Re}(E_n) \} $$
and define the following sets: 
\beq \label{ei.14}
L_{n, d} : = \{ E_n + de^{ i \pi/2 - \theta} - t e^{- \theta} : t \in \RR  \} \comma
\ene
\beq \label{ei.15}
L_n : = \cup_{d \geq 0} L_{n, d} \cap \mathcal{C}_n \period
\ene
By construction 
\beq \label{dist}
\dist(L_{n,d}, E_n) = d.
\ene
and
\beq \label{ei.13}
| z - E_n| \leq | z - E_n - r e^{- \theta}|, \; \forall z \in \tilde{\cE}_n \setminus L_n, \, \forall r \in [\sigma_{n + 1}, \infty) \period
\ene
Let $Z_{1, d} $ be the intersection of $L_{n, d}$ with the line $ E_n - \frac{i|\sin(\nu)|}{2} \sigma_{n + 1} + \RR $ and let
$ Z_{2, d} = L_{n, d}\cap (E_n  + \RR) $. We define furthermore $ Z_{3, d} = E_n +  de^{ i \pi/2 - \theta}$. It follows that 
\beq \label{beq} \begin{array}{l}
\sup_{Z \in L_{n, d} \cap \mathcal{C}_n } (|Z - E_n|) =
|Z_{1, d} - E_n|^2 = d^2 +|Z_{3, d} - Z_{1, d}| \\\\ = d^2 + (|Z_{3, d}-  Z_{2, d}|  
+ | Z_{2, d} - Z_{1, d}|)^2  = d^2 + (\frac{d}{|\tan(\nu)|} + \frac{1}{2} \sigma_{n + 1})^2
\end{array}
\ene
Eqs. (\ref{ei.12}), (\ref{dist}) and (\ref{beq}) imply that 
\beq \label{beq.1}
\frac{|z - E_n|}{|z - E_n - r e^{- \theta}| } \leq \Big(\big(\frac{\cos(\nu)}{\sin(\nu)}  + \frac{1}{\sin(\nu)}  \big)^2 + 1\Big)^{1/2} \leq  
\frac{2}{|\sin(\nu)|} \comma
\ene
for $r \in [\sigma_{n +1})$ and $z \in L_n$, which together with (\ref{ei.13}) implies that 
\beq \label{beq.2}
\frac{1}{|z - E_n - r e^{- \theta}| }  \leq  
\frac{2}{|\sin(\nu)|}\frac{1}{ |z - E_n|} \comma
\ene
for every $r \in [\sigma_{n +1})$ and $z \in \widetilde \cE_n$.

Eqs. (\ref{ei.12}) and (\ref{beq.2}) imply that
\beq \label{ei.18}\begin{array}{l}
\frac{1}{|z - E_n - re^{- \theta}|} \leq \frac{2}{(|\sin(\nu)|/2)} \frac{1}{ (|\sin(\nu)|/2) \sigma_{n + 1} + |z - E_n|  }, 
\; \forall z \in  \tilde{\cE}_n ,\\\\ \; \forall r \in [\sigma_{n+1}, \infty) \period
\end{array}
\ene 
Finally Eq. (\ref{L.ei.1.e1}) follows from (\ref{ei.11}) and (\ref{ei.18}).

\QED
In the next Lemma we use (\ref{eph.2}), (\ref{il.1})  and (\ref{il.15}). Remember that we omit the variable $\theta$.    

\begin{lemma}\label{L.ei.2}

Denote by $\cP^0 = 1$ and $\cP^1 = \cP$. For every $\iota \in \{ 0, 1 \}$,  $\rho > 0$, $ m > n $ (including $\infty$) and any
 $w \in \cW_{j}(\iota, m)$.   
\beq \label{L.ei.2.e1}
\begin{array}{l}
\| w (\hf^{\sigma_m, \sigma_n} + \rho)^{-j/2}  \cP^{1 -\iota} \| \leq C_{\ref{sym-e3}} \cG_\alpha(\sigma_n,\rho)^{j} 
\end{array}
\ene

\end{lemma}

\emph{Proof.}\\
We consider first the term $  (A^{\sigma_{m}, \sigma_n})^2   $ and $\iota = 0$.   By  (\ref{il.15}),
\beq \label{ei.21.1} \begin{array}{l}
(A^{\sigma_{m}, \sigma_n})^2    =
   a^*(\boG_\theta^{\sigma_m, \sigma_n}) \cdot  a^*(\boG_\theta^{\sigma_m, \sigma_n}) \\\\
+   a(\boG_{\overline{\theta}}^{\sigma_m, \sigma_n})\cdot  a(\boG_{\overline{\theta}}^{\sigma_m, \sigma_n})  +   
 a^*(\boG_\theta^{\sigma_m, \sigma_n})\cdot  a(\boG_{\overline{\theta}}^{\sigma_m, \sigma_n}) \\\\ +
a(\boG_{\overline{\theta}}^{\sigma_m, \sigma_n}) \cdot a^*(\boG_\theta^{\sigma_m, \sigma_n}) \period
\end{array}
\ene
We analyze the term $   a(\boG_{\overline{\theta}}^{\sigma_m, \sigma_n})\cdot a(\boG_{\overline{\theta}}^{\sigma_m, \sigma_n})   $, the 
other terms can be treated analogously. \\
By Lemma~\ref{L.ih.1} and Remark \ref{R.a.2} we have
\beq \label{ei.23}
\begin{array}{l}
\|    a(\boG_{\overline{\theta}, j}^{\sigma_m, \sigma_n})\cdot  a(\boG_{\overline{\theta}, j}^{\sigma_m, \sigma_n}) \frac{1}{\hf^{\sigma_m, \sigma_n} + \rho} \cP \|  \\\\ \leq
\|    a(\boG_{\overline{\theta}, j}^{\sigma_m, \sigma_n} (1 + |x|^2)^{-1/2} )\cdot  a(\boG_{\overline{\theta}, j}^{\sigma_m, \sigma_n}(1 + |x|^2)^{-1/2}) \frac{1}{\hf^{\sigma_{m}, \sigma_n} + \rho}\| \\\\ \cdot
\|  (1 + |x|^2)  \cP \|  \\\\ \leq
\|   \boG_{\overline{\theta}, j}^{\sigma_m, \sigma_n} (1 + |x|^2)^{-1/2}  \|_{\rho} \|   \boG_{\overline{\theta}, j}^{\sigma_m, \sigma_n} (1 + |x|^2)^{-1/2}  \|_{\rho}
 \\\\ \cdot \|  (1 + |x|^2)  \cP \| \leq \alpha^3 C_{\ref{R.a.2.1}}^2 \|  (1 + |x|^2)  \cP \| ((\sigma_n)^{2} \rho^{- 1/2} + (\sigma_n)^{3/2}  )^2 \period  
\end{array}
\ene
The desired result for this term follows from (\ref{ei.21.1})-(\ref{ei.23}) and similar estimates.\\

For the terms involving the commutator $ [A_{ j}^{\sigma_{m}, \sigma_n},  \vv_{\theta, q}^{\sigma_n} ] $, we  compute 
\beq \label{ei.23.1} \begin{array}{l}
 [A_{ j}^{\sigma_{m}, \sigma_n},  \vv_{\theta, q}^{\sigma_n} ]\\\\  =
 i e^{- \theta }a^*(  \frac{\partial}{\partial x_q}\boG_{\theta, j}^{\sigma_m, \sigma_n} ) + i e^{- \theta }
 a( \frac{\partial}{\partial x_q} \boG_{\overline{\theta}, j}^{\sigma_m, \sigma_n}) 
\end{array}
\ene
and estimate as before. The rest of the terms are analyzed similarly.

For $\iota = 0$ we proceed in a similar way using the following:
\beq \label{ei.23.1.0.1} \begin{array}{l}
 [ (1 + |x|^2)^{- 1/2} A_{ j}^{\sigma_{m}, \sigma_n},  \vv_{ q}^{\sigma_n} ]   =
 i e^{- \theta }a^*(  \frac{\partial}{\partial x_q}  (1 + |x|^2)^{- 1/2}  \boG_{\theta, j}^{\sigma_m, \sigma_n} ) \\\\ + i e^{- \theta }
 a( \frac{\partial}{\partial x_q} (1 + |x|^2)^{- 1/2} \boG_{\overline{\theta}, j}^{\sigma_m, \sigma_n}) 
\end{array}
\ene

\QED

\begin{lemma}\label{L.ei.2.1}

Let $h \in \CC$ satisfy the properties for $\theta$ in Assumptions~\ref{parameters-basis}. Suppose furthermore that 
 $\rho \leq e_0 -1 $ and $m > n$. For every $\iota \in \{ 0, 1 \}$ and any
 $w \in \cW_{j}(\iota, m)$,   
\beq \label{L.ei.2.1.e1} 
\begin{array}{l}
\| w  \frac{1}{ \overset{m}{H}_0(h) - \rho }  \| \leq 40 C_{\ref{sym-e3}}
 C_{\ref{an.0.16.2}}( 1 +  b_{1/4}) \big(\frac{\cG_\alpha(\sigma_n, 1)}{\sigma_n^{1 - \iota}}\big)^{j} 
\end{array}
\ene

\end{lemma}

\emph{Proof.}\\
First we notice that by (\ref{geh.5}),
\beq  \label{i.12prima} \begin{array}{l}
(3/4)\| ( \hf^{\sigma_m} - \rho - \Delta)\phi    \|_{\cH^{\sigma_m, \sigma_n}}  \leq \\\\
 \| ( \hf^{\sigma_m} - \rho - \Delta  + V(h) )\phi    \|_{\cH^{\sigma_m}} +  
 b_{1/4} \| \phi  \|_{\cH^{\sigma_m, \sigma_n}}  
\end{array}
\ene
and that 
\beq \label{i.13prima} \begin{array}{l}
\| ( \hf^{\sigma_m} - e^{- h}\hf^{\sigma_m}  - \Delta + e^{- 2 h}\Delta  )\phi \|_{\cH^{\sigma_m}} \\\\  \leq
(|1- e^{h}| + |1- e^{- 2 h}|  )\| ( \hf^{\sigma_m} - \rho - \Delta)\phi    \|_{\cH^{\sigma_m}}\period 
\end{array}
\ene
Eqs. (\ref{C.eeh.1.e1}), (\ref{i.12prima}) and (\ref{i.13prima}) imply
\beq \label{ei.23.1.0.3} 
\| (   - \Delta +  \hf^{\sigma_m} - \rho )  \frac{1}{ \overset{m}{H}_0(h) - \rho }   \|\leq 2 ( 1 + 8 b_{1/4} C_{\ref{an.0.16.2}}  )\period
\ene
Using functional calculus we prove that 
\beq \label{ei.23.1.0.2} 
\| (\hf^{\sigma_m, \sigma_n} - \rho) \frac{1}{ - \Delta +  \hf^{\sigma_m} - \rho }   \|\leq 1  \period
\ene
We conclude using (\ref{ei.23.1.0.3}), (\ref{ei.23.1.0.2}) and the proof of Lemma~\ref{L.ei.2}.  

\QED

\begin{lemma}\label{L.ei.3}
Suppose that $m > n$ and either 
 $ z \in \widetilde{\cE}_n \setminus \{ E_n \}$ (see \ref{ei.5}) for $ m = n + 1$ or $ z \in \cE_{(n, \infty)} $ (see \ref{in.6prima}) for $ m > n + 1$. 
 Suppose furthermore that $\rho > 0$ is such that $\rho \leq |z - E_n| $, then 
\beq \label{L.ei.3.e1}
\Big\| \frac{ \hf^{\sigma_m, \sigma_{n}} + \rho}{ e^{- \theta}\hf^{\sigma_m, \sigma_{n}} - (z - E_n ) }   \Big\| \leq \frac{10}{|\sin(\nu)|}  \period
\ene
\end{lemma}
\emph{Proof:}\\
We take $m = n+1$, the other cases are similar. \\
First we notice that 
\beq \label{ei.24}\begin{array}{l}
 \frac{ \hf^{\sigma_m, \sigma_{n}} + \rho}{ e^{- \theta}\hf^{\sigma_m, \sigma_{n}} - (z - E_n ) } = 
 e^{\theta} + e^{\theta} \frac{e^{- \theta} \rho + (z - E_n) }{  e^{- \theta}\hf^{\sigma_m, \sigma_{n}} - (z - E_n )    }  \comma
\end{array}
\ene
 then, by functional calculus, 
\beq \label{ei.25} \begin{array}{l}
\Big\| \frac{ \hf^{\sigma_m, \sigma_{n}} + \rho}{ e^{- \theta}\hf^{\sigma_m, \sigma_{n}} - (z - E_n ) }   \Big\| 
\leq 1 +   \sup_{r \in [\sigma_{m}, \infty)} \frac{2|z - E_n|}{ |  z - E_n - re^{- \theta}   |} 
\\\\ \leq  1 + \frac{4}{(|\sin(\nu)|/2)} \comma
\end{array}
\ene
where in the last inequality we used (\ref{beq.2}).

\QED

\begin{lemma}\label{L.ei.4}
Suppose that 
 $ z \in \widetilde{\cE}_n \setminus \{ E_n \}$ (see \ref{ei.5}). 
 Let $\cP^0 = 1$ and $\cP^1 = \cP$. Then for every $\iota \in \{ 0, 1 \}$ and any
 $w \in \cW_{j}(\iota, n+1)$

\beq \label{L.ei.4.e1}
\begin{array}{l}
\| w \cP^{1 -\iota} \tR^{n}(z) \| \leq C_{\ref{sym-e3}} \frac{30 \boC_{\ref{il.2.1}}^{n+1}}{|\sin(\nu)|}   \cG_\alpha(\sigma_n, |z - E_n|)^{j} 
|z - E_n|^{j/2 - 1}
\end{array}
\ene

The same estimates holds if we substitute $  \tR^{n}(z) $ by  $  \tR_\infty^{n}(z) $ (see (\ref{il.8.1})), $\tilde \cE^n$  by $\cE_{(n, \infty)}$ (see (\ref{in.6prima})) and $n+1$ for $\infty$.

\end{lemma}

\emph{Proof.}\\
We prove (\ref{L.ei.4.e1}) for $\iota = 1$ and $ w =  (A^{\sigma_{n + 1}, \sigma_n})^2 $, the other cases are similar. \\
Take $ \rho = | z - E_n  |  $ , we have that
\beq \label{ei.26} \begin{array}{l}
\|  (A^{\sigma_{n + 1}, \sigma_n})^2\cP \tR^n(z)   \| \leq \| (A^{\sigma_{n + 1}, \sigma_n})^2 (\hf^{\sigma_{n + 1}, \sigma_n} + \rho)^{-1} \cP  \| \\\\ \cdot
\|  \frac{ \hf^{\sigma_{n + 1}, \sigma_n} + \rho}{ e^{- \theta}\hf^{\sigma_{n+1},\sigma_n} - (z - E_n ) }  \| \| ( e^{- \theta}\hf^{\sigma_{n + 1}, \sigma_n} - (z - E_n )) \tR^n(z)  \| \\\\
\leq  C_{\ref{sym-e3}} \frac{10 }{|\sin(\nu)|}   \cG_\alpha(\sigma_n, |z - E_n|)^{j} 
\\\\ \cdot \| ( e^{- \theta}\hf^{\sigma_{n + 1}, \sigma_n} - (z - E_n )) \tR^n(z)  \| \comma
\end{array}
\ene
where we used Lemmata~\ref{L.ei.2} and \ref{L.ei.3}.\\
By functional calculus
\beq \label{ei.27} \begin{array}{l}
 \| ( e^{- \theta}\hf^{\sigma_{n + 1}, \sigma_n} - (z - E_n )) \tR^n(z)  \| \\\\
 \leq  \sup_{r \in \{ 0 \} \cup [\sigma_{n+1}, \infty) }   \| ( E_n - ( z -e^{- \theta} r) ) \overset{n}{R}(z -e^{- \theta} r )     \| \period
\end{array}
\ene
Next we choose a fixed number $r  \in \{ 0 \} \cup [\sigma_{n+1}, \infty) $. Eqs. (\ref{in.9}) and (\ref{in.8.1})   
 imply 
\beq \label{ei.28} \begin{array}{l}
 \| ( E_n - ( z -e^{- \theta} r) ) \overset{n}{R}(z -e^{- \theta} r )     \|   \leq
\| P_n   \| \\\\ 
+ \| ( E_n - ( z -e^{- \theta} r) ) \overset{n}{R}(z -e^{- \theta} r ) \bP_n    \|  \leq 2 + \boC_{\ref{il.2.1}}^{n+1} \leq 3\boC_{\ref{il.2.1}}^{n+1}
\end{array}
\ene
Eq. (\ref{L.ei.4.e1}) follows from (\ref{ei.26})-(\ref{ei.28}).  

For the case $ j = 1$ we use 
$$
\|  (\hf^{\sigma_{n + 1}, \sigma_n} + \rho)^{1/2}  \tR^n(z)  \|\leq  \|  (\hf^{\sigma_{n + 1}, \sigma_n} + \rho)  \tR^n(z)  \|^{1/2}
\|    \tR^n(z)  \|^{1/2}.
$$

\QED

Remember that we defined the velocity operator $v^{\sigma_n}$ in (\ref{il.1}).

\begin{lemma}\label{L.ei.5}

Suppose that $ h \in \CC $ satisfies the properties of $\theta$ in Assumptions~\ref{parameters-basis}. \\
For any $\rho \in (- \infty,   e_{0} - 1]  $, any $ m \geq n $ and any $ j, q \in \{1, 2, 3\}  $
\beq \label{L.ei.5.e1} \begin{array}{l}
\| (v^{\sigma_n}_{ j})^* v^{\sigma_n}_{ q} \frac{1}{ \overset{m}{H}_0(h) - \rho  }     \| \leq  20 C_{\ref{an.0.16.2}}(b_{1/4} + 1).  
\period
\end{array}
\ene
\end{lemma}
\emph{Proof:} \\
We use here that $\alpha^{3/2} \leq  \frac{1}{64(1 + 4 C_{\ref{an.0.16.2}} b_{1/4} )C_{\ref{R.a.2.1}}} $  and
$\alpha^{3/2} \leq \frac{1}{24 C_{\ref{i.13.4}} ( 1 +   8 C_{\ref{an.0.16.2}}  ) }  $ (see Assumptions~\ref{parameters-basis}).

We take $m = n$, the other cases follow in the same way.\\
We denote by 
\beq \label{ei.29.0.0}
p := (p_1, p_2, p_3) : = - i \nabla 
\ene
the momentum operator. \\
We calculate (remember that $\theta = i \nu $, $\nu \in \RR$),
\beq \label{ei.29} \begin{array}{l}
(v^{\sigma_n}_{ j})^* v^{\sigma_n}_{ q}  =  p_j p_q -  e^{- \overline{\theta}} ( a^*(p_j \boG^{\sigma_n}_{\theta,  q} ) - a(p_j  \boG^{\sigma_n}_{\overline{\theta}, q}) )\\\\ -  
e^{- \overline{\theta}} A_{ q}^{\sigma_n} (\theta) p_j - 
  A_{j}^{\sigma_n}(\overline{\theta}) e^{-\theta}p_q
+  A_{j}^{\sigma_n}(\overline{\theta}) A_{ q}^{\sigma_n}(\theta) \period
\end{array}
\ene 
We estimate separately the terms appearing in (\ref{ei.29}).  First 
we notice that 
\beq \label{ei.30}
\| p_j p_q \frac{1}{ \hf^{\sigma_n} - \rho - \Delta }     \| \leq 1 \period
\ene
Then we have as in (\ref{ei.23.1.0.3}) that,   
\beq \label{ei.31}
\|   p_j p_q    \frac{1}{ \overset{n}{H}_0(h) - \rho  }     \| \leq 2 (1 + 8 C_{\ref{an.0.16.2}} b_{1/4} ) \period  
\ene
Following the proof of (\ref{L.i.1.e1}), using (\ref{C.eeh.1.e1}) functional calculus and  our selection of $\rho$ we get for 
$\alpha^{3/2} \leq \frac{1}{24 C_{\ref{i.13.4}} ( 1 +   8 C_{\ref{an.0.16.2}}  ) }  $\\
\beq \label{ei.32}
\begin{array}{l}
\| \big( e^{- \overline{\theta}} A_{ q}^{\sigma_n}(\theta) p_j + 
  A_{j}^{\sigma_n}(\overline{\theta}) e^{-\theta}p_q
+  A_{j}^{\sigma_n}(\overline{\theta}) A_{ q}^{\sigma_n}(\theta) \big)  \frac{1}{ \overset{n}{H}_0(0) - \rho  }   \|  \leq \frac{1}{4} \period
\end{array}
\ene
Finally by Lemma~\ref{L.ih.1} and Remark \ref{R.a.2} 
\beq \label{ei.33}
\|( a^*(p_j \boG^{\sigma_n}_{\theta, q} ) + a(p_j \boG^{\sigma_n}_{\overline{\theta}, q}) ) \frac{1}{ \hf^{\sigma_n} - \rho}\| \leq 4 \alpha^{3/2} C_{\ref{R.a.2.1}} 
 \period 
\ene
Thus, using (\ref{ei.23.1.0.3}), (\ref{ei.23.1.0.2}) and that
 $\alpha^{3/2} \leq  \frac{1}{64(1 + 8 C_{\ref{an.0.16.2}} b_{1/4} )C_{\ref{R.a.2.1}}} $  we get 
\beq \label{ei.34} \begin{array}{l}
\|( a^*(p_j \boG^{\sigma_n}_{\theta, q} ) + a(p_j  \boG^{\sigma_n}_{\overline{\theta}, q}) ) \frac{1}{ \overset{n}{H}_0(h) - \rho  }   \|  \leq \frac{1}{4} 
\period
\end{array}
\ene
Eq. (\ref{L.ei.5.e1}) follows from (\ref{ei.29}), (\ref{ei.31}), (\ref{ei.32}), (\ref{ei.34}) and the fact that $C_{\ref{an.0.16.2}} > 1$.

\QED

\begin{lemma}\label{L.ei.6}

Suppose that $ h \in \CC $ satisfies the properties of $\theta$ in Assumptions~\ref{parameters-basis}. \\
For any $\rho \in (- \infty,   e_{0} - 1]  $, any $ m \geq n $ and any $ j \in \{1, 2, 3\}  $
\beq \label{L.ei.6.e1} \begin{array}{l}
\|  v^{\sigma_n}_{ j} \frac{1}{ \overset{m}{H}_0(h) - \rho  }     \| \leq  20 C_{\ref{an.0.16.2}}(b_{1/4} + 1)
 \period
\end{array}
\ene
\end{lemma}
\emph{Proof:} \\
We take $m = n$, the other cases are  similar. \\
By Lemma~\ref{L.ei.5}, 
\beq \label{ei.35} 
\|   |v^{\sigma_n}_{ j}|^2 \frac{1}{| \overset{n}{H}_0(h) - \rho|  }    \|   \leq   20 C_{\ref{an.0.16.2}}(b_{1/4} + 1) \period
\ene
As $| \overset{n}{H}_0(h) - \rho|  $ is positive (see Theorem X.8 of \cite{ReedSimonII1980}), 
\beq \label{ei.36}
 |v^{\sigma_n}_{ j}|^2 \leq 20 C_{\ref{an.0.16.2}}(b_{1/4} + 1) |(\overset{n}{H}_0(h) - \rho)| 
\ene
and then it follows that for any $\phi \in  \dom |\overset{n}{H}_0(h) - \rho | ^{1/2}$
\beq  \label{ei.37} \begin{array}{l}
\| v^{\sigma_n}_{ j}\phi \|_{\cH^{\sigma_n}} = \| |v^{\sigma_n}_{ j}| \phi \|_{\cH^{\sigma_n}} \\\\
 \leq ( 20 C_{\ref{an.0.16.2}}(b_{1/4} + 1))^{1/2} 
\| |\overset{n}{H}_0(h) - \rho|^{1/2} \phi \|_{\cH^{\sigma_n}} \period
\end{array}
\ene
Eq. (\ref{L.ei.6.e1}) follows from (\ref{ei.37}) and the fact that 
$ \|       \frac{1}{| \overset{n}{H}_0(h) - \rho |^{-1/2}  } \|  \leq 8^{1/2} (C_{\ref{an.0.16.2}})^{1/2} $, which
is a consequence of  (\ref{C.eeh.1.e1}) and

\beq \begin{array}{l}
\la   \frac{1}{| \overset{n}{H}_0(h) - \rho |^{-1/2}  } \phi  | \;    \frac{1}{| \overset{n}{H}_0(h) - \rho |^{-1/2}  } \phi  \ra \\\\
= \la \frac{1}{| \overset{n}{H}_0(h) - \rho |^{-1}  } \phi  | \;    \phi \ra \leq \|  | \overset{n}{H}_0(h) - \rho |^{-1}   \phi   \|  \| \phi \| \\\\
= \|  ( \overset{n}{H}_0(h) - \rho )^{-1}  \phi   \|  \| \phi \|\period 
\end{array}
\ene

\QED

\begin{lemma}\label{L.ei.7}
Suppose that $ z \in \tilde{\cE}_n \setminus \{ E_n \}$  

Take $\rho = 2e_0 - 1$, then the following estimate holds true,
\beq \label{L.ei.7.e1} \begin{array}{l}
\| (\overset{n}{H}_0(0) - \rho) \tR^n(z)   \| \leq  \frac{700}{|\sin(\nu)|} \boC_{\ref{il.2.1}}^{n + 1} (1 + |e_0|)(1 + \frac{1}{|z - E_n|})
\period
\end{array}
\ene
The same estimate is valid for $ z \in \cE_{(n, \infty)} $ replacing $\tR^n(z)  $ by $ \tR^n_\infty(z)  $ 	and 
$\overset{n}{H}_0(0)$ by $\overset{\infty}{H}_0(0)$.

\end{lemma}
\noindent \emph{ Proof: }\\
First we notice that 
\beq \label{ei.38} \begin{array}{l}
\| (\overset{n}{H}_0(0) - \rho) \tR^n(z)   \| \leq \| (\overset{n}{H}_0(0) - \rho) \tR^n(  e^{-\theta} \rho)    \|  \cdot \\\\
\|  (\tH^{n}(\theta) -  e^{-\theta} \rho )  \tR^n(z)  \| \period
\end{array}
\ene
Next we use functional calculus and Corollary~\ref{L.an.4} to estimate, 
\beq \label{ei.39} \begin{array}{l}
\| (\overset{n}{H}_0(0) - \rho) \tR^n(e^{-\theta} \rho )   \| \\\\ = \sup_{r \in \{ 0 \} \cup [\sigma_{n+1}, \infty)} 
\|  (\overset{n}{H}_0(0) - \rho) \frac{1}{\overset{n}{H}(\theta) - e^{ -\theta} (\rho - r)  }     \|
\leq 8 \period
\end{array}
\ene
By (\ref{in.8.1}), (\ref{ei.9}) and Lemma~\ref{L.ei.1}
\beq \label{ei.40} \begin{array}{l}
\|  \Big( \tH^{n}(\theta) - z + \big((z - E_n) - ( e^{-\theta} \rho - E_n ) \big) \Big)  \tR^n(z)  \| \\\\ \leq
 1 + \frac{2}{(|\sin(\nu)|/2)}(\boC_{\ref{il.2.1}}^{n + 1} + 2) + 2 \\\\ + 
 \frac{(|2e_{0} - 1| + |E_n|)  (\frac{2}{(|\sin(\nu)|/2)}(\boC_{\ref{il.2.1}}^{n + 1} + 2) + 2)}{ |z - E_n|  }  
\end{array}
\ene
Remark \ref{R.neri.1}, \ref{in.10.1.3} and Eq.  (\ref{in.5}) imply that 
\beq \label{ei.41}
| E_n - E_0  | \leq (|\sin(\nu)|/2) \sigma_0, \; \; \; \; | E_0 - e_{1} | \leq (|\sin(\nu)|/2) \sigma_0 
\ene
and therefore we have
\beq \label{ei.42}
| E_n | \leq |\sin(\nu)| \sigma_0 + |e_{0}| \period
\ene
Eq. (\ref{L.ei.7.e1}) follows from (\ref{ei.38})-(\ref{ei.40}) and (\ref{ei.42}) 
(notice that $\sigma_0\leq 1$ and $ \boC_{\ref{il.2.1}} \geq 1 $).

\QED

\begin{lemma}\label{L.ei.8}
Suppose that $ z \in \tilde{\cE}_n \setminus \{ E_n \}$.  \\
Denote by $\cP^0 = 1$ and $\cP^1 = \cP$. For every $\iota \in \{ 0, 1 \}$,  

\beq \label{L.ei.8.e1}  \begin{array}{l}
\| (1 + |x|^2)^{-\iota/2}  A^{\sigma_{n + 1}, \sigma_n} \cdot \vv^{\sigma_n} \cP^{1 - \iota} \tR^n(z)  \|  \leq 700 \boC_{\ref{il.2.1}}^{n + 1} \\\\ \Big(
C_{\ref{sym-e3}} \frac{30 }{|\sin(\nu)|^2}  \cdot   \cG_\alpha(\sigma_n, |z - E_n|)^{2} 
  C_{\ref{an.0.16.2}}(b_{1/4} + 1) \\\\ \cdot   (1 + |e_0|)(1 + \frac{1}{|z - E_n|})\Big)^{1/2} \period
\end{array} \ene 
\beq \label{L.ei.8.e2}  \begin{array}{l}
\|   (1 + |x|^2)^{-\iota/2}   \vv^{\sigma_n}\cdot  A^{\sigma_{n + 1}, \sigma_n}  \cP^{1 - \iota} \tR^n(z)  \| \leq \boC_{\ref{il.2.1}}^{n + 1} C_{\ref{sym-e3}}  \frac{90 }{|\sin(\nu)|} \\\\ \cdot    \cG_\alpha(\sigma_n, |z - E_n|) 
|z - E_n|^{-1/2}  + 
\| (1 + |x|^2)^{-\iota/2}  A^{\sigma_{n + 1}, \sigma_n} \\\\ \cdot \vv^{\sigma_n} \cP^{1 - \iota} \tR^n(z)  \| \period
\end{array} \ene 

The estimates are also valid for $ z \in \cE_{(n, \infty)} $ if we replace $ \tR^n(z)  $ for $  \tR_\infty^n(z)   $ and $n +1$ for $\infty$. 
\end{lemma}
\emph{Proof:}\\

Take $ \psi =\cP \tR^n(z) \phi $, for some $\phi \in \cH^{\sigma_{n + 1}}$. 
 A simple calculation 
 leads us to   
\beq \label{ei.44} \begin{array}{l}
\la  A^{\sigma_{n + 1}, \sigma_n} \cdot \vv^{\sigma_n}\psi   |\;  A^{\sigma_{n + 1}, \sigma_n} \cdot \vv^{\sigma_n} \psi  \ra_{\cH^{\sigma_{n+1}}} \\\\ =
\sum_{j, q \in  \{1, 2, 3 \} } \la  (v^{\sigma_n}_{ q})^* v^{\sigma_n}_{ j}  \psi |\;    
(A^{\sigma_{n + 1}, \sigma_n}_{ j})^*A^{\sigma_{n + 1}, \sigma_n}_{ q} \psi   \ra_{\cH^{\sigma_{n+1}}} \\\\ 
 +  \sum_{j, q \in  \{1, 2, 3 \} }  \la v^{\sigma_n}_{ j}  \psi |\;  
   (A^{\sigma_{n + 1}, \sigma_n}_{ j})^{*}  [ A^{\sigma_{n + 1}, \sigma_n}_{ q},  v^{\sigma_n}_{ q}   ] \psi  \ra_{\cH^{\sigma_{n+1}}}\\\\  
 +  \sum_{j, q \in  \{1, 2, 3 \} }  \la v^{\sigma_n}_{ j}  \psi |\;  
    [ (A^{\sigma_{n + 1}, \sigma_n}_{ j})^*,  v^{\sigma_n}_{ q} ]A^{\sigma_{n + 1}, \sigma_n}_{ q} \psi  \ra_{\cH^{\sigma_{n+1}}}
\period
\end{array}  \ene
Eq. (\ref{L.ei.8.e1}) follows from Lemmata~\ref{L.ei.4}, \ref{L.ei.5}, \ref{L.ei.6}, \ref{L.ei.7} and Eq. (\ref{ei.44}).
For (\ref{L.ei.8.e2}) we take the commutator and use Lemma~\ref{L.ei.4}. 

\QED

\begin{lemma}\label{L.ei.8.1} 
Suppose that $ z \in \tilde{\cE}_n \setminus \{ E_n \}$, $\iota \in \{ 0, 1 \}$.  \\
Denote by $\cP^0 = 1$ and $\cP^1 = \cP$. For every $\rho \leq e_{0} - 1$ and every  $h \in \CC$ satisfying 
the properties of $\theta$ in Assumptions~\ref{parameters-basis} 
\beq \label{L.ei.8.1.e1}  \begin{array}{l}
\|  (1 + |x|^2)^{-\iota}  A^{\sigma_{n + 1}, \sigma_n} \cdot \vv^{\sigma_n}    \frac{1}{\overset{n+1}{H}_0 (h) - \rho}  \| 
 \leq 200   C_{\ref{sym-e3}}
 C_{\ref{an.0.16.2}} \\\\ \cdot 
 ( 1 +  b_{1/4})\cG_\alpha(\sigma_n, 1) 
    \comma
\end{array} \ene 
\beq \label{L.ei.8.1.e2}  \begin{array}{l}
\| (1 + |x|^2)^{-\iota}  \vv^{\sigma_n}\cdot  A^{\sigma_{n + 1}, \sigma_n}  \frac{1}{\overset{n+1}{H}_0(h) - \rho}   \| 
  \leq  
 240 C_{\ref{sym-e3}} C_{\ref{an.0.16.2}} \\\\ ( 1 +  b_{1/4}) \cG_\alpha(\sigma_n, 1)  \period
\end{array} \ene 
\end{lemma}
The estimates are also valid for $ z \in \cE_{(n, \infty)} $ if we replace $ n+ 1  $ for $  \infty   $. 

\noindent \emph{Proof:}\\
The proof is similar as the one of Lemma~\ref{L.ei.8}, here we use Lemma~\ref{L.ei.2.1} (we used also that $C_{\ref{sym-e3}}\geq 1$ ).
\QED

\begin{theorem}\label{P.ei.1}

Denote by $\cP^0 = 1$ and $\cP^1 = \cP$. 
Suppose that $ z \in \tilde{\cE}_n \setminus \{ E_n \}$ and that $ |z - E_n | \geq \frac{(|\sin(\nu)|/2) \sigma_{n + 1}}{10 }  $. \\
Then for every  $\iota \in \{ 0, 1 \}$
\beq \label{P.ei.1.e1}  \begin{array}{l}
\|(1 + |x|^2)^{- \iota } W_{n + 1}^{n} \cP^{1 - \iota} \tR^n(z)  \| 
 \leq C_{\ref{P.ei.1.e2}} \frac{\alpha^{3/2}}{\mathcal{B}} \boC_{\ref{il.2.1}}^{n + 1} \sigma_n \comma 
\end{array} \ene 
\beq \label{P.ei.1.e1prima}  \begin{array}{l}
\| W_{n + 1}^{n} \cP^\iota \tR^n(z)  \| 
 \leq C_{\ref{P.ei.1.e2}} \frac{\alpha^{3/2}}{\mathcal{B}} \boC_{\ref{il.2.1}}^{n + 1} \comma
\end{array} \ene

where $  C_{\ref{P.ei.1.e2}}  $ is a constant independent of $n$ and $\mathcal{B}$, given by the formula 
\beq \label{P.ei.1.e2}\begin{array}{l}
C_{\ref{P.ei.1.e2}}: = 10^6 \frac{ C_{\ref{sym-e3}} C_{\ref{an.0.16.2}}(b_{1/4} + 1)  (1 + |e_0'|)}{\sin(\nu)^2} 
\end{array}
\ene
The estimates are also valid for $ z \in \cE_{(n, \infty)} $ if we replace $ \tR^n(z)  $ for $  \tR_\infty^n(z)   $ and $n +1$ for $\infty$. 

\end{theorem}

\emph{Proof:}\\
The proof of (\ref{P.ei.1.e1}) follows from (\ref{il.11}) and Lemmata~\ref{L.ei.4} and  \ref{L.ei.8}. We used also that the constants 
$ C_{\ref{sym-e3}}$ and $ C_{\ref{an.0.16.2}} $ are larger than $1$ and that $\sigma_0 \leq \frac{1}{2}$ (see also (\ref{sym-e4})). 
\\
The proof of (\ref{P.ei.1.e1prima}) is similar. \\
\QED

\subparagraph{Estimates for the Feshbach Map} $\\$

In this subparagraph we estimate the Feshbach map of our Hamiltonians. The key ingredient is Theorem~\ref{L.ei.10} which is the analogous result of Theorem~\ref{L.neri.2} and Theorem~\ref{P.ei.1}. Theorem~\ref{L.ei.10} (and Neumann expansions) imply the invertibility  of the Feshbach map applied to the Hamiltonian. 

\begin{proposition}\label{P.ei.2}
For any $z \in \tilde{\cE}_n $,
\beq \label{P.ei.2.e1}  \begin{array}{l}
\| \cP W_{n+1}^{n} \frac{1}{\bcP  \overset{n + 1}{H}(\theta)    \bcP - z  }  \| 
 \leq C_{\ref{P.ei.2.e2}} \alpha^{3/2} \sigma^{3/2}_n  \comma \\\\
\| \cP W_{n+1}^{n} \frac{1}{\bcP  \widetilde{H}^{n}    \bcP - z  }  \| 
 \leq C_{\ref{P.ei.2.e2}} \alpha^{3/2} \sigma^{3/2}_n \comma
\end{array} \ene 
where $ C_{\ref{P.ei.2.e2}}$, 
\beq \label{P.ei.2.e2}\begin{array}{l}
C_{\ref{P.ei.2.e2}}: =  10^6 \frac{ C_{\ref{sym-e3}} C_{\ref{an.0.16.2}}^2(b_{1/4} + 1)  (1 + |e_0'|)(1 + 1/\delta)}{|\sin(\nu)|} 
 \period 
\end{array}
\ene

The estimates are also valid for $ z \in \cE_{(n, \infty)} $ if we replace $ n+1  $ for $  \infty   $ in the first equation in (\ref{P.ei.2.e1}).
\end{proposition}

\emph{Proof:}\\
We prove the first inequality in (\ref{P.ei.2.e1}), the other can be proved in the same way.\\
The proof follows from (\ref{il.11}), Lemmata  \ref{L.il.fm.2.1},  \ref{L.ei.2.1},  \ref{L.ei.8.1} and the following estimate, 
\beq \begin{array}{l} \label{extra}
\| \cP W_{n + 1}^{n} \frac{1}{\bcP  \overset{n + 1}{H}    \bcP - z  }  \|  \leq ( \| \cP ( 1 + |x|^2)   \| + 1) \\\\ \cdot  
\|( 1 + |x|^2)^{-1}  W_{n + 1}^{n}   \frac{1}{  \overset{n + 1}{H}_0(\theta)     -( e_{0} - 1)  }  \| \\\\
\cdot \|  ( \overset{n+1}{H}_0(\theta)     - ( e_{0} - 1) )  \frac{ 1}{\bcP  \overset{n + 1}{H}    \bcP - z  }  \|\period
\end{array}
\ene
Notice that the term $  ( \| \cP ( 1 + |x|^2)   \| + 1) $ is already considered in the constant $C_{\ref{sym-e3}}$.

\QED

\begin{lemma}\label{L.ei.9}
Suppose that $ z \in \tilde{\cE}_n \setminus \{ E_n \}$ and that $ |z - E_n | \geq \frac{(|\sin(\nu)|/2) \sigma_{n + 1}}{10 }  $. \\
Then the following holds true.
\beq \label{L.ei.9.e1}
\|  W^{\sigma_{n+1}} \cP \tilde{R}^n(z)  \| \leq  C_{\ref{L.ei.9.e2}} \frac{ \alpha^{3/2}}{\mathcal{B}} 
\boC_{\ref{il.2.1}}^{n+1}\frac{1}{\sigma_n}
\ene
where
\beq \label{L.ei.9.e2} \begin{array}{l}
 C_{\ref{L.ei.9.e2}} : =  384 C_{\ref{i.13.4}} ( 1 + (2 + \frac{6}{(|\sin(\nu)|/2)})\\\\
 \cdot (1 + \frac{10(|\sin(\nu)|  + 3|e_{0}'| + 1)}{(|\sin(\nu)|/2) })  )
\end{array}
\ene

The estimate is also valid for $ z \in \cE_{(n, \infty)} $, if we replace $ \tR^n(z)  $ for $  \tR_\infty^n(z)   $ and $n +1$ for $\infty$. 

\end{lemma}
\emph{Proof:}\\

Take $\rho = 2 e_{0} - 1$ and use (\ref{L.i.1.e1}) to get (remember that $ H^{\sigma_{n + 1}}_{0}(0) $ is self-adjoint)
\beq \label{ei.45}
\|  W^{\sigma_{n + 1}} \cP \frac{1}{ \overset{n + 1}{H}_{0}(0) - \rho} \| \leq 24 C_{\ref{i.13.4}} \alpha^{3/2} \period  
\ene
As in the proof of (\ref{L.an.2.e1}) we have (notice that $ \widetilde{H}^n(0)  $ is self-adjoint and therefore we can use  functional calculus)
\beq \label{L.an.4.e1.0.0}
\|  (\overset{n + 1}{H}_{0}(0) - \rho)\frac{1}{ \widetilde{H}^n(0) - \rho }   \| \leq 2
\ene
It follows from the proof of Corollary~\ref{L.an.4} that
\beq \label{ei.46}
\|(\overset{n + 1}{H}_0(0) - \rho)\frac{1}{ \widetilde{H}^n(\theta) - \rho }  \| \leq 8 \period
\ene
We obtain finally 
\beq \label{ei.47.0.0}
\|  W^{\sigma_{n + 1}} \cP \frac{1}{ \widetilde{H}^n(\theta) - \rho }  \| \leq 384 C_{\ref{i.13.4}} \alpha^{3/2} \period  
\ene
Using (\ref{in.8.1}), Lemma~\ref{L.ei.1}, (\ref{ei.42}) and that $\sigma_{n + 1} = \mathcal{B} \sigma_n$ we obtain 
\beq \label{ei.48}\begin{array}{l}
\|  W^{\sigma_{n + 1}} \cP \frac{1}{ \widetilde{H}^n(\theta) - z }  \| \leq  
384 C_{\ref{i.13.4}} \alpha^{3/2} \\\\ \cdot  
(1 +  \| \frac{((z -E_n) + (E_n  - \rho)) (\widetilde{P_n} + \overline{ \widetilde{P_n} })}{ \widetilde{H}^n(\theta) - z }\|) \\\\
\leq 384 C_{\ref{i.13.4}} \alpha^{3/2} ( 1 + (2 + \frac{2}{(|\sin(\nu)|/2)} (\boC_{\ref{il.2.1}}^{n+1} + 2)) \\\\ 
\cdot (1 + \frac{10( |\sin(\nu)| \sigma_0 + 3|e_{0}'| + 1)}{(|\sin(\nu)|/2) \mathcal{B} \sigma_n})  )
\end{array}
\ene
Finally (\ref{L.ei.9.e1}) follows from (\ref{ei.48}).

\QED

\begin{theorem}\label{L.ei.10}
Suppose that $ z \in \tilde{\cE}_n \setminus \{ E_n \}$ and that $ |z - E_n | \geq \frac{(|\sin(\nu)|/2) \sigma_{n + 1}}{10 }  $. \\
Then the following holds true.\\

\beq \label{L.ei.10.e1}\begin{array}{l}
\|(\cF_\cP (\overset{n + 1}{H} - z) - \cF_\cP( \widetilde{H}^n - z  ))\frac{1}{ \cF_\cP ( \widetilde{H}^n - z    ) } \| \\\\ \leq
C_{\ref{L.ei.10.e2}} \sigma_n^{1/2}\frac{\alpha^{3/2}}{\mathcal{B}}\boC_{\ref{il.2.1}}^{n + 1} \comma
\end{array}
\ene
where
\beq \label{L.ei.10.e2}\begin{array}{l}
C_{\ref{L.ei.10.e2}} := 4 C_{\ref{P.ei.1.e2}}  +  
2 C_{\ref{P.ei.2.e2}}    C_{\ref{L.ei.9.e2}} 
   \\\\ +
8(1 + 4e^{-2}/\bob^2)   C_{\ref{T.eb.1.e2}}     
 C_{\ref{neri.15.2.1}}   
    C_{\ref{P.ei.2.e2}} 
  C_{\ref{L.ei.9.e2}}  \\\\  
+ 4 C_{\ref{neri.15.2.1}} C_{\ref{P.ei.1.e2}} 
\period
\end{array}
\ene

The estimate is also valid for $ z \in \cE_{(n, \infty)} $, if we replace $ \tH^n(z)  $ for $  \tH_\infty^n(z)   $ and $n +1$ for $\infty$.

\end{theorem}

\emph{Proof: }\\
First we notice that by (\ref{L.il.fm.3}),  $ \frac{1}{ \cF_\cP ( \widetilde{H}^n - z    ) } = \cP  \frac{1}{ \widetilde{H}^n - z     } \cP  $.\\
We compute 
\beq \label{ei.49}\begin{array}{l}
\|(\cF_\cP (\overset{n + 1}{H} - z) - \cF_\cP( \widetilde{H}^n - z  ))\cP  \frac{1}{ \widetilde{H}^n - z     } \cP\| \\\\ 
\leq 
\| \cP W_{n + 1}^{n}\cP  \frac{1}{ \widetilde{H}^n - z     } \cP  \|  + \|   \cP W_{n + 1}^{n}    \bcP   ( \bcP  \overset{n + 1}{H}  \bcP - z  )^{-1}  \|
\\\\  \cdot \|  W^{\sigma_{n + 1}} \cP   \frac{1}{ \widetilde{H}^n - z     } \cP \| \\\\ +
\|   \cP  e^{\beta  \la  x \ra}   \| \cdot 
 \|  e^{-\beta  \la  x \ra}   W^{\sigma_n}    \bcP   ( \bcP  \overset{n + 1}{H}  \bcP - z  )^{-1}  e^{\beta  \la  x \ra} \| \\\\ \cdot  
 \|   e^{- \beta  \la  x \ra}   W_{n + 1}^{n}     \frac{1}{ \bcP \widetilde{H}^n \bcP - z     }     \| \\\\  
 \cdot \|  W^{\sigma_{n + 1}} \cP   \frac{1}{ \widetilde{H}^n - z     } \cP \| +
 \|   \cP W^{\sigma_n}    \bcP    \frac{1}{ \bcP \widetilde{H}^n \bcP - z     }     \| 
\\\\  \cdot \|  W_{n + 1}^{n} \cP   \frac{1}{ \widetilde{H}^n - z     } \cP \|\\\\ 

\end{array}
\ene
\begin{equation*}
\begin{array}{l}
\leq 4 C_{\ref{P.ei.1.e2}} \frac{\alpha^{3/2}}{\mathcal{B}} \boC_{\ref{il.2.1}}^{n + 1} \sigma_n +  \\\\
2 C_{\ref{P.ei.2.e2}} \alpha^{3/2}\sigma_n^{3/2}   C_{\ref{L.ei.9.e2}} \frac{ \alpha^{3/2}}{\mathcal{B}} 
\boC_{\ref{il.2.1}}^{n+1}\frac{1}{\sigma_n}    \\\\ +
\|   \cP   e^{\beta  \la  x \ra}   \|      \\\\ \cdot 
 \|  e^{-\beta  \la  x \ra}  W^{\sigma_n}    \bcP   ( \bcP  \overset{n + 1}{H}  \bcP - z  )^{-1}  e^{\beta  \la  x \ra} \| \\\\ \cdot  
 \|   e^{- \beta  \la  x \ra} (1 + |x|^2) \|  \cdot  \|  (1 + |x|^2)^{-1}  W_{n + 1}^{n}    \frac{1}{ \bcP \widetilde{H}^n \bcP - z     }     \|\\\\
\cdot 2 C_{\ref{L.ei.9.e2}} \frac{ \alpha^{3/2}}{\mathcal{B}} 
\boC_{\ref{il.2.1}}^{n+1}\frac{1}{\sigma_n}   
\\\\
+ 4 \alpha^{3/2} C_{\ref{neri.15.2.1}}
\cdot C_{\ref{P.ei.1.e2}} \frac{\alpha^{3/2}}{\mathcal{B}} \boC_{\ref{il.2.1}}^{n + 1} \sigma_n  \comma
\end{array}
\end{equation*}
where we used (\ref{L.il.fm.1.e2}), (\ref{P.ei.1.e1}), (\ref{P.ei.2.e1}) and (\ref{L.ei.9.e1}).\\
We obtain (\ref{L.ei.10.e1}) from the proof of (\ref{P.ei.2.e1}) and noting that (see (\ref{geh.7.m1})) 
\beq \begin{array}{l} \label{ei.50}
\|  e^{- \beta  \la  x \ra} (1 +  |x|^2)\| \leq \|  e^{-  |x| } (1 +  |x|^2)  \| 
 \leq 1 + 4e^{-2}/\bob^2  \period  
\end{array}
\ene
and that 
\beq \label{ei.51}\begin{array}{l}
\|  e^{-\beta  \la  x \ra}  W^{\sigma_n}(\theta, \alpha, 0)    \bcP   ( \bcP  \overset{n + 1}{H}(\theta,\alpha, 0)  \bcP - z  )^{-1}  e^{\beta  \la  x \ra} \| \\\\
= \| W^{\sigma_n}(\theta, \alpha, \beta) (  \overline{P_{\at,1}}(\theta,\alpha, \beta)H^{\sigma_{n+1}}(\theta, \alpha, \beta) \overline{P_{\at,1}}(\theta, \alpha, \beta) - z   )\| \\\\
\leq 4 \alpha^{3/2} C_{\ref{neri.15.2.1}},
\end{array}
\ene
see the proof of (\ref{L.il.fm.1.e2}). Finally we take $\beta = \bob$ and use (\ref{T.eb.1.e2.1}).

\QED
 
\begin{corollary}\label{C.ei.1}
Suppose that $ z \in \tilde{\cE}_n \setminus \{ E_n \}$ and that $ |z - E_n | \geq \frac{(|\sin(\nu)|/2) \sigma_{n + 1}}{10 }  $. 
 Then
\beq \label{C.ei.1.e1} \begin{array}{l}
\|  \cF_\cP (\overset{n + 1}{H} - z)^{-1}   \| \leq  \frac{64}{ (|\sin(\nu)|/2) }\boC_{\ref{il.2.1}}^{n + 1}   \frac{1}{(|\sin(\nu)|/2) \sigma_{n + 1}  + |z - E_n|}
  \\\\\leq   \frac{64}{ (|\sin(\nu)|/2)^2 }\boC_{\ref{il.2.1}}^{n + 1}  \frac{1}{\sigma_{n + 1}} \period
\end{array}    
\ene
This estimate is also valid for $ z \in \cE_{(n, \infty)} $, if we replace $ n + 1  $ for $  \infty  $. 

\end{corollary}

\emph{Proof:}

From (\ref{il.2.1}) and (\ref{in.10.1.3}) it follows that $C_{\ref{L.ei.10.e2}} \sigma_n^{1/2}\frac{\alpha^{3/2}}{\mathcal{B}}\boC_{\ref{il.2.1}}^{n + 1} < \frac{1}{2}$.

We use Theorem~\ref{L.ei.10} and Neumann Series to obtain  
\beq \label{ei.52}\begin{array}{l}
\|  \cF_\cP (\overset{n + 1}{H} - z)^{-1}   \| \leq  2 \|   \cF_\cP (\widetilde{H}^n - z)^{-1}  \|  =
 2  \|   \cP (\widetilde{H}^n - z)^{-1} \cP  \| \\\\ \leq 
 \frac{16}{(|\sin(\nu)|/2)}(\boC_{\ref{il.2.1}}^{n + 1}  + 2) \frac{1}{(|\sin(\nu)|/2) \sigma_{n + 1}  + |z - E_n|} + 16 \frac{1}{ |z - E_n| }\\\\ \leq 
 \frac{16}{(|\sin(\nu)|/2)}(\boC_{\ref{il.2.1}}^{n + 1}  + 2) \frac{1}{(|\sin(\nu)|/2) \sigma_{n + 1}  + |z - E_n|}  \\\\ + 320  \frac{1}{(|\sin(\nu)|/2) \sigma_{n + 1}  + |z - E_n|  }
 \comma
\end{array}
\ene
where we used (\ref{e.p.eeh.3}), Lemma  \ref{L.il.fm.3}, (\ref{in.8.1}), (\ref{L.ei.1.e1}) and that $|\nu| \leq \frac{1}{120} $.  

\QED

\subparagraph{Estimates for the Resolvent} $\\$

\begin{theorem}\label{C.ei.2}

Suppose that $ z \in \tilde{\cE}_n \setminus \{ E_n \}$ and that $ |z - E_n | \geq \frac{(|\sin(\nu)|/2) \sigma_{n + 1}}{10 }  $.  Then

\beq \label{C.ei.2.e1}\begin{array}{l}
\|   (\overset{n + 1}{H} - z)^{-1}    \| \leq \\\\
\min \Big( (2 +   4 \alpha^{3/2} C_{\ref{neri.15.2.1}} )^2  \frac{64}{ (|\sin(\nu)|/2)^2 }\boC_{\ref{il.2.1}}^{n + 1}  \frac{1}{\sigma_{n + 1}} 
+ 6\frac{32 C_{\ref{an.0.16.2}}}{|\sin (\nu)| \delta}\\\\ 
,   (2 +   4 \alpha^{3/2} C_{\ref{neri.15.2.1}} )^2   \frac{64}{ (|\sin(\nu)|/2) }\boC_{\ref{il.2.1}}^{n + 1}   \frac{1}{(|\sin(\nu)|/2) \sigma_{n + 1}  + |z - E_n|} \\\\
+     6\frac{40 |e_{0}| C_{\ref{an.0.16.2}}}{|\sin (\nu)| \delta} 
\frac{16 |e_{0}|}{ \delta |\sin(\nu)|}  \frac{20}{ (|\sin(\nu)|/2) \sigma_{n + 1}  +  |z - E_n|}\Big)\period
\end{array}
\ene 

This estimate is also valid for $ z \in \cE_{(n, \infty)} $ if we replace $ n + 1  $ for $  \infty  $. 
\end{theorem}

\emph{Proof:}
We use (\ref{L.il.fm.1.e1}), (\ref{L.il.fm.1.e2}), (\ref{L.il.fm.3.e2}), Corollary~\ref{C.ei.1} and the following estimates.\\

Given two points $ z_1 $ and $z_2$ in the complex plane and a real number $\ell > 1$, the set of points $ z $
such that $ |z - z_1| = \frac{1}{\ell} | z - z_2 |$, is a circle with center at $ z_1 + \frac{1}{\ell^2 - 1} ( z_1 - z_2)  $ and
radius $ \frac{\ell}{\ell^2 - 1} |   z_1 - z_2 |  $.  It follows from (\ref{e.p.eeh.1}), (\ref{neri.0.1}), (\ref{in.10.1.3}), (\ref{in.6}), (\ref{in.7})  
 and (\ref{ei.5}) that $\dist(e_0, \tilde{\cE}_n ) \geq \delta |\sin(\nu)| /4  $, which implies that if $( \frac{\ell}{\ell^2 - 1} +   \frac{1}{(\ell^2 - 1)} )|e_{0}|  \leq \delta |\sin(\nu)| /4   $
then $ | z - E_n|  \leq \ell |z - e_{0}|$, for $z \in \tilde \cE_n$. We take $\ell = \frac{16 |e_{0}|}{ \delta |\sin(\nu)|}   $. We have
\beq   \label{ei.52.1} \begin{array}{l}
\frac{1}{ |z - e_{0}|} \leq \frac{16 |e_{0}|}{ \delta |\sin(\nu)|}  \frac{1}{ |z - E_n|} 
\leq \frac{16 |e_{0}|}{ \delta |\sin(\nu)|}  \frac{20}{ (|\sin(\nu)|/2) \sigma_{n + 1}  +  |z - E_n|} \period 
\end{array}
\ene

\QED

\begin{lemma}\label{L.ei.11}

Suppose that $ z \in \tilde{\cE}_n $ and that $ |z - E_n | \geq \frac{(|\sin(\nu)|/2) \sigma_{n + 1}}{10 }  $. \\
 Then
\beq \label{L.ei.11.e1} \begin{array}{l}
 \|(\overset{n + 1}{H} - z)^{-1}  -  (\widetilde{H}^{n} - z)^{-1} \| \\\\ \leq  C_{\ref{L.ei.11.e2}} \frac{ \alpha^{3/2}}{\mathcal{B}}
\boC_{\ref{il.2.1}}^{2n + 2} \frac{\sigma_n^{1/2}}{\sigma_{n + 1}}
+  6\frac{64 C_{\ref{an.0.16.2}} }{|\sin(\nu)|\delta}
\comma
\end{array}
\ene
where
\beq \label{L.ei.11.e2} \begin{array}{l}
 C_{\ref{L.ei.11.e2}} : = 
 2 \big( C_{\ref{T.eb.1.e2}}
  (1 + 4e^{-2}/\bob^2) 4  C_{\ref{neri.15.2.1}}+ 1 \big) C_{\ref{P.ei.2.e2}} 
\cdot ( 2 + 4  C_{\ref{neri.15.2.1}} )\cdot  \frac{64}{ (|\sin(\nu)|/2)^2 }   \\\\ 
 + ( 2 + 4  C_{\ref{neri.15.2.1}} )^2  \frac{64}{ (|\sin(\nu)|/2)^2 } 
C_{\ref{L.ei.10.e2}} \period
\end{array}
\ene

\end{lemma}

\emph{Proof:}
First we notice 
\begin{flushleft}

\end{flushleft}
that 
\beq \label{ei.53} \begin{array}{l}
\| \cP W^{\sigma_{n + 1}}(  \bcP \overset{n + 1}{H} \bcP - z)^{-1} - \cP W^{\sigma_n}(  \bcP \widetilde{H}^{n} \bcP - z)^{-1} \| \\\\ \leq 
\|\cP W_{n + 1}^{n}  ( \bcP \overset{n + 1}{H} \bcP - z)^{-1}  \| \\\\ +  \|   \cP  e^{\beta  \la  x \ra}   \| \cdot 
 \|  e^{-\beta  \la  x \ra}   W^{\sigma_n}    \bcP   ( \bcP  \overset{n + 1}{H}  \bcP - z  )^{-1}  e^{\beta  \la  x \ra} \| \\\\ \cdot  
 \|   e^{- \beta  \la  x \ra}   W_{n + 1}^{n}    \frac{1}{ \bcP \widetilde{H}^n \bcP - z     }     \| \\\\  \leq 
\big( \|  \cP   e^{\beta  \la  x \ra}   \|
  (1 + 4e^{-2}/\bob^2)  4 \alpha^{3/2} C_{\ref{neri.15.2.1}}+ 1 \big) C_{\ref{P.ei.2.e2}} \alpha^{3/2} \sigma_n^{3/2} \comma
\end{array}
\ene
Where we used  (\ref{P.ei.2.e1}) (and its proof, see (\ref{extra})), (\ref{ei.50}) and (\ref{ei.51}).
Similarly we have 
\beq \label{ei.54} \begin{array}{l}
\|(  \bcP \overset{n + 1}{H} \bcP - z)^{-1} W^{\sigma_{n + 1}}\cP -  (  \bcP \widetilde{H}^{n} \bcP - z)^{-1} W^{\sigma_n}\cP \| \\\\ \leq 
\big( \|  \cP   e^{\beta  \la  x \ra}   \|
  (1 + 4e^{-2}/\bob^2)  4 \alpha^{3/2} C_{\ref{neri.15.2.1}}+ 1 \big) C_{\ref{P.ei.2.e2}} \alpha^{3/2} \sigma_n^{3/2}\period
\end{array}
\ene

Using (\ref{L.il.fm.1.e1}), (\ref{L.il.fm.1.e2}), (\ref{L.il.fm.3.e2}), Corollary~\ref{C.ei.1}  and (\ref{ei.53}-\ref{ei.54})  we get
\beq \label{ei.55} \begin{array}{l}
\|(\overset{n + 1}{H} - z)^{-1}  -  (\widetilde{H}^{n} - z)^{-1} \| \\\\ \leq 
\| \big[  \cP - (\bcP \widetilde{H}^{n} \bcP - z)^{-1} W^{\sigma_n}\cP \big] 
\cdot (\cF_\cP (\overset{n + 1}{H} - z )^{-1}    - \cF_\cP (\widetilde{H}^{n} - z )^{-1} ) \\\\
 \cdot \big[ \cP -    (\bcP \overset{n + 1}{H} \bcP - z)^{-1} W^{\sigma_{n + 1}}\cP  \big] \| \\\\
+ 2\big( \|  \cP   e^{\beta  \la  x \ra}   \|
  (1 + 4e^{-2}/\bob^2)  4 \alpha^{3/2} C_{\ref{neri.15.2.1}}+ 1 \big) C_{\ref{P.ei.2.e2}} \alpha^{3/2} \sigma_n^{3/2}\\\\
\cdot ( 2 + 4 \alpha^{3/2} C_{\ref{neri.15.2.1}} )\cdot  \frac{64}{ (|\sin(\nu)|/2)^2 }\boC_{\ref{il.2.1}}^{n + 1}  \frac{1}{\sigma_{n + 1}} 
\\\\
+ 6  \frac{64 C_{\ref{an.0.16.2}} }{|\sin(\nu)|\delta}\period
\end{array}
\ene
Eq. (\ref{L.il.fm.1.e2}), Theorem~\ref{L.ei.10} Corollary~\ref{C.ei.1} and the second resolvent equation imply that
\beq \label{ei.56} \begin{array}{l}
\| \big[  \cP - (\bcP \widetilde{H}^{n} \bcP - z)^{-1} W^{\sigma_n}\cP \big] 
\cdot (\cF_\cP (\overset{n + 1}{H} - z )^{-1}    - \cF_\cP (\widetilde{H}^{n} - z )^{-1} ) \\\\
 \cdot \big[ \cP -    (\bcP \overset{n + 1}{H} \bcP - z)^{-1} W^{\sigma_{n + 1}}\cP  \big] \| \\\\
\leq ( 2 + 4 \alpha^{3/2} C_{\ref{neri.15.2.1}} )^2  \frac{64}{ (|\sin(\nu)|/2)^2 }\boC_{\ref{il.2.1}}^{n + 1}  \frac{1}{\sigma_{n + 1}} \\\\
C_{\ref{L.ei.10.e2}} \sigma_n^{1/2}\frac{\alpha^{3/2}}{\mathcal{B}}\boC_{\ref{il.2.1}}^{n + 1}\period
\end{array}
\ene
Eq. (\ref{L.ei.11.e1}) follows from (\ref{T.eb.1.e2.1}) and (\ref{ei.55}-\ref{ei.56}).

\QED

\subparagraph{Estimates for the Projection $P_{n +1}$. Proof of (\ref{in.10.1.2}) for the induction step}  $\\$

\begin{theorem} \label{T.ei.1} 
The following estimate holds true:
\beq \label{T.ei.1.e1}
\| P_{n + 1} - \tilde{P}_n  \| \leq  C_{ \ref{T.ei.1.e2} }  (\boC_{\ref{il.2.1}}^2)^{n + 1}\sigma_n^{1/2} 
\comma
\ene 
where

\beq \label{T.ei.1.e2}
C_{ \ref{T.ei.1.e2} } : = (|\sin(\nu)|/2) ( C_{\ref{L.ei.11.e2}}   
+  \frac{64 C_{\ref{an.0.16.2}} }{|\sin(\nu)|\delta}  )
\comma
\ene

\end{theorem}

\emph{Proof:}
We take $\tilde{\gamma}_n $ as in (\ref{ei.8}). We compute using Lemma~\ref{L.ei.11}: 
\beq \label{ei.57} \begin{array}{l}
\| P_{n + 1} - \tilde{P}_n  \| \leq \frac{1}{2 \pi} \int_{\tilde{\gamma}_n} |(\overset{n + 1}{R}(z) -  \tilde{R}^n(z) )| \\\\ \leq
 (|\sin(\nu)|/2) C_{\ref{L.ei.11.e2}} \frac{ \alpha^{3/2}}{\mathcal{B}}
\boC_{\ref{il.2.1}}^{2n + 2} \sigma_n^{1/2}
+  (|\sin(\nu)|/2) \frac{64 C_{\ref{an.0.16.2}} }{|\sin(\nu)|\delta} \sigma_{n+1} \period
\end{array}
\ene 
\QED
\begin{remark}\label{r.ei.1}
As $  C_{\ref{T.ei.1.e2}} \leq \boC_{\ref{il.2.1}}  $, Theorem~\ref{T.ei.1} establish the induction step for (\ref{in.10.1.2}). 
It also follows the existence of $E_{n + 1}$ and its non degeneracy (see (\ref{in.10.1.3})). Together with Theorem~\ref{C.ei.2}, Theorem~\ref{T.ei.1}
implies that $E_{n + 1}$ is the only eigenvalue in $\widetilde \cE_n$. 

\end{remark}


\begin{theorem} \label{L.ei.22} 

For every $z  \in \tilde{\mathcal{E}}_n $
\beq \label{L.ei.12.e1} \begin{array}{l}
\|\overset{n + 1}{R}(z) \overline{ P}_{n + 1}  \| \leq  C_{ \ref{L.ei.12.e2} }  \boC_{\ref{il.2.1}}^{n + 1}   \frac{1}{ (|\sin(\nu)|/2) \sigma_{n + 1} + |z - E_n| }  
\comma
\end{array}
\ene 
where
\beq \label{L.ei.12.e2}\begin{array}{l}
C_{ \ref{L.ei.12.e2} } : = 
  18(2 +   4 \alpha^{3/2} C_{\ref{neri.15.2.1}} )^2   \frac{64}{ (|\sin(\nu)|/2) }\boC_{\ref{il.2.1}}^{n + 1}  \\\\
+    18 \frac{40 |e_0'| C_{\ref{an.0.16.2}}}{|\sin (\nu)| \delta} 
\frac{16 |e_0'|}{ \delta |\sin(\nu)|} \cdot 20
\period
\end{array}
\ene

\end{theorem}
\emph{Proof:}
By (\ref{T.ei.1.e1}), there is only one eigenvalue of $  \overset{n + 1}{H} $ in $ \tilde{\mathcal{E}}_n $. Then we have 
that $   \overset{n + 1}{R}(z) \overline{ P}_{n + 1}  $ is analytic in $ \tilde{\mathcal{E}}_n   $. For $ |z - E_n  | \geq \frac{(|\sin(\nu)|/2) \sigma_{n + 1}}{10}$, 
the result follows from  (\ref{C.ei.2.e1}) (Remark \ref{r.ei.1} imply that $\| P_{n+1}\| \leq 2$). For  $ |z - E_n  | \leq \frac{(|\sin(\nu)|/2) \sigma_{n + 1}}{10}$ it follows from the maximum modulus 
principle.

\QED

\subparagraph{Estimates for the Energy. Proof of (\ref{in.5}) and (\ref{in.9})  for the induction step. Completion of the Induction Step} $\\$

For any vector $\phi  \in \cH^{\sigma_k}$ and every $ k  \leq   \tilde{k} \leq \infty $ we identify 
\beq \label{ee.0}
\phi : =  \phi \otimes \Omega^{\sigma_{\tilde{k}}, \sigma_{k}}\period
\ene
We take a unit eigenvector ($\phi_\at $) of  $ P_{\at, 1}(0, \alpha, 0)   $. We define foe every $m\leq n+1$
\beq \label{ee.3}
\psi_m(\theta) := P_m \phi_\at \period 
\ene
By (\ref{e.p.eeh.3}), (\ref{in.10.1.3}), (\ref{in.10.1.6tilde}) and(\ref{in.10.1.2}) we have that for $ m \leq n + 1$
\beq \label{ee.4}
\|   \psi_m(\theta) -  \phi_{\at}  \| \leq \frac{1}{8} + \frac{1}{120} + \frac{1}{100} \sum_{i = 0}^{n-1} (\frac{1}{100})^i\comma
\ene 
which implies that
\beq \label{ee.4.1} \begin{array}{l}
1 - \frac{1}{8} - \frac{1}{120} - \frac{1}{100}\sum_{i = 0}^{n-1} (\frac{1}{100})^i  \leq   \|   \psi_n(\theta)  \| \\\\ 
\leq 1 + \frac{1}{8} + \frac{1}{120} +\frac{1}{100} \sum_{i = 0}^{n-1} (\frac{1}{100})^i\comma
\end{array}
\ene 
and this in turn implies
\beq\label{ee.5} \begin{array}{l}
|\la P_{n+1} \psi_n(\theta) | \;    \phi_{\at} \ra | \geq - |\la (P_{n+1} - P_n)\psi_n(\theta) | \;    \phi_{\at} \ra |+
|\la  \psi_n(\theta) | \;    \phi_{\at} \ra | \\\\
 \geq 1 - \big(   \frac{1}{8} + \frac{1}{120} + \frac{1}{100}  \sum_{i = 0}^{m-1} (\frac{1}{100})^i  \big) - \frac{2}{100}
\end{array}
\ene
and 
\beq\label{ee.6}\begin{array}{l}
|\la  \psi_n(\overline{\theta}) | \; P_{n +1} \psi_{n}(\theta)   \ra| \\\\
\geq  |  \la   \phi_{\at}  |\; P_{n+1}  \psi_{n}(\theta)   \ra  | 
 -  |\la  \psi_n(\overline{\theta}) - \phi_{\at} |\;  P_{n+1} \psi_{n }(\theta)   \ra| \\\\
\geq  1 - \big(   \frac{1}{8} + \frac{1}{120} +  \frac{1}{100} \sum_{i = 0}^{n} (\frac{1}{100})^i  \big) - \frac{2}{100}\\\\ - 
(1+ \frac{1}{100}) \big(  1 +  \frac{1}{8} + \frac{1}{120} +\frac{1}{100} \sum_{i = 0}^{n} (\frac{1}{100})^i \big)\\\\ \cdot 
\big(  \frac{1}{8} + \frac{1}{120} + \frac{1}{100} \sum_{i = 0}^{n-1} (\frac{1}{100})^i  \big)\period
\end{array}
\ene
\begin{theorem}\label{T.ee.1}
The following inequality holds true:
\beq \label{T.ee.1.e1}
|E_{n + 1} - E_n| \leq C_{\ref{T.ee.1.e2}}  \sigma_n^2\boC_{\ref{il.2.1}}^{n + 1} \alpha^{3/2} 
\ene

\beq \label{T.ee.1.e2}
C_{\ref{T.ee.1.e2}}  : =   C_{\ref{P.ei.1.e2}} \frac{(|\sin(\nu)|/2) }{ 10} \cdot 2(1 + 4e^{-2}/\bob^2)  \cdot
4 \cdot  24 C_{\ref{T.eb.1.e2}}  \period
\ene

\end{theorem}

\emph{Proof:}
We have that 
\beq \label{ee.7} \begin{array}{l}
E_{n + 1} = \frac{\la \psi_n(\overline{\theta})   |\;  \overset{n + 1}{H} P_{n + 1}
 \psi_n(\theta)    \ra }
{  \la \psi_n(\overline{\theta})  |\;  
P_{n + 1}  \psi_n(\theta)    \ra   } \\\\ = 
 \frac{\la ( (\widetilde{H}^{n+1})^* + (W_{n + 1}^{n}(\theta))^* ) \psi_n(\overline{\theta})  |\;   P_{n + 1}
 \psi_n(\theta)   \ra }
{  \la \psi_n(\overline{\theta})  |\;  
P_{n + 1} \psi_n(\theta)    \ra   } \\\\ = 
E_n + \frac{\la e^{- \beta \la x  \ra }  W_{n + 1}^{n}(\overline{\theta})\psi_n(\overline{\theta})   |\;  
e^{ \beta \la x  \ra } P_{n + 1}
 \psi_n(\theta)   \ra }
{  \la \psi_n(\overline{\theta}) |\;  
 P_{n + 1} \psi_n(\theta)   \ra   } \period
\end{array}
\ene
now we estimate for any $z$ such that $|z - E_n(\overline{\theta})| = \frac{(|\sin(\nu)|/2) \sigma_{n + 1}}{ 10}$
\beq \label{ee.8} \begin{array}{l}
\| e^{- \beta \la x  \ra }   W_{n + 1}^{n}(\overline{\theta})\psi_n(\overline{\theta})    \| \\\\  =
| z - E_n(\overline{\theta}) | \| e^{- \beta \la x  \ra }   W_{n + 1}^{n}(\overline{\theta}) \frac{1}{ \tilde{H}^n(\overline{\theta})  - z  } 
\psi_n(\overline{\theta})   \| \\\\ \leq 
\frac{(|\sin(\nu)|/2) \sigma_{n + 1}}{ 10} \| e^{- \beta \la x  \ra }(1 + |x|^2) \| 
C_{\ref{P.ei.1.e2}} \frac{\alpha^{3/2}}{\mathcal{B}} \boC_{\ref{il.2.1}}^{n + 1} \sigma_n  \|   \psi_n(\overline{\theta})  \|  \\\\ \leq
\sigma_n^2 \boC_{\ref{il.2.1}}^{n + 1} \alpha^{3/2}    C_{\ref{P.ei.1.e2}} \frac{(|\sin(\nu)|/2) }{ 10} \cdot 2(1 + 4e^{-2}/\bob^2) \comma
\end{array}
\ene
where we used (\ref{P.ei.1.e1}), (\ref{ei.50}) and  (\ref{ee.4.1}). \\\\

(\ref{T.il.fm.1.e1}), (\ref{ee.4.1}), (\ref{ee.5}),  (\ref{ee.7}) and (\ref{ee.8}) imply 
\beq \label{ee.9} \begin{array}{l}
| E_{n + 1} - E_n | \leq  \sigma_n^2 \boC_{\ref{il.2.1}}^{n + 1} \alpha^{3/2}    C_{\ref{P.ei.1.e2}} \\\\ \cdot 
\frac{(|\sin(\nu)|/2) }{ 10} \cdot 2(1 + 4e^{-2}/\bob^2)  \cdot
4 \cdot  24 C_{\ref{T.eb.1.e2}} \period
\end{array}
\ene
Eq. (\ref{T.ee.1.e1}) follows from (\ref{ee.9}).

\QED

\begin{remark}\label{r.ee.1}
As $  C_{\ref{T.ee.1.e2}} < \boC_{\ref{il.2.1}}  $, Theorem~\ref{T.ee.1} establish the induction step for  (\ref{in.5}). 
\end{remark}

\begin{theorem} \label{T.Induction-Step}
For every $z  \in \tilde{\mathcal{E}}_n $
\beq \label{T.I.S.1} \begin{array}{l}
\| \overset{n+1}{R}(z) \overline{ P}_{n + 1}  \| \leq C_{ \ref{T.I.S.2} } \boC_{\ref{il.2.1}}^{n + 1}   \frac{1}{ (|\sin(\nu)|/2) \sigma_{n + 1} + |z - E_{n + 1}| }  
\comma
\end{array}
\ene 
where, 
\beq \label{T.I.S.2} 
C_{ \ref{T.I.S.2} } : =  \frac{10}{9} C_{ \ref{L.ei.12.e2} } \period
\ene

\end{theorem}

\emph{Proof:}
By (\ref{in.10.1.3}) and (\ref{T.ee.1.e1}) we have that
\beq \label{I.S.1}
|E_{n + 1} - E_n| \leq \frac{1}{10} (|\sin(\nu)|/2) \sigma_{n}\comma
\ene
thus
\beq \label{I.S.2}
 \frac{1}{ (|\sin(\nu)|/2) \sigma_{n + 1} + |z - E_{n }| }  \leq  \frac{10}{9} \frac{1}{ (|\sin(\nu)|/2) \sigma_{n + 1} + |z - E_{n +1 }| }   \period
\ene
(\ref{T.I.S.1}) follows from    (\ref{L.ei.12.e1})  and   (\ref{I.S.2}).

\QED

\begin{remark}\label{r.ee.2}
Notice that (\ref{in.10.1.3}) implies that $ \sigma_n^2 \boC_{\ref{il.2.1}}^{n +2 } \alpha^{3/2} \leq \frac{1}{10} \sin(\nu) \sigma_{n +1}$ and this in turn, together with (\ref{in.5}), implies that $\cE_{n+1} \subset \tilde \cE_n$ (see (\ref{in.6}) and (\ref{ei.5})).  

As $  C_{\ref{T.I.S.2}} < \boC_{\ref{il.2.1}}  $, Theorem~\ref{T.Induction-Step} establish the induction step for  (\ref{in.9}). 
This Remark together with Remark \ref{r.ei.1} and Remark \ref{r.ee.1} complete the proof of the induction step (see the Remarks on Section 
\ref{induction-hypothesis}).  

\end{remark}

\subsection{Existence of Resonances for the non-Regularized Hamiltonian}\label{Resonances} $\\$

In this section we suppose that Assumptions~\ref{parameters-basis} are valid and that Definition \ref{sigmas-sequence} is satisfied. \\
The resonances are eigenvalues of $H^{0}(\theta) $ (see \ref{icfh.1}) with imaginary part strictly negative. Here we prove the existence of such eigenvalues.  \\
We use notation (\ref{ei.1}) and identify  
$$H^0(\theta) = H^{0, \infty}(\theta) = \overset{\infty}{H}. 
$$

It is obvious from (\ref{in.10.1.3}) and (\ref{in.5}) that the sequence of eigenvalues 
$\{ E_n \}_{n \in \mathbb{N} }$ converges. We define 
\beq \label{Einfty}
E_\infty := \lim_{n \to \infty }E_n. 
\ene
Let $\phi_\at$ be a unit eigenvector of the atom Hamiltonian $H_\at(0, \alpha, 0)$ corresponding to the first excited state $e_1$.
We define the sequence of vectors 
\beq \label{psin}
\psi_n := P_n \phi_\at.
\ene
In the previous equation we identify vectors in  $\cH_\at$ and vectors in $\cH^{\sigma_n}$ with vectors in $\cH^{\sigma_\infty}$ by applying a tensor product with the corresponding vacuum state. \\
In this section we prove that the limit 
\beq \label{psiinfty}
\psi_\infty :=\lim_{n \to \infty} \psi_n
\ene
exists that it is an eigenvector of $\overset{\infty}{H}$ corresponding to the eigenvalue $E_\infty$.

We prove furthermore that $E_\infty$ is non-degenerate and that its imaginary part is strictly negative and that it is, thus, a resonance.  

\subsubsection{Existence a Resonant Eigenvalue $E_\infty$} \label{existence-resonant-eigenvalue}
\begin{theorem}
The complex number $E_\infty$ is an eigenvalue of $\overset{\infty}{H}$ with corresponding eigenvector $\psi_\infty$. 
  
\end{theorem}
\emph{Proof:}\\
We first prove that $E_\infty$ is  ein eigenvalue and that $\psi_\infty$ is a corresponding eigenvector.  
By  (\ref{in.10.1.3}) and (\ref{in.10.1.2}) 
\beq \label{r.3}
\|  \psi_{k + 1} - \psi_k  \| \leq \frac{1}{100} (\frac{1}{100})^k
\ene
and therefore the series $\{ \psi_k \}_{k\in \mathbb{N}}$ converges. \\
From (\ref{ee.0})-(\ref{ee.4.1}), it follows that
\beq \label{r.5}
 3/4  \leq  \|  \psi_{\infty}  \| \leq \frac{5}{4} 
\ene
and 
\beq\label{r.6}
|\la  \psi_\infty  | \;    \phi_{\at}    \ra | \geq  3/4 \period
\ene
We select some $z_n \in \cE_{(n, \infty)}$ with $|z_n- E_n| = \sigma_n$ (see (\ref{in.6prima})) and compute 
\beq \label{r.10}\begin{array}{l}
\overset{\infty}{H} \psi_n = (\tilde{H}^{n}_{\infty}  + W_\infty^n) 
\psi_n  \\\\
=  E_n \psi_n + (z_n - E_n) W_\infty^n  \tilde{R}_\infty^n(z) \psi_n \period
\end{array}
\ene
Thus we have that (See Corollary~\ref{C-Princ-Step}) 
\beq \label{r.11} \begin{array}{l}
\|  \overset{\infty}{H} \psi_n -   E_n \psi_n  \| \leq \boC_{\ref{il.2.1}}^{n +2 } \sigma_{n}  \comma
\end{array}
\ene
which implies by (\ref{in.10.1.3}) that 
\beq \label{r.12} 
\lim_{n \to \infty} \overset{\infty}{H} \psi_n = E_{\infty} \psi_\infty\period
\ene
As also
\beq \label{r.13} 
\lim_{n \to \infty}  \psi_n =  \psi_\infty 
\ene
and $  \overset{\infty}{H}  $ is closed, we conclude that $  \psi_\infty $ belongs to the 
domain of $  \overset{\infty}{H}  $ and that 
\beq \label{r.14} 
  \overset{\infty}{H} \psi_\infty  =  E_{\infty}  \psi_\infty   \comma
\ene
wich proves the statement. 

\QED

\subsubsection{Approximations For the Imaginary Part of $E_\infty$} \label{approximations-imaginary} $\\$
Here we give an explicit expression of the imaginary part of $E_\infty$ up to order $\alpha^3$. We prove that, under certain conditions, $E_\infty$ has a strictly negative imaginary part and it is, therefore, a resonance. The main result of this subsection is Theorem~\ref{imaginary-part-final}.

\begin{definition}
We utilize the symbols of Landau. Let $Y$ be a normed space. Suppose that  $F(\alpha,\cdot) $ is a $Y$- valued function that depends on $\alpha$ and another parameters. We say that 
\beq
F = \cO(\alpha^\mu) 
\ene
if there is a constant $C$ (independent of $\alpha$ and the other parameters) such that 
\beq
\|F(\alpha, \cdot)\| \leq C \alpha^\mu  
\ene
for sufficiently small $\alpha$. 
  
\end{definition}

The imaginary part of $E_\infty$ is given by $\tilde E_I$ (see (\ref{imaginary-tilde})) up to order $\alpha^3$ (see (\ref{r.14.3})) below. But unfortunately $\tilde E_I$ depends on $\alpha$. In the next lemma we extract an $\alpha-$independent quantity that equals $\tilde E_I$ up to order $\alpha^3$.   
\begin{lemma}
Let $\phi_\at$ we a unit eigenvector of $H_\at(0, 0, 0)$ corresponding to the eigenvalue $\boe_1$ (see Hypothesis \ref{hypothesis}). We define 
 (see Hypothesis \ref{hypothesis}, (\ref{e.p.eeh.1.1}), (\ref{e.p.eeh.2}) and (\ref{i.10.3.0}))
\beq \label{imaginary-tilde-0}\begin{array}{l}
 E_{Im} : =  \pi \int_{\mathbb{S}^2}  dS \| P_{\at, 0}(0, 0, 0)|\boe_1 - \boe_0| \\\\ \cdot \tilde w_{1,0}(0, 0, 0)(x, \frac{k}{|k|}|\boe_1 - \boe_0|)
 \phi_\at  \|^2, 
\end{array} 
\ene
where 
\beq \label{wtilde}
\tilde w_{1,0}(0, \alpha, 0) = \frac{1}{\alpha^{3/2}} w_{1,0}(0, \alpha, 0)
\ene
(see (\ref{i.10.3.0})).

It follows that 
\beq \label{imaginary-approx}
|E_{Im} + \frac{1}{\alpha^3}\tilde E_I | = \cO(\alpha). 
\ene

\end{lemma}
 
\emph{proof:} $\\$
We select a  $\beta > 0$  be satisfying Assumptions~\ref{parameters-basis}, we denote by 
$$\widehat k:= \frac{k}{|k|}
$$. 
We notice that (see (\ref{i.2}), (\ref{imaginary-tilde}) and (\ref{i.10.3.0}))  for $\psi_0$ as in (\ref{imaginary-tilde})
\beq \label{p.1} \begin{array}{l}
\|(\tilde w(0, \alpha, 0)(x, |e_1 - e_0|\widehat k) - \tilde w(0, 0, 0)(x, |e_1 - e_0|\widehat k)) \psi_0 \|_{\cH_\at}   \\\\
 = \| (2 G^0(0)(x, |e_1 - e_0|\widehat k) - 2G^0(0)(0, |e_1 - e_0|\widehat k))\cdot \nabla \psi_0 \|_{\cH_\at} \\\\
 =  \| (2 G^0(0)(x, |e_1 - e_0|\widehat k) - 2G^0(0)(0, |e_1 - e_0|\widehat k)) e^{- \beta \la x \ra} \cdot (e^{\beta \la x \ra} \\\\ 
 \nabla e^{- \beta \la x \ra})  |(2 e_0 - 1) - e_1 | \frac{1}{H(0,\alpha,- \beta) - (2e_0 -1)} e^{\beta \la x \ra} \psi_0 \|_{\cH_\at} 
 = \cO(\alpha),  
\end{array}
\ene 
where we used (\ref{an.0.10}) and (\ref{T.eb.1.e1}). It follows from (\ref{ep-1}) and (\ref{ep-2}) and standard perturbative arguments with Neumann series that 
\beq \label{p.2} 
\| P_{\at, 0}(0, \alpha, 0) - P_{\at, 0}(0, 0, 0) \| = \cO(\alpha^3),
\ene
which implies that for $i \in \{0, 1 \}$ 
\beq \label{p.3} 
\| e_i - \boe_i \| = \cO(\alpha^3).
\ene
Using (\ref{p.3}) we prove as in (\ref{p.1}) that
\beq \label{p.4} \begin{array}{l}
\|(\tilde w(0, 0, 0)(x, |e_1 - e_0|\widehat k) - \tilde w(0, 0, 0)(x, |\boe_1 - \boe_0|\widehat k)) \psi_0 \|_{\cH_\at}   \\\\
 = \cO(\alpha^3),  
\end{array}
\ene 
We conclude from (\ref{p.1})-(\ref{p.4}) that 
\beq\label{p.5} \begin{array}{l}
| \frac{1}{\alpha^3}\tilde E_I +   \pi \int_{\mathbb{S}^2}  dS \| P_{\at, 0}(0, 0, 0)|\boe_1 - \boe_0| \\\\ \cdot \tilde w_{1,0}(0, 0, 0)(x, |\boe_1 - \boe_0|\frac{k}{|k|})
 \psi_0  \|^2 | = \cO(\alpha).
 \end{array}
\ene
Now we take 
\beq \label{p.6}
\psi_0 = \frac{1}{\| P_{\at, 1}(0, \alpha, 0) \phi_\at \|} P_{\at, 1}(0, \alpha, 0) \phi_\at
\ene
and notice that by (\ref{p.2})
\beq \label{p.7}
\| \psi_0 - \phi_\at\| = \cO(\alpha^3). 
\ene
For any $h \in \cH_\at$ we have that 
\beq \label{p.8} \begin{array}{l}
|\la h |\; P_{\at, 0}(0, 0, 0)|\boe_1 - \boe_0|  \cdot \tilde w_{1,0}(0, 0, 0)(x,|\boe_1 - \boe_0| \frac{k}{|k|})
 (\psi_0 - \phi_\at)  \ra |
 \\\\ =
| \la  |\boe_1 - \boe_0|  (\tilde w_{1,0}(0, 0, 0)(x, |\boe_1 - \boe_0|\frac{k}{|k|}))^* \\\\ P_{\at, 0}(0, 0, 0) h | \; (\psi_0 - \phi_\at) \ra |
\leq C \| h \| \alpha^3, 
 \end{array}
\ene
for some constant $C$. Eq. (\ref{p.8}) implies that 
\beq \label{p.9} \begin{array}{l}
\| P_{\at, 0}(0, 0, 0)|\boe_1 - \boe_0|  \cdot \tilde w_{1,0}(0, 0, 0)(x, \frac{k}{|k|}|\boe_1 - \boe_0|)
 (\psi_0 - \phi_\at)  \| \\\\ = \cO(\alpha^3),
 \end{array}
\ene
which together with (\ref{p.5}) implies (\ref{imaginary-approx}).

\QED

\begin{theorem} \label{imaginary-part-final}$\\$
The imaginary part of $E_\infty$ is given by $- \alpha^3 E_{Im}$ up to order $\alpha^3$, i. e.  
\beq \label{p.10}
|{\rm Im} \; E_\infty - (-\alpha^3 E_{Im} ) | = \cO(\alpha^3 \alpha^{9/10}).
\ene
In particular if we assume that $  E_{Im}  $ is non-zero then the imaginary part of $ E_\infty$ is strictly negative for small $\alpha$ and it is therefore a resonance. Notice that $E_{Im}$ does not depend on $\alpha$.    
\end{theorem}

\emph{Proof:}

It follows from  (\ref{in.10.1.3}) and (\ref{in.5}) that
\beq \label{r.14.1}
| E_\infty - E_0 | \leq \frac{100}{99}\boC_{\ref{il.2.1}}^2 \alpha^{3/2} \sigma_0^2 =  \frac{100}{99}\boC_{\ref{il.2.1}}^2 \alpha^{3/2} \alpha^{2\upsilon }\comma 
\ene 
which implies together with (\ref{imaginary-basis}) that 
\beq \label{r.14.2}\begin{array}{l}
|{\rm Im}\; E_\infty - \tilde E_I  | \leq \frac{100}{99}\boC_{\ref{il.2.1}}^2 \alpha^{3/2}\alpha^{2\upsilon } \\\\
 + 4 C_{\ref{corr.16}} \alpha^3 
 (\alpha^\upsilon |\log(\alpha^\upsilon)| + \alpha^\upsilon + \alpha^{2\upsilon}  +  \alpha^{(3- \nu)/2} (1 + \alpha^{(3 - \upsilon)/2})^3).
\end{array}
\ene
Taking $\upsilon = \frac{6}{5} $ we get
\beq \label{r.14.3}\begin{array}{l}
|{\rm Im}\; E_\infty - \tilde E_I  | = \cO(\alpha^3 \alpha^{9/10})
\end{array}
\ene
Eq. (\ref{p.10}) follows from (\ref{imaginary-approx}) and (\ref{r.14.3}).

\QED

\subsubsection{Non Degeneracy of $E_\infty$}\label{non-degeneracy}

\begin{lemma}\label{L.nd.1}
For every $\psi \in \cH^{(0, \infty)} $
\beq \label{L.nd.1.e1}
\lim_{r \to \infty} 1_{\cH_\at \otimes \cF^{ (\sigma_r , \infty) }} \otimes (1 - P_{\Omega^{(0, \sigma_r)}}) \psi = 0 
\ene

\end{lemma}
\emph{Proof}
Let $\psi = ( \psi_n)_{n = 0}^\infty $, where for any $n \geq 1$, $\psi_n \in L^2( \RR^3 \times (\mathcal{K}^{(0,\infty)})^n) $ is symmetric on the 
variable belonging to $ (\mathcal{K}^{(0,\infty)})^n $, and 
$\psi_0 \in   L^2( \RR^3) $.\\\\
We identify the function $\psi_n$  with the element in $\cH^{(0, \infty)}$ such that its $n$-component is
equal to  $\psi_n$  and the others are equal to zero. 

As $\psi \in   \cH^{(0, \infty)} $ and   $  1_{\cH_\at \otimes \cF^{ (\sigma_r , \infty) }} \otimes (1 - P_{\Omega^{(0, \sigma_r)}}) $
is an orthogonal projection, for a given $\epsilon > 0$ we can choose $N$ such that 
\beq \label{nd.0.1} \begin{array}{l}
 \sum_{n = N}^{\infty} \|  1_{\cH_\at \otimes \cF^{ (\sigma_r , \infty) }} \otimes ( 1- P_{\Omega^{(0, \sigma_r)}})\psi_n \| ^2 \\\\ \leq 
 \sum_{n = N}^{\infty} \|     \psi_n \| ^2 < \epsilon \comma
\end{array}
\ene
uniformly in $r$. 
As $\sigma_r$ goes to zero, we have that there exist $R \in \NN $ such that for $r > R$
\beq \label{nd.0.2}
 \sum_{n = 0}^{N} \|    1_{\cH_\at \otimes \cF^{ (\sigma_r , \infty) }} \otimes (1 - P_{\Omega^{(0, \sigma_r)}})\psi_n  \| ^2 < \epsilon 
\ene
(\ref{nd.0.1}) and (\ref{nd.0.2}) imply (\ref{L.nd.1.e1}).

\QED

\begin{theorem}\label{T.nd.1}
The eigenvalue $   E_{\infty}  $ is non degenerate. 
\end{theorem}

\emph{Proof:}

Suppose that $\psi $ is such that $  \overset{\infty}{H} =   E_{\infty}  \psi $, we take $\gamma_n$ as in (\ref{in.8}) and $u \in \cE_{(n, \infty)} $ 
with $|u - E_n| = \sigma_n$ (see \ref{in.6prima}). We use Corollary~\ref{C-Princ-Step}.  

\beq \label{nd.1} \begin{array}{l}
P_n   1_{\cH_\at \otimes \cF^{ (\sigma_n , \infty) }} \otimes  P_{\Omega^{(0, \sigma_n)}} \psi  \\\\ = \frac{i}{2 \pi} \int_{\gamma_n} \overset{n}{R}(z)  1_{\cH_\at \otimes \cF^{ (\sigma_n , \infty) }} \otimes  P_{\Omega^{(0, \sigma_n)}}  \psi dz  \\\\ =
\frac{i}{2 \pi} \int_{\gamma_n}   \overset{n}{R}(z)   
 1_{\cH_\at \otimes \cF^{ (\sigma_n , \infty) }} \otimes  P_{\Omega^{(0, \sigma_n)}} \frac{  (\overset{\infty}{H} - z )  } { E_\infty - z } \psi  dz 
 \\\\ =
\frac{i}{2 \pi} \int_{\gamma_n}   \overset{n}{R}(z)   
 1_{\cH_\at \otimes \cF^{ (\sigma_n , \infty) }} \otimes  P_{\Omega^{(0, \sigma_n)}} \frac{  (\overset{n}{H} - z ) + W_\infty^n  } { E_\infty - z } \psi  dz  
\\\\ =
 \frac{i}{2 \pi} \int_{\gamma_n}  1_{\cH_\at \otimes \cF^{ (\sigma_n , \infty) }} \otimes  P_{\Omega^{(0, \sigma_n)}}   \overset{n}{R}(z)   
 \frac{  (\overset{n}{H} - z) + W_\infty^n  } { E_\infty - z } \psi  dz  
\end{array}
\ene
\begin{equation*}
\begin{array}{l}
 = \frac{i}{2 \pi} \int_{\gamma_n}  1_{\cH_\at \otimes \cF^{ (\sigma_n , \infty) }} \otimes  P_{\Omega^{(0, \sigma_n)}}   \overset{n}{R}(z)    
 \frac{  \overset{n}{H} - z } { E_\infty - z } \psi  dz 
 \\\\
 + \frac{i}{2 \pi} \int_{\gamma_n}  1_{\cH_\at \otimes \cF^{ (\sigma_n , \infty) }} \otimes  P_{\Omega^{(0, \sigma_n)}} (\overset{n}{R}(z)(z - u) + 1)\overset{n}{R}(u)    
 \frac{   W_\infty^n  } { E_\infty - z } \psi  dz 
\\\\
= \frac{i}{2 \pi} \int_{\gamma_n}  1_{\cH_\at \otimes \cF^{ (\sigma_n , \infty) }} \otimes  P_{\Omega^{(0, \sigma_n)}}  \frac{1} { E_\infty - z } \psi  dz 
 \\\\
+  \frac{i}{2 \pi} \int_{\gamma_n}  1_{\cH_\at \otimes \cF^{ (\sigma_n , \infty) }} \otimes  P_{\Omega^{(0, \sigma_n)}}  (\overset{n}{R}(z)(z - u) + 1)\overset{n}{R}(u)    
 \frac{   W_\infty^n  } { E_\infty - z } \psi  dz
\\\\
 =  1_{\cH_\at \otimes \cF^{ (\sigma_n , \infty) }} \otimes  P_{\Omega^{(0, \sigma_n)}}     
 \psi   \\\\
 +   
 \frac{i}{2 \pi} \int_{\gamma_n}  1_{\cH_\at \otimes \cF^{ (\sigma_n , \infty) }} \otimes  P_{\Omega^{(0, \sigma_n)}}     
  ((z-u)  \overset{n}{R}(z) + 1)   \tR_\infty^{n}(u)   \frac{   W_\infty^n e^{-\beta \la x \ra}  } { E_\infty - z }e^{\beta \la x \ra} \psi \comma
\end{array} 
\end{equation*}
where $\beta > 0$ satisfies Assumptions~\ref{parameters-basis}.  
Using Corollary~\ref{C-Princ-Step} we conclude that
\beq \label{nd.2} \begin{array}{l}
\lim_{n \to \infty} \frac{i}{2 \pi} \int_{\gamma_n}  1_{\cH_\at \otimes \cF^{ (\sigma_n , \infty) }} \otimes  P_{\Omega^{(0, \sigma_n)}}     
 \\\\ \cdot ((z-u)  \overset{n}{R}(z) + 1)   \tR_\infty^{n}(u)   \frac{   W_\infty^n e^{ - \beta \la x \ra}  } { E_\infty - z }e^{\beta \la x \ra} \psi = 0\comma
\end{array}
\ene
Thus by Lemma~\ref{L.nd.1}
\beq \label{nd.3} \begin{array}{l}
\lim_{n \to \infty }P_n   1_{\cH_\at \otimes \cF^{ (\sigma_n , \infty) }} \otimes  P_{\Omega^{(0, \sigma_n)}} \psi  = P_{\infty}\psi
= \\\\ \lim_{n \to \infty }1_{\cH_\at \otimes \cF^{ (\sigma_n , \infty) }} \otimes  P_{\Omega^{(0, \sigma_n)}}       \psi  = \psi
\end{array}
\ene
and we conclude that 
\beq \label{nd.4}
 P_{\infty}\psi = \psi \comma
\ene
which implies that $E_{\infty}$ is non degenerate.
 
\QED
 
\subsubsection{Estimates for the Resolvent} \label{estimates-resolvent}$\\$

\begin{theorem}\label{T.acs.1}
For any $z \in \cE_{(n, \infty)}$, z belongs to the resolvent set of $\overset{\infty}{H}$ and
\beq \label{T.acs.1.e1}
\| (\overset{\infty}{H} - z)^{-1}  \| \leq \boC_{\ref{il.2.1}}^{n + 2} \frac{1}{ (|\sin(\nu)|/2) \sigma_{n + 1} + |z- E_n| }.
\ene
In particular the set
\beq \label{T.acs.1.e2}
 \cE_\infty  := \begin{cases}  \cE^{\sigma_0}(\theta) \setminus (E_\infty + \RR  - i  [ 0, \infty )), & {\rm if} \; {\rm Im}(\theta) > 0 \comma \\
 \cE^{\sigma_0}(\theta) \setminus (E_\infty + \RR  + i [ 0, \infty )), & {\rm if} \; {\rm Im}(\theta) < 0  \period  \end{cases}      
\ene 
is contained in the resolvent set of  $\overset{\infty}{H}$.
\end{theorem}

\emph{Proof:}
It follows from
Theorem~\ref{C.ei.2}, that
\beq \label{acs.1prima}\begin{array}{l}
\| (\overset{\infty}{H} - z)^{-1}  \| \leq C_{\ref{acs.1}} \boC_{\ref{il.2.1}}^{n + 1}  \frac{1}{ (|\sin(\nu)|/2) \sigma_{n + 1}  +  |z - E_n|} \comma
\end{array}
\ene
where
\beq \label{acs.1}\begin{array}{l}
C_{\ref{acs.1}} : =  (2 +   4  C_{\ref{neri.15.2.1}} )^2   \frac{64}{ (|\sin(\nu)|/2) }   
+    6 \frac{40 |e_0'| C_{\ref{an.0.16.2}}}{|\sin (\nu)| \delta} 
\frac{16 |e_0'|}{ \delta |\sin(\nu)|} 20 
\period 
\end{array}
\ene 
Eq. (\ref{T.acs.1.e1}) follows from (\ref{il.2.1}). 

\QED

\section{ Proofs of Section~\ref{estimates-electron-hamiltonian}} \label{proofs-estimates-electron-hamiltonian}
\subsection{ Proof of Theorem~\ref{p.eeh.1}:}

\newtheorem*{p.eeh.1}{Theorem~\ref{p.eeh.1}}

\begin{p.eeh.1}
We suppose that $\alpha \leq \boa$, $|\beta| \leq \bob$ and that
\begin{equation} \tag{\ref{an.0.14}}
| \theta   | \leq \min\Big((32 C_{\ref{an.0.9}})^{-1}, \frac{1}{120}\Big)\comma
\end{equation}
the following holds true:
\begin{itemize}
\item 
There are only two points ($\{ e_0(\theta, \alpha, \beta),  e_1(\theta, \alpha, \beta) \}$) in the spectrum of $H_\at(\theta, \alpha, \beta)$ with real part less than $\boe_1 + \frac{15}{16} \delta_\at$ (see (\ref{delta-el})). 
They are simple eigenvalues and they do not depend on $\beta$ and $\theta$ (they are therefore real). They satisfy 
\beq \tag{\ref{e.p.eeh.1}}
| e_j(\theta,\alpha, \beta) - \boe_j | \leq \frac{\delta_\at}{16}, \: j \in \{0, 1\}.  
\ene
It follows that 
\beq  \tag{\ref{e.p.eeh.1.1}}
e_j(\theta,\alpha, \beta)  = e_{j}(0, \alpha, 0): = e_{j}(\alpha), \: j \in \{0, 1\}.  
\ene
We omit writing the dependence on $\alpha$ it is not required. 
\item Let 
\beq \tag{\ref{e.p.eeh.2}}
P_{\at, j}(\theta, \alpha, \beta),\:   j \in \{ 0, 1 \} 
\ene
be the projection into the eigen-space corresponding to $ e_j  $, $j  \in \{ 0, 1  \}$ respectively. 
It follows that 
\beq \tag{\ref{e.p.eeh.3}} \begin{array}{l}
\| P_{\at, j}(0, \alpha, \beta)  - P_{\at, j}(0, 0, 0) \| \leq \frac{1}{8}\comma \\\\
 \| P_{\at, j}(\theta, \alpha, \beta)  - P_{\at, j}(0, \alpha, \beta) \| \leq \frac{1}{8}.
\end{array}
\ene
\end{itemize}

\end{p.eeh.1}

We prove the assertions in several steps. 

\noindent {\bf Step 1:}

We prove that for any $\alpha \leq \boa$, any $|\beta| \leq \bob$ and any $z \in \CC$ such that 
$\dist(z, \sigma(H_\at(0,0,0)) ) \geq \frac{\delta_\at}{32} $ and ${\rm Re}\, z \leq |\boe_0|$,
\beq \label{geh.8}
\| (\alpha^3 V_{\pf}(\theta) +   \tilde V(\theta, \beta))\frac{1}{H_{\at}(0,0, 0) - z} \| \leq  \frac{1}{9}, \: \: \forall \theta \in \CC, |\theta| < \frac{1}{2}\period  
\ene
It follows from (\ref{geh.5}) that
\beq \label{geh.7.1} \begin{array}{l}
\| (1 - \Delta) \frac{1}{H_\at(0, 0, 0) - z}  \| \leq 2( 1 + |z| \| \frac{1}{H_\at(0, 0, 0) - z} \| \\\\ + (b_{1/2} + 1/2) \| 
\frac{1}{H_\at(0, 0, 0) - z} \|)\period
\end{array}
\ene
Eqs. (\ref{pft.15}) and (\ref{potential.1}) imply that
\beq \label{geh.2}
 |V_{PF}(\theta)(x)| \leq \min(|x|^2, \frac{1}{|x|}), \: \:  \: \forall x \in \mathbb{R}^3, \: \forall \theta, |\theta| <  \frac{1}{2} \comma
\ene
\beq \label{geh.3}
\| \tilde V(\theta, \beta)  \frac{1}{1 - \Delta}\| \leq 50 \beta, \: \: \forall \theta, |\theta| <  \frac{1}{2}.
\ene 

Eq. (\ref{geh.8}) follows from (\ref{geh.7.1}-\ref{geh.3}) and our assumptions on $\alpha$ and $\beta$.

\noindent {\bf Step 2:} 

We prove (\ref{e.p.eeh.1}) and (\ref{e.p.eeh.3}) for $\theta = 0$. \\

Using Neumann
series and (\ref{geh.8}) we prove that  for any $z \in \CC$ such that 
$\dist(z, \sigma(H_\at(0,0,0)) ) \geq \frac{\delta_\at}{32} $ and ${\rm Re}\, z \leq |\boe_0|$,
\beq \label{e.p.eeh.4} \begin{array}{l}
\| \frac{1}{ H_\at(0, \alpha, \beta) - z} \| \leq \frac{9}{8} \frac{1}{\dist(z, \sigma(H_\at(0, 0, 0)))}.
\end{array}
\ene

By theorem XIII.46 \cite{ReedSimonIV1978} the ground state $\boe_0$ defined in Hypothesis \ref{hypothesis} is non-degenerate and 
by hypothesis $\boe_1$ is non-generate.\\ 
The rank-one projections $ P_{\at, j}(0, 0, 0) $ ($j \in \{0, 1 \}$) are the integrals   
\beq \label{geh.9prima}
\begin{array}{l}
P_{\at, j}(0, 0, 0) = \frac{i}{2 \pi} \int_{\partial D_{\delta_\at /2}(\boe_j)}\frac{1}{ H_{\at}( 0, 0, 0) - z  } dz \period
\end{array}
\ene
Using again Neumann series and (\ref{geh.8}) we obtain that
\beq \label{geh.9.1}
\|   \frac{i}{2 \pi} \int_{\partial D_{\delta_\at /2}(\boe_j)}\frac{1}{ H_{\at}( 0, \alpha, \beta) - z  } dz  -  P_{\at, j}(0, 0, 0)  \| \leq \frac{1}{8}, 
\ene
wich implies that for any $j \in \{ 0, 1 \}$ there is only one eigenvalue -$e_j(0, \alpha, \beta)$ - of   $ H_{\at}( 0, \alpha, \beta) $ with
$| e_j(0, \alpha, \beta) - \boe_j | < \delta_\at/2$. Eq. (\ref{e.p.eeh.1}) follows from (\ref{e.p.eeh.4}). The first half of (\ref{e.p.eeh.3}) is a consequence of (\ref{geh.9.1}), since
\beq \label{geh.9}
\begin{array}{l}
P_{\at, j}(0, \alpha, \beta) = \frac{i}{2 \pi} \int_{\partial D_{\delta_\at /2}(\boe_j)}\frac{1}{ H_{\at}( 0, \alpha, \beta) - z  } dz \period
\end{array}
\ene

\noindent {\bf Step 3:} We prove (\ref{e.p.eeh.3}). \\

Using (\ref{geh.5}) and (\ref{e.p.eeh.4}) we obtain that for any $z \in \CC$ such that 
$\dist(z, \sigma(H_\at(0,0,0)) ) \geq \frac{\delta_\at}{32} $ and ${\rm Re}\, z \leq |\boe_0|$,
\beq \label{an.0.6}
  \|(1 - \Delta) \frac{1}{H_\at(0, \alpha, \beta) - z }  \| \leq \frac{4}{3}\Big(  1  + 2 \frac{|z| + b_{1/4} + 1}{\dist(z, 
  \sigma(H_\at(0, 0, 0) ))}  \Big) \comma
\ene
Lemma~\ref{L.a.2} implies that for $\theta \in \CC$ with $|\theta| \leq \frac{1}{120}$,
\beq \label{an.0.0.8}
\| ( H_\at(\theta, \alpha, \beta) - H_\at(0, \alpha, \beta)   ) \frac{1}{ 1 - \Delta}  \| \leq |\theta|
 (\frac{1}{2 (1/60 - 1/120)^2}C_{\ref{an.0.7}} + 2e^{1/15})\period
\ene
Then we have that for any $z \in \CC$ with ${\rm Re } (\, z) \leq |\boe_0|$ and  $\dist(z, \sigma(\boH_\at)) \geq \frac{\delta_\at}{16}$
\beq \label{an.0.8} \begin{array}{l}
\| ( H_\at(\theta,  \alpha, \beta) - H_\at(0, \alpha, \beta)   ) \frac{1}{H_\at(0, \alpha, \beta) - z }   \| \leq C_{\ref{an.0.9}} |\theta| \comma
\end{array}
\ene
It follows from (\ref{an.0.14}), (\ref{e.p.eeh.4}), (\ref{an.0.8}) and the Neumann series 
that for every $j \in \{ 0, 1  \}$, we can construct the projection 
$P_{\at, j}(\theta, \alpha, \beta)$ 
using the Dunford integral, 
\beq \label{eeh.2}
P_{\at, j}(\theta, \alpha, \beta) : = \frac{i}{2 \pi} \int_{|z- e_j| = \delta_\at/2 }
\frac{dz}{H_\at(\theta, \alpha, \beta) - z}
\ene
\noindent and that 
\beq \label{an.0.15}
\|P_{\at, j}(\theta, \alpha, \beta) - P_{\at, j}(0, \alpha, \beta)\| \leq \frac{1}{8} \period
\ene

\noindent {\bf Step 4:} We prove (\ref{e.p.eeh.1.1}).
First we notice that 
$$
e_j(\theta, \alpha, 0) = e_j(0, \alpha, 0)
$$
follows from Theorem III.36 \cite{ReedSimonIV1978}.

Suppose that $\phi$ is ein eigenvector of $ H_\at(\theta, \alpha, 0) $ corresponding to $e_j(\theta, \alpha, 0)$. 
It is easy to see that $ e^{- \beta \la x\ra} \phi \in \dom(H_\at (\theta, \alpha, \beta)) $ and that 
$$
H_\at (\theta, \alpha, \beta) e^{- \beta \la x\ra} \phi = e_j(0, \alpha, 0) e^{- \beta \la x\ra} \phi,
$$
wich implies that $ e_j(0, \alpha, 0) $ is an eigenvalue of $ H_\at (\theta, \alpha, \beta)$. 
Eq. (\ref{e.p.eeh.3}) implies that the range $ P_{\at, j}( \theta, \alpha, \beta ) $ is unidimensional. The only eigenvalue that can be associated to this projection is $ e_j(0, \alpha, 0) $.

\qed

\noindent
\subsection{\; Proof of Theorem~\ref{L.eeh.1}}

\newtheorem*{L.eeh.1}{Theorem~\ref{L.eeh.1}}

\begin{L.eeh.1} 
Suppose that $\theta$
 satisfies (\ref{an.0.14}). Suppose furthermore that $\alpha \leq \boa$ and $|\beta| \leq \bob$. 
Let $z \in \CC   $  with $\mathrm{Re (z)} < \boe_1 + \frac{7}{8}\delta_\at $. Then,  
\beq \tag{\ref{L.eeh.1.e1}}\begin{array}{l}
\|  (H_\at(\theta, \alpha, \beta) - z )^{-1} \overline{P}_{disc}(\theta, \alpha, \beta)  \|
\leq 54   \frac{1}{ |z - \boe_{\ref{an.0.16.1}}| } \\\\
\leq C_{\ref{an.0.16.2}} \frac{1}{| z - e_{0}| }     \comma
\end{array}
\ene
where
\beq \tag{\ref{an.0.16.1}}
\boe_{\ref{an.0.16.1}}  : =  \boe_1 + \frac{15}{16}\delta_\at      \comma
\ene
and 
\beq \tag{\ref{an.0.16.2}}
 C_{\ref{an.0.16.2}} := 972 \frac{|\boe_0|}{\delta_\at}     \period
\ene
\end{L.eeh.1}

Let $ \mathfrak{A} $ be the set $ \{  \mu  \in \CC : \mathrm{Re (\mu)} \leq \boe_{\ref{an.0.16.1}}   \} $
and   $ \mathfrak{F}  $ the function defined by the rule 
\beq \label{an.0.17}
\mathfrak{F}(\mu) : =    (H_{\at}(\theta, \alpha, \beta) - \mu )^{-1} \overline{P}_{disc}(\theta, \alpha, \beta)  \period
\ene 
$\mathfrak{F}$ is analytic.   \\ 
Given two points $ z_1 $ and $z_2$ in the complex plane and a real number $\ell > 1$, the set of points $ z $
such that $ |z - z_1| = \frac{1}{\ell} | z - z_2 |$, is a circle with center at $ z_1 + \frac{1}{\ell^2 - 1} ( z_1 - z_2)  $ and
radius $ \frac{\ell}{\ell^2 - 1} |   z_1 - z_2 |  $. \\
We take $ \ell = 4 $ and we define
 $ \mathcal{D} := D_{4/15| \boe_1 - \boe_{\ref{an.0.16.1}}| } ( \boe_1  + \frac{1}{15} ( \boe_1  -  \boe_{\ref{an.0.16.1}}   )  ) $ $ \cup $
$  D_{4/15| \boe_0 - \boe_{\ref{an.0.16.1}}| } (  \boe_0  + \frac{1}{15} ( \boe_0  -  \boe_{\ref{an.0.16.1}}  )  ) $. \\
For any $ z \in \CC \setminus \mathcal{D}  $, 
\beq \label{an.0.18} \begin{array}{l}
| z - \boe_{\ref{an.0.16.1}} | \leq 4 | z -  \boe_0  | \comma  \\\\
| z - \boe_{\ref{an.0.16.1}} | \leq 4 | z -  \boe_1  | \period
\end{array}
\ene
For any $ z \in D_{4/15| \boe_i - \boe_{\ref{an.0.16.1}}| } ( \boe_i  + \frac{1}{15} ( \boe_i  -  \boe_{\ref{an.0.16.1}}  )  ) $ and 
any $ \mu \in\partial  D_{ 4/15| \boe_i -  \boe_{\ref{an.0.16.1}}| } ( \boe_i  + \frac{1}{15} ( \boe_i  -  \boe_{\ref{an.0.16.1}}     )  ) $,  $i \in \{ 0, 1 \}$ 
\beq \label{an.0.19}
| z - \boe_{\ref{an.0.16.1}} | \leq \frac{ 1 + \frac{1}{\ell^2 - 1} + \frac{\ell}{\ell^2 - 1} }{ 
 1 + \frac{1}{\ell^2 - 1} - \frac{\ell}{\ell^2 - 1}   } |\mu - \boe_{\ref{an.0.16.1}}| \leq 1.67  |\mu - \boe_{\ref{an.0.16.1}}| \period
\ene
Eqs. (\ref{an.0.14}), (\ref{e.p.eeh.4}), (\ref{an.0.8}) and the Neumann series imply
\beq \label{an.0.5}
\|  \frac{1}{ H_{\at}(\theta, \alpha, \beta) - z }  \| \leq 2 \frac{1}{\dist(z, \sigma(H_\at(0, 0, 0)))}  \period
\ene
From (\ref{e.p.eeh.3}) follows  
\beq \label{an.0.16}
 1 - \frac{2}{8} \leq  \|P_{\at, j}(\theta, \alpha, \beta)\| \leq 1 + \frac{2}{8} \period
\ene

By (\ref{an.0.18}), (\ref{an.0.5}) and (\ref{an.0.16})
\beq \label{an.0.20}
\|\mathfrak{F}(z)\| \leq  32 \frac{1}{| z - \boe_{\ref{an.0.16.1}} |},\: z  \in \CC \setminus \mathcal{D}, \; {\rm Re}(\, z) \leq \boe_{\ref{an.0.16.1}} \period 
\ene
If $z \in \mathcal{D}$, it follows from the maximum modulus principle,  (\ref{an.0.19}) and (\ref{an.0.20}) that
\beq \label{an.0.21}
\|\mathfrak{F}(z)\| \leq  54 \frac{1}{| z - \boe_{\ref{an.0.16.1}}|}\period 
\ene
Now we take $z_1 =  \boe_{\ref{an.0.16.1}}$, $ z_2 = e_{0}  $  and
 $ \ell = 18 \frac{ |\boe_0|}{\delta_\at} $. 
We have that $  D_{ \ell/(\ell^2 - 1)| \boe_{\ref{an.0.16.1}} - e_{0}| } ( \boe_{\ref{an.0.16.1}}  + \frac{1}{\ell^2 -1}
 ( \boe_{\ref{an.0.16.1}}  -  e_{0}  )  )  $ is contained in the set $ \{z \in \CC : {\rm Re}(\, z) \geq \boe_1 + \frac{7}{8} \delta_\at    \}  $
and that 
\beq \label{an.0.22}
|z -  e_{0} | \leq \ell | z - \boe_{\ref{an.0.16.1}} |, \: z \in \CC,\,  {\rm Re}(\, z) \leq \boe_1 + \frac{7}{8} \delta_\at \period
\ene

\QED

\noindent
\subsection{\; Proof of Corollary~\ref{C.eeh.1}}
\newtheorem*{C.eeh.1}{Corollary \ref{C.eeh.1}}

\begin{C.eeh.1}  Suppose that $\theta$
 satisfies (\ref{an.0.14}). Suppose furthermore that $\alpha \leq \boa$ and $|\beta| \leq \bob$. 
Let $z \in \CC \setminus \{ e_{0}, e_{1}  \}   $  with $\mathrm{Re (z)} < \boe_1 + \frac{7}{8}\delta_\at $. The following estimates 
hold true,
\beq \tag{\ref{C.eeh.1.e1}}
\| (H_\at(\theta, \alpha, \beta) - z)^{-1}  \| \leq 4 C_{\ref{an.0.16.2}} ( \frac{1}{|z-e_{0}|} +   
\frac{1}{|z - e_{1}|} ) \period
\ene
\beq \tag{\ref{C.eeh.1.e2}}
\| (H_\at(\theta, \alpha, \beta) - z)^{-1} \overline{P}_{\at, 1}(\theta,  \alpha, \beta) 
 \| \leq  4 C_{\ref{an.0.16.2}}   \frac{1}{|z-e_{0}|}    \period
\ene

\end{C.eeh.1}

We have that 
\beq \label{eeh.4}\begin{array}{l}
(H_{\at}(\theta, \alpha, \beta)  - z)^{-1}P_{disc}(\theta, \alpha,  \beta) \\\\ = (e_{0} - z)^{-1}P_{\at, 0}(\theta, \alpha, \beta) +
 (e_{1} - z)^{-1}P_{\at, 1}(\theta, \alpha, \beta) \comma
\end{array}\ene
thus
\beq \label{eeh.5} \begin{array}{l}
\|(H_{\at}(\theta, \alpha, \beta)  - z)^{-1}P_{disc}(\theta, \alpha, \beta)\| \\\\ \leq |e_{0} - z|^{-1}\|P_{\at, 0}(\theta, \alpha, \beta)\| +
 |e_{1} - z|^{-1} \|P_{\at, 1}(\theta, \alpha, \beta)\| \period
\end{array}
\ene
Eqs. (\ref{an.0.16}) and (\ref{eeh.5}) imply
\beq \label{eeh.8}
\|(H_{\at}(\theta, \alpha, \beta) - z)^{-1}P_{disc}(\theta, \alpha, \beta)\| \leq 2 |e_{0}  - z|^{-1} +
 2|e_{1} - z|^{-1} \period
\ene
Eq. (\ref{C.eeh.1.e1}) follows form (\ref{eeh.8}) and Theorem 
\ref{L.eeh.1}.
Analogously we can probe  (\ref{C.eeh.1.e2}). 

\QED

\noindent
\subsection{\; Proof of Corollary~\ref{L.eeh.2}}

\newtheorem*{L.eeh.2}{Corollary \ref{L.eeh.2}}

\begin{L.eeh.2}
Suppose that $\theta$
 satisfies (\ref{an.0.14}). Suppose furthermore that $\alpha \leq \boa$ and $|\beta| \leq \bob$.  
Let $z, \; \mu \in \overline{B_{\delta_\at/6}(e_{1})}  $, it follows that
\beq \tag{\ref{L.eeh.2.e1}}
\|  (H_{\at}(\theta, \alpha, \beta)- z) \frac{\overline{P}_{\at, 1}(\theta, \alpha, \beta)}{ (H_{\at}(\theta, \alpha,\beta) - \mu)   } 
  \| \leq 4 + 4 C_{\ref{an.0.16.2}} \period 
\ene
\end{L.eeh.2}

by Corollary~\ref{C.eeh.1} and (\ref{an.0.16}) we  have that,
\beq \begin{array}{l}
 \frac{\overline{P}_{\at, 1}(\theta, \alpha, \beta)}{ (H_{\at}(\theta, \alpha, \beta) - \mu)   }     (H_{\at}(\theta, \alpha, \beta) - z) = 
\overline{P}_{\at, 1}(\theta, \alpha, \beta)(  1 + \frac{\mu - z}{ (H_{\at}(\theta, \alpha, \beta) - \mu)   }  ) \\\\
\leq 4 + 4 C_{\ref{an.0.16.2}} \period
\end{array}
\ene


\QED

\subsection{\; Proof of Theorem~\ref{T.ebe.1}}
\newtheorem*{T.ebe.1}{Theorem~\ref{T.ebe.1}}

\begin{T.ebe.1}
Suppose that $\theta$
 satisfies (\ref{an.0.14}). Suppose furthermore that $\alpha \leq \boa$ and $|\beta| \leq \bob$.  Then 
the range of $ P_{\at, 1}(\theta, \alpha, 0)  $ is contained in the domain of $e^{\beta \la x \ra}$ (see (\ref{geh.7.m1})) 
and
\beq \tag{\ref{T.eb.1.e1}} \begin{array}{l}
 \|  e^{\beta \la x \ra}  P_{\at, 1}(\theta, \alpha, 0) \|  \leq
 8 \| e^{\beta \la x\ra}  \phi   \|_{\cH_\at}  \comma
\end{array}
\ene
In particular the following uniform bounds hold true
\beq \tag{\ref{T.eb.1.e3}} \begin{array}{l}
 \|  e^{\beta \la x \ra}  P_{\at, 1}(\theta, \alpha, 0) \| \leq  C_{\ref{T.eb.1.e2}} \comma
\end{array}
\ene
and 
\beq \tag{\ref{T.eb.1.e4}} \begin{array}{l}
 \| (1 + |x|^2) P_{\at, 1 }(\theta, \alpha, 0) \| \leq  C_{\ref{T.eb.1.e2.1}}  \comma
\end{array}
\ene
where
\beq \tag{\ref{T.eb.1.e2}}
C_{\ref{T.eb.1.e2}}  = 8 \| e^{\bob \la x \ra}  \phi   \|_{\cH_\at}   \comma
\ene
\beq \tag{\ref{T.eb.1.e2.1}}
C_{\ref{T.eb.1.e2.1}}  =   8 \|  e^{\bob \la x \ra}  \phi   \|_{\cH_\at} (1 + e^{-2}(1 + 4 /\bob^2) )  \comma
\ene

and $  \phi $ is any unit eigenvector of $  P_{\at, 1, 0, 0}(0)  $.  

\end{T.ebe.1}

By (\ref{potential.1}), $ P_{\at, 1}(\theta, \alpha, \beta)  $ is an analytic operator valued function with respect to $\beta$. 
For purely imaginary $\beta$, the operator $ e^{\beta \la x \ra} $ is unitary and therefore for such $\beta$
$$
  P_{\at, 1}(\theta, \alpha, \beta) =   e^{- \beta \la  x \ra }P_{\at, 1}(\theta, \alpha, 0)e^{\beta \la x\ra }, \: {\rm Re\, \beta} = 0. 
$$
By analyticity it follows that for any $\phi, \psi \in C_0^{\infty}(\mathbb{R}^3)  $,
$$
 \la \psi  |\; P_{\at, 1}(\theta, \alpha, \beta) \phi\ra_{\cH_\at} = 
 \la e^{- \overline{\beta} \la  x \ra }\psi |\;  P_{\at, 1}(\theta, \alpha, 0)e^{\beta \la x\ra }\phi \ra_{\cH_\at}, \: |\beta| < \frac{1}{2}. 
$$

As $C_0^\infty (\RR^3)$ is dense in $\cH_\at$ and  the range of $ P_{\at, 1}(\theta, \alpha, 0) $ is one dimensional, 
$  P_{\at, 1}(\theta, \alpha, 0)[ C_0^\infty (\RR^3) ] $ covers the range of  $ P_{\at, 1}(\theta, \alpha, 0) $ and therefore for any $\phi$ in the range of 
 $ P_{\at, 1}(\theta, \alpha, 0) $ there is a constant $C$ such that,
$$
\la e^{- \overline{\beta} \la  x \ra }\psi |\; \phi \ra_{\cH_\at}  \leq C \|\psi \|_{\cH_\at},  \forall \psi \in  C_0^\infty (\RR^3), 
$$ 
which implies that 
$$
\| e^{ 1/4 \la  x \ra }\phi \|_{\cH_\at}  \leq C^{1/2} \|\phi \|^{1/2}_{\cH_\at},  
$$
thus, the range of $ P_{\at, 1}(\theta, \alpha, 0)  $ is contained in the domain of $e^{1/4 \la x\ra}$. Related results can be found in 
Lemma, page 196, \cite{ReedSimonIV1978}). \\

Let $ \phi_{0, 0} \in \cH_\at(0, 0, 0)$ be such that $ P_{\at, 1}(0, 0, 0) \phi_{0, 0}  =  \phi_{0, 0} $ and 
 $\|  \phi_{0, 0} \|_{\cH_\at} = 1$, and take $   \phi_{\alpha, \theta} : =  P_{\at, 1}(\theta, \alpha, 0) \phi_{0, 0}  $. 
By (\ref{an.0.15}) and  (\ref{an.0.16}) the following estimates hold true  
\beq \label{eb.0}\begin{array}{l}
\|    P_{\at, 1}(\overline{\theta}, \alpha, 0) (  1 -  P_{\at, 1}(\theta, \alpha, 0)   )   \| = \\\\
\|   ( P_{\at, 1}(\overline{\theta}, \alpha, 0)  -  P_{\at, 1}(\theta, \alpha, 0) ) (  1 -  P_{\at, 1}(\theta, \alpha, 0)   )   \| \leq \frac{3}{4}\comma
\end{array}
\ene
\beq \label{eb.0.1}\begin{array}{l}
 1 - \frac{2}{8} 
 \leq  \|    \phi_{\alpha, \theta}  \|_{\cH_\at} \\\\ = \|    ( P_{\at, 1}(\theta, \alpha, 0)   -    P_{\at, 1}(0, 0, 0)) \phi_{0, 0} + \phi_{0, 0}  \|
\leq  1 +   \frac{2}{8}\period
\end{array}
\ene

Now we prove that if $ \phi_{\alpha, \theta} \ne 0$ is such that $ P_{\at, 1}(\theta, \alpha, 0) \phi_{\alpha, \theta}  =  \phi_{\alpha, \theta} $
then 
\beq \label{eb.1} \begin{array}{l}
\| e^{\beta \la  x  \ra }   P_{\at, 1}(\theta, \alpha, 0) \| \\\\ \leq \frac{ \|    e^{\beta \la  x  \ra }  
 \phi_{\alpha, \theta} \|_{\cH_\at}  }{   
\| \phi_{\alpha, \theta} \|_{\cH_\at}  (1 - \|    P_{\at, 1, \alpha, 0}(\overline{\theta}) (  1 -  P_{\at, 1}(\theta, \alpha, 0)   )   \|^2)^{1/2} } \period
\end{array}
\ene
Suppose that $\phi \ne 0$, as the range of $   P_{\at, 1}(\theta, \alpha, 0)  $ is one dimensional we can write
\beq \label{eb.2}
\phi = \lambda_\phi \phi_{\alpha, \theta} + \bar{\phi} \comma
\ene
where $ (1 -  P_{\at, 1}(\theta, \alpha, 0) )\bar{\phi} = \bar{\phi}  $. \\
A simple computation yields 
\beq \label{eb.3} \begin{array}{l}
\| \phi \|_{\cH_\at}^2 \geq 
 |\lambda_\phi|^2 \|  \phi_{\alpha, \theta}  \|_{\cH_\at}^2  + \|   \bar{\phi}   \|_{\cH_\at}^2 \\\\ 
- 2 |\lambda_\phi | | \la  P_{\at, 1}(\theta, \alpha, 0)   \phi_{\alpha, \theta}   |\;  (1 -  P_{\at, 1}(\theta, \alpha, 0) )  \bar{\phi} \ra | \\\\
\geq  |\lambda_\phi|^2 \|  \phi_{\alpha, \theta}  \|_{\cH_\at}^2  + \|   \bar{\phi}   \|_{\cH_\at}^2 \\\\ 
- 2 |\lambda_\phi  \| \phi_{\alpha, \theta} \|_{\cH_\at}   \|\bar{\phi} \|_{\cH_\at}   
\|   P_{\at, 1}(\overline{\theta}, \alpha, 0)   (1 -  P_{\at, 1}(\theta, \alpha, 0) )\| \\\\  \geq
 |\lambda_\phi|^2 \|  \phi_{\alpha, \theta}  \|_{\cH_\at}^2  (1 - \|   P_{\at, 1}(\overline{\theta}, \alpha, 0) 
   (1 -  P_{\at, 1}(\theta, \alpha, 0))\|^2  ) \period
\end{array}
\ene
Then we have that 
\beq \label{eb.4} \begin{array}{l}
\|   e^{\beta \la  x  \ra}   P_{\at, 1}(\theta, \alpha, 0)  \|= 
\sup_{\phi \ne 0, \lambda_\phi \ne 0} \frac{ \|    e^{\beta \la  x  \ra }   P_{\at, 1}(\theta, \alpha, 0) \phi  \| }{ \| \phi \|  } \\\\
\leq \sup_{\phi \ne 0, \lambda_\phi \ne 0} \frac{ |\lambda_\phi|   \|    e^{\beta \la  x  \ra }  \phi_{\alpha, \theta}  \|_{\cH_\at} }
{ |\lambda_\phi|    \| \phi_{\alpha, \theta} \|_{\cH_\at} (1 - \|   P_{\at, 1}(\overline{\theta}, \alpha, 0)   
(1 -  P_{\at, 1}(\theta, \alpha, 0) )\|^2  )^{1/2} }\comma
\end{array}
\ene
which implies (\ref{eb.1}). 
 
We calculate 
\beq \label{eb.5}\begin{array}{l}
\|     e^{ \beta\la  x  \ra}   \phi_{\alpha, \theta}  \|_{\cH_\at}  =  \|  e^{\beta\la  x  \ra}  P_{\at, 1}(\theta, \alpha, 0)    \phi_{0,0}  \|_{\cH_\at} 
\\\\ 
 =  \|    P_{\at, 1}(\theta, \alpha, - \beta)    e^{\beta \la  x  \ra } \phi_{0,0}  \|_{\cH_\at}  \comma
\end{array}
\ene
Using (\ref{an.0.16}), (\ref{eb.0})-(\ref{eb.1}) and (\ref{eb.5}) we obtain
\beq \label{eb.6}\begin{array}{l}
\|  e^{\beta \la x \ra}  P_{\at, 1}(\theta, \alpha, 0) \| \leq \frac{ \|  P_{\at, 1}(\theta, \alpha, - \beta) \|  \|   e^{\beta \la  x  \ra } \phi_{0,0}  \|_{\cH_\at}  }{   
\| \phi_{\alpha, \theta} \|_{\cH_\at}  (1 - \|    P_{\at, 1}(\overline{\theta}, \alpha, 0) (  1 -  P_{\at, 1}(\theta, \alpha, 0))   \|^2)^{1/2} }  \\\\
 \leq 8  \|   e^{\beta \la  x  \ra } \phi_{0,0}  \|_{\cH_\at}   \period
\end{array}
\ene

To prove (\ref{T.eb.1.e4}) we notice that 
\beq \begin{array}{l} \label{ei.50prima}
\|  e^{- \bob  \la  x \ra} (1 +  |x|^2)\| \leq \|  e^{- \bob   |x| } (1 +  |x|^2)  \| \\\\
 \leq (1 + e^{-2}(1 + 4 /\bob^2) )  \period  
\end{array}
\ene

\QED

\section{ \; Proofs of Section~\ref{s.rbapha}} \label{rbapha-proofs}

\subsection{\; Proof of Lemma~\ref{L.i.1}}

\newtheorem*{L.i.1}{Lemma~\ref{L.i.1}}

\begin{L.i.1}
 Let $ s' \leq s  < t$ be positive real or infinite numbers. Suppose that $\alpha \leq \boa $ and $|\beta| \leq \bob $ ( see \ref{geh.6.1}). 

For every $\rho > 0$ and every $\theta, \eta \in \CC$ with $|\theta| \leq \frac{1}{120}$ and $ |\theta + \eta| \leq \frac{1}{120} $ there exist constants $C_{\ref{i.13.4}}$ and 
$C_{\ref{i.17}}$ such that the following  
estimates hold true. 
\beq\tag{\ref{L.i.1.e1}}
\begin{array}{l}
\| W^{s,t}(\theta) \phi \|_{\cH^{s',t}} \leq   C_{\ref{i.13.4}} ( (1 + \frac{1}{\rho^{1/2} })\alpha^{3/2} +  (1 + \frac{1}{\rho^{1/2} })^2\alpha^{3})\cdot \\\\ 
(\|(H_0^{s',t}(\theta) + \rho)\phi \|_{\cH^{s',t}} +  \| \phi \|_{\cH^{s',t}}) \comma
\end{array}
\ene

more over, if $-\rho $ belongs to the resolvent set of $H_{0}^{s,t}(0, \alpha, \beta)$,
\beq
\begin{array}{l} \tag{\ref{L.i.1.e2}}
\big\| \big(W^{s,t}(\theta + h) - W^{s,t}(\theta)\big) \frac{1}{ H^{s',t}(0) + \rho}   \big\| 
\leq |h|  C_{\ref{i.17}} \big( (1 + \frac{1}{\rho^{1/2} })\alpha^{3/2} +  \\\\ (1 + \frac{1}{\rho^{1/2} })^2\alpha^{3}\big) 
(1 + \|(H_0^{s',t}(0) + \rho)^{-1} \| ) \period
\end{array}
\ene
and the operator-valued function $\theta \to W^{s,t}(\theta)
\frac{1}{ H^{s',t}(0) + \rho} $ is analytic for $|\theta| < \frac{1}{120}$.\\
Explicit values for the constants are given in (\ref{i.13.4}) and (\ref{i.17}). 
\end{L.i.1}

\noindent We introduce the functions
\beq \label{pft.1.1} \begin{array}{l}
 \boQ^{s,t}(\theta)(x, k)  :=  G^{s,t}(\theta)(0, k) \cdot \Big( \eta\Big( |x||k|  \Big)  e^{\theta} x \Big) \comma \\\\
 Q^{s,t}(\theta)(x, k ) : =  2(   G^{s,t}(\theta)(x, k) -  e^{- \theta}\nabla  \boQ^{s,t}(\theta)(x, k) ) \\\\ \cdot
(i e^{- \theta} \beta \frac{x}{ (1 + |x|^2)^{1/2} }  ) 
+ i e^{- \theta}|k| \boQ^{s,t}(\theta)(x, k)  \period
\end{array}
\ene
and the operator
\beq \label{i.6.2}
\begin{array}{l}
\tilde b^{s,t}(\theta)   
: =  a^*( Q^{s,t}(\theta)) +  a( Q^{s,t}(\overline{\theta})  ) \period
\end{array}
\ene

From Eqs. (\ref{eph})-(\ref{icfh.1}) we can write 
\beq \label{i.8} \begin{array}{l}
W^{s,t}(\theta) : =  e^{- \theta }A^{s,t}(\theta) \cdot (i\nabla) \otimes  \one_{\cH_{\cF^{s,t}}}  + 
e^{-\theta} (i\nabla) \otimes  \one_{\cH_{\cF^{s,t}}}  \cdot  A^{s,t}(\theta) \\\\ + A^{s,t}(\theta) \cdot A^{s,t}(\theta)  
+ \tilde b^{s,t}(\theta) \period
\end{array}
\ene 

Below we do not write the operator $\one_{\cH_{\cF^{s,t}}}$, in the understanding that we identify $ i \nabla $ with $ (i\nabla) \otimes  \one_{\cH_{\cF^{s,t}}}  \ $.

By Lemma~\ref{L.ih.1} (see also (\ref{eph.2})), for every $\phi \in \dom(\hf^{s',t}) \cap
\dom(- \Delta )$
\beq \label{i.11}
\begin{array}{l}
\| e^{- \theta} A^{s,t}(\theta) \cdot \nabla \phi  \|_{\cH^{s',t}} \leq 2 e^{ 1/30} \\\\ \cdot \sum_{j = 1}^{3} \sup_{ |\theta| \leq \frac{1}{30}} 
\| G^{s,t}(\theta)_j  + e^{ - \theta} \partial_j \boQ^{s,t}(\theta)  
 \|_{\rho}
\| (\hf^{s,t} + \rho)^{1/2} (- i \partial_{j}) \phi \|_{\cH^{s',t}} \\\\
\leq e^{1/30} ( \sum_{j = 1}^{3} \sup_{|\theta| \leq \frac{1}{30} } \|  G^{s,t}(\theta)_j  + e^{ - \theta} \partial_j \boQ^{s,t}(\theta) 
 \|_{\rho}  ) \\\\ \cdot 
\| ( \hf^{s,t} + \rho - \Delta)\phi    \|_{\cH^{s',t}} \comma \\\\
\|  A^{s,t}(\theta) \cdot A^{s,t}(\theta)  \phi  \|_{\cH^{s',t}} \\\\ \leq  4
 ( \sum_{j = 1}^{3} \sup_{|\theta| \leq \frac{1}{30}} (\| G^{s,t}(\theta)_j +   e^{- \theta} \partial_j  \boQ^{s,t}(\theta)  
 \|_{\rho}) ^2 )  \\\\ \cdot 
\| ( \hf^{s',t} + \rho - \Delta)\phi    \|_{\cH^{s',t}} \period
\end{array}
\ene
By (\ref{geh.5}),
\beq  \label{i.12} \begin{array}{l}
(1 - \epsilon)\| ( \hf^{s',t} + \rho - \Delta)\phi    \|_{\cH^{s',t}}  \leq \\\\
 \| ( \hf^{s',t} + \rho - \Delta  + V(\theta) )\phi    \|_{\cH^{s',t}} +  
 b_{\epsilon} \| \phi  \|_{\cH^{s',t}}   \period
\end{array}
\ene
We have also that, 
\beq \label{i.13} \begin{array}{l}
\| ( \hf^{s',t} - e^{-\theta}\hf^{s',t}  - \Delta + e^{- 2\theta}\Delta  )\phi \|_{\cH^{s',t}} \\\\  \leq
(|1- e^{\theta}| + |1- e^{- 2\theta}|  )\| ( \hf^{s',t} + \rho - \Delta)\phi    \|_{\cH^{s',t}} 
\end{array}
\ene
Now we use Remark \ref{R.a.1} to estimate 
\beq \label{i.13.1} \begin{array}{l}
\| G^{s,t}(\theta)_j +   e^{- \theta} \partial_j  \boQ^{s,t}(\theta)  
 \|_{\rho} \\\\ \leq (1 + \frac{1}{\rho^{1/2}}) C_{\ref{R.a.1.1}} (1 + (1 + 6 \|  \eta'  \|_\infty)   ) \alpha^{3/2}\comma 
 \end{array}
\ene
where we used (\ref{pft.1}),  (\ref{pft.1.1}). Notice that $ \eta(|x||k|) = 0$ if $ |x||k|\geq 2 $ and thus we can bound
$  |x| $ by $\frac{2}{|k|} $ when we estimate the integrals.

 Analogously we get
\beq \label{i.13.2}
\begin{array}{l}
\|\Delta  e^{- 2\theta}  \boQ^{s,t}(\theta)  
 \|_{\rho} \leq 3 e^{1/15} (1 + \frac{1}{\rho^{1/2}})C_{\ref{R.a.1.1}} 
( 4 \| \eta'   \|_{\infty}  + 2\| \eta''   \|_{\infty} ) \alpha^{3/2} \comma\\\\
\|   Q^{s,t}(\theta)  
 \|_{\rho} \leq 6 e^{1/15} (1 + \frac{1}{\rho^{1/2}})C_{\ref{R.a.1.1}} \alpha^{3/2} \\\\
+  12\beta (1 + \frac{1}{\rho^{1/2}})C_{\ref{R.a.1.1}} \alpha^{3/2}(1 + 2\| \eta' \|_\infty)\period 
\end{array}
\ene

Using (\ref{i.11})-(\ref{i.13.2}), that 
\beq \label{i.13.3}
 \| (\hf^{s,t} + \rho)^{1/2}  \phi \|_{\cH^{s',t}}  \leq \frac{1}{2} ( \| (\hf^{s,t} + \rho)  \phi \|_{\cH^{s',t}} + \|  \phi \|_{\cH^{s',t}} )\comma
\ene
the fact that $|1- e^{\theta}| + |1- e^{- 2\theta}| \leq 1/4 $  
 for $ |\theta| \leq \frac{1}{30} $  and similar 
computations, we deduce Eq. (\ref{L.i.1.e1}) with
\beq \label{i.13.4} \begin{array}{l}
C_{\ref{i.13.4}} : = 30(1 + b_{1/4})\Big[( 3 +   6 \| \eta'   \|_{\infty}  + 2\| \eta''   \|_{\infty} )^2 ( C_{\ref{R.a.1.1}} + C_{\ref{R.a.1.1}}^2 ) \\\\
+ 24\bob C_{\ref{R.a.1.1}} (1 + 2\| \eta' \|_\infty)\Big]\period  
\end{array}
\ene

The function $G^{s,t}(\theta)(x, k)$  in Eq. (\ref{i.2}) is analytic in 
$\theta $ when we leave $x$ and $k $ fixed, we denote by $G^{s,t}(\theta)(x, k)'$ 
its derivative with respect to $\theta$. We denote by
$ A^{s,t}(\theta)'  $ the operator that we obtain after writing $ G^{s,t}(\theta)(x, k)'  $
instead of $ G^{s,t}(\theta)(x, k) $ in (\ref{i.6}), (\ref{lambda-pf}) and (\ref{eph.2}). 
In the same way we define the formal derivative with
respect to $\theta$ of the operator $ \tilde b^{s,t}(\theta) $ introduced 
in Eq. (\ref{i.6.2}) and we denote it by  $ \tilde b^{s,t}(\theta)' $.    
The formal derivative of the interaction is the operator 
\beq \label{i.14}
\begin{array}{l}
W^{s,t}(\theta)' :=   e^{- \theta} A^{s,t}(\theta)' \cdot (i \nabla) +  e^{- \theta}(i \nabla) \cdot A^{s,t}(\theta)' \\\\
+  e^{- \theta} A^{s,t}(\theta) \cdot (-i \nabla) +  e^{- \theta}(-i \nabla) \cdot A^{s,t}(\theta) 
\\\\ +   A^{s,t}(\theta) \cdot (A^{s,t}(\theta))' + 
(A^{s,t}(\theta) )'\cdot A^{s,t}(\theta) +  \tilde b^{s,t}(\theta)'  \period 
\end{array}
\ene

\noindent By Lemma~\ref{L.ih.1}, 
we obtain, 
\beq \label{i.15}
\begin{array}{l}
\| e^{- \theta} (\frac{A^{s,t}(\theta + h) - A^{s,t}(\theta) }{h} - (A^{s,t}(\theta))')
 \cdot \vnabla \phi  \|_{\cH^{s',t}}  \\\\ \leq 2 e^{1/30}  \Big( \sum_{j = 1}^{3}
 \| \frac{G^{s,t}(\theta+ h)_j - G^{s,t}(\theta)_j}{h} - (G^{s,t}(\theta)_j)' 
\\\\
 +     \frac{  e^{-\theta - h}  \partial_j   \boQ^{s,t}(\theta + h)          - e^{-\theta}  \partial_j   \boQ^{s,t}(\theta)}{h} 
- (e^{-\theta}  \partial_j   \boQ^{s,t}(\theta))'     \|_{\rho}  \Big) \cdot \\\\
 \| ( \hf^{s,t} + \rho - \Delta)\phi    \|_{\cH^{s',t}} \comma
\end{array}
\end{equation}

\beq \label{i.15.1}
\begin{array}{l}
\| ( \frac{ A^{s,t}(\theta + h)\cdot A^{s,t}(\theta + h) - 
A^{s,t}(\theta)\cdot A^{s,t}(\theta)}{h}  -  (A^{s,t}(\theta)\cdot A^{s,t}(\theta)' +
A^{s,t}(\theta)'\cdot A^{s,t}(\theta) )) \phi   \|_{\cH^{s',t}} \\\\
\leq 4 \Big( \sum_{j = 1}^{3}
2 \|  \frac{G^{s,t}(\theta+ h)_j - G^{s,t}_{\theta , j}}{h} - (G^{s,t}(\theta)_j)' 
   \\\\ +     \frac{  e^{-\theta - h}  \partial_j   \boQ^{s,t}(\theta + h)          - e^{-\theta}  \partial_j   \boQ^{s,t}(\theta)}{h} 
- (e^{-\theta}  \partial_j   \boQ^{s,t}(\theta))'                  \|_{\rho} \\\\  \cdot \| G^{s,t}_{\theta , j} + e^{-\theta}  \partial_j  
 \boQ^{s,t}(\theta)    \|_{\rho}  + \\\\
\|  \frac{G^{s,t}(\theta+ h)_j - G^{s,t}_{\theta , j}}{h} + 
  \frac{  e^{-\theta - h}  \partial_j   \boQ^{s,t}(\theta + h)     
     - e^{-\theta}  \partial_j   \boQ^{s,t}(\theta)}{h}  \|_{\rho} \\\\ \cdot
\| G^{s,t}(\theta + h)_j - G^{s,t}_{\theta , j} +   e^{-\theta - h}  \partial_j   \boQ^{s,t}(\theta + h)   
 - e^{-\theta}  \partial_j   \boQ^{s,t}(\theta)  \|_{\rho} \Big)  \\\\ \cdot
 \| ( \hf^{s,t} + \rho - \Delta)\phi    \|_{\cH^{s',t}} \period
\end{array}
\ene 
Using (\ref{L.a.2.e0}), it follows from (\ref{i.12}- \ref{i.13.4}),  (\ref{i.15}), (\ref{i.15.1}) and similar estimates that 
\beq \label{i.16}  \begin{array}{l}
\| (\frac{W^{s,t}(\theta + h) - W^{s,t}(\theta)}{h} - W^{s,t}(\theta)' \phi \|_{\cH^{s',t}} \leq 
 (\frac{1}{(1/60 -1/120)^3 } +   \frac{1}{4(1/60 -1/120)^4 })|h|  C_{\ref{i.13.4}}\cdot  \\\\
( (1 + \frac{1}{\rho^{1/2} })\alpha^{3/2} +  (1 + \frac{1}{\rho^{1/2} })^2\alpha^{3}) 
(\|(H^{s',t}_{0}(0) + \rho)\phi \|_{\cH^{s',t}} +  \| \phi \|_{\cH^{s',t}})     \comma 
\end{array}
\ene

and therefore $ W^{s,t}(\theta)  \frac{1}{ H^{s',t}_{0}(0) + \rho  }  $ is analytic
 in for $|\theta| < \frac{1}{120} $, whenever $  \rho $ belongs to the resolvent set of $  H^{s',t}_{0}(0)  $. \\
Finally (\ref{L.i.1.e2}) can be proved as in (\ref{i.16}) with 
\beq \label{i.17}
C_{\ref{i.17}} : =  (\frac{1}{(1/60 -1/120)^3 } +   \frac{1}{4(1/60 -1/120)^4 }) C_{\ref{i.13.4}} \period   
\ene

\QED

\subsection{\; Proof of Theorem~\ref{L.inh.1}}

\newtheorem*{L.inh.1}{Theorem~\ref{L.inh.1}}

\begin{L.inh.1} 

For every $\rho > 0$ such that $- \rho$ belongs to the resolvent set 
of $H^{s,t}_0(0)$, the function $\theta  \to H^{s,t}( \theta) \frac{1}{H^{s,t}(0) + 
\rho}$ is an operator-valued analytic function for $|\theta| < \frac{1}{120}$.
 Moreover, for every 
$\theta, h \in \CC$ such that $  |\theta| < \frac{1}{120}, |\theta + h| < \frac{1}{120} $
 the following estimate holds

\beq \tag{\ref{L.inh.1.e1}} 
\|(H^{s,t}(\theta + h) - H^{s,t}(\theta )) \frac{1}{H_0^{s,t}(0) + \rho}\| \leq
C_{\ref{L.inh.1.e2}}  |h|(1 + \| \frac{1}{H_0^{s,t}(0) + \rho}  \|) \comma
\ene
where
\beq \tag{\ref{L.inh.1.e2}}
\begin{array}{l}
C_{\ref{L.inh.1.e2}} : =  C_{\ref{L.inh.1.e2}}(\alpha,\rho) : = \big( C_{\ref{i.17}} ( (1 + \frac{1}{\rho^{1/2} })\alpha^{3/2} \\\\ +  (1 + \frac{1}{\rho^{1/2} })^2\alpha^{3}) 
 + \frac{4}{3} (  1 + b_{1/4} )(\frac{C_{\ref{an.0.7}}}{2(1/60 - 1/120)^2} +  3 e^{1/15} )\big) 
\end{array}
\ene

\end{L.inh.1}

By (\ref{i.12}) we have that, 
$ \| \frac{ 1- \Delta + \rho + \hf^{s,t} }{H^{s,t}_{0}(0) + \rho}\|  \leq \frac{1}{1-1/4}
( 1 + (b_{1/4} + 1 )\|  \frac{1}{ H^{s,t}_{0}(0) + \rho}   \|)  $. Then (\ref{L.inh.1.e1} )
follows from (\ref{L.i.1.e2}), (\ref{L.a.2.e0}), functional calculus and
 the following,

\beq \label{inh.4} 
\begin{array}{l}
\|(V(\theta + h, \alpha, \beta) - V(\theta, \alpha, \beta) + e^{- 2\theta - 2 h}\Delta- e^{- 2\theta} \Delta
+ e^{-\theta - h}\hf^{s,t} - e^{-\theta}\hf^{s,t}) \frac{1}{H^{s,t}_{0}(0) + \rho}\| \\\\
\leq \Big( \| (V(\theta + h,\alpha, \beta) - V(\theta, \alpha, \beta))\frac{1}{1 - \Delta }   \|
\|  \frac{1 - \Delta }{ 1 - \Delta + \rho + \hf^{s,t} }  \|     \\\\ + 
 \| e^{-2 (\theta + h)} - e^{- 2 \theta}  \| \| \frac{-\Delta}{ 
1 - \Delta + \rho + \hf^{s,t}  }     \|  +  \| e^{-  (\theta + h)} -
 e^{-   \theta}  \| \| \frac{\hf^{s,t}   }{ 
1 - \Delta + \rho + \hf^{s,t}  }     \| \Big) \\\\
 \| \frac{ 1- \Delta + \rho + \hf^{s,t} }{H^{s,t}_{0}(0) + \rho} \| \period
\end{array}
\ene
The proof for analyticity can be done analogously.

\QED

\subsection{\; Proof of Corollary~\ref{L.an.1}}

\newtheorem*{L.an.1}{Corollary \ref{L.an.1}}

\begin{L.an.1}
Suppose that $\alpha \leq \boa $ and $|\beta| \leq \bob $ (see \ref{geh.6.1}). Suppose furthermore that $\alpha^{3/2} \leq \frac{1}{128 C_{\ref{i.17}} }$ $z \in \CC$ is such that $ \rm{Re}(z) \leq \frac{3}{2} e_0 - \frac{3}{4}  $. Then 
$H^{s,t}(0, \alpha, \beta) - z$ is invertible and 
\beq \tag{\ref{L.an.1.e1}}
\|  \frac{1}{H^{s,t}(0, \alpha, \beta) - z}  \| \leq \Big(\frac{4}{3}\Big)^2 \frac{1}{| z - \boe_0  |} \comma
\ene
\beq \tag{\ref{L.an.2.e1}}
\| (H_{0}^{s,t}(0, \alpha, \beta) - z)  \frac{1}{H^{s,t}(0, \alpha, \beta) - z}  \| \leq  2 \period
\ene
Suppose furthermore that $\theta, h \in \CC$ are such that $|\theta| < \frac{1}{120}$ and $|\theta + h| < \frac{1}{120}$, then
\beq \tag{\ref{L.an.3.e1}} \begin{array}{l}
\| (H^{s,t}(\theta + h, \alpha, \beta) - H^{s,t}(\theta, \alpha, \beta))  \frac{ 1  }{H^{s,t}(0, \alpha, \beta) - z}  \| \\\\ \leq 
|h|\cdot 2 \Big( 1 +  \Big( \frac{4}{3}\Big)^2 \Big) \big( \frac{1}{16} +   \frac{4}{3} (  1 + b_{1/4} )(\frac{C_{\ref{an.0.7}}}{2(1/60 - 1/120)^2} +  3 e^{1/15} )\big)  \period
\end{array}
\ene
In particular if  
$$ |h|  \leq   \frac{1}{4}    \Big( 2 \Big( 1 +  \Big( \frac{4}{3}\Big)^2 \Big) \cdot \Big( 
  \frac{1}{16} +   \frac{4}{3} (  1 + b_{1/4} )(\frac{C_{\ref{an.0.7}}}{2(1/60 - 1/120)^2} +  3 e^{1/15} ) \Big) \Big)^{-1},  
$$ it follows that 
\beq \tag{\ref{L.an.3.1.e1}}
\|  \frac{ 1  }{H^{s,t}(\theta, \alpha, \beta) -  z}  \| \leq \Big(\frac{4}{3 }\Big)^3 \cdot \frac{1}{| z - \boe_0  |}   \period
\ene
\end{L.an.1}

\noindent{\bf Proof of (\ref{L.an.1.e1}): }\\
By (\ref{geh.8}), for any $z \in \CC$ with ${\rm Re( \, z) \leq \boe_0 - \frac{\delta_\at}{16}}$
\beq \label{an.m1.1}
\|  \frac{1}{H_{\at}(0, \alpha, \beta) - z }\|  \leq \frac{4}{3}\frac{1}{|z - \boe_0|}\period
\ene
By functional calculus, 
\beq \label{an.m1.2}
\|  \frac{1}{H_{0}^{s,t}(0, \alpha, \beta) - z }   \| \leq \max_{r > 0 }\|  \frac{1}{H_{\at}(0, \alpha, \beta) - z + r }\| \leq  
\frac{4}{3}\frac{1}{|z - \boe_0|}\period 
\ene
Then we use (\ref{L.i.1.e1}) to show that 
for our choice of $\alpha$ (notice that $  C_{\ref{i.17}} \geq   C_{\ref{i.13.4}} \geq 1$)
\beq \label{an.1} \begin{array}{l}
\| W^{s,t}(0)  \frac{1}{ H_{0}^{s,t}(0, \alpha, \beta)  - {\rm Re} (\, z)  }   \| \leq \frac{1}{16} \comma \\\\
\| W^{s,t}(0)  \frac{1}{ H_{0}^{s,t}(0, \alpha, \beta)  -   z  }   \| 
 \leq \frac{1}{16} \|   \frac{     H_{0}^{s,t}(0, \alpha, \beta)  - {\rm Re} (\, z)       }{ H_{0}^{s,t}(0, \alpha, \beta)  -  z  }   \| \leq  \frac{1}{16} (1 + \frac{4}{3})
\leq \frac{1}{4}
\period
\end{array}
\ene
Now we use the Neumann series to compute $   \frac{1}{H^{s,t}(0, \alpha, \beta) -  z  }   $: 
\beq \label{an.2}
 \frac{1}{H^{s,t}(0, \alpha, \beta)  -  z  }  =
  \frac{1}{ H_{0}^{s,t}(0, \alpha, \beta)   - z  }  \sum_{j=0}^{\infty}
 \big[ -    W^{s,t}(0)  \frac{1}{ H_{0}^{s,t}(0, \alpha, \beta)  - z  }     \big]^j \period
\ene
Eq. (\ref{an.1}) assures the convergence of the Neumann series and the bound (\ref{L.an.1.e1}).  

\noindent{\bf Proof of (\ref{L.an.2.e1}): }\\
Using (\ref{L.i.1.e1}) and taking on account our selection of $\alpha$ we obtain,
\beq \label{an.3}
\| W^{s,t}(0) \phi    \|_{\cH^{s,t}} \leq \frac{1}{32}(\|(H_{0}^{s,t}(0, \alpha, \beta) - z) \phi \|_{\cH^{s,t}} +
 (1 + |   {\rm Im} (\, z) | ) \| \phi \|_{\cH^{s,t}})\period 
\ene
Then we have
\beq \label{an.4}\begin{array}{l} 
(1 - \frac{1}{32})\|(H_{0}^{s,t}(0, \alpha, \beta) - z)\phi \|_{\cH^{s,t}} \\\\ \leq \|(H^{s,t}(0, \alpha, \beta) - z)\phi \|_{\cH^{s,t}} +   
\frac{1}{32}  (1 + |  {\rm Im} (\, z) | )   \| \phi \|_{\cH^{s,t}}  \period
\end{array} \ene
Using  \ref{L.an.1.e1} we get, 
\beq \label{an.5} \begin{array}{l}
\|(H^{s,t}_0(0, \alpha, \beta) - z) \frac{1}{H^{s,t}(0, \alpha, \beta) - z} \| \leq \frac{32}{31} 
( 1  +   \frac{1}{16}  \Big(\frac{4}{3} \Big)^2    )  
\period
\end{array}
\ene

\noindent{\bf Proof of \ref{L.an.3.e1} } \\
Eq. (\ref{L.an.3.e1}) follows from Theorem~\ref{L.inh.1} and (\ref{L.an.1.e1})-(\ref{L.an.2.e1}).

\noindent{\bf Proof of \ref{L.an.3.1.e1} }\\

We use (\ref{L.an.1.e1}), (\ref{L.an.3.e1}) and the Neumann series.

\QED

\subsection{ \; Proof of Corollary~\ref{L.an.4}}

\newtheorem*{L.an.4}{Corollary \ref{L.an.4}}

\begin{L.an.4}
Suppose that $\alpha \leq \boa $ and $|\beta| \leq \bob $ ( see \ref{geh.6.1}). Suppose furthermore that $\alpha^{3/2} \leq \frac{1}{128 C_{\ref{i.17}} }$. Then
for every $\theta \in D_{1/120}(0)$, $h \in D_{1/120}(0) \cap(i \RR)$   with 

\beq
\begin{array}{l}\tag{\ref{theta}} 
 |\theta| <  \frac{1}{16}    \Big( 2 \Big( 1 +  \Big( \frac{4}{3}\Big)^2 \Big) \\\\ \cdot
 \Big( \frac{1}{16} +   \frac{4}{3} (  1 + b_{1/4} )(\frac{C_{\ref{an.0.7}}}{2(1/60 - 1/120)^2} +  3 e^{1/15} )\Big)  \Big)^{-1}  $$
\end{array}
\ene
 the following estimate holds

\beq \tag{\ref{L.an.4.em1}}
\|  (H^{s,t}_0(0, \alpha, \beta) -    \rho )\frac{1}{  H^{s,t}(\theta, \alpha, \beta) - \rho   }    \| \leq 2 \comma
\ene

\beq \tag{\ref{L.an.4.e1}}
\|  (H^{s,t}_0(0, \alpha, \beta) -    \rho )\frac{1}{  H^{s,t}(\theta, \alpha, \beta) - e^{h} (\rho + r)  }    \| \leq 8 \comma
\ene

where $r \in \RR $ is any negative number and $\rho \leq 2 e_0 - 1 $.  
\end{L.an.4}

Let $z = x+ i y$ with $x \leq \rho$.  From (\ref{L.i.1.e1}) and its proof we obtain that 
\beq \label{pen.1}
\|  W^{s,t}(\theta) \phi\|_{\cH^{s,t}} \leq \frac{1}{128} (\|  (H_0^{s,t}(0) - z) \phi  \|_{\cH^{s,t}} +(1+ |y|) \| \phi\|_{\cH^{s,t}}) 
\ene
where we used that $C_{\ref{i.17}} \geq 60 C_{\ref{i.13.4}} \geq 1 $. 

From the proof of Theorem~\ref{L.inh.1}, in particular from (\ref{inh.4}) it follows that 
\beq \label{pen.2}
\| (H_0^{s,t}(\theta) - H_0^{s,t}(0)) \frac{1}{H_0^{s,t}(0) - z}\|_{\cH^{s,t}} \leq \frac{1}{16},
\ene
where we used the assumptions on $\alpha$ and $\theta$ (see the proof of (\ref{L.an.3.e1})).  \\
Using (\ref{pen.1}) and (\ref{pen.2}) we conclude that 
\beq \label{pen.3}\begin{array}{l}
(1 - \frac{1}{128} - \frac{1}{16}) \|  (H_0^{s,t}(0) -  z)\phi\|_{\cH^{s,t}} \\\\ 
\leq  \|  (H^{s,t}(\theta) -  z)\phi\|_{\cH^{s,t}} 
+ \frac{1}{128} (1 + |y|) \| \phi\|_{\cH^{s,t}}
\end{array}
\ene
We obtain finally from (\ref{L.an.3.1.e1}) 
\beq \label{pen.4}
\|(H_0^{s,t}(0) -  z)\frac{1}{H^{s,t}(\theta) -  z}  \| \leq \frac{1}{ (1 - \frac{1}{128} - \frac{1}{16}) } (1 + \frac{2}{128}(4/3)^3),
\ene
which proves (\ref{L.an.4.em1}).

Now we prove (\ref{L.an.4.e1}). We compute 
\beq \label{pen.5} \begin{array}{l}
\|  (H^{s,t}_0(0, \alpha, \beta) -    \rho )\frac{1}{  H^{s,t}(\theta) - e^{h} (\rho + r)  }    \|
 \leq  \frac{1}{ (1 - \frac{1}{128} - \frac{1}{16}) } (1 + \frac{2}{128}(4/3)^3) 
\\\\ + \|  (e^{h} (\rho + r)- \rho )\frac{1}{  H^{s,t}(\theta) - e^{h} (\rho + r)  }    \| \leq  \frac{1}{ (1 - \frac{1}{128} - \frac{1}{16}) } (1 + \frac{2}{128}(4/3)^3) \\\\
+ (\frac{4}{3})^3 + (\frac{4}{3})^3  \frac{|\boe_0 - \rho|}{ |\boe_0 -  e^{h} (\rho + r)    | } \leq 8,
\end{array}
\ene
where we used (\ref{L.an.3.1.e1}). This last equation proves (\ref{L.an.4.e1}).

\QED

\section{ $\:$ Proofs of Section~\ref{s.rirh}} \label{s.rirh.p}

\subsection{ $\;$ Proof of Lemmata~\ref{L.neri.1} and \ref{L.neri.2}}

Before starting our estimates we write down, for the convenience of the reader, the explicit expression of the functions 
$w_{m,n}^{s,t}(\theta, \alpha, \beta)$ (see (\ref{i.10.4}) and (\ref{i.10.3})). We use notations (\ref{pft.1.1})-(\ref{i.8}).

\beq \label{i.10.3.0}
\begin{array}{l}
w^{s,t}_{1, 0}(\theta, \alpha, \beta)(x, k) := w^{s,t}_{1, 0}(\theta, \alpha, \beta)(k)  \\\\ : =
  2 \big(  G^{s,t}(\theta)(x, k)  -  e^{- \theta}( \nabla  \boQ^{s,t}(\theta)(x,k)) \big) \cdot (-i e^{- \theta} \nabla) \\\\  
+ i e^{- 2 \theta} \Delta  \boQ^{s,t}(\theta)(x,k) +   Q^{s,t}(\theta)(x,k)  
 \period
\end{array}
\ene

\beq \label{i.10.3.1}
\begin{array}{l}
w^{s,t}_{0, 1}(\theta, \alpha, \beta)(x, k) := w^{s,t}_{0, 1}(\theta, \alpha, \beta)(k)  \\\\ : =
 2 \big(  \overline{G^{s,t}}(\overline{\theta})(x, k)  -  e^{- \theta}( \nabla \overline{ \boQ}^{s,t}(\overline{\theta})(x,k)) \big) \cdot (-i e^{- \theta} \nabla)  \\\\  
+ i e^{- 2 \theta} \Delta \overline{ \boQ}^{s,t}(\overline{\theta})(x,k) +   \overline{Q}^{s,t}(\overline{\theta})(x,k)   
\period
\end{array}
\ene

\beq \label{i.10.3.2}
\begin{array}{l}
w^{s,t}_{2, 0}(\theta, \alpha, \beta)(x, k, k') := w^{s,t}_{2, 0}(\theta, \alpha, \beta)(k, k') \\\\ :=
 \big(  G^{s,t}(\theta)(x, k)  -  e^{- \theta}( \nabla  \boQ^{s,t}(\theta)(x,k)) \big) \\\\ \cdot 
  \big(  G^{s,t}(\theta)(x, k')  -  e^{- \theta}( \nabla  \boQ^{s,t}(\theta)(x,k')) \big)
 \comma
\end{array}
\ene

\beq \label{i.10.3.3}
\begin{array}{l}
w^{s,t}_{0, 2}(\theta, \alpha, \beta)(x, k, k') := w^{s,t}_{2, 0}(\theta, \alpha, \beta)(k, k') \\\\ :=
 \big(  \overline{G^{s,t}}(\overline{\theta})(x, k)  -  e^{- \theta}( \nabla \overline{ \boQ}^{s,t}(\overline{\theta})(x,k)) \big) \\\\ \cdot
 \big(  \overline{G^{s,t}}(\overline{\theta})(x, k')  -  e^{- \theta}( \nabla \overline{ \boQ}^{s,t}(\overline{\theta})(x,k')) \big)\comma
\end{array}
\ene

\beq \label{i.10.3.5}
\begin{array}{l}
w^{s,t}_{1, 1}(\theta, \alpha, \beta)(x, k, k') := w^{s,t}_{1, 1}(\theta, \alpha, \beta)(k, k') \\\\ := 
2 \big(  G^{s,t}(\theta)(x, k)  -  e^{- \theta}( \nabla  \boQ^{s,t}(\theta)(x,k)) \big) \\\\ \cdot 
\big(  \overline{G^{s,t}}(\overline{\theta})(x, k')  -  e^{- \theta}( \nabla \overline{ \boQ}^{s,t}(\overline{\theta})(x,k')) \big)\comma 
\end{array}
\ene

\beq \label{i.10.3.6}
\begin{array}{l}
w^{s,t}_{0, 0}(\theta, \alpha,\beta)(x):= w^{s,t}_{0, 0}(\theta, \alpha,\beta)\\\\  : = 
\big\la   \big(  \overline{G^{s,t}}(\overline{\theta})(x, k)  -  e^{- \theta}( \nabla \overline{ \boQ}^{s,t}(\overline{\theta})(x,k)) \big)   \; \\\\
 \big| 
  \big(  G^{s,t}(\theta)(x, k)  -  e^{- \theta}( \nabla  \boQ^{s,t}(\theta)(x,k)) \big)   \big\ra \period
\end{array}
\ene

\noindent
\subsubsection{ $\;$ Proof of Lemma~\ref{L.neri.1}}

\newtheorem*{L.neri.1}{Lemma~\ref{L.neri.1}}

\begin{L.neri.1}
Suppose that $\theta =\pm i  \nu $, where  $\nu \in \RR$ is different
from zero and that $\alpha \leq \boa$, $|\beta| \leq \bob$. Suppose furthermore that it satisfies (\ref{an.0.14}) and (\ref{theta}). Then,  
for every $\mu \geq 0$ and every $z \in \cE^{s}(\theta) \setminus \{e_{1}\}$, 
\beq \tag{\ref{L.neri.1.e0}}
\|   R^{s,t}_{0}(\theta, \alpha, \beta)(z)     \| 
\leq \frac{ 50 |e_0| C_{\ref{an.0.16.2}}  }{\delta |\sin(\nu)|} 
 \frac{1}{|z - e_{1}| }  \comma 
\ene
\beq \tag{\ref{L.neri.1.e1}}
\| |  R^{s,t}_{0}(\theta, \alpha, \beta)(z)|  (\hf^{s,t} + \mu)   \| 
\leq \frac{50 |e_0| C_{\ref{an.0.16.2}}}{\delta|\sin(\nu)|} (1  +
 \frac{\mu}{|z - e_{1}| }  ) \comma
\ene
\beq \tag{\ref{L.neri.1.e2}}
\|  | R^{s,t}_{0}(\theta, \alpha, \beta)(z)|  ( H_\at \pm i \delta)   \| \leq 
\frac{  C_{\ref{L.neri.1.e3} } }{|\sin(\nu)|} ( 1 +
  \frac{1}{|z- e_{1}|}  )) \period
\ene
For any $ z \in \cE^{(1/2 \delta_\at, \infty)} $,
\beq \tag{\ref{L.neri.1.e0.1}}
\| \overline{P}_{\at, 1}(\theta, \alpha, \beta)  R^{s,t}_{0}(\theta, \alpha, \beta)(z)     \| 
\leq \min \Big(  \frac{ 16 C_{\ref{an.0.16.2}}  }{|\sin(\nu)|\delta} ,
\frac{ 20 |e_{0}| C_{\ref{an.0.16.2}}  }{|\sin(\nu)|\delta} \frac{1}{|z - e_{0}|} \Big)  \comma 
\ene

where 
\beq \tag{\ref{L.neri.1.e3}}
C_{\ref{L.neri.1.e3} } :=    2\big(1+ 8 C_{\ref{an.0.16.2}}  + \frac{8 C_{\ref{an.0.16.2}}  }{|\sin(\nu)|}(1 + \frac{|e_0'|}{\delta}) + 
 \frac{8C_{\ref{an.0.16.2}} }{| \sin{\nu} |}( \delta + |e_1'|)\big)
 \period 
\ene

\end{L.neri.1}

First we stress the following: Suppose that $O_1, O_2$ are closed operators, densely defined on a Hilbert space. Suppose 
furthermore that $O_1$ and $ O_1 O_2 $ are bounded and that $ O_1 O_2 =  O_2 O_1 $, then 
\beq \label{neri.4.0}
\|  O_1 O_2 \| =\|  |O_1| O_2 \|= \| | O_1 O_2 | \| \period 
\ene 
We prove now (\ref{L.neri.1.e1}). By the functional calculus and Corollary~\ref{C.eeh.1}, 
\beq \label{neri.4}
\begin{array}{l}
\|   R^{s,t}_{0}(\theta, \alpha, \beta)(z)  (\hf^{s,t} + \mu) \| \\\\   \leq \sup_{r \in \{0\}\cup [s, \infty)}
(\| (H_{\at}(\theta, \alpha, \beta) - (z - e^{ -i\nu}r ))^{-1} \| (r + \mu) ) \\\\
 \leq 4 C_{\ref{an.0.16.2}} \sup_{r \in \{0\}\cup [s, \infty)} ( \frac{r + \mu}{|z - e^{-i\nu}r - e_{0}|}  
+ \frac{r + \mu}{| z - e^{-i\nu}r - e_{1} |}  ) \period
\end{array}
\ene
It is geometrically clear that  for $r \geq s$, 
$$  |z - e^{-i\nu}r - e_{1}| \geq s |\sin(\nu)|/2 .  
$$  
It is also easy to prove that for $r \geq s$ and for
$$z \in \mathcal{C}_\nu : = \{ z = z_1 + i z_2 \in  \mathcal{E}^{s,t}(\theta):
 z_1 > e_{1} \}, 
$$
\beq 
|z- e_1| \leq s \big( |\tan(\nu)/4|  + 
| \sin(\nu)/2 |   + \sqrt{ (1/4)^2 + (\sin(\nu)/2)^2} \big)\period
\ene 
Therefore for $z \in \mathcal{C}_v$ and $r \geq s$  
\beq \begin{array}{l}
\frac{|z - e_1|}{|z - e^{-i\nu}r - e_{1}|} \leq \frac{  |\tan(\nu)/4|  + 
| \sin(\nu)/2 |  + \sqrt{ (1/4)^2 + (\sin(\nu)/2)^2  } }
 {|\sin(\nu)|/2}. 
\end{array}
\ene
As  $|\nu | \leq \frac{1}{120} $ and   
$$    |z - e^{-i\nu}r - e_{1}| \geq | z - e_{1} | $$ 
for $ z \in \mathcal{E}^{s,t}(\theta)\setminus \mathcal{C}_\nu   $ we have that 
\beq \label{neri.5}
\frac{ | z - e_{1} | }{|z - e^{-i\nu}r - e_{1}| } \leq \frac{1}{|\sin(\nu)|} , \; 
\forall z \in \mathcal{E}^{s,t}(\theta)\setminus \{ e_{1} \}, \: \forall r \in \{0\} \cup[s, \infty) \period
\ene
Similarly we obtain that 
\beq \label{neri.6}
\begin{array}{l}
\frac{ 1 }{|z - e^{-i\nu}r - e_{0}| } 
\leq \min\Big( \frac{1}{\delta( 3/4 - \cos(\nu)/2 )|\sin(\nu)| } ,  \frac{ 5 |e_{0}| }{|\sin(\nu)|\delta} \frac{1}{|z - e_{0}|}
\Big)  \comma
\end{array}
\ene
for all  $ z \in \mathcal{E}^{s,t}(\theta)   $.\\
Now we denote by 
$$ \mathcal{C}_{\nu, 1} :  
 = \{ z = z_1 + i z_2  \in \mathcal{E}^{s,t}(\theta) : z_1 > e_{0} \} . $$ 
 For every 
$z \in \mathcal{C}_{\nu, 1}$, 
$$ |  z- e_{1} | \leq |e_0| 
(( |\tan(\nu)|5/4 + |\sin(\nu)|/2)^2 + 
(5/4)^2 )^{1/2}  \leq  2 |e_0|  $$ 
so, by (\ref{neri.6}), 
$$ \frac{1}{| z - e^{-i\nu}r - e_{0}  | } \leq  
 \frac{8|e_0|}{\delta | z- e_{1}  | |\sin(\nu)|}, $$
  for $z \in 
\mathcal{C}_{\nu, 1}$. \\
For $z = z_1 + i z_2 \in \mathcal{E}^{s,t}(\theta) 
\setminus \mathcal{C}_{\nu, 1} $ we have that
$$ |z  - e^{-i\nu}r  - e_0 | \geq |z_2| \geq
 (1/5)^{1/2} |\sin(\nu)| |z- e_1|.  
 $$ 
Then we have, 
\beq \label{neri.6.1}
\frac{1}{|z  - e^{-i\nu}r  -e_{0}|} \leq  \frac{8 |e_0|}{\delta |\sin(\nu)|}\frac{1}{|z-e_{1}|}
\; \forall  z \in \mathcal{E}^{s,t}(\theta) \setminus\{ e_{1} \} \period
\ene
Eq. (\ref{L.neri.1.e1})   follows from (\ref{neri.4}) - (\ref{neri.6.1}).\\
To estimate (\ref{L.neri.1.e2}) we proceed as in the proof of Lemma 3.12
of \cite{BachFroehlichSigal1999}: 
By the functional calculus, 
\beq \label{neri.7}
\| |R^{s,t}_{0}(\theta, \alpha, \beta)(z)| (H_{\at}(0, \alpha, \beta) \pm i \delta ) 
 \| \leq \sup_{r \in \{ 0 \} \cup [s, \infty)} 
\| Y_{\pm}  \| \comma 
\ene
where, 
\beq \label{neri.8}
Y_{\pm}:=  (H_{\at}(\theta, \alpha, \beta) - z + e^{-i \nu} r)^{-1} (H_{\at}(0, \alpha, \beta) \pm i\delta )\period
\ene
The identity, 
\beq \label{neri.9} \begin{array}{l}
Y_\pm = 1 - Y_\pm (H_{\at}(0, \alpha, \beta) \pm i \delta )^{-1} (H_{\at}(\theta, \alpha, \beta) - H_{\at}(0, \alpha, \beta) ) + \\\\
 (\pm  i \delta + z - e^{-i \nu} r ) (H_{\at}(\theta, \alpha, \beta) - z + e^{-i \nu} r)^{-1}
\end{array}
\ene
yields 
\beq \label{neri.10} \begin{array}{l}
\sup_{r \in \{ 0 \} \cup [t, \infty)} 
\| Y_{\pm}  \| \leq \frac{1}{1 - C_{\ref{an.0.9}}|\theta| }( 1+ 8 C_{\ref{an.0.16.2}} + \frac{8 C_{\ref{an.0.16.2}}  }{|\sin(\nu)|}
(1 + \frac{|e_{0}|}{\delta})\\\\ + 
 \frac{8C_{\ref{an.0.16.2}} }{| \sin{\nu} |}( \delta + |e_{1}|)
  \frac{1}{|z- e_{1}|}  )   \comma 
\end{array}  \ene
where we used (\ref{an.0.8}), Corollary~\ref{C.eeh.1}, (\ref{neri.5}) and 
(\ref{neri.6}). Eq. (\ref{L.neri.1.e2}) follows from (\ref{neri.10}). 

Finally  (\ref{L.neri.1.e0}) follows from
the functional calculus, Corollary~\ref{C.eeh.1} and (\ref{neri.5}, \ref{neri.6.1}).  (\ref{L.neri.1.e0.1})
is proved in the same way.

\QED

\subsubsection{ $\;$ Proof of Lemma~\ref{L.neri.2}}
\newtheorem*{L.neri.2}{Lemma~\ref{L.neri.2}}

\begin{L.neri.2}

Suppose that $ s' \leq  s < t  $ and that $\theta = \pm i \nu $, with $\nu \in \RR \setminus \{0 \}$.
Suppose furthermore that  $ \theta  $ satisfies (\ref{an.0.14}) and (\ref{theta}) and that  $\alpha \leq \boa$, $|\beta| \leq \bob$. For every 
$z \in \mathcal{E}^{s,t}(\theta) \setminus \{ e_{1} \}$ there is a constant $  C_{\ref{neri.15.2}} $ such that  
\beq \tag{\ref{L.neri.2.e0}}\begin{array}{l}
\|  |R^{s,t}_{0}(\theta, \alpha, \beta)(z)|^{1/2} W^{s,t}_{m, n}(\theta, \alpha, \beta)   |R^{s,t}_{0}(\theta, \alpha, \beta)(z)|^{1/2}  \| \\\\
\leq   \frac{C_{\ref{neri.15.2}}}{|\sin(\nu)|^{2}}\alpha^{3(m+n)/2} (1 + \frac{1}{|z- e_{1}|})^{1/2}  \comma m + n \geq 1\comma
\end{array}
\ene
\beq \tag{\ref{L.neri.2.e0.1}}\begin{array}{l}
\|  |R^{s,t}_{0}(\theta, \alpha, \beta)(z)|^{1/2} W^{s,t}_{0, 0}(\theta, \alpha, \beta)   |R^{s,t}_{0}(\theta, \alpha, \beta)(z)|^{1/2}  \| \\\\
\leq   \frac{C_{\ref{neri.15.2}}}{|\sin(\nu)|}\alpha^{3} (1 + \frac{1}{|z- e_{1}|})  \comma
\end{array}
\ene
\beq \tag{\ref{L.neri.2.e0.2}}\begin{array}{l}
\|  |R^{s,t}_{0}(\theta, \alpha, \beta)(z)|^{1/2} W^{s,t}(\theta, \alpha, \beta)   |R^{s,t}_{0}(\theta, \alpha, \beta)(z)|^{1/2}  \| \\\\
\leq    \frac{C_{\ref{neri.15.2}}}{|\sin(\nu)|^{2}}\alpha^{3/2} (1 + \frac{1}{|z- e_{1}|})^{1/2}   +    
 \frac{C_{\ref{neri.15.2}}}{|\sin(\nu)|}\alpha^{3} (1 + \frac{1}{|z- e_{1}|})  \comma
\end{array}
\ene
For any $z \in \cE^{(1/2 \delta_\at, \infty)}$,
\beq \tag{\ref{L.neri.2.e0.2.1}}\begin{array}{l}
\|  W^{s,t}(\theta, \alpha, \beta)  
 \overline{P}_{\at}(\theta, \alpha, \beta)R^{s',t}_{0}(\theta, \alpha, \beta)(z)  \| \\\\
\leq    C_{\ref{neri.15.2.1}}\alpha^{3/2}     \period
\end{array}
\ene
The explicit value of the constants  $  C_{\ref{neri.15.2}} $ and $ C_{\ref{neri.15.2.1}} $  are given in (\ref{neri.15.2}) and ({\ref{neri.15.2.1}}). 
\end{L.neri.2}

\begin{lemma}\label{L.neri.1.1}
Suppose that  $\theta = \pm i \nu $, with $\nu \in \RR \setminus \{0 \}$.
Suppose furthermore that  $ \theta  $ satisfies (\ref{an.0.14}) and (\ref{theta}) and that  $\alpha \leq \boa$, $|\beta| \leq \bob$.
Let $\mu \in B_{(|\sin(\nu)|/2) s}(e_{1})$ and $z \in \CC$ be such that $|z - e_{1}|  = (|\sin(\nu)|/2) s $ then 
\beq \label{L.neri.1.1.e1}
\| (H_0^{s,t}(\theta, \alpha, \beta) - z) \frac{\overline{P}_{\at, 1}(\theta, \alpha, \beta) + \overline{P}_{\Om^{s,t}} }{ H_0^{s,t}(\theta, \alpha, \beta) - \mu }  \| \leq
100 (1 +   C_{\ref{an.0.16.2}})  \comma
\ene
where $    \overline{P}_{\Om^{s,t}} : 1- P_{\Om^{s,t}}  $, $ P_{\Om^{s,t}}  $ being the projection over the vacuum $\Om^{s,t}$.
\end{lemma}
\noindent
\emph{Proof:}\\
By the functional calculus
\beq \label{neri.10.1} \begin{array}{l}
\| (H^{s,t}_{0}(\theta, \alpha,\beta) - z) \frac{\overline{P}_{\at, 1}(\theta, \alpha, \beta) + \overline{P}_{\Om^{s,t}} }{ H_{0}^{s,t}(\theta, \alpha, \beta)) - \mu }  \| \\\\ 
\leq \sup_{r \in  \{ 0 \} \cup [s, \infty)}  \| (H_{\at}(\theta, \alpha, \beta) - (z  - r e^{-i \nu}  )) 
\frac{\overline{P}_{\at, 1}(\theta, \alpha, \beta) }{ H_{\at}(\theta, \alpha, \beta) - (\mu   - r e^{-i \nu}   ) }  \|     \\\\
+  \sup_{r \in   [s, \infty)}  \| (H_{\at}(\theta, \alpha, \beta) - (z  - r e^{-i \nu}  )) 
\frac{1}{ H_{\at}(\theta, \alpha, \beta) - (\mu   - r e^{-i \nu}   ) }  \|  \\\\
\leq    \| (H_{\at}(\theta, \alpha, \beta) - z  ) 
\frac{\overline{P}_{\at, 1}(\theta, \alpha, \beta) }{ H_{\at}(\theta, \alpha, \beta) - \mu     }  \| \\\\
+   \sup_{r \in   [s, \infty)}  \| (H_{\at}(\theta, \alpha, \beta) - (z  - r e^{-i \nu}  )) 
\frac{1}{ H_{\at}(\theta, \alpha, \beta) - (\mu   - r e^{-i \nu}   ) }  \| \\\\
 \cdot (1 + \|  \overline{P}_{\at, 1}(\theta, \alpha, \beta)  \|) \period    
\end{array}
\ene
As in (\ref{neri.5}) we prove that
\beq \label{neri.10.1.1}
\frac{ | z - e_{1} | }{|\mu - e^{-i\nu}r - e_{1}| } \leq 1 \comma \; r \geq s \period
\ene
using Corollary~\ref{C.eeh.1} and (\ref{neri.6}) we obtain (remember that $s \leq \frac{1}{2}\delta_\at \leq  \delta $)
\beq \label{neri.10.2}\begin{array}{l}
\| (H_{\at}(\theta, \alpha, \beta) - (z  - r e^{-i \nu}  )) 
\frac{1}{ H_{\at}(\theta, \alpha, \beta) - (\mu   - r e^{-i \nu}   ) }  \|  \\\\ \leq 1
 +\| \frac{|\mu - z|}{  H_{\at}(\theta, \alpha, \beta) - (\mu   - r e^{-i \nu}   ) } \| \leq 1 + 
4 C_{\ref{an.0.16.2}}( 2 + \frac{ 8 (|\sin(\nu)|/2) s }{\delta |\sin(\nu)|}) \\\\
\leq   1 +  30 C_{\ref{an.0.16.2}}    \period
\end{array}
\ene
Eq. (\ref{L.neri.1.1.e1}) follows from (\ref{an.0.16}), Corollary~\ref{L.eeh.2}, (\ref{neri.10.1}) and (\ref{neri.10.2}).

\QED

{\bf Proof of Lemma~\ref{L.neri.2}:}

First remember that $\om(k) = |k|$.\\
The proof follows from the proof of Lemma 3.13 of 
\cite{BachFroehlichSigal1999}, using Lemma~\ref{L.neri.1} instead of 
Lemma 3.12 of \cite{BachFroehlichSigal1999}.
For the reader's convenience we sketch the proof.\\
We use the expression (\ref{i.10.4}), (\ref{i.10.3}) for the interaction. 
First we bound the term $ \|  |R^{s,t}_{0}(\theta,\alpha, \beta)(z)|^{1/2} W^{s,t}_{0,1}(\theta, \alpha, \beta)   |R^{s,t}_{0}(\theta, \alpha, \beta)(z)|^{1/2}  \|   $.
By Eq. (3.118) \cite{BachFroehlichSigal1999}
\beq \label{neri.11}\begin{array}{l}
\|  |R^{s,t}_{0}(\theta,\alpha, \beta)(z)|^{1/2} W^{s,t}_{0,1}(\theta, \alpha, \beta)   |R^{s,t}_{0}(\theta, \alpha, \beta)(z)|^{1/2}  \| \\\\
 \leq \Big( \int_{\cK^{s,t}} \frac{dk}{\om(k)} 
\|  |R^{s,t}_{0}(\theta, \alpha, \beta)(z)|^{1/2} ((\hf)^{s,t} + \om(k))^{1/2}  \|^2 \\\\ \cdot \|   w_{0,1}(\theta, \alpha, \beta)| H_\at(0,\alpha,\beta) \pm i \delta |^{-1/2}  \|^2 \\\\ 
\cdot \| | H_{\at}(0, \alpha, \beta) \pm i \delta |^{1/2}   |R^{s,t}_{0}(\theta,\alpha, \beta)(z - e^{- i \nu} \om (k))|^{1/2}   \|^2 \Big)^{1/2} \\\\
\leq  \Big(  \frac{50|e_0| C_{\ref{an.0.16.2}} C_{\ref{L.neri.1.e3}}}{\delta |\sin(\nu)|^2  }  \int_{\cK^{s,t}}\frac{dk}{\om(k)} 
( 1 + \frac{\om (k)}{|z- e_{1}|}  ) \\\\ \cdot (  1 + \frac{1}{|z- e_{1} - e^{- i \nu} \om (k) |}   ) 
 \Big(  2 \big(  |G^{s,t}(\theta)  -   e^{- \theta} \nabla  \boQ^{s,t}(\theta) | \big)\\\\\cdot 3^{1/2} | (-i e^{- \theta} C_{\ref{an.0.11}})| 
 +| i e^{- 2 \theta} \Delta  \boQ^{s,t}(\theta)| +   |Q^{s,t}(\theta)    | \Big)^2 \Big)^{1/2}\comma
\end{array} 
\ene
where we used (\ref{an.0.10}) and Lemma~\ref{L.neri.1}. \\
By (\ref{neri.5}), for $ \om(k) \leq 2 |z - e_{1}| $,
$( 1 + \frac{\om (k)}{|z- e_{1}|}  )(  1 + \frac{1}{|z- e_{1} - e^{- i \nu} \om (k) |}   ) $
  $ \leq  \frac{3}{|\sin(\nu)|} (1 + \om(k))(  1 + \frac{1}{|z- e_{1}|}   )   $. For
$   \om(k) > 2  |z - e_{1}|  $ we have that 
$  ( 1 + \frac{\om (k)}{|z- e_{1}|}  )(  1 + \frac{1}{|z- e_{1} - e^{- i \nu} \om (k) |}   )  $
$ \leq (2 \frac{\om(k)}{|z-e_{1}|})( 1 + \frac{2}{\om(k)}  )  $
$\leq  4   (1 + \om(k))(  1 + \frac{1}{|z- e_{1}|}   )     $.  Using  (\ref{i.13.1}), (\ref{i.13.2}) (with $\rho = 1$) 
and the triangle inequality  we obtain, 
\beq \label{neri.12}
\begin{array}{l}
\|  |R^{s,t}_{0}(\theta,\alpha, \beta)(z)|^{1/2} W^{s,t}_{0, 1}(\theta, \alpha, \beta)   |R^{s,t}_{0}(\theta, \alpha, \beta)(z)|^{1/2}  \| \leq \\\\ 
2 \frac{(50 |e_0| C_{\ref{an.0.16.2}} C_{\ref{L.neri.1.e3}})^{1/2}}{ \delta^{1/2} |\sin(\nu)|^{3/2} } \Big(
 12  C_{\ref{an.0.11}}     C_{\ref{R.a.1.1}} (1  +  ( 1 + 6 \|  \eta'  \|_{\infty} ))\alpha^{3/2}  \\\\ + 
 6 e^{1/15} C_{\ref{R.a.1.1}} 
( 4 \| \eta'   \|_{\infty}  + 2\| \eta''   \|_{\infty} ) \alpha^{3/2}  +  12 e^{1/15} C_{\ref{R.a.1.1}} \alpha^{3/2} \\\\
+  24\beta C_{\ref{R.a.1.1}} \alpha^{3/2}(1 + 2\| \eta' \|_\infty)
\Big) (  1 + \frac{1}{|z- e_{1}|}   )^{1/2} \period 
\end{array}
\ene

Now we analyze the term containing $ W_{0, 2}(\theta, \alpha, \beta) $. By (3.121) and (3.122) of \cite{BachFroehlichSigal1999},

\beq \label{neri.13}\begin{array}{l}
\|  |R^{s,t}_{0}(\theta, \alpha, \beta)(z)|^{1/2} W^{s,t}_{0,2}(\theta, \alpha, \beta)   |R^{s,t}_{0}(\theta, \alpha, \beta)(z)|^{1/2}  \|^2 \\\\ \leq \int_{(\cK^{s,t})^2}
 \frac{dk dk'}{\om(k) \om(k')} \|   |R^{s,t}_{0}(\theta, \alpha, \beta)(z)|^{1/2} (\hf^{s,t} + \om(k) + \om(k') )^{1/2}  \|^2 \\\\
\cdot  \|   |R^{s,t}_{0}(\theta, \alpha, \beta)(z - (  \om(k) + \om(k') )e^{-i \nu}  )|^{1/2} (\hf^{s,t} + \om(k) + \om(k') )^{1/2}  \|^2 \\\\ \cdot
| w^{s,t}_{0, 2}(\theta, \alpha, \beta) (k, k'; \theta )|^2  \\\\
\leq  \frac{(50 |e_0| C_{\ref{an.0.16.2}})^2}{ |\delta \sin(\nu)|^2 }   \int_{(\cK^{s,t})^2}
 \frac{dk dk'}{\om(k) \om(k')} (1 + \frac{\om(k) + \om(k')}{|z-e_{1}|})
( 1 + \frac{\om(k) + \om(k')}{ |z-e_{1} - (\om(k) + \om(k'))e^{-i\nu} |  }  )\\\\
\cdot | w^{s,t}_{0, 2}(\theta, \alpha, \beta) (k, k'; \theta )|^2 \comma  
\end{array} 
\ene
where we used Lemma~\ref{L.neri.1}. \\
By (\ref{neri.5}), for $ r \in [s, \infty) $, 
\beq \label{neri.14} \begin{array}{l}
\frac{r}{| z-e_{1} - re^{-i\nu}|} \leq    \frac{  |z-e_{1} - re^{-i\nu  }|}{| z-e_{1} - re^{-i\nu}|} + \frac{ | z-e_{1}|}
{| z-e_{1} - re^{-i\nu}|} \\\\ \leq 
1 + \frac{1}{|\sin(\nu)|}\period 
\end{array}
\ene
By (\ref{i.13.1}), (\ref{neri.13}) and (\ref{neri.14}), 
\beq \label{neri.15} \begin{array}{l}
\|  |R^{s,t}_{0}(\theta, \alpha, \beta)(z)|^{1/2} W^{s,t}_{0,2}(\theta, \alpha, \beta)   |R^{s,t}_{0}(\theta, \alpha, \beta)(z)|^{1/2}  \| \\\\
\leq \frac{1000 |e_0| C_{\ref{an.0.16.2}}}{ \delta |\sin(\nu)|^{3/2} }    C_{\ref{R.a.1.1}}^2 (1  +  ( 1 + 6 \|  \eta'  \|_{\infty} ))^2\alpha^{3} 
  (1 + \frac{1}{|z - e_{1} |})^{1/2}\comma
\end{array}
\ene
where we used that $ \frac{1}{\omega(k)} \frac{1}{\omega(k')}  + \frac{1}{\omega(k)} \frac{1}{|z - e_{1} |} +
   \frac{1}{\omega(k')} \frac{1}{|z - e_{1} |}  $   $ \leq (
  \frac{1}{\omega(k)} \frac{1}{\omega(k')}  + \frac{1}{\omega(k)}  +
   \frac{1}{\omega(k')})( 1 + \frac{1}{|z - e_{1} |}  )   $
$\leq ( 1 +  \frac{1}{\omega(k')}  )  ( 1 +  \frac{1}{\omega(k)}  )  ( 1 + \frac{1}{|z - e_{1} |}  )  $. \\
Similarly we can analyze the term $ W^{s,t}_{1, 1}(\theta, \alpha, \beta)   $ and obtain the bound 
\beq \label{neri.13.2}\begin{array}{l}
\|  |R^{s,t}_{0}(\theta,\alpha, \beta)(z)|^{1/2} W^{s,t}_{1,1}(\theta, \alpha, \beta)   |R^{s,t}_{0}(\theta,\alpha, \beta)(z)|^{1/2}  \| 
 \\\\ \leq \big( \int_{(\cK^{s,t})^2}
 \frac{dk dk'}{\om(k) \om(k')} \|   |R^{s,t}_{0}(\theta,\alpha, \beta)(z -   \om(k) e^{-i \nu} )|^{1/2} (\hf^{s,t} + \om(k)  )^{1/2}  \|^2 \\\\
\cdot  \|   |R^{s,t}_{ 0}(\theta,\alpha, \beta)(z -  \om(k') e^{-i \nu}  )|^{1/2} (\hf^{s,t}  + \om(k') )^{1/2}  \|^2 
 \cdot  |w_{1, 1}(\theta, \alpha, \beta)(k, k'; \theta)|^2 \big)^{1/2} \\\\
\leq \big( \frac{(50 |e_0| C_{\ref{an.0.16.2}})^2}{\delta|\sin(\nu)|^2}  \int_{(\cK^{s,t})^2}
 \frac{dk dk'}{\om(k) \om(k')}( 1 + \frac{\om(k) }{ z-e_1 - \om(k)e^{-i\nu}   }  )
( 1 + \frac{ \om(k')}{ z-e_{1} -  \om(k')e^{-i\nu}   }  )\\\\
\cdot   |w_{1, 1}(\theta, \alpha, \beta)(k, k'; \theta)|^2    \big)^{1/2} \\\\ \leq 
\frac{1000 |e_0| C_{\ref{an.0.16.2}}}{\delta |\sin(\nu)|^{2} }    C_{\ref{R.a.1.1}}^2 (1  + 1 ( 1 + 6 \|  \eta'  \|_{\infty} ))^2\alpha^{3} 
  \comma  
\end{array}
\ene

We also have
\beq \label{neri.15.1} \begin{array}{l}
\|  |R^{s,t}_{0}(\theta, \alpha, \beta)(z)|^{1/2} W^{s,t}_{0, 0}(\theta, \alpha, \beta)   |R^{s,t}_{0}(\theta, \alpha, \beta)(z)|^{1/2}  \| \\\\
\leq \frac{50 |e_0| C_{\ref{an.0.16.2}}}{ \delta |\sin(\nu)| }    C_{\ref{R.a.1.1}}^2 (1  +  ( 1 + 6 \|  \eta'  \|_{\infty} ))^2\alpha^{3} 
  (1 + \frac{1}{|z - e_{1} |})\comma
\end{array}
\ene
Now we define 
\beq \label{neri.15.2} \begin{array}{l}
C_{\ref{neri.15.2}} : = 5 \frac{|e_0'|}{\delta}\Big( 1000 C_{\ref{an.0.16.2}}    C_{\ref{R.a.1.1}}^2 (1  +  ( 1 + 6 \|  \eta'  \|_{\infty} ))^2 \\\\
+ 2 (50 C_{\ref{an.0.16.2}} C_{\ref{L.neri.1.e3}})^{1/2} \big(
 12  C_{\ref{an.0.11}}     C_{\ref{R.a.1.1}} (1  +  ( 1 + 6 \|  \eta'  \|_{\infty} ))  \\\\ + 
 6 e^{1/15} C_{\ref{R.a.1.1}} 
( 4 \| \eta'   \|_{\infty}  + 2\| \eta''   \|_{\infty} )   +  12 e^{1/15} C_{\ref{R.a.1.1}}  \\\\
+  24\bob C_{\ref{R.a.1.1}} (1 + 2\| \eta' \|_\infty)\big)\Big)
\end{array}
\ene
Finally other terms in (\ref{i.10.4}) can treated as in
(\ref{neri.12}) and (\ref{neri.15}). \\\\

To prove (\ref{L.neri.2.e0.2.1}) we use (\ref{L.i.1.e1}) with $ \rho = 1 - e_{0}$,
(\ref{L.neri.1.e0.1}), (\ref{an.0.16}) and 
\beq  \label{neri.15.2.1}
C_{\ref{neri.15.2.1}} : =  6 C_{\ref{i.13.4}}  
( 3 + \frac{16 (2 + 2|e_0'|) C_{\ref{an.0.16.2}} }{|\sin(\nu) |\delta}   +  \frac{20 |e_0'|  C_{\ref{an.0.16.2}} }{|\sin(\nu) |\delta}  )\period 
\ene

\QED

\subsection{ $\;$ Proof of Theorem~\ref{T.rrp}}\label{S.rrp.p}

\subsubsection{ $\;$  The Feshbach Map}

We define the projection operator on $\cH^{s,t}$
\beq \label{re.1}
\p^{s,t}(\theta): = \p : = P_{\at, 1}(\theta, \alpha, \beta) \otimes P_{\Om^{s,t}}, \; \; \bp^{s,t}(\theta)=: \bp : = 1 - \p  \comma 
\ene
where $  P_{\at, 1}(\theta, \alpha, \beta) $ is defined in (\ref{e.p.eeh.2}) and 
$P_{\Om^{s,t}}$ is the projection on the vacuum vector of $\cF^{s,t}$.\\
We denote by 
\beq \label{re.2}
  H^{s,t}_{\p}(\theta) : =  \p   H^{s,t}(\theta, \alpha, \beta)  \p,  \; \;    H^{s,t}_{\bp}(\theta) : =  
\bp   H^{s,t}(\theta, \alpha, \beta)  \bp \period  
\ene  
We prove below that for any $z \in \cE^{s}(\theta)$ (see (\ref{neri.0.1})) $ H^{s,t}_{\bp}(\theta) - z $ is invertible 
on the range of $ \bp $ and that 
\beq \label{re.3} \begin{array}{l}
\|    \bp (H_{\bp}^{s,t}(\theta) - z)^{- 1} \bp H^{s,t}(\theta) \p \| < \infty\comma \\\\
\| \p  H^{s,t}(\theta) \bp (H_{\bp}^{s,t}(\theta) - z)^{- 1} \bp    \| < \infty \period 
\end{array}
\ene
We define the Feshbach map, $f_\p(H^{s,t}(\theta)) - z$, by
\beq \label{re.4} \begin{array}{l}
\cF_\p := \cF_{\p}(H^{s,t}(\theta) - z)  : = \p (H^{s,t}(\theta) - z) \p \\\\ -  \p H^{s,t}(\theta) 
\bp  (H_{\bp}^{s,t}(\theta) - z)^{- 1} \bp H^{s,t}(\theta) \p \comma
\end{array}
\ene
on the range of $\p$.\\
The Feshbach map is discussed in detail in  \cite{BachFroehlichSigal1998a} and \cite{BachFroehlichSigal1998b}, 
in this text we use it to estimate the value of $E^{s,t}(\theta)$. The remaining of this section 
is devoted to prove the invertibility  of $ H^{s,t}_{\bp}(\theta) - z $ and (\ref{re.3}).

\begin{lemma}\label{L.fm.1}
Suppose that  $\theta = \pm i \nu $, with $\nu \in \RR \setminus \{0 \}$, and that  $ \theta  $ satisfies (\ref{an.0.14}) and (\ref{theta}).
Suppose furthermore that $\alpha $ satisfies (\ref{L.rrp.1.m10}) and that  $\alpha \leq \boa$, $|\beta| \leq \bob$. 
Then $ \mathcal{E}^{s}(\theta)$ is contained in the resolvent set of $H_{\bp}^{s,t}(\theta)$ and  for every $\mu \in   \mathcal{E}^{s}(\theta)$,
\beq \label{L.fm.1.e1}
  \|  (H_{\bp}^{s,t}(\theta) - \mu)^{-1} \bp \|  \leq 150 (1 + C_{\ref{an.0.16.2}}) \frac{50 |e_0| C_{\ref{an.0.16.2}}}{\delta |\sin(\nu)| }
 \frac{1}{ (|\sin(\nu)|/2) s} \period
\ene
\end{lemma}

\emph{Proof.}\\
Let $ \mu \in  \mathcal{E}^{s,t}(\theta)\setminus \{ e_{1} \}  $. We construct $    (H_{\bp}^{s,t}(\theta) - \mu)^{-1} \bp  $ by a norm-convergent  Neumann series 
\beq \label{re.5}
  (H_{\bp}^{s,t}(\theta) - \mu)^{-1}\bp   = \sum_{n= 0}^{\infty} \frac{\bp}{ H^{s,t}_{0}(\theta, \alpha,\beta) - \mu } 
\big[- W^{s,t}(\theta, \alpha, \beta) \frac{\bp}{ H^{s,t}_{0}(\theta, \alpha,\beta) - \mu } \big]^n
\ene
Next we estimate the $n^{th}$ order term. We take  $  z = \mu  $ if $ | \mu - e_{1} | \geq (|\sin(\nu)|/2) s   $, and
$  z = z_0 $ with $|z_0 - e_{1}| = (|\sin(\nu)|/2) s  $ if $    | \mu - e_{1} | < (|\sin(\nu)|/2) s $.  \\ 
Using the proof of Lemma~\ref{L.neri.1.1} and Lemma~\ref{L.neri.2}  we obtain,  

\beq \label{re.6} \begin{array}{l}
 \big\|   \frac{\bp}{ H^{s,t}_{0}(\theta, \alpha,\beta) - \mu }  \big[- W^{s,t}(\theta, \alpha, \beta)  \frac{\bp}{ H^{s,t}_{0}(\theta, \alpha,\beta) - \mu } \big]^n \big\| \leq 
\|   |  H^{s,t}_{0}(\theta, \alpha,\beta) - z  |   \frac{\bp}{ H^{s,t}_{0}(\theta, \alpha,\beta) - \mu }   \|^{n+1} 
\\\\ \cdot \|   |  H^{s,t}_{0}(\theta, \alpha,\beta) - z  |^{-1}   \| \cdot  
\| |  H^{s,t}_{0}(\theta, \alpha,\beta) - z  |^{-1/2}  W^{s,t}(\theta, \alpha, \beta)  |  H^{s,t}_{0}(\theta, \alpha,\beta) - z  |^{-1/2}    \|^n \leq \\\\
100 (1 + C_{\ref{an.0.16.2}})    \|   R^{s,t}_{0}(\theta, \alpha, \beta)(z)   \| \frac{1}{5^n} \period 
\end{array}
\ene
This proves the convergence of the Neumann series and, by Lemma~\ref{L.neri.1}, the bound
(\ref{L.fm.1.e1}). \\
If $\mu = e_{1}$, the same estimate follows because the right hand side of Eq. (\ref{re.5}) is analytic  
in $  \mathcal{E}^{s}(\theta) \setminus \{ e_{1} \} $ and bounded in a neighbourhood of $ e_{1} $, so it is analytic in 
$   \mathcal{E}^{s}(\theta) $.   

\QED

\begin{lemma}\label{L.fm.1.1}
Suppose that  $\theta = \pm i \nu $, with $\nu \in \RR \setminus \{0 \}$, and that  $ \theta  $ satisfies (\ref{an.0.14}) and (\ref{theta}).
Suppose furthermore that $\alpha $ satisfies (\ref{L.rrp.1.m10}) and that  $\alpha \leq \boa$, $|\beta| \leq \bob$. Then  for every $ \mu \in   \mathcal{E}^{s}(\theta)$ and every $z $ with $ |z - e_{1}| = (|\sin(\nu)|/2) s $,  
\beq \label{L.fm.1.1.e1} \begin{array}{l}
  \|   |  H^{s,t}_{0}(\theta, \alpha, \beta) - z  |^{1/2}  (H_{\bp}^{s,t}(\theta) - \mu)^{-1} \bp   |  H^{s,t}_{0}(\theta, \alpha, \beta) - z  |^{1/2}   \| \\\\
 \leq  150 (1 + C_{\ref{an.0.16.2}})   \period
\end{array}
\ene
\end{lemma}

\emph{Proof.}\\
It can be proved using similar estimates as in the proof of Lemma~\ref{L.fm.1}. 
\QED

\begin{lemma}\label{L.fm.2}
Suppose that  $\theta = \pm i \nu $, with $\nu \in \RR \setminus \{0 \}$, and that  $ \theta  $ satisfies (\ref{an.0.14}) and (\ref{theta}).
Suppose furthermore that $\alpha $ satisfies (\ref{L.rrp.1.m10}) and that  $\alpha \leq \boa$, $|\beta| \leq \bob$.  
 Then
\beq \label{L.fm.1.e2} \begin{array}{l}
\|    \bp (H_{\bp}^{s,t}(\theta) - z)^{- 1} \bp H^{s,t}(\theta, \alpha, \beta) \p \| \leq  
150 (1 + C_{\ref{an.0.16.2}})  \\\\  \cdot \Big(\frac{C_{\ref{neri.15.2}}}{|\sin(\nu)|^{2}}\alpha^{3/2} (1 + \frac{1}{(|\sin(\nu)|/2) s /4  })^{1/2}   +    
 \frac{C_{\ref{neri.15.2}}}{|\sin(\nu)|}\alpha^{3} (1 + \frac{1}{ (|\sin(\nu)|/2) s /4 })\Big)  \cdot 2 ( \frac{50 |e_0| C_{\ref{an.0.16.2}}}{ \delta |\sin{\nu}| }  )^{1/2}
     \comma \\\\ 
\| \p  H^{s,t}(\theta, \alpha, \beta) \bp (H_{\bp}^{s,t}(\theta) - z)^{- 1} \bp    \| \leq  
150 (1 + C_{\ref{an.0.16.2}})  \\\\  \cdot \Big(\frac{C_{\ref{neri.15.2}}}{|\sin(\nu)|^{2}}\alpha^{3/2} (1 + \frac{1}{(|\sin(\nu)|/2) s /4  })^{1/2}   +    
 \frac{C_{\ref{neri.15.2}}}{|\sin(\nu)|}\alpha^{3} (1 + \frac{1}{ (|\sin(\nu)|/2) s /4 })\Big)  \cdot 2 ( \frac{50 |e_0| C_{\ref{an.0.16.2}}}{ \delta |\sin{\nu}| }  )^{1/2}
   \period 
\end{array} \ene
\end{lemma}

\emph{Proof.}\\ The proof of (\ref{L.fm.1.e2}) is similar to the one of (\ref{L.fm.1.e1})(see also (\ref{e.1}) below), here we use
Lemmata    \ref{L.neri.1} and \ref{L.neri.2}   instead of Lemma~\ref{L.neri.1} its self. We make use of (\ref{an.0.16}) also. 

\QED

Lemmata~\ref{L.fm.1} and \ref{L.fm.2} prove the existence of the Feshbach map (\ref{re.4}).
Theorem IV.1 \cite{BachFroehlichSigal1998a} implies that $\cF_\p$ is invertible in 
$\cE^{s,t}(\theta) \setminus \{ E^{s,t}(\theta) \}$. An algebraic manipulation leads us to the identity
\beq \label{re.7} \begin{array}{l}
(H^{s,t}(\theta, \alpha, \beta) - z)^{-1} = \\\\ \big[  \p - \bp (H_{\bp}^{s,t}(\theta) - z)^{-1} W^{s,t}(\theta, \alpha, \beta) \p  \big] 
\cdot \cF_\p^{-1} \big[ \p - \p W^{s,t}(\theta, \alpha, \beta) \bp   (H_{\bp}^{s,t}(\theta) - z)^{-1}   \big] \\\\
+ \bp   (H_{\bp}^{s,t}(\theta) - z)^{-1}  \bp \comma
\end{array}
\ene
for every $z \in \cE^{s,t}(\theta) \setminus \{ E^{s,t}(\theta)  \}$.

\subsubsection{ $\;$ Estimations for $E^{s,t}(\theta)$}

As the annihilation operator applied to the vacuum is zero and thus the
the creation operator followed by the projection on the vacuum is zero, 
the following is clear. 

\begin{remark}\label{R.e.1}
For any $m$ and $n$ belonging to $\{ 0, 1, 2 \}$
\beq  \label{R.e.1.e1}
\p W^{s,t}_{m,n}(\theta, \alpha, \beta) \p = 0 \comma \: 1 \leq m + n \leq 2 \period
\ene
Furthermore
\beq  \label{R.e.1.e2}
W^{s,t}_{0,2}(\theta, \alpha, \beta) \p = 0 =  \p  W^{s,t}_{2,0}(\theta, \alpha, \beta)  
\ene
and
\beq  \label{R.e.1.e3}
W^{s,t}_{0,1}(\theta, \alpha, \beta)\p = 0 = \p W^{s,t}_{0,1}(\theta, \alpha, \beta)\period
\ene
\end{remark}

In the next Lemma we should remember  (\ref{neri.0.1}). 
 
\begin{lemma}\label{L.e.1}
Suppose that  $\theta = \pm i \nu $, with $\nu \in \RR \setminus \{0 \}$, and that  $ \theta  $ satisfies (\ref{an.0.14}) and (\ref{theta}).
Suppose furthermore that  $\alpha \leq \boa$, $|\beta| \leq \bob$. 

 For any $z \in \cE^{s,t}(\theta)\setminus \{ e_{1} \}$, 
\beq \label{L.e.1.e1} \begin{array}{l}
\|  |R^{s,t}_{0}(\theta,\alpha, \beta)(z)|^{1/2}   W^{s,t}_{m,n}(\theta, \alpha, \beta) \p \| \\\\ \leq 
2 \frac{C_{\ref{neri.15.2}}}{|\sin(\nu)|^2}\alpha^{3(m + n)/2} (1 + |z- e_{1}|)^{1/2} \comma \: \: m+n \geq 1
\end{array}
\ene
\beq \label{L.e.1.e2}\begin{array}{l}
\| \p   W^{s,t}_{m,n}(\theta, \alpha, \beta) |R^{s,t}_{0}(\theta,\alpha, \beta)(z)|^{1/2}  \| \\\\  \leq  
2 \frac{C_{\ref{neri.15.2}}}{|\sin(\nu)|^2}\alpha^{3(m + n)/2} (1 + |z- e_{1}|)^{1/2} \comma  \: \: m +n \geq 1  \period
\end{array}
\ene

\beq \label{L.e.1.e3} \begin{array}{l}
\|  |R^{s,t}_{0}(\theta,\alpha, \beta)(z)|^{1/2}   W^{s,t}_{0,0}(\theta, \alpha, \beta) \p \| \\\\ \leq 
2 \frac{C_{\ref{neri.15.2}}}{|\sin(\nu)|^2}\alpha^{3} (   |z- e_{1}|^{ 1/2} + |z- e_{1}|^{-1/2}) \comma
\end{array}
\ene
\beq \label{L.e.1.e4}\begin{array}{l}
\| \p   W^{s,t}_{0,0}(\theta, \alpha, \beta) |R^{s,t}_{0}(\theta,\alpha, \beta)(z)|^{1/2}  \| \\\\  \leq  
2 \frac{C_{\ref{neri.15.2}}}{|\sin(\nu)|^2}\alpha^{3} (  |z- e_{1}|^{ 1/2}  + |z- e_{1}|^{- 1/2}) \period
\end{array}
\ene

\end{lemma}

\emph{Proof.}\\
We use  (\ref{an.0.16}), Lemma~\ref{L.neri.2} and the bound
\beq \label{e.1}
\| |R^{s,t}_{0}(\theta,\alpha, \beta)(z)|^{1/2} \p \| \leq 2 |  z - e_{1} |^{1/2} \period 
\ene

\QED

\begin{lemma}\label{L.e.2}
Suppose that  $\theta = \pm i \nu $, with $\nu \in \RR \setminus \{0 \}$, and that  $ \theta  $ satisfies (\ref{an.0.14}) and (\ref{theta}). Suppose furthermore that  $\alpha \leq \boa$, $|\beta| \leq \bob$.
 Then
\beq \label{L.e.2.e1}\begin{array}{l}
\p  (W^{s,t}_{0, 1}(\theta, \alpha, \beta) + W^{s,t}_{1, 0}(\theta, \alpha, \beta) )  \frac{\bp}{ H_0^{s,t}(\theta, \alpha, \beta) - z   } 
 (W^{s,t}_{0, 1}(\theta, \alpha, \beta) \\\\ + W^{s,t}_{1, 0}(\theta, \alpha, \beta) )  \p  = 
\int_{\cK^{s,t}} P_{\at, 1}(\theta, \alpha, \beta) w_{0,1}(\theta, \alpha, \beta)(k)\\\\ \frac{1}{  H_{\at}^{s,t}(\theta, \alpha, \beta) - (z - e^{-i \nu} |k| )  }  
 w_{1,0}(\theta, \alpha, \beta)(k) P_{\at, 1}(\theta, \alpha, \beta) P_{\Om^{s,t}} 
 \period  
\end{array}
\ene
\end{lemma}
\emph{Proof.}\\
By Remark \ref{R.e.1}, 
\beq \label{e.2}\begin{array}{l}
\p  (W^{s,t}_{0, 1}(\theta, \alpha, \beta) + W^{s,t}_{1, 0}(\theta, \alpha, \beta) )  \frac{\bp}{ H_{0}^{s,t}(\theta,\alpha, \beta) - z   }  
(W^{s,t}_{0, 1}(\theta, \alpha, \beta) \\\\ + W^{s,t}_{1, 0}(\theta, \alpha, \beta) )  \p  = 
\p  W^{s,t}_{0, 1}(\theta, \alpha, \beta)   \\\\ \frac{1}{ H^{s,t}_{0}(\theta, \alpha, \beta) - z   }   W^{s,t}_{1, 0}(\theta, \alpha, \beta)   \p \period
\end{array}
\ene
As $\p$ projects on the vacuum photon space, it is enough to apply $ \p $  $  W^{s,t}_{0, 1}(\theta, \alpha, \beta) $  $  
 \frac{1}{ H^{s,t}_{0}(\theta, \alpha, \beta) - z   }  $  $  W^{s,t}_{1, 0}(\theta, \alpha, \beta)  $ to a function  $\psi : \RR^3 \to 
\cF^{s,t} = \CC \Om^{s,t}$. The function   
\beq \label{e.2.1}
 \psi_1 : =  W^{s,t}_{1, 0}(\theta)  \psi    
\ene
is a function  form $\RR^3$ to $\cF^{s,t}$:
\beq \label{e.2.2}
\psi_1(x)(k) =   w_{1,0}(\theta, \alpha, \beta)(x,k)\psi (x)\period   
\ene
Therefore 
\beq \label{e.3}
 \frac{1}{ H^{s,t}_{0}(\theta, \alpha, \beta) - z   } \psi_1 =  \frac{1}{ H_{\at}(\theta , \alpha, \beta) - (z - e^{- \theta} |k|)} \psi_1 \period 
\ene
We obtain that 
\beq \label{e.4}\begin{array}{l}
(W^{s,t}_{0, 1}(\theta, \alpha, \beta) \frac{1}{ H^{s,t}_{0}(\theta, \alpha, \beta) - z   } \psi_1) (x) \\\\ = \int_{\cK^{s,t}} dk \;  w_{0,1}(\theta, \alpha, \beta)(x,k)  \frac{1}{ H_{\at}(\theta , \alpha, \beta) - (z - e^{- \theta} |k|)} \psi_1(x, k)\period 
\end{array}
\ene
Eqs. (\ref{e.2.2}) and (\ref{e.4}) imply
\beq \label{e.4.1}\begin{array}{l}
\p  (W^{s,t}_{0, 1}(\theta, \alpha, \beta) + W^{s,t}_{1, 0}(\theta, \alpha, \beta) )  
\frac{\bp}{ H^{s,t}_{0}(\theta, \alpha, \beta) - z   }  (W^{s,t}_{0, 1}(\theta, \alpha, \beta) \\\\ + W^{s,t}_{1, 0}(\theta, \alpha, \beta) )  \p  = 
\int_{\cK^{s,t}} \p w_{0,1}(\theta,\alpha, \beta)(k)\\\\ \frac{1}{  H_{\at}^{s,t}(\theta, \alpha, \beta) - (z - e^{-i \nu} |k| )  }   
w_{1,0}(\theta, \alpha, \beta)(k)\p \period  
\end{array}
\ene
To obtain (\ref{L.e.2.e1}) we notice (see (\ref{re.1})) that
 $   w_{1,0}(\theta, \alpha, \beta)(k)\p = $   $   w_{1,0}(\theta, \alpha, \beta)(k)  $  $(P_{\at, 1}(\theta, \alpha, \beta)\otimes 1) $ $( 1 \otimes P_{\Om^{s,t}}) $
  $ =  P_{\Om^{s,t}} $ $  w_{1,0}(\theta, \alpha, \beta)(k) $ 
$ P_{\at, 1}(\theta, \alpha, \beta)  $.

\QED

On the next lemma we should remember the definition of the Feshbach map (\ref{re.4}). 
\begin{lemma}\label{L.e.3}
Suppose that  $\theta = \pm i \nu $, with $\nu \in \RR \setminus \{0 \}$, and that  $ \theta  $ satisfies (\ref{an.0.14}) and (\ref{theta}).
Suppose furthermore that $\alpha $ satisfies (\ref{L.rrp.1.m10}) and that  $\alpha \leq \boa$, $|\beta| \leq \bob$. 
We assume additionally that $ (|\sin(\nu)|/2) s\leq 1 $.  
Then for every $z \in \CC$ with $|z -e_{1}| \leq (|\sin(\nu)|/2) s $,
\beq \label{L.e.3.e1}
\begin{array}{l}
\cF_\p  = (e_{1} - z) \p - \int_{\cK^{s,t}} P_{\at, 1}(\theta, \alpha, \beta) 
w_{0,1}(\theta,\alpha, \beta)(k) \\\\ \cdot \frac{1}{  H_{\at}^{s,t}(\theta, \alpha, \beta) - (z - e^{-i \nu}|k| )  }  
   w^{s,t}_{1,0}(\theta, \alpha,\beta)(k) P_{\at}(\theta, \alpha, \beta) P_{\Om^{s,t}}  \\\\ +  \p  W_{0, 0}(\theta, \alpha,  \beta)\p
+ Rem_0 + Rem_1 \comma
\end{array}
\ene
where 
\beq \label{L.e.3.e2}
\| Rem_0 \| \leq C_{Rem_0}  \frac{\alpha^{9/2}}{((|\sin(\nu)|/2) s)^{1/2}}  (5 + \frac{\alpha^{3/2}}{ ((|\sin(\nu)|/2) s)^{1/2} })^3  \comma
\ene

\beq \label{L.e.3.e3}
\| Rem_1 \| \leq C_{Rem_1}    \frac{4 \alpha^{3}}{ ((|\sin(\nu)|/2) s )^{1/2} } (  2 \alpha^{3/2}  +    \frac{4 \alpha^{3}}{ ((|\sin(\nu)|/2) s )^{1/2} } ) 
   \comma  
\ene

and 
\beq \label{L.e.3.e4}
C_{Rem_0} := 4 (150(1 + C_{\ref{an.0.16.2}}))^{2} \Big( \frac{4C_{\ref{neri.15.2}}}{|\sin(\nu)|^2} \Big)^3 2^3 
\ene

\beq \label{L.e.3.e5}
C_{Rem_1} :=   8 (150(1 + C_{\ref{an.0.16.2}})) \Big( \frac{4C_{\ref{neri.15.2}}}{|\sin(\nu)|^2} \Big)^2 2^2 
\ene

\end{lemma}
\emph{Proof.}\\
By (\ref{re.1}), (\ref{re.4}) and (\ref{R.e.1.e1}),
\beq \label{e.5}
\begin{array}{l}
\cF_\p =  (e_{1}  - z )\p  +  \p  W_{0, 0}(\theta, \alpha,  \beta)   \p 
\\\\  -  \p W^{s,t}(\theta, \alpha, \beta) \bp  (H_{\bp}^{s,t}(\theta) - z)^{- 1} \bp W^{s,t}(\theta, \alpha, \beta) \p \period
\end{array}
\ene
Next we use Lemma~\ref{L.e.2} to obtain, 
\beq\label{e.6}
\begin{array}{l}
 \p W^{s,t}(\theta, \alpha, \beta) \bp  (H_{\bp}^{s,t}(\theta) - z)^{- 1} \bp W^{s,t}(\theta, \alpha, \beta) \p  \\\\ =
\int_{\cK^{s,t}} \p w_{0,1}(\theta, \alpha, \beta)(k)\frac{1}{  H_{\at}^{s,t}(\theta, \alpha, \beta)
 - (z - e^{-i \nu}|k| )  }  w_{1, 0}(\theta, \alpha,  \beta)(k)\p \\\\ +
Rem_0 + Rem_1 \comma
\end{array}
\ene
where
\beq\begin{array}{l} \label{e.7}
\Rem_0 :=  \p W^{s,t}(\theta, \alpha, \beta) \bp  ((H_{\bp}^{s,t}(\theta) - z)^{- 1} \\\\ - ( \bp H^{s,t}_{0}(\theta, \alpha, \beta)\bp - z)^{- 1}     )
 W^{s,t}(\theta, \alpha, \beta) \p 
\end{array}
\ene
and 
\beq \label{e.8}
\begin{array}{l}
\Rem_1 :=  \p W^{s,t}(\theta, \alpha, \beta) \bp   (H^{s,t}_{0}(\theta, \alpha, \beta) - z)^{- 1}   \bp W^{s,t}(\theta, \alpha, \beta) \p \\\\
-  \p (     W^{s,t}_{1, 0}(\theta, \alpha, \beta)        +   W_{0, 1}(\theta, \alpha, \beta) ) 
 \bp   (H^{s,t}_{0}(\theta, \alpha, \beta) - z)^{- 1} \\\\ \cdot    
\bp (  W^{s,t}_{1, 0}(\theta, \alpha, \beta)      +   W_{0, 1}(\theta, \alpha, \beta) )  \p \period 
\end{array}
\ene
To estimate $Rem_0$ we use the second resolvent equation, 
\beq \label{e.9}
\begin{array}{l}
Rem_0 = - \p W^{s,t}(\theta, \alpha, \beta) \bp  (H_{\bp}^{s,t}(\theta) - z)^{- 1} \bp  
W^{s,t}(\theta, \alpha, \beta)  \bp \\\\ \cdot ( \bp H^{s,t}_{0}(\theta, \alpha, \beta) \bp - z)^{- 1}      \bp W^{s,t}(\theta, \alpha, \beta) \p 
\\\\
 = -

  \p W^{s,t}(\theta, \alpha, \beta)  |  H^{s,t}_{0}(\theta, \alpha, \beta)  - \tilde{z}|^{- 1/2} \\\\  \cdot

|  H^{s,t}_{0}(\theta, \alpha, \beta)  - \tilde{z}|^{ 1 /2} (H_{\bp}^{s,t}(\theta) - z)^{- 1} \bp   |  H^{s,t}_{0}(\theta, \alpha, \beta)  - \tilde{z}|^{ 1/2} \\\\ \cdot

\bp | H^{s,t}_{0}(\theta, \alpha, \beta)  - \tilde{z}|^{- 1/2}  W^{s,t}(\theta, \alpha, \beta)  | H^{s,t}_{0}(\theta, \alpha, \beta)  - \tilde{z}|^{- 1/2} \\\\ \cdot \bp

  | H^{s,t}_{0}(\theta, \alpha, \beta) - \tilde{z}|^{ 1/2} (  H^{s,t}_{0}(\theta, \alpha, \beta)  - z)^{- 1}  | H^{s,t}_{0}(\theta, \alpha, \beta) - \tilde{ z} |^{ 1/2} \\\\ \cdot

 |  H^{s,t}_{0}(\theta, \alpha, \beta)  - \tilde{z}|^{- 1/2}   W^{s,t}(\theta, \alpha, \beta) \p \period \\\\ 

\end{array}
\ene 
where $\tilde{z}  \in \CC $ is such that $| \tilde{z} - e_{1} | = (|\sin(\nu)|/2) s$.  We used that $\bp$ commutes with $  H_{0}^{s,t}(\theta,\alpha, \beta) $,  
 that the range of $    (H_{\bp}^{s,t}(\theta) - z)^{- 1}  $ is 
contained in the range of $\bp$.\\
Eq. (\ref{L.e.3.e2}) follows from (\ref{L.neri.1.1.e1}), Lemma~\ref{L.neri.2}, (\ref{L.fm.1.1.e1}),  Lemma  \ref{L.e.1} 
 and  (\ref{e.9}). \\
Eq. (\ref{L.e.3.e3}) is deduced similarly, there we only use Lemma~\ref{L.e.1}.

\QED

\begin{theorem}\label{T.e.1}
Suppose that  $\theta = \pm i \nu $, with $\nu \in \RR \setminus \{0 \}$, and that  $ \theta  $ satisfies (\ref{an.0.14}) and (\ref{theta}).
Suppose furthermore that $\alpha $ satisfies (\ref{L.rrp.1.m10}) and that  $\alpha \leq \boa$, $|\beta| \leq \bob$. We assume additionally that $ (|\sin(\nu)|/2) s\leq 1 $.  
Then for every $z \in \CC$ with $|z -e_{1}| \leq (|\sin(\nu)|/2) s $,

\beq \label{L.e.3.e1.0.0}
\begin{array}{l}
\cF_\p  = (e_{1} - z) \p   - \alpha^{3} Z^{s,t}_d(\theta) -  \alpha^{3} Z^{s,t}_{od}(\theta) \\\\
 + Rem_0 + Rem_1 + Rem_2  +   \p  W_{0, 0}(\theta, \alpha,  \beta)\p \comma
\end{array}
\ene
where, 
\beq \label{T.e.1.e2}\begin{array}{l}
Z^{s,t}_{od}(\theta) := \frac{1}{ \alpha^{3}}  \int_{\cK^{s,t}} dk  P_{\at, 1}(\theta, \alpha, \beta)  w_{0,1}(\theta, \alpha, \beta)(k) P_{\Om^{s,t}} \\\\ \cdot
\frac{\overline{P}_{\at, 1}(\theta, \alpha, \beta)}{  H_{\at}(\theta,\alpha, \beta) - (e_{1} - e^{-i \nu}|k| )  }   w_{1,0}(\theta, \alpha, \beta)(k)
 P_{\at, 1}(\theta) P_{\Om^{s,t}}  \comma 
\end{array}
\ene
\beq \label{T.e.1.e2.0} \begin{array}{l}
Z^{s,t}_d(\theta) := \frac{1}{ \alpha^{3}}  \int_{\cK^{s,t}} \frac{dk}{ e^{-i \nu} |k|} P_{\at, 1}(\theta)  w_{0,1}(\theta, \alpha, \beta)(k)
P_{\at, 1}(\theta)  \\\\ \cdot  w_{1,0}(\theta, \alpha, \beta)(k) P_{\at, 1}(\theta) P_{\Om^{s,t}} \comma
\end{array}
\ene
\beq \label{T.e.1.e3} \begin{array}{l}
Rem_2 \leq  C_{\ref{T.e.1.e3}}    \alpha^3 s (|\log(s)| + 4),  \comma  \\\\ 
C_{\ref{T.e.1.e3}} : =   4 C_{\ref{R.a.1.1}}^2 \Big( 24 | (|e_1'| + \delta +1)^{1/2}C_{\ref{an.0.11}} 
\big( 12 + 6 \bob + 42 \|  \eta' \|_\infty 
\\\\
+ 18 \|  \eta' \|_\infty  \bob + 18 \| \eta'' \|_\infty   \big)\Big)^2
 \big(1 + \frac{3}{2} (4 C_{\ref{an.0.16.2}})^2 \big( \frac{4}{\delta |\sin{\nu}|} + 1  \big)^2\big)
\comma
\end{array}
\ene
and $Rem_0$ and $Rem_1$ are defined in Lemma~\ref{L.e.3} and $\bob$ is defined in (\ref{geh.6.1}).  

\end{theorem}

\emph{Proof.}\\
We use Lemmata~\ref{L.e.2} and \ref{L.e.3} to obtain, 
\beq \label{e.10} \begin{array}{l}
\cF_\p  = (e_{1} - z) \p - \alpha^3 Z^{s,t}_{od}(\theta) 
 -     \int_{\cK^{s,t}} dk P_{\at, 1}(\theta, \alpha, \beta)  \\\\ \cdot w_{0,1}(\theta, \alpha, \beta)(k)
 P_{\at, 1}(\theta, \alpha, \beta)  \frac{1}{ e_{1} - z + e^{-i \nu} |k|}  w_{1, 0}(\theta, \alpha,  \beta)(k)
 \\\\ \cdot P_{\at, 1}(\theta, \alpha, \beta)  P_{\Om^{s,t}} \\\\
-     \int_{\cK^{s,t}} dk P_{\at, 1}(\theta, \alpha, \beta)  w_{0,1}(\theta, \alpha, \beta)(k)
\overline{P}_{\at, 1}(\theta, \alpha, \beta) \\\\ 
\big( \frac{1}{ H_{\at}(\theta, \alpha, \beta) - z + e^{-i \nu} |k|} 
-\frac{1}{ H_{\at}(\theta, \alpha, \beta) - e_{1} + e^{-i \nu} |k|}  \big)w_{1, 0}(\theta, \alpha,  \beta)(k)  \\\\ \cdot  P_{\at, 1}(\theta, \alpha, \beta) P_{\Om^{s,t}}

 +   \p  W_{0, 0}(\theta, \alpha,  \beta)\p  
+ Rem_0 + Rem_1   \comma
\end{array}
\ene
where we used that $ 1 =P_{\at, 1}(\theta, \alpha, \beta) +   \bP_{\at, 1}(\theta, \alpha, \beta) $.
Next note that for $ |k| \geq s $ and $|   e_{1} - z  | \leq \frac{s}{2}  $, 
\beq \label{e.11}
| \frac{1}{ e_{1} - z + e^{-i \nu} |k|} - \frac{1}{ e^{-i \nu} |k|} | \leq  \frac{  s }{|k|^2} \period
\ene
Using the second resolvent equation and (\ref{C.eeh.1.e2}), (\ref{neri.6}) we get
\beq \label{e.11.1} \begin{array}{l}
\overline{P}_{\at, 1}(\theta, \alpha, \beta)  
\big( \frac{1}{ H_{\at}(\theta, \alpha, \beta) - z + e^{-i \nu} |k|} 
-\frac{1}{ H_{\at}(\theta, \alpha, \beta) - e_{1} + e^{-i \nu} |k|}  \big) \\\\
\leq \frac{3}{2} (4 C_{\ref{an.0.16.2}})^2 \big( \frac{4}{\delta |\sin{\nu}|}  \big)^2 s
\end{array}
\ene
For fixed $k$, the operator $  w_{1,0}(\theta, \alpha, \beta)(k)   $ is an operator on $\cH_\at$, 
we have by (\ref{i.10.3.0}) and (\ref{an.0.10}) that 
\beq \label{e.12} \begin{array}{l}
\|   w_{1,0}(\theta, \alpha, \beta)(k) P_{\at, 1}(\theta, \alpha, \beta)    \| \leq \\\\
12 | (|e_{1}| + \delta)^{1/2}C_{\ref{an.0.11}}    \big(  G^{s,t}  +    \vnabla  \boQ^{s,t}(\theta)  \big) )| \\\\  
+ 2| e^{- 2 \theta} \Delta \boQ^{s,t}(\theta)| +  2| Q^{s,t}(\theta) |  \\\\ 
\leq 
24  (|e_{1}| + \delta +1)^{1/2}C_{\ref{an.0.11}} \big( 12 + 6  |\bob| + 42 \|  \eta' \|_\infty 
\\\\
+ 18 \|  \eta' \|_\infty  |\beta| + 18 \| \eta'' \|_\infty   \big) (1 + |k|) \max_{j \in \{ 1, 2, 3 \}}
| G_{\theta, j}  |\period 
\end{array}
\ene 
and a similar bound holds for $  w_{0,1}(\theta, \alpha, \beta)(k) P_{\at, 1}(\theta, \alpha, \beta)   $.
 Then we obtain (\ref{L.e.3.e1.0.0}) by (\ref{an.0.16}),
(\ref{e.10}-\ref{e.12}) and Remark \ref{R.a.1}.

\QED
\noindent 
\subsubsection{ $\;$ Proof of Theorem~\ref{T.rrp}}

\newtheorem*{T.rrp}{Theorem~\ref{T.rrp}}

\begin{theorem} 
Suppose that  $\theta = \pm i \nu $, with $\nu \in \RR \setminus \{0 \}$, and that  $ \theta  $ satisfies (\ref{an.0.14}) and (\ref{theta}).
Suppose furthermore that $\alpha \leq \boa$ satisfies (\ref{L.rrp.1.m10}) and that  $ (|\sin(\nu)|/2) s \leq 1 $ and $ s = \alpha^{\upsilon}$ for some 
$\upsilon \in (0, 2)$. We assume that $\beta = 0$. 

Let (see (\ref{i.10.3.0}))
\beq \tag{\ref{imaginary-tilde}}\begin{array}{l}
\tilde E_I : = - \pi \int_{\mathbb{S}^2}  dS \| P_{\at, 0}(0, \alpha, 0)|e_1 - e_0| \\\\ \cdot  w_{1,0}(0, \alpha, 0)(x, \frac{k}{|k|}|e_1 - e_0|)
 \psi_0  \|^2, 
\end{array} 
\ene
where $\mathbb{S}^2$ is the sphere and $\psi_0$  is a unit eigenvector of $ P_{\at, 1}(0, \alpha, 0)  $ (see (\ref{corr2})). 

There is a constant $C_{\ref{corr.16}}$ such that 
\beq \tag{\ref{imaginary-basis}} \begin{array}{l}
|\tilde E_I - { \, Im} (E^{s,t}(\theta)) | \leq 4 C_{\ref{corr.16}} \alpha^3 
 (\alpha^\upsilon |\log(\alpha^\upsilon)| + \alpha^\upsilon  +  \alpha^{2\upsilon} \\\\ +  \alpha^{(3- \upsilon)/2} (1 + \alpha^{(3 - \upsilon)/2})^3)
\end{array}
\ene

The explicit value of the constant $C_{\ref{corr.16}}$ is written in (\ref{corr.16}).  

\end{theorem}

The dilation operator (\ref{d1}) is a one parameter group of unitary operators when $\theta$ is real. The set of analytic vectors for the generator of the group is dense in $L^2(\mathbb{R}^3)$. We select an analytic vector $\psi$ such that 
\beq \label{corr3prima}
\psi_0 : = P_{\at, 1}(0, \alpha, 0) \psi \ne 0,   \: \: \| \psi_0 \|_{\cH_\at} = 1.
\ene
We define 
\beq \label{corr2}
\psi_\theta : = P_{\at, 1}(\theta, \alpha, 0) u(\theta)\psi.
\ene
It follows that 
\beq \label{corr3}
\la \psi_{\overline{\theta}} |\:  \psi_\theta \ra_{\cH_\at} 
\ene
is analytic in $\theta$  for any theta satisfying (\ref{an.0.14}) and (\ref{theta}). As $u(\theta)$ is unitary for real $\theta$, (\ref{corr3}) is constant for real theta, thus 
\beq \label{corr3primaprima}
\la \psi_{\overline{\theta}} |\:  \psi_\theta \ra_{\cH_\at} = 1. 
\ene
Eq. (\ref{corr3primaprima}) implies that the operator  
$$
| \psi_\theta \ra \la \psi_{\overline{\theta}} |
$$
is a projection in $L^2(\mathbb{R}^3)$. The range of $ | \psi_\theta \ra \la \psi_{\overline{\theta}} | $ equals the range of $P_{\at, 1}(\theta, \alpha, 0)$. As the null space of $ P_{\at, 1}(\theta, \alpha, 0) $ is the orthogonal complement of the range of  
$  P_{\at, 1}(\theta, \alpha, 0)^* =  P_{\at, 1}(\overline{\theta}, \alpha, 0)  $  we conclude that the null space of both projections coincide and therefore 
\beq \label{corr4}
 P_{\at, 1}(\theta, \alpha, 0) = | \psi_\theta \ra \la \psi_{\overline{\theta}} |.
\ene
Eq. (\ref{corr4}) implies that
\beq \label{corr5}
\begin{array}{l}
Z^{s,\infty}_\od(\theta) = z_\od^{s, \infty}(\theta)  P_{\at, 1}(\theta, \alpha, 0) \otimes P_{\Omega^{s, \infty}}, \\\\ 
Z^{s,\infty}_d(\theta) = z_d^{s, \infty}(\theta)  P_{\at, 1}(\theta, \alpha, 0)\otimes P_{\Omega^{s, \infty}},\\\\
\p  W_{0, 0}(\theta, \varsigma, 0)  \p = w^{s, \infty}(\theta)   P_{\at, 1}(\theta, \alpha, 0)\otimes P_{\Omega^{s, \infty}} \comma 
\end{array}
\ene
where
\beq \label{corr6}
\begin{array}{l}
z_\od^{s, \infty}(\theta) = \la \psi_{\overline{\theta}}\otimes P_{\Omega^{s, \infty}} |\:   Z^{s,\infty}_\od(\theta) 
\psi_\theta \otimes P_{\Omega^{s, \infty}}    \ra_{\cH^{s, \infty}}  , \\\\ 
z_d^{s, \infty}(\theta) = \la \psi_{\overline{\theta}}\otimes P_{\Omega^{s, \infty}} |\: Z^{s,\infty}_d(\theta) \psi_\theta \otimes P_{\Omega^{s, \infty}}    \ra_{\cH^{s, \infty}}    ,\\\\
 w^{s, \infty}(\theta) = \la \psi_{\overline{\theta}} \otimes P_{\Omega^{s, \infty}}|\:  W_{0, 0}(\theta, \alpha, \beta)    \psi_\theta \otimes P_{\Omega^{s, \infty}}    \ra_{\cH^{s, \infty}}    \period
\end{array}
\ene
It follows from  (\ref{i.10.3.6}),  (\ref{T.e.1.e2}) and (\ref{T.e.1.e2.0}) that
\beq \label{corr7} \begin{array}{l}
z_\od^{0, \infty}(\theta) - z_\od^{s, \infty}(\theta) = z_\od^{0, s}(\theta), \: z_d^{0, \infty}(\theta) - 
z_d^{s, \infty}(\theta) = z_d^{0, s}(\theta),\\\\
w^{0, \infty}(\theta) - w^{s, \infty}(\theta) = w^{0, s}(\theta).
\end{array}
\ene
Using (\ref{e.p.eeh.3}), (\ref{C.eeh.1.e2}), (\ref{neri.6}), (\ref{e.12}), (\ref{corr2}) and remark \ref{R.a.1} we obtain
\beq \begin{array}{l} \label{corr8}
|z_\od^{0, s}(\theta)| \leq 4C_{\ref{an.0.16.2}} \frac{4}{\delta |\sin{\nu}|} \Big(  24  (|e_{1}| + \delta +1)^{1/2}C_{\ref{an.0.11}} \big( 12 + 6  |\bob| \\\\
 + 42 \|  \eta' \|_\infty 
+ 18 \|  \eta' \|_\infty  |\bob| + 18 \| \eta'' \|_\infty   \big)\Big)^2 \\\\ \cdot \sup_{|\theta|\leq \frac{1}{120}} 16 \| u(\theta)\psi \|^2
C_{\ref{R.a.1.1}}^2 s^2
\end{array}
\ene
Similarly we get
\beq \begin{array}{l}\label{corr9}
|z_d^{0, s}(\theta)| \leq 2 \Big(  24 (|e_{1}| + \delta +1)^{1/2}C_{\ref{an.0.11}} \big( 12 + 6  |\bob| \\\\
 + 42 \|  \eta' \|_\infty 
+ 18 \|  \eta' \|_\infty  |\bob| + 18 \| \eta'' \|_\infty   \big)\Big)^2 \\\\ \cdot \sup_{|\theta|\leq \frac{1}{120}} 16 \| u(\theta)\psi \|^2
C_{\ref{R.a.1.1}}^2 s
\end{array}
\ene  
and   
\beq \begin{array}{l}\label{corr10}
|w^{0, s}(\theta)| \leq 4 \Big(  24 C_{\ref{an.0.11}} \big( 12 +  
 + 42 \|  \eta' \|_\infty 
 + 18 \| \eta'' \|_\infty   \big)\Big)^2 \\\\ \cdot \sup_{|\theta|\leq \frac{1}{120}} 16 \| u(\theta)\psi \|^2
C_{\ref{R.a.1.1}}^2 s^2
\end{array}
\ene

The numbers $  z_\od^{0, \infty}(\theta)  $,  $z_d^{0, \infty}(\theta)$ and $ w^{0, \infty}(\theta) $ do not depend on $\theta$.  We show this for the case of 
$  z_d^{0, \infty}(\theta)$ to present the argument: We define the functions $ \theta \to f_\theta $,  $ \theta \to g_\theta $ by
\beq \label{corr11} \begin{array}{l}
f_\theta(x, k) : = \frac{1}{e^{- i \nu}|k|}P_\at (\theta, \alpha, 0) w_{1, 0}(\theta, \alpha, 0)\psi_\theta, \\\\
g_\theta(x, k) : = (w_{1, 0}(\theta, \alpha, 0))^*\psi_{\overline{\theta}} 
\end{array}
\ene
with values in $L^2(\mathbb{R}^3 \times \cK^{0, \infty})$. \\
It is clear that 
\beq \label{corr12}
z_d^{0, \infty}(\theta) = \la   g_\theta |  f_\theta \ra_{L^2(\mathbb{R}^3 \times \cK^{0, \infty})}. 
\ene
For real theta 
\beq \label{corr13} \begin{array}{l}
\la   g_\theta |  f_\theta \ra_{L^2(\mathbb{R}^3 \times \cK^{0, \infty})} \\\\
=  \la   u(\theta) \otimes u(- \theta)  g_0 | u(\theta) \otimes u(- \theta) f_0 \ra_{L^2(\mathbb{R}^3 \times \cK^{0, \infty})} \\\\ 
=   \la   g_0 |  f_0 \ra_{L^2(\mathbb{R}^3 \times \cK^{0, \infty})} \comma
\end{array}
\ene
since $u(\theta)$ is unitary (see \ref{d1}). As $ z_d^{0, \infty}(\theta) $ is analytic, we conclude that it is constant. Similarly we prove that 
$  z_\od^{0, \infty}(\theta)  $ and $ w^{0, \infty}(\theta) $ are constant. We denote by $  z_\od^{0, \infty}(0)  $ the limit $\theta \to 0$ of 
$  z_\od^{0, \infty}(\theta)  $. From (\ref{T.e.1.e2.0}) taking $\theta$ to zero we conclude that 
$ z_d^{0, \infty}(\theta)  $ is real and similarly we conclude that $ w^{0, \infty}(\theta)  $ is real. We compute now the imaginary part of 
$  z_\od^{0, \infty}(\theta)  $.

\beq \label{T.e.1.e2tilde.0.0}\begin{array}{l}
{\rm  Im}\,  z^{0,\infty}_\od(\theta)    =\lim_{\tilde{\nu} \to 0}  \frac{1}{ 2 \alpha^{3} i}  \int_{\cK^{0,\infty}} dk  
\la \psi_{\overline{i\tilde \nu}} |  w_{0,1}(i\tilde{\nu}, \alpha, 0)(k)  \\\\ \cdot
\frac{P_{\at, 0}(i\tilde{\nu}, \alpha, 0)  + 
 (1 -P_{\at, 0}(i\tilde{\nu}, \alpha, 0) - P_{\at, 1}(i\tilde{\nu}, \alpha, 0) )
}{  H_{\at}(i\tilde{\nu}, \alpha, 0) - (e_{1} - e^{-i \tilde{\nu}}|k| )  } 
  w_{1,0}(i\tilde{\nu}, \alpha, 0)(k)
      \psi_{i \tilde \nu} \ra_{\cH_\at}                       \\\\  -
\frac{1}{ 2 i \alpha^{3}}  \int_{\cK^{0,\infty}} dk  
 \la \psi_{i\tilde \nu} | w_{0,1}(\overline{i\tilde{\nu}}, \alpha, 0)(k)  \\\\ \cdot
\frac{P_{\at, 0}(\overline{i\tilde{\nu}}, \alpha, 0)  + 
 (1 -P_{\at, 0}(\overline{i\tilde{\nu}}, \alpha, 0) - P_{\at, 1}(\overline{i\tilde{\nu}}, \alpha, 0)
}{  H_{\at}(\overline{i\tilde{\nu}}, \alpha, 0) - \overline{(e_{1} - e^{-i \tilde{ \nu}}|k| )}  } 
  w_{1,0}(\overline{i\tilde{\nu}}, \alpha, 0)(k)
 \psi_{\overline{i \tilde \nu}} \ra_{\cH_\at}  \\\\
= \lim_{\tilde{\nu} \to 0}  \frac{1}{ 2 \alpha^{3} i}  \int_{\cK^{0,\infty}} dk  
 \la \psi_{0} |  w_{0,1}(0, \alpha, 0)(k)  
\big( \frac{P_{\at, 0}(0,\alpha, 0)
}{  e_{0}   - (e_{1} - e^{-i \tilde{\nu}}|k| )  } \\\\ -  
  \frac{P_{\at, 0}(0,\alpha, 0)
}{  e_{0}   - (e_{1} - e^{i\tilde{ \nu}}|k| )  }
\big)  w_{1,0}(0, \alpha, 0)(k)
 \psi_{0} \ra_{\cH_\at}   \\\\
= \lim_{R \to \infty} \lim_{\tilde{\nu} \to 0}  \frac{1}{ 2 \alpha^{3} i} \int_{\mathbb{S}^2} dS \int_{0}^{R} dr  |e_0 - e_1|^2  
 \\\\ \la \psi_{0} |    w_{0,1}(0, \alpha, 0)( |e_{0} - e_{1}|\frac{k}{|k|} )  
\big( e^{-i\tilde \nu}\frac{P_{\at, 0}(0,\alpha, 0)
}{  e_{0}   - e_{1} + e^{-i \tilde{\nu}}r   } \\\\ - 
  e^{i\tilde \nu}\frac{P_{\at, 0}(0,\alpha, 0)
}{  e_{0}   - e_{1} + e^{i\tilde{\nu}} r } \big)   w_{1,0}(0, \alpha, 0)(|e_{0} - e_{1}| \frac{k}{|k|}   )
 \psi_{0} \ra_{\cH_\at}  
 \comma 
\end{array}
\ene
where in the last step we use the exponential decay of  $   w_{1,0}(0, \alpha, 0) $, $ \mathbb{S}^2  $ is the 
$2$-sphere, $dS$ is the volume element in  $ \mathbb{S}^2  $ and $  \frac{k}{|k|} \in S $. \\ 
To compute the radial integral we use the analyticity of the function 
$  \tilde{z} \to  \frac{ 1 } { e_{0}   - e_{1} + \tilde{z}  }   $, then we have for large $R$
\beq \label{T.e.1.e2tilde.1.0} \begin{array}{l}
\lim_{\tilde{\nu} \to 0} \frac{1}{2 \pi i} \int_0^R \big( \frac{e^{-i \tilde{\nu}}
}{  e_{0}   - e_{1} + e^{-i \tilde{\nu}}r   } - 
  \frac{ e^{i \tilde{\nu}}
}{  e_{0}   - e_{1} + e^{i\tilde{\nu}} r } \big) \\\\ =  \frac{1}{2 \pi i}\int_{\mathbb{S}^1 -  e_{0}   + e_{1} }  \frac{ 1 } { e_{0}   - e_{1} + \tilde{z}  } 
 d \tilde{z} =  1 \comma
\end{array}
\ene
Then we have 
\beq \label{T.e.1.e2tilde.2} \begin{array}{l} 
{\rm Im} z_{\od}^{0,\infty}(\theta) =   \frac{\pi}{ \alpha^{3}}  \int_{\mathbb{S}^2} dS   \||e_1 - e_0|
P_{\at, 0}(0, \alpha, 0)  \\\\ w_{1,0}(0, \alpha, 0)(\frac{k}{|k|}|e_{0} -e_{1}  |)   \psi_0 \|^2 \comma
\end{array}
\ene
where $ \mathbb{S}^2  $ is the 
$2$-sphere, $dS$ is the volume element in  $ \mathbb{S}^2  $ and $  \frac{k}{|k|} \in S $.

It follows from Lemma~\ref{L.e.3}, Theorem~\ref{T.e.1} and Eqs. (\ref{corr8}) to (\ref{corr10}) that
\beq \label{corr14}
\cF_\boP = (e_1 - z + w^{0, \infty}(0) - \alpha^3 z_\od^{0, \infty}(0) - \alpha^3 z_d^{0, \infty}(0))\boP + Rem, 
\ene
where 
\beq \label{corr.15}
\| Rem \| \leq C_{\ref{corr.16}} \alpha^3(s |\log{s}| + s + s^2 + \frac{\alpha^{3/2}}{s^{1/2}} (1 + \frac{\alpha^{3/2}}{s^{1/2}})^3)
\ene
and
\beq \label{corr.16}
 C_{\ref{corr.16}} : = 5^3 \frac{C_{Rem_0} + C_{Rem_1}}{(|\sin(\nu)|/2)^{2}} + 10 C_{\ref{T.e.1.e3}} (1 + \sup_{|\theta| \leq 1/2} \|u(\theta)\psi\| )^2.
\ene
Now we take
\beq \label{corr.17}
s = \alpha^{\upsilon}, \: \upsilon \in (0, 2). 
\ene
It follows from 
the fact that $e_1, w^{0, \infty}(0) $ and $  z_d^{0, \infty}(0)) $ are real that if 
\beq \label{corr.18}
\frac{2 C_{\ref{corr.16}}\alpha^3 (\alpha^\upsilon |\log(\alpha^\upsilon)| + \alpha^{\upsilon} + \alpha^{2\upsilon} 
+ \alpha^{(3- \upsilon)/2} (1 + \alpha^{(3 - \upsilon)/2})^3) }{  | {\rm Im}(z + \alpha^3 z_\od^{0, \infty}(0))  |} 
< \frac{1}{2}
\ene
then $\cF_\boP$ is invertible and by (\ref{re.7}) $H^{s,t}(\theta, \alpha, 0)$ is also invertible. We conclude by remark \ref{R.neri.1} that 
\beq \label{corr18}\begin{array}{l}
| {\rm Im} (E^{s,t}(\theta) + \alpha^3 z_\od^{0, \infty}(0)) |  \leq 4 C_{\ref{corr.16}} \alpha^3 
 (\alpha^\upsilon |\log(\alpha^\upsilon)| + \alpha^{\upsilon} + \alpha^{2\upsilon}  \\\\ +  \alpha^{(3-\upsilon)/2} (1 + \alpha^{(3 - \upsilon)/2})^3),
\end{array}
\ene
which proves (\ref{imaginary-basis}). 
\QED

\appendix
\secct{Relevant Integrals for the Interaction Terms} 
\label{sec-II.A}
In this appendix we estimate some integrals and define some constants that are used frequently in most parts of the text.

\begin{lemma}\label{L.a.1}for any positive real numbers $s, t$ with $ s > t$, the following estimates 
hold (see (\ref{extra.5}) and (\ref{i.2})),
\beq \label{L.a.1.e1} 
\begin{array}{l}
\sup_{|\theta|\leq 1/30, \; \vx \in \RR^3, \; j \in \{1, 2, 3\}}\int_{ s \geq |k| \geq t  } 
 |G(\theta)_j(x, k) |^{2} dk \leq \alpha^3 \min( C_1^2, \:  C_2^2 s(s- t) )
\comma \\\\  
 \sup_{|\theta|\leq 1/30, \; \vx \in \RR^3, \; j \in \{1, 2, 3\}}\int_{ s \geq |k| \geq t } 
 \frac{|G(\theta)_j(x, k) |^{2}}{|k|} dk \leq \alpha^3
\min( C_{1, \om}^2, \: C_{2}^2(s - t))   
 \comma \\\\

 \sup_{|\theta|\leq 1/30, \; \vx \in \RR^3, \; j \in \{1, 2, 3\}}\int_{ |k| \geq s  } 
 \frac{|G(\theta)_j(x, k) |^{2}}{|k|^2} dk \leq \alpha^3      C_2^2 (|\log(s)| + 1) \comma \\\\

\sup_{|\theta|\leq 1/30, \; \vx \in \RR^3, \; j \in \{1, 2, 3\}}\int_{ s \geq |k| \geq t  } 
 ||k|G(\theta)_j(x, k) |^{2} dk \leq \alpha^{3} \min(  C_1^2 \frac{e}{2} ,C_2^2 s^3(s - t) )
\comma \\\\

\sup_{|\theta|\leq 1/2, \; \vx \in \RR^3, \; j \in \{1, 2, 3\}}\int_{ s \geq |k| \geq t  } 
 \frac{||k|G(\theta)_j(x, k) |^{2}}{|k|} dk \leq \alpha^{3} \min(C_1^2, C_2^2 s^2 (s - t) )
\comma \\\\

\end{array}
\ene

where
\beq \label{L.a.2.e2}
\begin{array}{l}
C_1 : =  2 \frac{ e^{3/4} }{2 (2 \pi)  }, \; \;  
C_2 :=   2^{1/2} \frac{e^{1/4} }
{ 2 \pi }  \comma      \\\\
C_{1, \om} : =  2^{1/2} \frac{ e^{1/2}  }{2^{1/4} (2 \pi)  }
\frac{\pi^{1/4}}{2^{1/2}} 
\period
\end{array}
\ene
\end{lemma} 

\noindent
\emph{Proof.}

By (\ref{i.2}), using spherical coordinates we get, 
\beq \label{a.0} 
\begin{array}{l}
\int_{ s \geq |k|  \geq t }  |G(\theta)_j(x, k) |^{2} dk \leq  2 \frac{\alpha^3 2 \pi }
{ (2 \pi)^3  } e^{1/2} \int_{t}^{s} \exp(-2e^{-1} r^2)r \\ \leq  
\min(2 \frac{\alpha^3 2 \pi }{ (2 \pi)^3  } e^{1/2} s(s-t), 
 2\frac{\alpha^3 2 \pi }{4 (2 \pi)^3  } e^{3/2} ) \period
\end{array}
\ene
The second inequality in (\ref{L.a.1.e1}) is estimated similarly. \\
For the third inequality we use, 
\beq
\begin{array}{l}
\int_{s}^{\infty}  \exp(-2e^{-1} r^2)\frac{1}{r}   \leq  |\log(s)| \exp(-2e^{-1} s^2) + \int_{1}^\infty dt \log(t^2)  \exp(-2e^{-1} t^2) 4e^{-1} t \\\\
\leq | \log(s)| +   2e^{-1}  \int_1^\infty dt \log(t)  \exp(-2e^{-1} t) \leq |\log(s)| + \int_1^\infty dt \frac{1}{t}  \exp(-2e^{-1} t) \\\\
\leq  |\log(s)| +   \int_1^\infty dt  \exp(-2e^{-1} t) \leq  |\log(s)| + 1   \comma
\end{array}
\ene
where in the first and third inequalities we used integration by parts. \\
The forth integral is calculated similarly, here we either compute the integral $ \int_0^{\infty}  \exp(-2e^{-1} r^2)r^3   $ or
use the bound $ \int_{t}^{s} \exp(-2e^{-1} r^2)r^3 \leq s^2 \int_{t}^{s} \exp(-2e^{-1} r^2)r   $. The fifth  
integral is estimated similarly, additionally we have to compute the variance of the Gaussian  $\exp(-2e^{-1} r^2)$.

\QED

\begin{remark}\label{R.a.1}{\rm
We define 
\beq \label{R.a.1.1}
C_{\ref{R.a.1.1}} : =  \sqrt{\frac{e}{2}} \max (C_1, C_{1, \om}, C_2) \period 
\ene
From the previous lemma it follows that}
\beq \begin{array}{l}
\sup_{|\theta|\leq 1/30, \; \vx \in \RR^3, \; j \in \{1, 2, 3\}}\int_{ s \geq |k| \geq t  } 
 |k|^m|G^{s,t}(\theta)_j(x, k) |^{2} dk \\\\ \leq \alpha^3 \min(C_{\ref{R.a.1.1}}^2, \: C_{\ref{R.a.1.1}}^2 s^{m + 2} ), \, \, \, m \in \{-1, \cdots 2\} \comma \\\\  
\sup_{|\theta|\leq 1/30, \; \vx \in \RR^3, \; j \in \{1, 2, 3\}}\int_{  |k| \geq t  } 
 |k|^{-2}|G^{s,t}(\theta)_j(x, k) |^{2} dk \\\\ \leq \alpha^3 C_{\ref{R.a.1.1}}^2 (|\log(t)| + 1)\period 
\end{array}
\ene

\end{remark}

Remember that the functions $ \boG^{s,t}_{\theta, j} $ are defined in (\ref{il.14}), 
\begin{remark}\label{R.a.2}{\rm
We define 
\beq \label{R.a.2.1}
C_{\ref{R.a.2.1}} : = 3^{1/2}C_{\ref{R.a.1.1}} (4 + 4 \| \eta' \|_\infty  + 2  \| \eta'' \|_\infty)  \period 
\ene
From the previous lemma it follows that
\beq \begin{array}{l} \label{R.a.2.1.1}
\sup_{|\theta|\leq 1/30, \; \vx \in \RR^3, \; j \in \{1, 2, 3\}}\int_{ s \geq |k| \geq t  } 
 |k|^m|\boG^{s,t}_{\theta, j}(x, k)( 1 + |x|^2)^{-1/2} |^{2} dk \\\\ \leq \alpha^3  C_{\ref{R.a.2.1}}^2 s^{m + 4}, \, \, \, m \in \{-1, \cdots 2\} \period 
\end{array}
\ene
\beq \begin{array}{l}\label{R.a.2.2.1}
\sup_{|\theta|\leq 1/30, \; \vx \in \RR^3, \; j \in \{1, 2, 3\}}\int_{ s \geq |k| \geq t  } 
 |k|^m|(\frac{\partial}{\partial x_q } \boG^{s,t}_{\theta, j}(x, k) |^{2} dk \\\\ 
\leq \alpha^3  C_{\ref{R.a.2.1}}^2 s^{m + 4}, \, \, \, m \in \{-1, \cdots 2\} \period 
\end{array}
\ene

}

\end{remark}

\noindent{\it Proof:} 
The proof of (\ref{R.a.2.1.1}) follows from the proof of Lemma~\ref{L.a.1}, Remark \ref{R.a.1} and the following.  
By (\ref{i.2}), (\ref{il.14}) and (\ref{pft.1.1}) we have that 
\beq \label{a.0.0.1} \begin{array}{l}
|\boG^{s,t}_{\theta,j}(x,k) (1 + |x|^2)^{-1/2} |  \leq \alpha |k| |  G^{s,t}(\theta)_j(0, k)  |  \frac{|x|}{(1 + |x|^2 )^{1/2}}
\\\\  + |  G^{s,t}(\theta)_j(0, k)  | \cdot |1 - \eta(|x||k|)|  \frac{1}{(1 + |x|^2 )^{1/2}}
 \\\\  + | |k|  G^{s,t}(\theta)(0, k) \cdot x (1 + |x|^2)^{-1/2} | \| \eta' \|_\infty \\\\  \leq 
|k| |   G^{s,t}(\theta)(0, k) |  (2 + \| \eta' \|_\infty ) \comma
\end{array}
\ene
where we used that   $|1 - \eta(|x||k|)|  = 0$ for $ |x||k| \leq 1  $ and therefore    
$|1 - \eta(|x||k|)| \leq |1 - \eta(|x||k|)| \cdot |x| |k| $.   

To prove (\ref{R.a.2.2.1}) we notice that 
\beq \label{a.0.0.2} \begin{array}{l}
|\frac{\partial}{\partial x_q } \boG^{s,t}_{\theta, j}(x, k) | \leq 
|  G^{s,t}(\theta)(0, k)| \cdot |k|  \big( \alpha + 4 \| \eta' \|_\infty  + |x||k|\cdot |\eta''(|x||k|)|   \big) 
\end{array}
\ene
and that $  |x||k|  |\eta''(|x||k|)| \leq 2|\eta''(|x||k|)| $.

\QED

\begin{lemma}\label{L.a.2}
Let $r \in (0, \frac{1}{2}] $
and $\iota$ be a vector-valued analytic function defined 
on a neighbourhood the closed ball 
 $ \overline{D_{r}^{\CC}(0)}: = \{ \theta \in \CC : |\theta | \leq r   \} $ 
and let $\gamma$ be the curve $\gamma: = \{ r e^{i t} : t \in [0, 2 \pi] 
 \}$. \\
 For any $ s < r $, every $\theta \in D_s(0)$, 
every $h \in  \CC $ such that $\theta + h \in  D_s(0) $
 the following estimations hold. 

\beq \label{L.a.2.e0}
\begin{array}{l}
\| \frac{ \iota(\theta + h) - \iota(\theta) }{h} -  
\iota'(\theta) \| \leq 
  \frac{|h|}{2(r - s)^3  } \max_{z \in \gamma} \| \iota(z) \|  \comma
 \\\\
\|  \iota(\theta + h) - \iota(\theta)  \| \leq 
 \frac{|h|}{2(r - s)^2  }    \max_{z \in \gamma}
  \| \iota(z) \|   \period
\end{array}
\ene

In particular for every $j \in \{ 1, 2, 3  \}$ and every  $\theta \in D_{1/120}(0)$, 
every $h \in \CC $ such that $\theta + h \in  D_{1/120}(0) $,
\beq \label{L.a.2.e1}
\begin{array}{l}
\|  \frac{G^{s,t}(\theta + h)_j - G^{s,t}(\theta)_j }{h} -  
G^{s,t}(\theta)_j' \|^{s,t}_{\rho} \leq \alpha^{3/2}
  \frac{|h|}{2(1/60 - 1/120)^3  }C_{\ref{R.a.1.1}} (1 + \frac{1}{\rho^{1/2}})   \comma
 \\\\
\|  G^{s,t}(\theta + h)_j - G^{s,t}(\theta)_j  \|^{s,t}_{\rho} \leq  \alpha^{3/2}
 \frac{|h|}{2(1/60 - 1/120)^2  } C_{\ref{R.a.1.1}}   (1 + \frac{1}{\rho^{1/2}}) \comma
\end{array}
\ene
where $C_{\ref{R.a.1.1}} $  is defined in (\ref{R.a.1.1}).

\end{lemma}

\noindent{\it Proof:}

We estimate the first inequality in (\ref{L.a.2.e0}), the second 
is estimated similarly. \\
 
We use the Cauchy's integral formula the contour 
$\gamma: = \{ r e^{i t} : t \in [0, 2 \pi]  \}$ to get, 
\beq \label{a.2}
\begin{array}{l}
| \frac{\iota(\theta + h) - \iota(\theta)}{h} -  \iota'(\theta) |  = 
\frac{|h|}{2 \pi }| \int_{\gamma} \frac{\iota(z)}
{(z- \theta)^2(z-(\theta + h))} dz | \\
 \leq  \frac{|h|}{2(r - s)^3  } \sup_{z\in \gamma} | \iota (z)  | \period
\end{array}
\ene

Now we choose $\iota$ to be given by
\beq \label{a.1}
\iota(\theta)  :=   G^{s,t}(\theta)_j(x,k) \period  
\ene
Eqs. (\ref{L.a.2.e1}) follows from (\ref{L.a.2.e0}) 
and Remark  \ref{R.a.1}.

\QED

\section{The Pauli-Fierz Transformation}\label{ap-b}

The operator $ \lambda_{PF}^0(0) $ 
defined in (\ref{lambda-pf}) 
given by (see(\ref{pft.1.1prima}))
$$
\lambda_{PF}^0(0) = a^*(\boQ^0(0)) + a(\boQ^0(0))
$$

 is self-adjoint in
$\mathcal{H}^{0}$ and therefore the operator 
\beq \label{pft.3}
e^{- i\lambda_{PF}^0(0) }
\ene
is unitary.   \\
Assumptions (\ref{i.1}) imply that 
the components of the magnetic potential (see (\ref{i.6})) commute between each other: 
\beq \label{pft.5}
[\boA^{0}(x)_\nu,   \boA^{0}(y)_\mu] = 0, \, \, \, \nu, \mu \in \{ 1, 2, 3 \}  \period
\ene  
By (\ref{pft.5}), 
\beq \label{pft.6}
e^{- i\lambda_{PF}^0(0) } \boA^{0}e^{ i\lambda_{PF}^0(0) } = \boA^{0} \period
\ene
We have that, 
\beq\label{pft.6.1}
\begin{array}{l}
\frac{\partial}{\partial t} e^{- i\lambda_{PF}^0(0) }(i \nabla)  e^{ i\lambda_{PF}^0(0) } \\\\
=   (\nabla \lambda_{PF}^0(0)  )  \period  
\end{array}
\ene
Since this last expression does not depend on $t$, we can integrate with respect to $t$ to obtain: 
\beq \label{pft.7}
e^{- i\lambda_{PF}^0(0) } (-i\nabla) e^{- i\lambda_{PF}^0(0) } = -i \nabla +  (\nabla \lambda_{PF}^0(0)  )  \period
\ene
It is easy to prove that  
\beq \label{pft.8}
[ \hf^0  , a^*(\boQ^0(0))] = a^*( |k| \boQ^{0}(0)  ) \period
\ene
Taking adjoints we get
\beq \label{pft.9}
[ \hf^0  , a(\boQ^0(0))] = - a(|k| \boQ^{0}(0)) \period
\ene

Using (\ref{pft.8}) and (\ref{pft.9}) we get (see also (\ref{i.6})), 
\beq \label{pft.10}
\begin{array}{l}
\frac{\partial}{\partial t}  e^{- i t  \lambda_{PF}^0(0)  } \hf^0  e^{ i t   \lambda_{PF}^0(0)  } \\\\ = 
  e^{- i t    \lambda_{PF}^0(0)  } i(  a^*( |k| \boQ^{0}(0) )  - a( |k| \boQ^{0}(0) ) )    e^{ i t \lambda_{PF}^0(0)   } 
 \period
\end{array}
\ene
By the commutation relations (\ref{ih.3}) we have for $\nu, \mu \in \{1, 2, 3\}$, 
\beq \label{pft.11}
\begin{array}{l}
\frac{\partial}{\partial t} e^{- i t    \lambda_{PF}^0(0)  } i(  a^*( |k| \boQ^{0}(0) )  - a( |k| \boQ^{0}(0) ) )    e^{ i t \lambda_{PF}^0(0)   } \\\\ 
=  \la \boQ^{0}(0)  |\; |k| \boQ^{0}(0) \ra +  \la |k| \boQ^{0}(0)  |\;    \boQ^{0}(0)  \ra
\period 
\end{array}
\ene
Since this last quantity does not depend on $t$, we can integrate and obtain 
\beq \label{pft.12}
\begin{array}{l}
 e^{- i t    \lambda_{PF}^0(0)  } i(  a^*( |k| \boQ^{0}(0) )  - a( |k| \boQ^{0}(0) ) )    e^{ i t \lambda_{PF}^0(0)   } \\\\  =   
  i(  a^*(|k| \boQ^{0}(0))  - a(|k| \boQ^{0}(0)) )  
  +   2 t \la \boQ^{0}(0)   |\; |k| \boQ^{0}(0) \ra   \period
\end{array}
\ene
Integrating  (\ref{pft.10}) we get, 
\beq \label{pft.13}
\begin{array}{l}
 e^{- i   \lambda_{PF}^0(0)  } \hf^0  e^{ i    \lambda_{PF}^0(0)  }    = \hf^0    +  i(  a^*(|k| \boQ^{0}(0))  - a(|k| \boQ^{0}(0)) )  
\\\\  +    \la  \boQ^{0}(0)  |\; |k| \boQ^{0}(0) \ra \period  
\end{array}
\ene
Eqs. (\ref{pft.15}), (\ref{pft.7}) and (\ref{pft.13}) imply (\ref{eph.1}).

\end{document}